\documentclass[
	aps,
	onecolumn,
	a4paper,
	10pt,
	amsmath,
	amssymb,
	preprintnumbers,
	tightenlines,
	notitlepage,
	floatfix,
	nofootinbib,
	longbibliography,
	superscriptaddress
	]{revtex4-2}\pdfoutput=1\pdfsuppresswarningpagegroup=1
\usepackage[utf8]{inputenc}
\usepackage{graphicx}
\usepackage{color}
\usepackage[colorlinks=true]{hyperref}
\usepackage{aas_macros}
\usepackage{booktabs}
\usepackage[nonumberlist]{glossaries}                             
\usepackage{xparse}
\usepackage[inline]{enumitem}
\usepackage{bm}
\usepackage{multirow}
\usepackage{xspace} 
\usepackage[normalem]{ulem}
\usepackage{tabularray}
\usepackage{orcidlink}

\setglossarystyle{index}

\graphicspath{{figs/}}

\allowdisplaybreaks[0]

\glsdisablehyper
\newglossarystyle{custom_acronyms}
{
    \setglossarystyle{long3colheader}%
}

\input{CQGformat.input}

\newcommand{\edit}[1]{{\color{black}{#1}}}

\newcommand{\nwip}[2]{\langle #1 | #2 \rangle}

\renewcommand{\arraystretch}{0.5}

\makeatletter
\def\cat@comma@active{\catcode`\,12}%
\makeatother

\newcommand{\Ms}{\ifmmode{\textrm{M}_\odot}\else{$M_\odot$}\fi}

\def\apj{Astrophysical Journal }                
\def\apjl{Astrophysical Journal Letters}             

\def\apjs{ApJS}              
\def\mnras{MNRAS}            
\def\nat{Nature}              

\def\aap{Astronomy \& Astrophysics}

\ExplSyntaxOn
\NewDocumentCommand{\makeabbrev}{mm}
 {
  \func_makeabbrev:nnn { #1 } { #2 }
 }

\cs_new_protected:Npn \func_makeabbrev:nnn #1 #2
 {
  \clist_map_inline:nn { #2 }
   {
    \cs_new_protected:cpn { #1 } { \gls{ ##1 } \xspace}
	\cs_new_protected:cpn { #1 s} { \glspl{ ##1 } \xspace}
   }
 }
\ExplSyntaxOff

\makeglossaries

\newacronym{AAK}{AAK}{Augmented Analytic Kludge}
\newacronym{AGN}{AGN}{active galactic nuclei}
\newacronym{BBH}{BBH}{binary black hole}
\newacronym{BH}{BH}{black hole}
\newacronym{BHPT}{BHPT}{black hole perturbation theory}
\newacronym{BHNS}{BHNS}{binary black hole-neutron star}
\newacronym{BNS}{BNS}{binary neutron star}
\newacronym{BWD}{BWD}{binary white dwarf}
\newacronym{CHE}{CHE}{close hyperbolic encounters}
\newacronym{CO}{CO}{compact object}
\newacronym{CPU}{CPU}{central processing unit}
\newacronym{CS}{CS}{cosmic string}
\newacronym{EFT}{EFT}{effective field theory}
\newacronym{EM}{EM}{electromagnetic}
\newacronym{EMRI}{EMRI}{extreme mass-ratio inspiral}
\newacronym{EOB}{EOB}{effective one-body}
\newacronym{FEW}{FEW}{Fast EMRI Waveforms}
\newacronym{FD}{FD}{frequency domain}
\newacronym[longplural={Galactic binaries}]{GB}{GB}{Galactic binary}
\newacronym{GR}{GR}{general relativity}
\newacronym{GPU}{GPU}{graphical processing unit}
\newacronym{GSF}{GSF}{gravitational self-force}
\newacronym{GW}{GW}{gravitational wave}
\newacronym{IMRI}{IMRI}{intermediate mass-ratio inspiral}
\newacronym{IMBH}{IMBH}{intermediate mass black hole}
\newacronym{ISCO}{ISCO}{innermost stable circular orbit}
\newacronym{LISA}{LISA}{Laser Interferometer Space Antenna}
\newacronym{LVK}{LVK}{LIGO-Virgo-KAGRA Collaboration}
\newacronym{MBH}{MBH}{massive black hole}
\newacronym[longplural={massive black hole binaries}]{MBHB}{MBHB}{massive black hole binary}
\newacronym{MHD}{MHD}{magnetohydrodynamic}
\newacronym{NR}{NR}{numerical relativity}
\newacronym{NS}{NS}{neutron star}
\newacronym{ODE}{ODE}{ordinary differential equation}
\newacronym{PBH}{PBH}{primordial black holes}
\newacronym{PDE}{PDE}{partial differential equation}
\newacronym{PN}{PN}{post-Newtonian}
\newacronym{PM}{PM}{post-Minkowskian}
\newacronym{QCD}{QCD}{quantum chromodynamics}
\newacronym{QNM}{QNM}{quasi-normal mode}
\newacronym{ROM}{ROM}{reduced-order model}
\newacronym{SDSS}{SDSS}{Sloan Digital Sky Survey}
\newacronym{SNR}{SNR}{signal-to-noise ratio}
\newacronym{SPA}{SPA}{stationary phase approximation}
\newacronym[longplural={stellar-origin black hole binaries}]{SOBHB}{SOBHB}{stellar-origin black hole binary}
\newacronym{SOBH}{SOBH}{stellar-origin black hole}
\newacronym{TD}{TD}{time domain}
\newacronym{TS}{TS}{transient source}
\newacronym{UCXB}{UCXB}{ultra-compact X-ray binary}
\newacronym{WDBH}{WDBH}{white dwarf-black hole}
\newacronym{WDNS}{WDNS}{white dwarf-neutron star}
\newacronym{XMRI}{XMRI}{extremely-large mass-ratio inspiral}

\makeabbrev{#1}{AAK, AGN, BBH, BH, BHNS, BNS, BHPT, BWD, CHE, CO, CPU, CS, EFT, EM, EMRI, EOB, FD, FEW, GB, GPU, GR, GSF, GW, IMRI, IMBH, ISCO, LISA, LVK, MBH, MBHB, MHD, NR, NS, ODE, PBH, PDE, PN, PM, QCD, QNM, ROM, SDSS, SNR, SOBHB, SOBH, SPA, SSO, TD, TS, UCXB, WDBH, WDNS, XMRI}

\begin{document}

\title{Waveform Modelling for the Laser Interferometer Space Antenna\\ \vskip 0.5cm \normalfont \normalsize  LISA Consortium Waveform Working Group}


\date{\today}
\begin{abstract}
LISA, the Laser Interferometer Space Antenna, will usher in a new era in gravitational-wave astronomy. As the first anticipated space-based gravitational-wave detector, it will expand our view to the millihertz gravitational-wave sky, where a spectacular variety of interesting new sources abound: from millions of ultra-compact binaries in our Galaxy, to mergers of massive black holes at cosmological distances; from the beginnings of inspirals that will venture into the ground-based detectors’ view to the death spiral of compact objects into massive black holes, and many sources in between. Central to realising LISA's discovery potential are waveform models, the theoretical and phenomenological predictions of the pattern of gravitational waves that these sources emit. This white paper is presented on behalf of the Waveform Working Group for the LISA Consortium. It provides a review of the current state of waveform models for LISA sources, and describes the significant challenges that must yet be overcome.
\end{abstract}

\vspace*{\fill}
\hspace*{\fill}Corresponding email: wav-wg-chairs@lisamission.org



\newcommand{\aei}{\affiliation{Max Planck Institute for Gravitational Physics (Albert Einstein Institute), D-14476 Potsdam, Germany}}
\newcommand{\ucd}{School of Mathematics and Statistics, University College Dublin, Belfield, Dublin 4, Ireland}
\newcommand{\UIUC}{Department of Physics and Illinois Center for Advanced Studies of the Universe, University of Illinois at Urbana-Champaign, Urbana, Illinois 61801, USA}
\newcommand{\NBIA}{Niels Bohr International Academy, Niels Bohr Institute, Blegdamsvej 17, 2100 Copenhagen, Denmark}
\newcommand{\Southampton}{School of Mathematical Sciences and STAG Research Centre, University of Southampton, Southampton, SO17 1BJ, United Kingdom}
\newcommand{\Perimeter}{Perimeter Institute for Theoretical Physics, Waterloo, Ontario N2L 2Y5, Canada}
\newcommand{\uib}{Departament de F\'isica, Universitat de les Illes Balears, IAC3 -- IEEC, Crta. Valldemossa km 7.5, E-07122 Palma, Spain}

\newpage
\author{Niayesh Afshordi}
\affiliation{\Perimeter}

\author{Sarp Ak\c{c}ay~\orcidlink{0000-0003-2216-421X}}
\affiliation{\ucd}

\author{Pau Amaro Seoane~\orcidlink{0000-0003-3993-3249}}
\affiliation{Universitat Politècnica de València, València, Spain}
\affiliation{Max Planck Institute for Extraterrestrial Physics, Garching, Germany}
\affiliation{Higgs Centre for Theoretical Physics, Edinburgh, United Kingdom}
\affiliation{Kavli Institute for Astronomy and Astrophysics, Beijing 100871, China}

\author{Andrea \surname{Antonelli}}
\affiliation{William H. Miller III Department of Physics and Astronomy, Johns Hopkins University, Baltimore, Maryland 21218, USA}

\author{Josu C. Aurrekoetxea~\orcidlink{0000-0001-9584-5791}}
\affiliation{Astrophysics, University of Oxford, Denys Wilkinson Building, Keble Road, Oxford OX1 3RH, United Kingdom}

\author{Leor Barack~\orcidlink{0000-0003-4742-9413}}
\affiliation{\Southampton}

\author{Enrico Barausse~\orcidlink{0000-0001-6499-6263}}
\affiliation{SISSA, Via Bonomea 265, 34136 Trieste, Italy and INFN Sezione di Trieste\\IFPU - Institute for Fundamental Physics of the Universe, Via Beirut 2, 34014 Trieste, Italy}

\author{Robert Benkel}
\aei

\author{Laura \surname{Bernard}~\orcidlink{0000-0003-2856-1662}}
\affiliation{Laboratoire Univers et Th{\'e}ories, CNRS, Observatoire de Paris, Universit{\'e} PSL, Universit{\'e} Paris Cit{\'e}, 92190 Meudon, France}

\author{Sebastiano \surname{Bernuzzi}}
\affiliation{Theoretisch-Physikalisches Institut, Friedrich-Schiller-Universit{\"a}t Jena, 07743, Jena, Germany}

% ...

\collaboration{Coordinators and contributors}\noaffiliation


\author{Angelica \surname{Albertini}~\orcidlink{0000-0002-9556-1323}}
\affiliation{Astronomical Institute of the Czech Academy of Sciences, Bo\v{c}n\'{i} II 1401/1a, CZ-141 00 Prague, Czech Republic}
\affiliation{Institute of Theoretical Physics, Faculty of Mathematics and Physics, Charles University, CZ-180 00 Prague, Czech Republic}

\author{K. G. \surname{Arun}}
\affiliation{Chennai Mathematical Institute, Siruseri, 603103, India}

\author{Miguel \surname{Bezares}}
\affiliation{Nottingham Centre of Gravity, University of Nottingham, University Park, Nottingham, NG7 2RD, United Kingdom}
\affiliation{School of Mathematical Sciences, University of Nottingham, University Park, Nottingham, NG7 2RD, United Kingdom}

\author{Alexander Bonilla~\orcidlink{0000-0002-7001-0728}}
\affiliation{Observat\'orio Nacional, Rua General Jos\'e Cristino 77, S\~{a}o Crist\'ov\c{c}o, 20921-400 Rio de Janeiro, RJ, Brazil}

\author{Christian Chapman-Bird~\orcidlink{0000-0002-2728-9612}}
\affiliation{SUPA, University of Glasgow, Glasgow G128QQ, United Kingdom}

\author{Bradley Cownden~\orcidlink{0000-0002-5831-1384}}
\affiliation{\ucd}

\author{Kevin Cunningham~\orcidlink{0000-0002-1094-2267}}
\affiliation{\ucd}

\author{Chris Devitt~\orcidlink{0000-0002-7650-5828}}
\affiliation{\ucd}

\author{Sam Dolan~\orcidlink{0000-0002-4672-6523}}
\affiliation{Consortium for Fundamental Physics, School of Mathematics and Statistics, University of Sheffield, Hicks Building, Hounsfield Road, Sheffield S3 7RH, United Kingdom}

\author{Francisco Duque}
\aei

\author{Conor Dyson}
\affiliation{\NBIA}

\author{Chris L.\ Fryer}
\affiliation{Center for Theoretical Astrophysics, Los Alamos National Laboratory, Los Alamos, NM 87545, USA}

\author{Jonathan R.~Gair~\orcidlink{0000-0002-1671-3668}}
\aei

\author{Bruno \surname{Giacomazzo}}
\affiliation{Dipartimento di Fisica ``G. Occhialini'', Universit\'a degli Studi di Milano-Bicocca, Piazza della Scienza 3, 20126 Milano, Italy}
\affiliation{INFN, Sezione di Milano-Bicocca, Piazza della Scienza 3, 20126 Milano, Italy}
\affiliation{INAF, Osservatorio Astronomico di Brera, Via E. Bianchi 46, I-23807 Merate, Italy}

\author{Priti Gupta}
\affiliation{Department of Physics, Indian Institute of Science, Bangalore 560012, India}

\author{Wen-Biao \surname{Han}~\orcidlink{0000-0002-2039-0726}}
\affiliation{Shanghai Astronomical Observatory, CAS, 200030, Shanghai, China}

\author{Roland \surname{Haas}~\orcidlink{0000-0003-1424-6178}}
\affiliation{University of Illinois}

\author{Eric W. \surname{Hirschmann}~\orcidlink{0000-0003-2164-8001}}
\affiliation{Department of Physics and Astronomy, Brigham Young University, Provo, UT 84602, USA}

\author{E.~A.~\surname{Huerta}~\orcidlink{0000-0002-9682-3604}}
\affiliation{Data Science and Learning Division, Argonne National Laboratory, Lemont, Illinois 60439, USA}
\affiliation{Department of Computer Science, The University of Chicago, Chicago, Illinois 60637, USA}
\affiliation{Department of Physics, University of Illinois at Urbana-Champaign, Urbana, Illinois 61801, USA}

\author{Philippe \surname{Jetzer}}
\affiliation{Department of Physics, University of Z\"urich, Winterthurerstrasse 190, 8057 Z\"urich, Switzerland}

\author{Bernard \surname{Kelly}}
\affiliation{Center for Space Sciences and Technology, University of Maryland Baltimore County, 1000 Hilltop Circle Baltimore, MD 21250, USA}
\affiliation{Gravitational Astrophysics Lab, NASA Goddard Space Flight Center, Greenbelt, MD 20771, USA}
\affiliation{Center for Research and Exploration in Space Science and Technology, NASA Goddard Space Flight Center, Greenbelt, MD 20771, USA}

\author{Mohammed \surname{Khalil}}
\affiliation{\Perimeter}

\author{Jack \surname{Lewis}~\orcidlink{0000-0001-8345-3176}}
\affiliation{\Southampton}

\author{Nicole  \surname{Lloyd-Ronning}}
\affiliation{Computational Physics and Methods Group, Los Alamos National Lab, Los Alamos, NM 87544, USA}

\author{Sylvain \surname{Marsat}}
\affiliation{Laboratoire des 2 Infinis - Toulouse (L2IT-IN2P3), Universit\'e de Toulouse, CNRS, UPS, F-31062 Toulouse Cedex 9, France}

\author{Germano \surname{Nardini}~\orcidlink{0000-0002-3523-0477)}}
\affiliation{Faculty of Science and Technology, University of Stavanger, 4036 Stavanger, Norway}

\author{Jakob \surname{Neef}~\orcidlink{0000-0002-3215-5694}}
\affiliation{\ucd}

\author{Adrian \surname{Ottewill}~\orcidlink{0000-0003-3293-8450}}
\affiliation{\ucd}

\author{Christiana Pantelidou~\orcidlink{0000-0002-7983-8636}}
\affiliation{\ucd}

\author{Gabriel Andres Piovano~\orcidlink{0000-0003-1782-6813}}
\affiliation{\ucd}

\author{Jaime \surname{Redondo-Yuste}}
\affiliation{\NBIA}

\author{Laura \surname{Sagunski}~\orcidlink{0000-0002-3506-3306}}
\affiliation{Institute for Theoretical Physics, Goethe University, 60438 Frankfurt am Main, Germany}

\author{Leo C.\ \surname{Stein}~\orcidlink{0000-0001-7559-9597}}
\affiliation{Department of Physics and Astronomy, University of Mississippi, University, MS 38677, USA}

\author{Viktor \surname{Skoup\'{y}}}
\affiliation{Institute of Theoretical Physics, Faculty of Mathematics and Physics, Charles University, CZ-180 00 Prague, Czech Republic}
\affiliation{Astronomical Institute of the Czech Academy of Sciences, Bo\v{c}n\'{i} II 1401/1a, CZ-141 00 Prague, Czech Republic}

\author{Ulrich Sperhake~\orcidlink{0000-0002-3134-7088}}
\affiliation{Department of Applied Mathematics and Theoretical Physics, University of Cambridge, Wilberforce Road, Cambridge CB3 0WA, United Kingdom}

\author{Lorenzo \surname{Speri}~\orcidlink{0000-0002-5442-7267}}
\aei

\author{Thomas F.M.~\surname{Spieksma}}
\affiliation{\NBIA}

\author{Chris \surname{Stevens}}
\affiliation{School of Mathematics and Statistics, University of Canterbury, Christchurch 8041, New Zealand}

\author{David Trestini~\orcidlink{0000-0002-4140-0591}}
\affiliation{CEICO, Institute of Physics of the Czech Academy of Sciences, Na Slovance 2, 182 21 Praha 8, Czechia}

\author{Alex \surname{Va\~n\'o-Vi\~nuales}~\orcidlink{0000-0002-8589-006X}}
\affiliation{CENTRA, Departamento de F\'{\i}sica, Instituto Superior T\'ecnico -- IST, Universidade de Lisboa -- UL, Avenida Rovisco Pais 1, 1049-001 Lisboa, Portugal}

\collaboration{Endorsers}\noaffiliation

\maketitle

\newpage
\setcounter{tocdepth}{2}
\tableofcontents
\setlength{\parindent}{0pt}
\setlength{\parskip}{6pt}

\clearpage

\section*{Notation}

\begin{table}[!htbp]
	\centering

\begin{tabular}{lll} \toprule
	$G$					& gravitational constant						&								\\
	$c$					& speed of light								&								\\
	$m_1$			   	& mass of the primary		 					&	$m_1 > m_2$					\\
	$m_2$				& mass of secondary								&	$m_2 < m_1$					\\
	$M$					& total mass									&	$m_1 + m_2$					\\
	$\mu$				& reduced mass 									&	$m_1 m_2/M$		\\
	$q$					& large mass ratio      						&	$m_1/m_2$					\\
	$\epsilon$			& small mass ratio								&	$m_2/m_1$					\\
	$\nu$				& symmetric mass ratio 							&	$m_1 m_2/M^2$ 		\\
	$S_i$				& spin angular momentum of body $i$ 					&								\\
	$a_i$				& Kerr spin parameter of body $i$ 		& 	$S_i/(m_i c)$					\\
	$\chi_i$			& dimensionless spin of body $i$		& 	$a_i c^2/ (G m_i) \equiv S_i c / (G m_i^2)$					\\
	$e$					& orbital eccentricity							&								\\
	$\Omega$			& orbital frequency								&								\\
	$\mu_S$				& string tension								&								\\
\bottomrule
\end{tabular}
	\caption{Frequently used symbols throughout this whitepaper, $G$ and $c$ are set to 1 throughout.}
    \label{tab:notation}
\end{table}

\section*{List of Acronyms}
\noindent
\begin{tabular}{@{}p{0.15\textwidth} p{0.8\textwidth}@{}}
\textbf{AAK}   & Augmented Analytic Kludge \\
\textbf{AGN}   & active galactic nuclei \\[\medskipamount]

\textbf{BBH}   & binary black hole \\
\textbf{BH}    & black hole \\
\textbf{BHNS}  & binary black hole-neutron star \\
\textbf{BHPT}  & black hole perturbation theory \\
\textbf{BNS}   & binary neutron star \\
\textbf{BWD}   & binary white dwarf \\[\medskipamount]

\textbf{CHE}   & close hyperbolic encounters \\
\textbf{CO}    & compact object \\
\textbf{CPU}   & central processing unit \\
\textbf{CS}    & cosmic string \\[\medskipamount]

\textbf{EFT}   & effective field theory \\
\textbf{EMRI}  & extreme mass-ratio inspiral \\
\textbf{EOB}   & effective one-body \\[\medskipamount]

\textbf{FD}    & frequency domain \\
\textbf{FEW}   & Fast EMRI Waveforms \\[\medskipamount]

\textbf{GB}    & Galactic binary \\
\textbf{GPU}   & graphical processing unit \\
\textbf{GR}    & general relativity \\
\textbf{GSF}   & gravitational self-force \\
\textbf{GW}    & gravitational wave \\[\medskipamount]

\textbf{IMBH}  & intermediate mass black hole \\
\textbf{IMRI}  & intermediate mass-ratio inspiral \\
\textbf{ISCO}  & innermost stable circular orbit \\[\medskipamount]

\textbf{LISA}  & Laser Interferometer Space Antenna \\
\textbf{LVK}   & LIGO-Virgo-KAGRA Collaboration \\[\medskipamount]

\textbf{MBH}   & massive black hole \\
\textbf{MBHB}  & massive black hole binary \\[\medskipamount]

\textbf{NR}    & numerical relativity \\
\textbf{NS}    & neutron star \\[\medskipamount]

\textbf{ODE}   & ordinary differential equation \\[\medskipamount]

\textbf{PBH}   & primordial black holes \\
\textbf{PDE}   & partial differential equation \\
\textbf{PM}    & post-Minkowskian \\
\textbf{PN}    & post-Newtonian \\[\medskipamount]

\textbf{QCD}   & quantum chromodynamics \\
\textbf{QNM}   & quasi-normal mode \\[\medskipamount]

\textbf{ROM}   & reduced-order model \\[\medskipamount]

\textbf{SDSS}  & Sloan Digital Sky Survey \\
\textbf{SNR}   & signal-to-noise ratio \\
\textbf{SOBH}  & stellar-origin black hole \\
\textbf{SOBHB} & stellar-origin black hole binary \\
\textbf{SPA}   & stationary phase approximation \\[\medskipamount]

\textbf{TD}    & time domain \\[\medskipamount]

\textbf{UCXB}  & ultra-compact X-ray binary \\[\medskipamount]

\textbf{WDBH}  & white dwarf-black hole \\
\textbf{WDNS}  & white dwarf-neutron star \\[\medskipamount]

\textbf{XMRI}  & extremely-large mass-ratio inspiral \\
\end{tabular}
\newpage

\section{Introduction}


Waveforms, theoretical predictions for \GW signals, play a vital role in \GW astronomy. 
Most \GW signals are buried deep in instrumental noise. 
By using waveforms as matched filters, such signals can be detected. 
Once a \GW is found the properties of its source can be inferred by further comparing the system to theoretical waveforms. 
The \LISA will open a window on a new frequency band of \GWs in the mHz regime. 
In this band we will encounter a slew of new sources of \GWs ranging from local Galactic white-dwarf binaries to distant mergers of \MBHs, and many sources in between. 
Realizing the science potential of \LISA detection and measurement of \GW signals will require new waveform models, which will need to cover a much wider range of sources while being significantly more accurate than the models used currently in ground based \GW observations by LIGO, Virgo, and KAGRA~\cite{Purrer:2019jcp,Ferguson:2020xnm}.

A large class of \GW sources expected for \LISA feature the inspiral of a binary of compact objects decaying under the emission of \GWs, often leading to a merger at the end. 
Producing theoretical waveforms for such events requires solving the relativistic two body problem. 
This is a notoriously difficult problem with no known closed form solution for radiating binaries.  
There are \edit{four} main approaches to obtain approximate solutions to the relativistic two-body problem from first principles. 
First, \NR takes the non-linear \PDEs of \GR, puts them on a grid, and evolves them on a supercomputer. 
\edit{
Second, one can make a weak field approximation and obtain a perturbative analytical solution.
This is known as the \PM approximation, or -- when an additional simultaneous slow motion approximation is introduced -- the \PN approximation.
} 
\edit{Third,} the \GSF{}  \edit{formalism} expands the two-body dynamics in powers of the (small) ratio of the masses of the two bodies. 
\edit{Finally, \BHPT uses homogeneous perturbations of isolated \BHs and associated \QNM calculations to model the post-merger behaviour.}

\begin{figure}
	\includegraphics[width=0.6\textwidth]{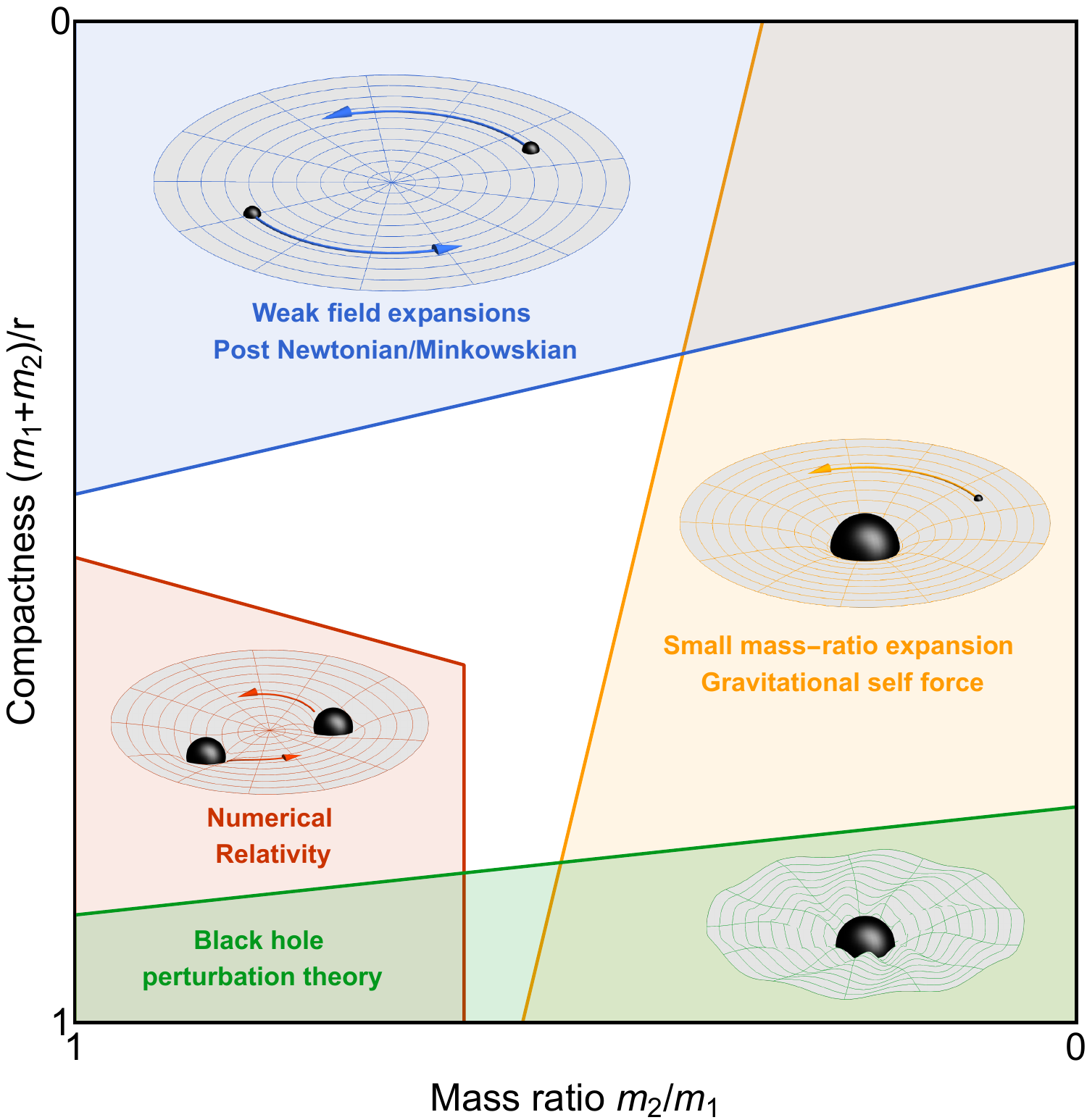}
	\caption{
		A sketch of the natural domains of applicability of the four main approaches to solving the relativistic two-body problem. 
		The approaches are largely complementary and building waveforms for \LISA will require input from all four. 
		The solid lines shown are illustrative of the reach of each approach. 
		The precise reach of each region depends on the source type and the accuracy requirements of the model. 
		In addition, the Effective-One-Body and Phenomenological models absorb information from the four main approaches to produce more global models that can compute waveforms sourced by binaries across large portions of the parameter space. 
	}	\label{fig:cartoon}
\end{figure}

Each of these four approaches has its own natural domain of applicability. 
Due to its computational cost \NR is limited to a relatively low number of orbits before merger and to systems with comparable masses. 
The \edit{\PN and \PM weak-field approximations limit their} application to the early inspiral, while \GSF theory is at its best when dealing with binaries with relatively disparate mass-ratios. 
The post-merger phase of binaries containing a \BH can be modelled with \BHPT.
As illustrated in Fig.~\ref{fig:cartoon}, the natural domains of applicability of the four first-principle approaches to the relativistic two-body problem are largely complementary, and building waveforms for \LISA will require input from all four --- sometimes for modelling a single source. 
Effective waveform models like \EOB and the Phenom-family combine inputs from the different approaches to provide waveforms that can straddle these domains. 

The LISA Waveform Working Group (WavWG) serves as an interface between the LISA Science Group --- tasked with realizing LISA's scientific mission --- and the wider scientific community developing waveform models and studying the relativistic two-body problem. 
It serves to prioritize waveform development, inform the wider community of LISA's waveform modelling needs, and as a recruiting pool for waveform related tasks and projects in the LISA Science Group.  

In this white paper, the WavWG discusses the current status of waveform modelling and what further development is needed in realize LISA's science. 
It is organized in five main sections. 
In $\S$\ref{sec:sources} we take a brief inventory of the sources LISA is expected to observe along with their expected parameters. 
This sets the primary goals for waveform model accuracy and parameter space coverage. 
$\S$\ref{sec:modelling_requirements} discusses what requirements LISA data analysis puts on waveform models in terms of accuracy, efficiency, and format. 
The main approaches to modelling waveforms from compact binaries are described in $\S$\ref{sec:modelling_binaries}, discussing both their status to date, and challenges to be overcome to meet LISA's waveform requirements. 
$\S$\ref{sec:waveform_accel} discusses methods for accelerating the production of waveform models. 
Finally, $\S$\ref{sec:modling_beyond_GR} covers the modeling for beyond GR, beyond Standard Model, and cosmic strings sources. 
We note that while stochastic signals are of considerable interest, this paper does not address them as the focus  here is on individual detectable sources. For the reader's convenience, we have provided a table of notation used throughout the paper in Table~\ref{tab:notation} and a list of commonly used acronyms on the page proceeding this introduction.  

\hrulefill

\textit{The writing of this white paper was coordinated by the co-chairs of the LISA Waveform Working Group: Maarten van de Meent, Deirdre Shoemaker, Niels Warburton, and Helvi Witek. Additional coordination was provided by the co-chairs of the LISA Waveform Work Package: Leor Barack, Anna Heffernan, and Harald Pfeiffer.
\\
\\
The coordinators and contributors to individual sections of the white paper are listed at the start of the each section.
}

\newpage
\section{LISA Sources}\label{sec:sources}
LISA will be sensitive to a wide range of \GW sources in the mHz regime. In this section, we take an inventory of anticipated sources that need to be modelled to extract useful scientific results {\it and} are well-enough understood at this stage to be modelled in some detail.
A more extensive survey of LISA's expected sources and the science that can be done with them can be found in the whitepaper produced by the LISA Astrophysics Working Group \cite{LISA:2022yao}. The short summaries here mainly focus on the expected parameter ranges to provide a context for the modelling approaches discussed in the rest of the white paper.  Conventional long-lived astronomical sources include \MBHBs, \EMRIs, \IMRIs, \GBs and \SOBHBs.
These sections (\ref{sec:sources_MBHBs}--~\ref{sec:sources_SOBHBs}) present LISA sources in \GR and the Standard Model.
Theoretically interesting but empirically speculative sources include \CSs and sources based on physics beyond \GR and beyond the Standard Model are discussed in Secs.~\ref{sec:sources_cosmicstrings} and~\ref{sec:sources_beyond_GR}. These sources, and resulting science, will also be applicable to other mHz gravitational wave detectors \cite{TianQin:2015yph, Hu:2017mde}.

A key metric for the detectability of sources is their \SNR.
This is defined via the standard formula
\edit{\begin{equation}
	\text{SNR}= \sqrt{\nwip{h}{h}},
\end{equation}
where $h$ is the \GW strain and $\nwip{\cdot}{\cdot}$ is the noise weighted inner product,
\begin{equation}\label{eq:nwip}
\nwip{h_1}{h_2} = 4 \Re \int \frac{\tilde{h}_1^{*}(f)\tilde{h}_2(f)}{S_n(f)}\mathrm{d}f,
\end{equation}
with $\tilde{h}_i(f)$ denoting the Fourier transform of the strain, the star superscript (${^{*}}$) denoting complex conjugation, and $S_n(f)$ the noise power spectral density of the detector \cite{LISA:2017pwj,Robson:2018ifk}.}

Unless otherwise stated this is computed for an observation period of up to 4 years.

\subsection{Massive black hole binaries (MBHBs)}
\label{sec:sources_MBHBs}

Coordinator: Enrico Barausse \\
Contributors: M.~Bonetti, J.~Garcia-Bellido

\subsubsection{Description}
In the local Universe, \MBHs
are observed in the centers of virtually all massive galaxies~\cite{1984ApJ...278...11G, Kormendy:1995er,Heckman:2014kza},
as well as in many low mass dwarf galaxies~\cite{Reines:2011na, Reines:2013pia, Baldassare:2019yua}.
Their cosmic evolution is inextricably intertwined with that of their host galaxies.
The latter provide \MBHs with gas out of which they grow by accretion (thereby shining as
quasars or \AGN), and 
\MBHs influence the evolution
of their hosts by injecting energy into their surroundings (\AGN feedback)~\cite{Croton:2005hbr,2019ApJ...884..180D, 2020ApJ...897..103S}.
This co-evolution of \MBHs and galaxies is reflected in the scaling 
relations between \MBH mass and galactic properties~\cite{Ferrarese:2000se,Gebhardt:2000fk,Barger:2004xw,Kormendy:2013dxa,McConnell:2012hz,Schramm:2012jt}, although the detailed physical processes leading to their emergence are still not fully understood.

Present day galaxies/dark matter haloes are believed to have formed hierarchically from
the merger of smaller systems. Likewise, ``seed'' \MBHs are expected to inhabit
at least a fraction of the high redshift progenitors of today's galaxies (see \cite{Latif:2016qau} for a review). These seed \BHs
are expected to grow by a combination of gas accretion and mergers \cite{Volonteri:2007ax,Volonteri:2012tp,Lupi:2014vza,Mayer:2014nva,2016MNRAS.457.3356V,Regan:2017vre,2019Natur.566...85W}. Their coalescences, as well as those of the later generations of \MBHs that they give rise to, are indeed a prime target for \LISA~\cite{Sesana:2004sp,Sesana:2004gf,Klein:2015hvg,Barausse:2020mdt,Barausse:2023yrx}.

By detecting \MBH mergers, LISA could confirm or disprove this hierarchical scenario for the evolution of \MBHs. 
Indeed, considerable uncertainties exist about the timescale on which \MBHs form a bound binary
and eventually coalesce, after two galaxies have merged. The main uncertainty is whether
\MBH pairs can efficiently make their way from separations of hundreds of pc down to the
sub-pc scales on which gravitational waves dominate the dynamics~\cite{Begelman:1980vb,Yu:2001xp,Milosavljevic:2002bn,Merritt:2004gc,Sesana:2007sh,Lodato:2009qd,Khan:2011gi,Preto:2011gu,Colpi:2014poa,2015ApJ...810...49V,Sesana:2015haa,Dosopoulou:2016hbg,2018MNRAS.475.4967T,Bortolas:2020uib,Barausse:2020mdt}. First dynamical friction and then stellar and/or gaseous hardening are the main drivers of the evolution of \MBHs on their path to coalescence. Triple or quadruple \MBH interactions,
which naturally arise in models where the evolution of \MBH pairs/binaries is slow or even stalls~\cite{Bonetti:2018tpf,Ryu:2018yhv,Mannerkoski:2021lal},
should in any case ensure LISA detection rates of at least a few per year~\cite{Bonetti:2018tpf,Barausse:2020mdt}. 

Conversely, the observed detection rate will
inform us about the efficiency with which \MBH binaries come together and merge in nature~\cite{Barausse:2020mdt}, even though 
that information may be partly degenerate with the number of \MBH seeds formed at high redshift. In that respect, 
it is worth pointing out the possibility that there may be massive \PBH
formed during the radiation era that may act as seeds for the 
\MBH population~\cite{Clesse:2015wea,Garcia-Bellido:2019tvz}\edit{, resulting in high redshift \MBHB mergers.} 
\edit{\PBHs of mass $\sim 10^6 M_\odot$ are believed to form at the electron-positron annihilation epoch ($\sim 1$ second after the Big Bang) \cite{Carr:2023tpt}. They could start accreting gas very soon after recombination ($\sim 400, 000$ years after the Big Bang) and grow to become supermassive ($10^9 M_\odot$) around  $z \sim 10$ (470 Myr after the Big Bang). Close binaries of million-solar-mass \PBHs could form very early and merge in a few million years. If LISA detects \MBHBs at $z=25$ (130 Myr after the Big Bang) we should be confident they are PBH at $ >99.97\%$ confidence interval.}

\subsubsection{Expected source parameters}\label{ssec:MBHB_expected_source_parameters}
Further information on the underlying astrophysics can be gained by measuring the parameters of \MBH binaries.
Measurements of the component masses will allow for constraining the initial mass function
of the seed black holes at high redshift~\cite{Sesana:2004sp,Sesana:2004gf,Sesana:2010wy,Klein:2015hvg,Barausse:2020mdt}. Indeed, the latter may form as relatively low-mass
black holes ($\sim  100-1000 M_\odot$), e.g. from the collapse of the first generation of stars~\cite{Madau:2001sc}, or
they may be born already as higher mass seeds of $\sim10^4-10^5 M_\odot$ (e.g. from
the collapse of massive quasi-stars~\cite{Begelman:2009ty}, runaway instabilities in stellar clusters~\cite{PortegiesZwart:1999nm,Stone:2016ryd}, instabilities of
protogalactic disks~\cite{Volonteri:2007ax}, etc.). 
Spin measurements, together with mass measurements, will clarify
the properties of the accretion process and its importance in the evolution of \MBHs relative to mergers~\cite{Berti:2008af,Barausse:2012fy,Sesana:2014bea}.
Measurements of distance will allow for placing the detected system at the right epoch in the history of
the universe. Further help in this respect will naturally be provided by sufficiently accurate/precise
sky localization, which might allow for the identification of an electromagnetic counterpart; and, therefore,
a direct measurement of photometric or spectroscopic redshift~\cite{Tamanini:2016zlh,Mangiagli:2020rwz}. In the presence of independent
redshift and distance measurements, one may even attempt to construct a \GW-based cosmography,
potentially measuring the Hubble constant or the density of matter/dark energy in our universe~\cite{Tamanini:2016zlh}.
Finally, measurements of the eccentricity and mass ratio will provide insight on the
mechanism driving \MBH pairs to separations at which \GW emission is enough to lead to a merger within a Hubble time~\cite{Bonetti:2018tpf,Barausse:2020mdt}.

In more detail, current expectations on the parameters of \MBHBs  and their seeds 
from astrophysical models are generally as follows:
%
\begin{itemize}
	\item Source frame total masses range from a few $10^2$ to a few $10^8 M_\odot$, with possible peaks around  $10^3$ and
	$10^5 M_\odot$ for respectively light and heavy seed formation in some models. See, e.g., Fig.~2 of Ref.~\cite{Barausse:2020mdt}.
	
	\item Comparable mass binaries with mass ratio $\epsilon\equiv m_2/m_1$ between 0.1 and 1 are expected to make up the bulk of LISA detections, although tails
	extending down to $\epsilon\sim$ a few $\times 10^{-3}$ may be present (cf.~Fig.~3 of  \cite{Barausse:2020mdt}). 
	
	\item Large uncertainties generally affect the prediction for spin magnitudes and orientations. Moderate to large spins $\gtrsim 0.7$
	are expected in models calibrated to electromagnetic measurements (i.e. iron K$\alpha$) of \MBH spins in local \AGNs~\cite{Sesana:2014bea}, but the spins
	of the first generations of \MBH mergers are relatively unknown. Similarly, spins may be approximately parallel at the peak
	of the star formation $z\sim 2-3$, where gas torques may align them with the orbital angular momentum to within $10-30$ degrees~\cite{Dotti:2009vz}, but the
	situation may be different at higher and lower redshift.
	
	\item Eccentricity may be very significant and evolving in the LISA window ($e> 0.9$ upon entrance into the LISA band and up to 0.1 at plunge)
	if triple/quadruple \MBH interactions are more efficient than gas and stellar interactions to drive the binary's evolution at $\sim $pc scales~\cite{Bonetti:2018tpf}.
	
	\item Event rates and \SNRs  can vary significantly according to the underpinning astrophysical model~\cite{Sesana:2010wy,Klein:2015hvg,Barausse:2020mdt}: in light seed scenarios
	rates may be as low as a few per year, especially if feedback from
	supernova explosions is included~\cite{Barausse:2020mdt}, which results in most events having small total masses and thus low \SNRs (below 100). More generally, the number of light seed detections may be threshold limited~\cite{Bonetti:2018tpf,Barausse:2020mdt}. This is again
	especially true  if supernova feedback is included,
	as most events have \SNRs of a few tens in that case. In heavy seed scenarios, instead, irrespective of whether supernova feedback is included or not, rates and \SNRs tend to be larger, i.e. roughly a few tens of detections in 4 yr and \SNRs up to several thousands.
	
	\item Environmental effects from gas and stars generally have negligible
	impact in the LISA observation window~\cite{Barausse:2014tra}, except under certain circumstances \cite{Garg:2022nko, Caputo:2020irr}. In addition, they might leave a recognizable imprint in the eccentricity distribution of detected events \cite{Roedig:2011rn}. The same applies to \MBH triple/quadruple interactions, which are expected to cause the aforementioned very high ``relic'' eccentricity when the binaries enter the LISA band~\cite{Bonetti:2018tpf}.
	
	\item When delays between galaxy and \MBH mergers are accounted for, the highest number of detections is expected around $z\sim 2-4$~\cite{Barausse:2020mdt}.
\end{itemize}

In addition to the large parameter space described above, waveform templates for MBHBs will also need to accurately represent the waveform for many hundreds or thousands of cycles before merger --- see, e.g., Fig.~16 of Ref.~\cite{Mangiagli:2020rwz}.

\begin{table}[!htbp]
	\centering
\begin{tabular}{lll} \toprule
	Parameter   	& Notation		& Astrophysically relevant range    	
\\ \midrule
	Total mass in the detector frame	& $M$                   & \qquad \qquad $10^5-10^7 M_\odot$									
\\
	Mass ratio $(>1)$ & $q$			& \qquad \qquad $1-10$		
\\
	Dimensionless spin &	$|\chi_i|$	& \qquad \qquad $0-0.998$			
\\ 
	Eccentricity entering LISA band & $e_\text{init}$			& \qquad \qquad $0-0.99$		
		\\
	Eccentricity at last stable orbit & $e_\text{merge}$	& \qquad \qquad $<0.1$	
		\\
	Signal to noise ratio & \SNR		& \qquad \qquad $10-10^4$			
								\\ \bottomrule
\end{tabular}
\caption{Summary of the anticipated parameters for LISA \MBHB sources. 
The total mass of a massive binary in the source frame ranges from $10^2-10^8 M_\odot$ while moving to the detector 
frame picks up a factor of $\sim10$ giving the range $10^3-10^9 M_\odot$. 
This is squeezed on both ends by another factor of $10$ (SNR and LISA band constrain the low and high limits respectively) 
giving a possible range of $10^4-10^8 M_\odot$ with the majority of sources expected between $10^5-10^7 M_\odot$~\cite{Barausse:2020mdt}. 
The mass ratio, $q$ is expected to range from $1-1000$, again with the majority of sources expected to be in 
the $1-10$ range~\cite{Barausse:2020mdt}. \edit{The maximum spin of 0.998 comes about because
if a \MBH accretes gas,  the  radiation emitted by the disk and falling into  the hole  has negative angular momentum, preventing
a spin up past $\chi=0.998$~\cite{Thorne:1974ve}.}}
\label{tab:SMBBHparams}
\end{table}


\newpage

\subsection{Extreme mass-ratio inspirals (EMRIs)}
\label{sec:sources_EMRIs}

Coordinator: Huan Yang\\
Contributors: S.~Akcay, P.~Amaro Seoane, S.~Bernuzzi, J.~Gair, A.~Heffernan, T.~Hinderer, S.~A.~Hughes,  S.~Isoyama, G.~Lukes-Gerakopoulos, A.~Nagar, Z.~Pan, Z.~Sam, C.~F.~Sopuerta, V.~Witzany

\subsubsection{Description}

\EMRIs are binaries in which one member is substantially more massive than the other.  
The canonical \EMRI is expected to be a stellar mass compact body (typically a black hole of 10--100$\,M_\odot$) captured by a \BH of $10^5$--$10^7\,M_\odot$ by multibody scattering processes in the core of a galaxy \cite{Amaro-Seoane:2012lgq}. 
\EMRIs can also form via the capture of other compact objects such as neutron stars or white dwarfs, though the small mass of these objects reduces their detectability with LISA \cite{Hopman:2006xn}.
A consequence of the large mass-ratio of \EMRIs is the quite slow inspiral of the smaller body, executing $10^4$--$10^5$ of orbits as it moves through the black hole's strong field \cite{Finn:2000sy, Berry:2016bit}.  
Binaries involving a \MBH orbited by a substellar object, such as a brown dwarf, are called \XMRIs and these are potential LISA sources if they form around the \MBH in the centre of the Milky Way \cite{Amaro-Seoane:2019umn,Gourgoulhon:2019iyu}.
Measuring the gravitational waves generated by the strong-field orbits of E/XMRIs will make it possible to map the properties of massive black holes with great precision \cite{Babak:2017tow,Vazquez-Aceves:2022dgi}.

\subsubsection{Expected source parameters}

The expected LISA detection rate of \EMRIs depends on several astrophysical ingredients \cite{Babak:2017tow}, 
including the mass/spin function of \MBHs at different redshifts, 
the fraction of \MBHs living in stellar cusps, 
stellar-mass \COs capture rate per \MBH, the characteristic mass of stellar-mass \COs and their orbit eccentricities.
The mass function of \MBHs is usually modelled as a power law $dn/d\log M = n_0\left(M/3\times10^6 M_\odot\right)^{(-0.3,0.3)}$ \cite{Gair:2010yu}, with $n$ being the MBH number density and $n_0\in(0.002,0.005)$ Mpc$^{-3}$.
MBHs can be extremely spinning with $a\approx0.998$, slowly spinning with $\chi_1 \approx 0$ or 
of an extending spin distribution in the range $0\le \chi_1 \le 1$, 
depending on their growth channels (accretion and/or mergers) \cite{Dotti:2012qw,Sesana:2014bea,Pan:2020dze}.
The fraction of \MBHs living in stellar cusps is determined by how frequently MBHs merge, when the stellar cusps are destroyed,  
and their regrowth time after mergers. \COs in stellar cusps captured by \MBHs generically reside in low angular momentum orbits with large eccentricities, which subsequently decay to the range of $0< e < 0.2$ at the final plunge \cite{Babak:2017tow}.  The capture rate depends on the cusp relaxation time and the density profile of the cusp \cite{Amaro-Seoane:2010dzj}.
Taking into account of all these uncertainties with semi-analytic models, \citet{Babak:2017tow} forecasted that there will be several to thousands of EMRIs detected by LISA per year, assuming the contributing \COs are BHs with mass $10$ or $30$ $M_\odot$, and a detection threshold of \SNR $=20$. 
 On the other hand, recent rate studies predicted that accretion disk-assisted \EMRI formation may be more common for LISA detection \cite{Pan:2021ksp,Pan:2021oob, Derdzinski:2022ltb}, thanks to the high efficiency of disks in transporting stellar-mass black holes towards galaxy centers. These disk-assisted \EMRIs tend to have negligible eccentricity comparing to the \EMRIs formed by gravitational capture. 
 
For \XMRIs their extremely slow evolution means they will spend millions of years in the LISA band. 
Although the event rate of an XMRI forming is quite low, their long duration in band means in the Galactic centre there could be a few dozen sources, some of which may be highly eccentric.
Due to the short distance between the centre of the Galaxy and the solar system, the \SNRs  can range from 10 to $10^4$ \cite{Amaro-Seoane:2019umn,Gourgoulhon:2019iyu}.

\subsubsection{Science with \EMRIs}

Constraining model parameters of \EMRIs from their \GW signals benefits from the large number of cycles completed in the LISA band \cite{Babak:2017tow,Huerta:2008gb,Pan:2020dze}: all intrinsic parameters of an \EMRI (the redshifted masses,  the \MBH spin magnitude, and the \CO orbit eccentricity at plunge) can be measured with fractional uncertainty $\sim 10^{-6}-10^{-4}$; external parameters are mainly determined by the \GW amplitude and its modulation with time, where the luminosity distance $D_{\rm L}$, the source sky location $\Omega_{\rm s}$ and the MBH spin direction $\Omega_{a}$ can be measured with fractional uncertainty $\sigma(\ln D_{\rm L})\approx \text{\SNR}^{-1}$, median solid angle uncertainties  $\sigma(\Omega_{\rm s})\approx 0.05 (\text{\SNR}/100)^{-5/2}$deg$^2$ and $\sigma(\Omega_{\rm a})\approx 30 \sigma(\Omega_{\rm s})$, respectively. 
In light of the source location constraints in both radial  and transverse directions, a fraction of low-redshift \EMRIs
are expected to be traced back to their host galaxies without the aid of any electromagnetic counterparts~\cite{Pan:2020dze}.

To identify the nature of the secondary in an \EMRI, in the case of a light black hole or a neutron star, the only potentially detectable parameter beyond its mass will be the component of its spin \edit{parallel to its orbital angular momentum} \cite{Witzany:2019nml,Skoupy:2023lih}. 
\edit{The detectability in the aligned-spin case has been discussed in \citet{Piovano:2021iwv} while more generic waveform model is being developed to address it in the  general precessing-spin scenario \cite{Drummond:2023wqc}.} For less compact objects, such as white dwarfs or brown dwarfs, there is a possibility of gaining some information about their quadrupoles \cite{Rahman:2021eay}. 
However, all of the parameters of the secondary beyond its mass may be poorly constrained and partially or wholly degenerate with parameters controlling other sub-leading effects.

Detection of EMRIs will provide critical information with which to understand the growth history of \MBHs and their environment. 
Different growth mechanisms, i.e., via mergers and/or via gas accretion, predict significantly different mass and spin distribution for \MBHs \citep{King:2008au,Berti:2008af,Barausse:2012fy,Sesana:2014bea}. 
With a population of \EMRI events, the co-evolution of \MBH masses and spins over cosmic time can be measured to further reveal how they are formed and how they grew \citep{Berti:2008af, Barausse:2012fy,Sesana:2014bea, Pan:2020dze}. 
N-body calculations have shown that \EMRI rates sensitively depend on stellar distribution 
at galactic centers, so that the rate inferred from observation will also be able to constrain various distribution models \citep{Babak:2017tow}. 
Environmental effects, including the interaction with possible accretion disk around the \MBH and close stellar objects near the \EMRI system, 
may induce sizeable phase shifts to the \EMRI waveform \cite{Barausse:2014tra,Bonga:2019ycj, Derdzinski:2018qzv, Derdzinski:2020wlw}.  
The accretion channel has illustrated that \MBHs with \AGNs will account for a significant number of \EMRIs \cite{Pan:2021oob}. 
Estimates of the resulting \GW phase correction vary over many orders of magnitude ($0.1 - 10^4$ rad / year), depending on the assumed disk model \cite{Yunes:2011ws, Barausse:2014pra}.
 Tidal resonances induced by the tidal field of nearby stars and stellar-mass black holes can generate a phase correction of $\mathcal{O}(1)$ in a significant fraction of the phase space~\cite{Gupta:2022fbe}. The main uncertainty comes from the population of nearby stellar-mass objects, which may come from the (disk)-migrating stellar-mass objects in the \AGN or leftovers from the previous \AGN life cycle (Fig.~6 of \cite{Pan:2021xhv}), or simply from the mass segregation effect (\cite{Emami:2019mzi, Emami:2019uty}). 

By precisely mapping the background spacetime geometry of a \MBH \cite{Ryan:1995wh,Barack:2006pq,Barausse:2020rsu}, \EMRIs can be used to check whether the Kerr metric description is accurate and \GR holds in the strong-field regime \cite{Barausse:2020rsu}. At the \ISCO of a typical \MBH, the curvature scale is higher than Solar-system curvatures but lower than curvatures of some well-observed compact-binaries \citep{Yunes:2016jcc}. Consequently, modified gravities satisfying existing observational constraints generically predict very small deviations from \GR in the standard \EMRI scenario. However, since \EMRIs are expected to be highly sensitive to small departures from the standard Kerr black hole background paradigm \citep{Bambi:2011mj,Cardoso:2019rvt}, \EMRIs will allow us to further tightly constrain any deviations caused by an additional scalar or vector channel for radiation in the long-term dissipation of energy and angular momentum \citep{Chamberlain:2017fjl}. In fact, even in the case where the field of the primary is just a generic stationary axisymmetric \GR vacuum field, we should be able to determine the matter multipoles of the primary \citep{Ryan:1995wh} and check possible violations of the no-hair theorem \cite{Carter:1971zc}, e.g., a deviation from the Kerr solution in the quadrupole moment \cite{Glampedakis:2005cf,Barack:2006pq}. 

In addition, as is true for all extragalactic LISA sources, various cosmological dispersion and propagation effects can be tightly constrained from \EMRI signals \cite{Chamberlain:2017fjl}.

\begin{table}[!htbp]
	\centering
\begin{tabular}{lll} \toprule
	Parameter   	& Notation		& Astrophysically relevant range     	
\\ \midrule
	Total mass in the detector frame	& $M$ 	& \qquad \qquad $10^5 - 10^7 M_{\odot}$    
								\\
	Mass ratio $(>1)$ & $q$	          & \qquad \qquad $10^3 - 10^6 $					
								\\
	Dimensionless spin &	${\rm max}|\chi_i|$	& \qquad \qquad  $ 0 - 0.998$				
									\\ 
	Eccentricity entering LISA band & $e_\text{init}$  & \qquad \qquad  $0 - 0.8$	
									\\
	Eccentricity at last stable orbit  & $e_\text{merge}$      & \qquad \qquad  $0 - 0.2$	
							\\
	Signal to noise ratio & \SNR     &\qquad \qquad   $ 20 - 100 $					
								\\ \bottomrule
\end{tabular}
\caption{Summary of the anticipated parameters for LISA \EMRI sources. Total masses of $10^4-10^7 M_{\odot}$ can be seen by LISA~\cite{Gair:2010yu} with those greater than $10^7M_{\odot}$ \edit{ending} outside the LISA band, \edit{although a rapidly spinning \MBH may be detectable with a few $10^7 M_{\odot}$  as the \ISCO is closer to the \MBH} \cite{Amaro-Seoane:2012lgq}. Regardless of formation channel, the smaller compact body \edit{as} a \edit{typical $\mathcal{O}(10) M_{\odot}$} stellar mass black hole has an expected mass ratio of $10^3 - 10^6$, or even $10^4-10^7$ if neutron star \EMRIs are included. Using \MBH of $3 \times 10^6 M_{\odot}$ as a basis, the ``loss-cone'' \EMRIs are expected to have high eccentricities, capped at $\sim0.8$ entering the LISA band~\cite{Hopman:2005vr}, while the accretion channel tends to have smaller eccentricities~\cite{Pan:2021oob}. Evolving a large sample of compact bodies from capture results in a flat eccentricity distribution for the last stable orbit in the range $0<e<0.2$~\cite{Babak:2017tow}. A small tail of outlying higher eccentricities is possible. There is also the possibility of \GW bursts from unbound / hyperbolic systems~\cite{Berry:2013poa,Berry:2013ara}. 
Ref.~\cite{Babak:2017tow} predicts the loudest \EMRIs to have \SNR$\sim100$ \edit{for an observation time of a year}.}
\label{Tab:EMRIparameters}
\end{table}




\newpage

\subsection{Intermediate mass-ratio inspirals (IMRIs)}
\label{sec:sources_IMRIs}

Coordinator: Carlos F.~Sopuerta\\
Contributors: M.~Dhesi, M.~Hannam, T.~Hinderer, D.~Neilsen, H.~Pfeiffer, H.~Sundar, N.~Warburton

\subsubsection{Description}
An \IMRI is a binary system with mass ratio in the range $10 \lesssim q \lesssim 10^4$ placing it between comparable mass binaries and \EMRIs.
IMRIs can come in two flavors, both of which contain an \IMBH with mass in the range $10^2$-$10^4 M_\odot\,$. 
There is little observational evidence for \BHs in the \IMBH mass range, mainly because their main formation channel is inside globular clusters, the interiors of which are very difficult to observe \cite{Greene:2019vlv}.  
However, recent \GW observations with ground-based detectors have found a population of \BHs formed from \BBH mergers with a total mass greater than $10^{2}\,M_\odot$ \cite{LIGOScientific:2021usb,KAGRA:2021vkt}.
The largest confidently observed remnant (of GW190521) has a mass of $\sim 150 M_\odot$~\cite{LIGOScientific:2020iuh,LIGOScientific:2020ufj}.
At the other end of the mass range, we have evidence of \BHs with masses in the range $10^4$-$10^5 M_\odot$ from observations of dwarf galaxies out to redshift $\sim 2.4\,$~\cite{2018MNRAS.478.2576M,Greene:2019vlv}. 
These may correspond to an interpolation of the mass function observed for low \MBH mass~\cite{Reines:2013pia}.

The two flavors in which \IMRIs are expected to appear are: 
(i) Light \IMRIs. 
A \SOBH, or another sufficiently massive and compact object, inspiraling into an \IMBH. 
For instance, dwarf galaxies or globular clusters may contain an \IMBH which could capture an \SOBH \cite{Arca-Sedda:2020lso}. 
In this case, it is quite likely that the merger occurs outside the LISA frequency band, at higher frequencies.
It is thus possible that an \IMRI can be observed during its inspiral with LISA and have its merger seen by ground-based \GW observatories (only for light \IMBHs, with mass $\lesssim10^3\,M_\odot$~\cite{Gair:2010dx}).
(ii) Heavy \IMRIs. 
Dynamical friction can produce the orbital decay of globular clusters into a galactic nucleus allowing the formation of an IMBH–MBH binary system, with the \MBH in the LISA mass range ($10^5$-$10^7 M_\odot$) \cite{Matsubayashi:2005eg,Arca-Sedda:2018kne}.   
Another possible channel to form heavy \IMRIs is the merger of a dwarf galaxy satellite with its main galaxy \cite{Bellovary:2018gbb}. 
A third  possibility is the formation of an \IMBH in a galactic nuclei via hierarchical mergers \cite{Gerosa:2021mno} via migration traps in \AGN \cite{Bellovary:2015ifg}.
Interestingly, GW190521 may have occurred in such an environment \cite{Graham:2020gwr}.

For more information on possible formation channels, and the uncertain event rates for these binaries, see the LISA Astrophysics Working Group White Paper~\cite{LISA:2022yao} and~\cite{Fragione:2018nnl,Jani:2019ffg}, or the \IMRI reviews~\cite{Amaro-Seoane:2012lgq,Amaro-Seoane:2018gbb}.
The nature of the different formation channels for \IMRIs (both light and heavy) may lead to distinct \IMRI dynamics so that LISA detections may shed some light on what are the most viable formation scenarios.

\subsubsection{Science with IMRIs}

As we only have evidence for IMBHs at the two ends of the mass range that defines these objects, the search for IMRIs with LISA is especially relevant with huge discovery potential.
Event rates and parameter estimates for \IMRIs can provide valuable information about the formation and growth of \IMBHs in globular clusters, as well as details of stellar dynamics in those systems.
This information is often difficult to glean from electromagnetic observations \cite{Greene:2019vlv}.

For heavy \IMRIs the masses and spin of the \MBH can be measured to within an relative error of a fraction of a percent \cite{Huerta:2011zi}.
Unlike EMRIs, where the spin on the secondary can be difficult to constrain \cite{Piovano:2021iwv,Burke:2023lno}, the spin of the \IMBH can be constrained to within $\sim 10\%$ \cite{Huerta:2011zi}.
\IMRIs spend a long time orbiting in the strong field and due to their high \SNR (relative to \EMRIs) this makes then uniquely precise probes of \GR and the Kerr hypothesis \cite{Miller:2004va}.
They can also be sensitive to modifications to Einstein’s equations of \GR or the presence of dark matter \cite{Sopuerta:2009iy,Dai:2021olt}.

Finally, {\it multiband} \GW astronomy observations of light \IMRIs consisting of observing the inspiral in the LISA detector and the merger in ground-based detectors \cite{LIGOScientific:2014pky,VIRGO:2014yos,Sathyaprakash:2012jk,Reitze:2019iox} offers a range of increased science potential \cite{Jani:2019ffg,Amaro-Seoane:2009vjl,Sesana:2016ljz}.
In particular, observing the inspiral with LISA can place tight constraints on the sky location and the merger time which can allow for targeted alerts for electromagnetic counterparts to the merger.
Conversely, knowledge of the source parameters from ground-based observations of the merger can allow researchers to perform targeted searches for \IMRIs in the archived LISA data \cite{Wong:2018uwb}.
Multiband observations can also enhance the potential for performing tests of \GR by breaking degeneracies between some parameters in the waveform models~\cite{Datta:2020vcj}.

\begin{table}[!htbp]
%
	\centering
\begin{tabular}{llll} \toprule
	Parameter   & Notation			& Heavy \IMRIs    	&   Light \IMRIs  \\ \midrule
	Binary mass & $M$				& $10^4 - 10^7 M_\odot$	&   $10^2 - 10^4 M_\odot$	\\
	Mass ratio ($>1$) & $q$			& $10 - 10^4$			& 	$10 - 10^4 $			\\
	Dimensionless spin & ${\rm max}|\chi_i|$	 	& $0 - 0.998$			&   $0 - 0.998$ 	 		\\ 
	Eccentricity entering LISA band & $e_\text{init}$	& $0 - 1$	&	$0 - 0.9995$ 			\\
	Eccentricity at last stable orbit or leaving LISA band& $e_\text{merge}$	& $0 - 0.9$		&	$0 - 0.9$			    \\
Signal to noise ratio &\SNR					& $10-10^2$				&	$10-10^3$		\\ \bottomrule

\end{tabular}
\caption{
Summary of the anticipated parameters for LISA \IMRI sources. By defintion, light \IMRIs' primary \BH is from $10^2 M_\odot$ to $10^4 M_\odot$, i.e., an \IMBH, while heavy \IMRIs' primary \BH is a \MBH~\cite{LISA:2022yao}. 
Again the definition of an \IMRI has a mass ratio from comparable ($10$) up to $10^5$, above which the binary is defined as an \EMRI. 
Eccentricity can be quite high for light \IMRIs entering the LISA band ~\cite{Amaro-Seoane:2018gbb}.
The initial eccentricity for heavy IMRIs is largely unknown. 
\SNRs are taken from Table 10 of the Astrophysics Working Group White Paper~\cite{LISA:2022yao}. 
}
\label{Tab:IMRIparameters}
\end{table}


\newpage

\subsection{Galactic Binaries (GBs)}
\label{sec:sources_GBs}

%
Coordinator: Milton Ruiz\\
Contributors: V. Korol, H. Lim, V. Paschalidis, S. L. Shapiro, A. Tsokaros

\subsubsection{Description}
Our Galaxy is home to a variety of stellar binaries formed by white dwarfs, neutron stars, and black holes. Approximately a million years 
before their anticipated merger, these compact binaries transition into the millihertz \GW frequency band accessible with LISA. Given such long evolution timescales, these systems will manifest as nearly-monochromatic sources of gravitational waves. Furthermore, these compact binaries will be detectable in large numbers by LISA, potentially emerging as the most numerous \GW source within the millihertz band \cite{Postnov:2014tza,Wagg:2021cst,Lamberts:2019nyk,Marsh:2011yj}.

Galactic compact binaries represent a key source class for the LISA mission. 
Firstly, their existence as \GW sources is confirmed, as several have already been identified and characterised with electromagnetic telescopes (see~e.g.~\cite{Kupfer:2023nqx,Digman:2022zyh,Carvalho:2022pst,Johnson:2021grx,Yu:2020bic,Biscoveanu:2022sul,Wolz:2020sqh}). 
Gravitational radiation from some of these known sources has already been measured indirectly by monitoring binary orbital contraction over extended periods \cite{Hermes:2012us, Burdge:2019hgl, Munday:2022wtz}. Secondly, as we can anticipate their \GW signatures before the mission, these binaries have been suggested as ``verification sources'' for LISA (e.g. \cite{Stroeer:2006rx,Finch:2022prg,Littenberg:2024bso,Shah:2012vc,Kilic:2021dtv}). Most notably, Galactic compact object binaries accessible to LISA promise a wealth of insights into stellar and binary evolution. This encompasses understanding the nature of compact objects, unravelling the physical processes that govern binary interactions, and exploring their role in the formation and evolution of the Milky Way -- see the LISA Astrophysics white paper \cite{LISA:2022yao} for a review. Finally, it is noteworthy that a substantial portion of these binaries will not be individually resolvable by LISA, contributing to an unresolved stochastic foreground that will act as an additional source of noise for the instrument. Therefore, accurate characterization of this foreground is crucial to ensuring the precise characterization of other LISA sources.

Here we consider Galactic double compact objects, which we refer to as \GBs, specifically those where at least one component is a compact object other than a black hole (binaries with two stellar origin black holes are treated in the next section). These can be categorised into various subtypes: \BWDs, with the subset of accreting \BWDs, known as AM Canum Venaticorum or AM CVns in literature; \BNSs; and mixed systems. The mixed systems encompass \WDNSs, which can emerge as \UCXBs in electromagnetic radiation, \WDBHs, and \BHNSs.

\subsubsection{Masses, mass-ratios, eccentricities, and known LISA verification sources}
Observing \GB using electromagnetic observatories poses a significant challenge due to their inherently small size and dimness (in some cases the entire absence of electromagnetic emissions). These features, when combined with selection effects and incompleteness of dedicated electromagnetic surveys, has limited our ability to know the true distributions of their parameters such as orbital separations, component masses, mass ratios, and eccentricities. Thus, much of our understanding of these binaries primarily hinges upon population synthesis studies. However, electromagnetic observations using \SDSS and Gaia have recently improved our understanding~\cite{Postnov:2014tza, Maoz:2018epf, 2020ApJ...889...49B,Kilic:2021dtv}

Binary population synthesis studies (e.g., Table 6 in~\cite{LISA:2022yao}) show that for frequencies less than approximately 2 mHz, the \BWDs form an unresolved foreground for LISA (see also~\cite{Evans:1987qa,1987A&A...176L...1L,Hils:1990vc,Ruiter:2007xx}). Binaries with frequencies greater than approximately 2 mHz and/or closer than a few kpc do not overlap with this background~\cite{Evans:1987qa, Nelemans:2001hp, LISA:2022yao,Ruiter:2007xx,Korsakova:2024sut} and are called ``resolvable". However, not all resolvable \GBs are detectable, and unresolved \GBs can also be detectable. Only binary systems with a significant \SNR may be detected by LISA within a 4-year mission. Theoretical estimates suggest  that among the hundreds of millions of binaries in the Galaxy, the number of resolvable binaries are about 6000-10000 \BWDs, 100-300 \WDNSs, 2-100 \BNSs, 0-3 \WDBHs and 0-20 \BHNSs \cite{LISA:2022yao,Ruiter:2007xx,Korsakova:2024sut}. However, the number of detectable binaries above the noise are about 6000 \BWDs, 100 \WDNSs, 30 \BNSs, 3 \WDBHs, 3 \BHNSs and 2 hot subdwarf binaries~\cite{Nelemans:2001hp,Kupfer:2023nqx}.

The chirp mass, a key distinctive parameter among compact binary classes, varies significantly across different binary systems. For instance, \BWDs typically peak at a chirp mass of around $0.25 M_\odot$ (e.g.~\cite{Korol:2021pun,Li:2020voo,Edlund:2005cg}), while \BNS tend to have a chirp mass around $1.2 M_\odot$~\cite{Lau:2019wzw,Korol:2020irk}. Systems comprising \BHs and \NSs exhibit even higher chirp masses. However, it is important to note that LISA's limited ability to measure frequency derivatives for $f < 2$ mHz, and thus the chirp mass may introduce potential classification ambiguities~\cite{Lau:2019wzw,Korol:2020irk}. For example, a nearby \BWD system may be misidentified as a more distant \BNS. Another discriminative feature could be the eccentricity measurements~\cite{Lau:2019wzw, Andrews:2019vou}
. \BWDs, typically formed via isolated binary evolution, are expected to be circularised due to recurrent mass transfer episodes. Contrastingly, \BNS systems detectable in the LISA band might present measurable eccentricities due to natal kicks imparted by the supernova explosions that birth the \NSs. Nevertheless, some rare eccentric \BWD can form in globular clusters or via triple interactions (e.g.~\cite{Kremer:2018tzm,Willems:2007xe}).

There are approximately 30 \GBs, identified through electromagnetic detection, that have been confirmed as resolvable verification targets for the LISA mission. Notably, 18 of these are expected to be detectable within merely 3 months of scientific operations~\cite{Kupfer:2023nqx}, proving invaluable in the early stages of the mission by assisting in the validation of LISA's performance relative to pre-launch expectations. These \GBs are characterised by orbital periods ranging from $\sim300$ to $\sim6000$ seconds, equivalent to \GW frequencies approximately between 0.5 and 6 mHz. Their total mass typically lies between $0.5$ to $1.0 M_\odot$, with mass ratios roughly between 0.1 and 0.7. Their estimated \SNRs can exceed 100 after four years of integration~\cite{Finch:2022prg,Liang:2024ulf,Kupfer:2018jee}.

Certain types of binaries can be sources of low and high frequencies, and hence can straddle both the LISA and ground-based detector bands. This is true for mixed binaries when unstable mass transfer leads to merger. Gravitational waves from the inspiral and merger of \WDNSs may go from LISA to ground-based from potential oscillations of the \NS after merger and/or its eventual collapse to a \BH~\cite{Paschalidis:2009zz,Paschalidis:2010dh,Paschalidis:2011ez,Sun:2018puk,Kang:2023rux}. The same straddling of LISA/ground-based frequency bands holds for stellar-mass black hole binaries~\cite{Sesana:2016ljz,Toubiana:2020cqv,Moore:2019pke}.

\subsubsection{Modeling requirements \& methods to improve/remove the Galactic background}
\GBs can be classified as detached or interacting (mass-transferring), depending on whether significant mass transfer occurs between the components. This classification is crucial for modelling their \GW signals accurately. Although these systems are practically monochromatic to first order, the frequency does evolve with time. In particular, there is a difference in how the frequency evolves for detached binaries and interacting binaries. The orbital evolution of detached binaries is driven by the emission of gravitational waves, causing a gradual inward spiral of the binary (manifested as a negative frequency derivative in the \GW signal). On the other hand, in accreting binaries, the redistribution of mass between the components results in an increase in their separation (leading to a positive frequency derivative in the \GW signal)~\cite{Postnov:2014tza,Paschalidis:2009zz}. This leads to different requirements for waveform modelling. Note also that the shortest period binaries can be tidally-locked and corotating~\cite{Bildsten:1992my}. For binaries near the Roche separation, tidal effects may have a non-negligible contribution to the phase evolution of the wave~\cite{Broek:2012rs,Yu:2021hwl}. Understanding the frequency evolution will help determine if mass transfer is taking place~\cite{Breivik:2017jip,Littenberg:2020bxy} and/or tidal interactions are at play. A large number of mass-transferring systems may constrain the physics of mass transfer and the efficiency of angular momentum removal from the disk/companion system and its reinjection back into the orbit~\cite{Paschalidis:2009zz}. 

Some of the \GW sources that LISA will observe may be part of triple or higher-order multiple systems (e.g. \cite{2016ComAC...3....6T, Rajamuthukumar:2022rhl}). This includes sources in the Galactic disc that form through isolated triple evolution, as well as those in dense environments. When a \BWD system --- as the most common Galactic LISA source --- is part of a larger system with an additional stellar or even a substellar object, the gravitational interaction with the substellar object introduces a modulation in the observed \GW frequency due to the Doppler and related effects. This is a consequence of the motion of the \BWD around the centre of mass of the three-body system. As the orbit of the tertiary must be larger than that of the inner binary for the system to remain stable, the modulation timescale will be longer than the \GW frequency produced by the inner \BWD. However, to detect this modulation with LISA, it should also be shorter than the mission's lifetime \cite{Seto:2008di, Tamanini:2018cqb, Katz:2022izt}.

Signals from many \GBs will be below the \BWD confusion
background, which
is generated both by detached and 
interaction/mass-transferring \BWDs. 
Efforts have been made to remove this background in order to resolve more binaries~(see e.g.~\cite{Crowder:2006eu,Littenberg:2011zg,Bouffanais:2015sya,Littenberg:2020bxy}).
Notice that the modeling of this confusion background may provide insight about the \BWD
  population in the Milky Way (and nearby satellite galaxies).

\begin{table}[!htbp]
	\centering
\begin{tabular}{lllll} \toprule
	Parameter   &                    	Notation		& \BWD                  	& \WDNS             		& \BNS	\\ \midrule
	Total Chirp Mass &             	$M$	 		&$0.1-1 M_\odot$	&$0.4-1.2 M_\odot$	&$1- 2.2 M_\odot$	\\
	Mass Ratio ($>1$) &           	$q$	     		& $1-10$	          	&$1-5$        			&   $1-1.6$	\\
	Eccentricity in LISA band &	$e_\text{init}$	&0				&$0-1 $				& $0-1$         	\\
	Signal to Noise Ratio 		&\SNR		& $< 1000$		&$<1000	$			&$<1000$  \\ \bottomrule
\end{tabular}
\caption{Summary of the anticipated parameters for LISA \GB sources. \BWDs can have a chirp mass between $0.1-1 M_\odot$ \cite{Korol:2017qcx}, mass ratios of $1-10$ (which can reach 100 for AM CVns systems) and spins ranging from seconds to hours \cite{2022MNRAS.509L..31P, 2021ApJ...923L...6K}. \WDNSs and \BNSs will have chirp masses in the range $0.4-1.2 M_\odot$ \cite{Korol:2023jfz} and $1.0-2.2  M_\odot$ \cite{Vigna-Gomez:2018dza,Wagg:2021cst,2020A&A...639A.123K} respectively. These binaries should exhibit spin periods ranging from seconds down to milliseconds, consistent with those observed in current systems, 
  and near any eccentricity \cite{Vigna-Gomez:2018dza, Lau:2019wzw}. As these sources are considered nearly monochomatic in the LISA band, their eccentricity is expected to evolve only a negligible amount inband.}
\end{table}


\newpage

\subsection{Stellar origin black hole binaries (SOBHBs)}
\label{sec:sources_SOBHBs}
Coordinators:  Antoine Klein and Ilya Mandel \\
Contributor: D.~Gerosa

\subsubsection{Description}

Stellar origin \BH binaries are \BBHs with component masses ranging from a few solar masses up to about $\sim 50$-$100 M_\odot$, where models of pair-instability supernovae predict a mass gap to appear \cite{Woosley:2016hmi, Farmer:2019jed, Woosley:2021xba}. Those systems are in a mass range such that they can be observed by both LISA and by ground-based detectors, as they sweep through a few decades in frequency during the last stage of their inspiral.  This broad coverage of the \GW frequency spectrum makes it possible to probe the evolutionary history of such binaries \cite{Sesana:2016ljz,Mandel:2017pzd}. 

A circular \BBH with two 45 M$_\odot$ components will merge in about 10 years from a \GW frequency of $\sim10$ mHz \cite{Peters:1964zz,Sesana:2016ljz}.  
It is thus possible to track its evolution from the LISA band to the ground-based detector band, taking advantage of LISA's ability to better measure some of the system properties at low frequency and the improved measurements of parameters such as the remnant spin from high-frequency data.  Of particular interest for LISA is the sky localisation capability, which is typically limited to tens of square degrees for \BBHs in current ground-based detector data \cite{LIGOScientific:2021usb, KAGRA:2021vkt}.  The long duration of observations in the LISA band means that the detector will complete multiple orbits around the Sun, and thus effectively act as an instrument with a baseline of order the size of the orbit.  The accuracy of sky localisation can be estimated as \citep{Mandel:2017pzd}
\begin{equation}
	\sigma_\theta \sim 0.025 \frac{0.01\mathrm{Hz}}{f} \frac{8}{\text{\SNR}} \frac{\mathrm{AU}}{\mathrm{baseline}},
\end{equation}
where the \GW frequency $f$ should be replaced by the detector bandwidth for a source that evolves out of the band during the observation, and \SNR is the signal-to-noise ratio.  
For a 2 AU baseline, this would yield sub-degree localization, though the exact localization accuracy would depend on the source location in the orbital plane.  
Another example could involve the measurements of \BH spins and spin-orbit misalignments, which store information about formation scenarios \cite{Vitale:2015tea,Stevenson:2017dlk,Zevin:2017evb,Farr:2017uvj,Gerosa:2021mno}.  
While we may expect that high-frequency ground-based observations are best suited to measure mass ratios and spin-orbit couplings, which enter the waveform at higher orders in the orbital frequency, \LVK observations to date have demonstrated the challenge of making such measurements precisely \citep{KAGRA:2021vkt}.  
LISA could observe $\gtrsim 10^5$ orbital cycles and therefore assist in making precise constraints for the handful of systems that will be observed by both LISA and ground-based detectors \cite{Klein:2022rbf}.

Beyond individual sources, LISA could track changes in the source population as binaries evolve in frequency.  For example, \LVK observations indicate that the majority of black holes in merging \BBHs have masses $\lesssim 40 M_\odot$ \citep{KAGRA:2021duu}, low to moderate spins \cite{Roulet:2020wyq,Galaudage:2021rkt,Callister:2021fpo}, and most have circularized by the time of merger \cite{Romero-Shaw:2019itr}.  On the other hand, LISA could observe \BBHs while they are still eccentric, which would likely indicate dynamical formation \cite{LIGOScientific:2016vpg,Romero-Shaw:2020siz}.  The appearance of sources at higher frequencies would on its own indicate high birth eccentricity \cite{McNeill:2022jti}.

Another possibility worth mentioning is the detectability of a third body, particularly a massive black hole (MBH), through its impact on the binary's \GW signature.  Nuclear clusters around an MBH have been proposed as a possible \BBH merger site, with possible assistance from gas in an active galactic nucleus \cite{OLeary:2008myb,Bellovary:2015ifg,Tagawa:2019osr}.  Ref.~\cite{Mandel:2017pzd} argued that even when the orbital period of a binary of mass $M_\mathrm{bin}$ around an MBH of mass $M_\mathrm{MBH} \gg M_\mathrm{bin}$ is much longer than the observation duration $T_\mathrm{obs} $,  the orbital acceleration of the binary due to the MBH could still be detectable provided the \GW frequency $f_\mathrm{GW}$ exceeds
\begin{equation}
f_\mathrm{GW} \gtrsim 0.02\, \mathrm{Hz} \left(\frac{M_\mathrm{MBH}}{10^6\ M_\odot}\right)^{-1}\, \left(\frac{a}{\mathrm{pc}}\right)^{2} \left(\frac{T_\mathrm{obs}}{5 \mathrm{yr}}\right)^{-2} \left(\frac{\text{\SNR}}{8}\right)^{-1},
\end{equation}
where $a$ is the distance from the binary to the MBH.  A \BBH merger in a massive globular cluster of a similar mass could carry a comparable signature.

\subsubsection{Expected source parameters}

Judging by evidence from ground-based observations, the majority of \BBHs observed so far have chirp masses between 5 and $\sim 40 M_\odot$ \citep{KAGRA:2021duu}, though BH masses could extend down to the maximum neutron star mass and up to the IMBH mass range, especially if hierarchical mergers in dense dynamical environments fill the mass gap from pair-instability supernovae \cite{Yang:2019cbr,Mapelli:2020xeq,Gerosa:2021mno}.  LISA will be particularly sensitive to more massive \BBHs.

Most observed \BBHs have mass ratios $q \lesssim 4$ \cite{KAGRA:2021duu}, which is consistent with the bulk of model predictions, but systems with more extreme mass ratios are possible, such as GW190814, which involved a $\sim 23 M_\odot$ \BH and a $\sim 2.6 M_\odot$ companion \cite{LIGOScientific:2020zkf}.  As mentioned above, most observed \BBHs have low to moderate companion spins, though it remains unclear whether this is a generic feature of stellar-mass \BHs \cite{Mandel:2020lhv,Fishbach:2021xqi}.  Lastly, high eccentricities and generic spin-orbit misalignments could be a telltale sign of dynamical formation in dense stellar environments or in hierarchical triples (see Refs.~\cite{Mandel:2018hfr,Mapelli:2018uds} for reviews of formation channels).  

\BBHs emit the bulk of their orbital energy above the LISA frequency band; therefore, moderate SNRs are expected except for fortuitously nearby sources \citep{Sesana:2016ljz}. The minimal SNR for detection of BBHs will be influenced by the technical challenges associated with searching for signals with a complicated morphology and many in-band cycles \citep{Moore:2019pke}.  Furthermore, beyond an SNR threshold for detection, signals must also show evidence of frequency evolution in order for masses to be measurable, so that BBHs can be identified among the much larger population of signals from double white dwarfs \cite{Lau:2019wzw}.

\begin{table}[!htbp]
	\centering
\begin{tabular}{lll} \toprule
	Parameter   	& Notation		& range for majority of sources    
\\ \midrule
	Chirp Mass & $M$						& 5--40					
\\
	Mass Ratio ($>1$) & $q$				& 1--4					
\\
	Effective Spin &  ${\rm max}|\chi_\mathrm{eff}|$			& 0--0.3					
\\ 
	Eccentricity entering LISA band &   $e_\text{init}$		& 0--1		
\\
	Eccentricity at last stable orbit  & $e_\text{merge}$	& out of band		
\\
	Signal to Noise Ratio &   \SNR			& $<50$					
\\ \bottomrule
\end{tabular}
\caption{Summary of the anticipated parameters for LISA \SOBHB sources. The most recent data release from \LVK indicates a mass distribution of BBHs centred around $5-40M_{\odot}$~\cite{KAGRA:2021duu}, however their masses could theoretically extend down to neutron star mass ($2 M_{\odot}$) and up towards IMBH ($100 M_{\odot}$) \cite{Yang:2019cbr,Mapelli:2020xeq,Gerosa:2021mno, Woosley:2021xba}. \LVK observations also show a mass ratio and effective spin magnitude distribution between $1-4$ and $0-0.3$ respectively~\cite{KAGRA:2021duu}, with more extreme mass ratio binaries observed~\cite{LIGOScientific:2020zkf} and the possibility of higher spins theoretically possible~\cite{Mandel:2020lhv,Fishbach:2021xqi}.  Binary eccentricity decays with increasing gravitational-wave frequency roughly as $e \propto f^{-18/19}$ \cite{Peters:1964zz}, so sources that appear in the LISA band would circularise by the time they reach the \LVK detector band; therefore, existing \LVK observations do not constrain LISA band eccentricies. If we consider a SNR threshold for detection of 12, then $1-(12/50)^3\approx 98.6$\% of sources in an isotropic homogeneous Universe would have $\text{\SNR}<50$.}
\end{table}


\newpage

\subsection{Cosmic strings}
\label{sec:sources_cosmicstrings}
Coordinators: Barry Wardell  \\
Contributors: D.~Chernoff, J.~Wachter

\subsubsection{Description}

Strings are effectively one-dimensional stress energy sources. 
If a network of strings is generated at
early times then it can have many cosmological consequences including
the production of gravitational waves (see~\cite{Vilenkin:2000jqa} for a general review) that are potentially observable by LISA.
String sources at energy scales comparable to those postulated in Grand Unified Theories (GUTs) have been ruled out by observations of the CMB (see~\cite{Planck:2013mgr} for Planck-derived limits; the energy scale of the string-forming phase transition is $<4.7 \times 10^{15}$ GeV, somewhat less than the charateristic GUT energy scale $\sim 10^{16}$ GeV).
However,
String Theory suggests new sources ---fundamental strings and D-branes
wrapped on small dimensions--- in the context of certain inflationary
string theory scenarios
~\cite{Witten:1998cd,Sen:1998ii,Horava:1998jy,Dvali:1998pa,Sen:1998tt,Giddings:2001yu,Jones:2002cv,Sarangi:2002yt,Kachru:2003aw,Copeland:2003bj,Jones:2003da,Kachru:2003sx,Dvali:2003zj,HenryTye:2006uv}.
Macroscopic cosmic strings (CSs) are created and stretched to
superhorizon scales by inflation (see \cite{Chernoff:2014cba,Baumann:2014nda} for a review). Extensions to the Standard Model that introduce new symmetry breakings in-between the GUT and electro-weak scale can also produce viable strings~\cite{Dror:2019syi,Dunsky:2021tih}.

Irrespective of the detailed microscopic origin, these CSs may evolve to generate a network with some common features: after the Universe enters
its radiation dominated phase the long, horizon-crossing strings begin
to collide, break, reconnect and form small, sub-horizon scale loops. All
string elements are dynamical, radiating gravitational waves and
possibly other
quanta \cite{Brandenberger:1986vj,Srednicki:1986xg,Bhattacharjee:1991zm,Damour:1996pv,Wichoski:1998kh,Peloso:2002rx,Vachaspati:2009kq,Sabancilar:2009sq,Long:2014mxa}.
Isolated loops, for example, radiate their entire rest mass energy and
eventually disappear. Together the string elements generate a
stochastic \GW background
~\cite{Vilenkin:1981bx,Vachaspati:1984gt,Hogan:1984is,Accetta:1988bg,Bennett:1990ry,Caldwell:1991jj,Siemens:2006yp,DePies:2007bm,Olmez:2010bi,Sanidas:2012ee,Sanidas:2012tf,Binetruy:2012ze,Kuroyanagi:2012jf,Kuroyanagi:2012wm,
Sousa:2013aaa,Blanco-Pillado:2013qja,Sousa:2014gka,Blanco-Pillado:2017oxo,Chernoff:2017fll,Ringeval:2017eww,Blanco-Pillado:2017rnf,Cui:2018rwi,Jenkins:2018nty}
and \GW bursts
\cite{Damour:2000wa,Damour:2001bk,Siemens:2006yp,Binetruy:2009vt,Olmez:2010bi,Regimbau:2011bm,Kuroyanagi:2012wm, Kuroyanagi:2012jf,LIGOScientific:2013tfe,Ringeval:2017eww,Chernoff:2017fll,LIGOScientific:2017ikf,Jenkins:2018nty}. Detection and measurement of the
string-generated gravitational waves by LISA will be informative for cosmology
and high energy physics (see the whitepaper from the LISA Cosmology Working Group for a summary~\cite{LISACosmologyWorkingGroup:2022jok}).

The physics of CSs
is sensitive to (1) the set of fields to which the
string couples, (2) whether the strings are global or local, (3) the
ratio of the characteristic string width to curvature scale. Here, we
assume that the strings are minimally coupled (they interact with each
other but only radiate gravitational waves), local and well-described
in the classical limit by the Nambu-Goto action (see \cite{Aoyama2004} for a brief description and \cite{Zwiebach:2004tj} for string-theory applications). We refer to this as
the ``minimally coupled-string network''. There are additional
possibilities but this defines a wide, interesting arena for this
whitepaper.

Average properties of the minimally coupled string network are
encapsulated in the Velocity One Scale (VOS) model
\cite{Martins:1996jp,Martins:2000cs,Martins:2003vd,Sousa:2013aaa}.
It turns out that all quantitative features depend primarily
on the string tension $\mu_S$, or $G \mu_S/c^2$ in dimensionless terms. In
particular, the total density in string components is parametrically
small when the tension is small.

The VOS model is a valuable guide for forecasting the observations
LISA may make. For specific numerical estimates below we assume
minimally coupled local strings and adopt the following secondary
parameters: intercommutation probability $\sim 1$ (intersecting field theory
strings break and reconnect with probability
of order unity \cite{Shellard:1987bv} whereas
string theory strings do so with smaller probability
\cite{Jackson:2004zg}), number of string
species $1$ (multiple species exist in realistic string
constructions, for a review see \cite{Chernoff:2014cba}),
fraction $\sim 0.2$ of long length strings chopped into
loops of size $\sim 0.1$ of the horizon (for review of
the small and large components inferred from simulation see
\cite{Chernoff:2009tp,LISACosmologyWorkingGroup:2022jok})
and rate of gravitational energy loss $dE/dt = \Gamma G\mu_S^2c$ implied
by dimensionless parameter
$\Gamma=50$ \cite{Vachaspati:1984gt,Burden:1985md,Garfinkle:1987yw,Blanco-Pillado:2017oxo}.
Broadly speaking, changes to these adopted secondary values do not
qualitatively change the network properties predicted by the VOS
model. Many network properties are {\it not} included in VOS and have
not yet been addressed in simulations. For example, isolated loops
should evolve under the force of radiative backreaction
(see \cite{Wachter:2016rwc,Blanco-Pillado:2018ael,Chernoff:2018evo,Blanco-Pillado:2019nto})
but that process is not included in current numerical simulations and
so it is difficult to accurately incorporate into statistical
descriptions. These model-dependent, as opposed to
parameter-dependent, uncertainties are important systematic
deficiencies in our understanding and hard to quantify.

\subsubsection{Tension limits}

The most well-studied modern scenario involves Type IIB string theory
and low tension strings produced at the end of brane inflation.
Strings --- not monopoles nor domain walls --- are produced when a brane and
anti-brane pair annihilate and initiate the Big Bang cosmology (\cite{Jones:2002cv,Sarangi:2002yt}). The
primary parameter, string tension, cannot currently be calculated a
priori from theory. Instead we must turn to observations.

Empirical upper bounds on $G \mu_S/c^2$ have been derived from null
results for experiments involving lensing \cite{Vilenkin:1981zs,Hogan:1984unknown,Vilenkin:1984ea,deLaix:1997dj,Bernardeau:2000xu,Sazhin:2003cp,Sazhin:2006fe,Christiansen:2008vi}, 
\GW background and bursts \cite{Vachaspati:1984gt,Economou:1991bc,Battye:1997ji,Damour:2000wa,Damour:2001bk,Damour:2004kw,Siemens:2006vk,Hogan:2006we,Siemens:2006yp,LIGOScientific:2009bje,LIGOScientific:2009qal,LIGOScientific:2013tfe,LIGOScientific:2017ikf,LIGOScientific:2019ppi,KAGRA:2021kbb}, pulsar
timing \cite{Bouchet:1989ck,Caldwell:1991jj,Kaspi:1994hp,Jenet:2006sv,DePies:2007bm,Blanco-Pillado:2017oxo,Blanco-Pillado:2017rnf}, 
cosmic microwave background radiation \cite{COBE:1992syq,Bennett:1996ce,Pogosian:2003mz,Pogosian:2004ny,Tye:2005fn,Wyman:2005tu,Pogosian:2006hg,Seljak:2006bg,WMAP:2006bqn,Bevis:2007qz,Fraisse:2006xc,Pogosian:2008am,Planck:2013mgr}. It has long been recognized
\cite{Chernoff:2014cba} that
all such bounds are model-dependent and typically involve observational and
astrophysical uncertainties. Constraints from the CMB power spectrum rely on
well-established gross properties of large-scale string networks and
are relatively secure. Limits from optical lensing in fields of
background galaxies rely on the theoretically well-understood deficit
angle geometry of a string in spacetime but require a precise
understanding of optical selection effects.  Bounds from big bang
nucleosynthesis rely on changes to the expansion rate from extra
gravitational energy density but only constrain the strings formed
prior to that epoch. Roughly speaking, these limits imply
$G \mu_S/c^2 \lesssim 3 \times 10^{-8} - 3 \times 10^{-7}$
(see \cite{Siemens:2006yp} for comparisons of limits).  More stringent
bounds on tension generally invoke additional
assumptions \cite{Battye:2010xz}.  Gravitational wave experiments
\cite{Siemens:2006yp,LIGOScientific:2009qal,LIGOScientific:2017ikf,LIGOScientific:2019vic}
can monitor the occurrence of bursts. In particular, the \LVK set a bound $G\mu_S/c^2 \lesssim 4\times
10^{-15}$ based on non-detection of assumed cusp-like bursts \cite{LIGOScientific:2021nrg}. 
Long-term pulsar timing searches for a stochastic background have set the bound of $G \mu_S/c^2 \lesssim 1.5 \times 10^{-11}$~\cite{Jenet:2006sv,Ringeval:2017eww,Blanco-Pillado:2017oxo,Blanco-Pillado:2017rnf}.
In the future LISA may achieve limits as low as $G \mu_S \sim 10^{-17}$ for Nambu-Goto strings \cite{LISACosmologyWorkingGroup:2022jok,LISA:2022kgy}.


\subsubsection{Loop sources for LISA}

The VOS model for the minimally coupled string network generates a loop size distribution weighted towards small sizes \cite{Chernoff:2017fll}.
%
For string tensions $G \mu_S/c^2 \ll 10^{-7}$ the string loops are the most important elements of the network for \GW science \cite{LISACosmologyWorkingGroup:2022jok}. 
Small tensions imply weak \GW damping.
The undamped string is a non-linear oscillator with a
fundamental period $T=\ell/(2c)$ and frequency $f=1/T$. The Fourier
transform of its motion yields power in all harmonics $n f$ for $n \ge
1$. A survey of the loop dynamics reveals large scale motions and
distinctive small scale feature: cusps (infinitesimal bits of the
string that move at the speed of light twice per fundamental period)
and kinks (discontinuous changes of slope that perpetually
circumnavigate the loop).  Gravitational wave emission is sourced not
only by the large scale oscillations but also by cusps and kinks.  All
long-lived loops are expected to possess cusps or kinks else they
intercommute and produce kinks. Cusp-generated power decays with
harmonic $n$ asymptotically $\propto n^{-4/3}$, kink-generated power
$\propto n^{-5/3}$ and kink-kink collisions $\propto
n^{-2}$~\cite{Vachaspati:1984gt,Damour:2000wa,Damour:2001bk,Binetruy:2009vt}.
At high frequencies, cusps dominate if they are present.  Conversely,
the period-averaged area of the sky illuminated by gravitational
radiation increases from cusps to kinks to large scale modes.

There are three scenarios for LISA detections. (1) Loop decay creates
a stochastic \GW background from a large number of
unresolved sources \cite{LISACosmologyWorkingGroup:2022jok}.  (2) Specific cusp or kink containing sources
produce bursts of emission that stand above the general
background~\cite{LISACosmologyWorkingGroup:2022kbp}. (3) A few nearby loops, possibly associated with the Galaxy, produce emission that is strong,
smooth and always on~\cite{DePies:2009mf,Chernoff:2009tp,Khakhaleva-Li:2020wbr,Jain:2020dct}.

\subsubsection{Science with Cosmic Strings}

(1) Interesting fundamental results are destined to emerge whether or
not LISA detects evidence of gravitational radiation from strings. A
positive result for the stochastic background will allow the inference
of the string tension; a negative result will provide upper limits on
the tension of any string component that might be present \cite{LISACosmologyWorkingGroup:2022jok}. 
If the network has a String Theory origin then either determination helps
guide progress towards a realistic model scenario for String Theory
that incorporates the Standard Model.

%

(2) A positive detection (either background, burst or nearby loop) is
also fundamentally significant for cosmology because loops of macroscopic
size are created during an epoch of inflation, supporting the inflationary
paradigm \cite{Guth:1980zm,Linde:1981mu,Albrecht:1982wi}. The
universe's precise inflationary scenario remains a profound problem
for cosmology and for fundamental physics.  The almost scale-invariant
density perturbation spectrum predicted by inflation is strongly
supported by cosmological observations, in particular the cosmic
microwave background radiation \cite{Planck:2013mgr}.

A negative detection is also very informative.  The production of
string-like structures is a rather generic theoretical prediction
whenever inflation does occur \cite{Jeannerot:2003qv,Dunsky:2021tih}. A negative result might be explained if
the strings are unstable and/or couple to additional fields that
promote their decay.  This sort of result will guide the search for
models that allow such interactions.


\newpage

\subsection{Beyond GR and beyond Standard Model sources}
\label{sec:sources_beyond_GR}
Coordinator: Paolo Pani, Helvi Witek \\
Contributors:
N.~Afshordi,
R. Benkel,
G. Bozzola,
R. Brito,
A.~C\'{a}rdenas-Avenda\~{n}o,
E.~Maggio,
M. Okounkova,
V. Paschalidis,
C. Sopuerta

\subsubsection{Introduction}
\BHs and compact binaries, and their \GW emission
have tremendous potential to probe for new physics beyond the Standard Model
in the strong-field, nonlinear regime of gravity.
LISA is likely to detect loud sources, such as \MBHBs and \EMRIs,
which will allow us to test the nature of \BHs,
the validity of \GR in the strong-field, highly-dynamical regime of gravity,
or the presence of additional fundamental fields with unprecedented precision.
These observations can help us address fundamental questions such
as~\cite{Arvanitaki:2010sy,Yunes:2013dva,Berti:2015itd,Barack:2018yly,Cardoso:2019rvt,Barausse:2020rsu}:
\begin{enumerate*}[label={(\roman*)}]
\item What is the nature of dark matter, and how can \BH detections with LISA aid the
search for new particles?
\item What is the nature of gravity?
         Are there new fundamental fields and \GW polarizations, as predicted by some
extensions of \GR and of the Standard Model?
\item How do gravitational waves propagate over cosmological distances?
\item Are the massive objects observed at galactic centers consistent with the rotating
\BHs predicted by \GR?
\item Do exotic compact objects other than \BHs and neutron stars exist in the universe?
\end{enumerate*}
This complex topic in the context of LISA science is discussed in depth in the White paper of LISA's Fundamental Physics Working Group~\cite{LISA:2022kgy}.
Here we briefly summarize the most relevant sources for probing fundamental physics with LISA
as a guide for the modelling of gravitational waveforms in \GR and beyond.

\subsubsection{Black holes and ultralight fields}
\paragraph{Description}
Ultralight bosonic fields such as axions, axion-like particles or dark photons are
predicted in several particle and theoretical physics models.
A remarkable example is the string axiverse scenario, which predicts a multitude of
axion-like particles
emerging naturally from string theory compactifications~\cite{Arvanitaki:2009fg}.
These ultralight particles play a crucial role in diverse areas of physics
and have been proposed
\begin{enumerate*}[label={(\roman*)}]
\item as a solution to the strong CP problem in \QCD~\cite{Peccei:1977hh},
\item as compelling dark matter candidates~\cite{Hui:2016ltb,Ferreira:2020fam}
and \item in cosmology~\cite{Marsh:2015xka}.
\end{enumerate*}
Excitingly, we can employ \BHs to search for (or constrain) ultralight bosons
in a mass range that is complementary to traditional particle colliders or direct detection experiments~\cite{Arvanitaki:2010sy,Brito:2015oca,Bertone:2019irm,Arvanitaki:2014wva,Brito:2017zvb,Brito:2017wnc}.
This surprising connection between \BHs and particle physics is provided by the
superradiant instability of \BHs~\cite{Press:1972zz,
Starobinsky:1973aij,Bekenstein:1973mi,Teukolsky:1974yv,
Detweiler:1980uk,Shlapentokh-Rothman:2013ysa,Brito:2015oca}:
low-frequency bosonic fields scatter off a rotating \BH superradiantly,
thereby extracting mass and angular momentum from the \BH.
Fields of mass-energy $\mu_{\rm B}$ are efficiently confined in the vicinity of a \BH with mass $M$
if the gravitational coupling
$M\mu_{\rm B}\lesssim
0.4$~\cite{Dolan:2007mj,Witek:2012tr,Pani:2012vp,East:2017mrj,Wang:2022hra},
corresponding to $\mu_{\rm B}\lesssim 10^{-16}(M/10^6 M_\odot) \,{\rm eV}$.
In this case they efficiently form a bosonic condensate (``cloud'') around the \BH.
An alternative formation scenario involves accretion of such ultra-light fields onto
\BHs~\cite{Clough:2019jpm,Bamber:2020bpu} in the same mass range.
If the bosonic field is complex, this process gives rise to hairy \BHs~\cite{Herdeiro:2015waa,Herdeiro:2014goa,Herdeiro:2016tmi,Santos:2020pmh}.

The details of the cloud's formation depends
on the initial parameters such as the \BH spin
and the gravitational coupling between \BH and bosonic field~\cite{Brito:2014wla,Ficarra:2018rfu,East:2017ovw}.
Once formed, the condensate dissipates by emitting a
quasi-monochromatic \GW signal~\cite{Yoshino:2013ofa,Arvanitaki:2014wva,Brito:2014wla,Okawa:2014nda,Zilhao:2015tya,Brito:2017zvb,Brito:2017wnc,East:2017ovw,East:2018glu,Siemonsen:2019ebd,Zhu:2020tht,Brito:2020lup}.
The presence of boson clouds can also significantly affect the dynamics of binaries,
e.g., through
``dragging'' of the cloud~\cite{Zhang:2019eid}, dynamical
friction~\cite{Traykova:2021dua} or tidal effects~\cite{Cardoso:2020hca}.
\edit{Numerical simulations of comparable-mass \BH binaries interacting with a
high-density bosonic cloud have shown that
scalar clouds may condense around the binary to form ``gravitational
molecules''~\cite{Ikeda:2020xvt}, that yield scalar radiation.
This interaction may yield a \GW phase shift~\cite{Bamber:2022pbs}
that is also present for low-eccentricity initial data~\cite{Cheng2023InPrep}.
The post-merger quasinormal ringdown frequency are changed in the presence of a scalar
cloud~\cite{Choudhary:2020pxy}.
}
In \EMRIs the \GW signal is modified relative to the vacuum
case~\cite{Macedo:2013qea,Wong:2020qom}
and the presence of a secondary
\BH yields resonances
~\cite{Baumann:2018vus,Baumann:2019eav,Baumann:2019ztm,Berti:2019wnn}.
The detection of \EMRIs can be used to infer the boson's
mass~\cite{Hannuksela:2018izj,Hannuksela:2019vip}.
%

\paragraph{Expected source parameters}
All ``traditional'' \GW sources in the LISA band,
including
compact, comparable-mass \BH binaries, \EMRIs, and isolated spinning massive \BHs,
are potentially affected by bosonic clouds.
Therefore, they are also good sources to act as cosmic laboratories for ultralight fields.
Because the underlying mechanism only relies on the gravitational coupling,
but is independent from the coupling of the bosons to the Standard Model of particle
physics,
they probe all types of bosons, i.e., (pseudo-) scalars such as axion-like
particles~\cite{Peccei:1977hh,Arvanitaki:2009fg},
ultralight dark matter~\cite{Hui:2016ltb,Ferreira:2020fam},
ultralight vector~\cite{Goodsell:2009xc,Baryakhtar:2017ngi}
and tensor~\cite{Brito:2020lup,Dias:2023ynv} fields.
LISA sources are particularly well suited for detecting or constraining
ultra-light bosons in the mass range $\mu_{\rm B}\in\left[10^{-19},10^{-15} \right]\,{\rm eV}$~\cite{Brito:2017zvb,Brito:2017wnc,Isi:2018pzk,Stott:2020gjj,Brito:2020lup} (and even wider for massive spin-2 fields~\cite{Dias:2023ynv}),
and they are suited for multi-wavelength searches in combination with ground-based
instruments~\cite{Ng:2020jqd}.

\subsubsection{Binary black holes as probes of the nature of gravity}\label{sssec:BBHinModifiedGravity}

\paragraph{Description}
Does \GR, our Standard Model of gravity, truly describe gravitational phenomena at all
scales? It is expected to break down at high-energy scales as signaled by the presence of
singularities inside \BHs or at the Big Bang.
At these scales a more complete theory of quantum gravity is needed that consistently combines
gravity and quantum mechanics.
However, \GR cannot be quantized with standard approaches and it is not renormalizable.
Therefore, \GR (or a quantized version thereof) is not a viable candidate for quantum
gravity.
While a complete theory of quantum gravity remains elusive,
most candidates
predict similar extensions to \GR such as
higher curvature corrections or (non-minimial) coupling to new fields.
\BHs provide an ideal probe to search for such beyond-\GR theories
because, e.g.,
the presence of additional fields
may endow \BHs with scalar hair (or ``charges''), thus violating the no-hair
theorems of \GR~\cite{Hawking:1972qk,Bekenstein:1996pn}.
More specifically, new fundamental fields arise in several low-energy effective field
theories of gravity~\cite{Berti:2015itd},
e.g., in the low-energy limit of  quantum gravity, in the Horndeski class of scalar-tensor
theories~\cite{Horndeski:1974wa,Deffayet:2009wt}
tensor-vector-scalar theories~\cite{Moffat:2005si},
and in theories with quadratic curvature corrections~\cite{Kanti:1995vq,Alexander:2009tp,
Kanti:1995cp,Cano:2021rey}.
Such fields also arise in extensions of the Standard Model of particle physics,
e.g.,
\edit{hidden U(1) fields (including ``dark photons'') in mini-charged dark matter models~\cite{Cardoso:2016olt}
},
primordial magnetic monopoles~\cite{Preskill:1984by}, darkly charged dark-matter~\cite{Fan:2013yva},
and the aforementioned bosonic clouds formed around BHs due to the superradiant instability.

In comparable-mass compact binaries,
the coalesence of such ``hairy'' \BHs generates additional scalar radiation that accelerates
the inspiral and yields a \GW phase shift.
Furthermore, new polarization channels can exist in modified gravity theories.
The detection or absence of such extra polarizations will be an important probe for new physics.
In addition to modifications to the background solutions and the \GW emission,
modified theories of gravity may also change the physical properties of the gravitational waves
once they are emitted, e.g., by changing the dispersion relation, the polarization
and the way they interact with matter and with the detector~\cite{Yunes:2013dva}.
As modifications to the propagation of gravitational waves accumulate with the distance traveled,
and the capability to put constraints on the mass of the graviton
(Compton GW wavelength) scales with the chirp mass,
comparable-mass binaries are the most
effective systems for measuring these effects~\cite{Barausse:2020rsu}.

\EMRIs will provide an excellent probe of the multipolar stucture of its primary object
and, thus, test if the primary is consistent with the Kerr \BH metric predicted by \GR.
\GW measurements will
reveal details of both the conservative (time-symmetric) and the dissipative
(time-asymmetric) sectors of the gravitational theory.
Additional degrees of freedom, such as dynamical scalar or vector fields,
will introduce modifications to the motion of bodies,
and additional sources of \GW energy and angular momentum emission.
Given that in most modified theories the gravitational field is described by a
spin-2 metric tensor field and by additional fields~\cite{Berti:2015itd},
the interaction between matter and the new fields may give rise generally
to an effective ``fifth force'', leading to deviations from the universality of
free fall~\cite{Nordtvedt:1968qr}, or in other words,
to violations of the ``strong'' equivalence principle.
How well a beyond-\GR theory may be constrained depends on the relative \PN order
at which the correction enters and on the dimensions of the extra couplings~\cite{Cardenas-Avendano:2019zxd,Yunes:2016jcc,Chamberlain:2017fjl}.
\edit{
For example, the best \GW tests of theories with higher-order curvature invariants, such as the Gauss-Bonnet
invariant~\cite{Metsaev:1987zx,Campbell:1991kz,Kanti:1995vq} or the Pontryagin
density~\cite{Alexander:2009tp}, involve small-mass \BH{s} whose curvature
is larger than that of supermassive \BH{s}.
Therefore, the best probes of higher curvature corrections are \IMRIs and \EMRIs in the
LISA band~\cite{Sopuerta:2009iy,Gair:2012nm}
or stellar-mass \BBHs detected with ground-based \GW instruments.
}

\paragraph{Expected source parameters}
Both \MBHBs and \EMRIs can be employed to test
gravity and the nature of compact objects with LISA;
\edit{see the LISA Fundamental Physics whitepaper~\cite{LISA:2022kgy} for details.}
.
For example, with nearly equal-mass \BHs,
one can perform null tests with inspiral-merger-ringdown consistency and BH spectroscopy,
as well as searching for specific deviations with parametrized inspiral and ringdown
waveforms;
\edit{see Refs.~\cite{Berti:2015itd,Franchini:2023eda,Colleoni:2024lpj}.}
\edit{
Given the different nature and mass range of the gravitational wave sources that will be
detected with LISA,
tests of gravity with LISA will be complementary to the suite of tests that has been
performed by the \LVK
collaboration~\cite{LIGOScientific:2019fpa,LIGOScientific:2020tif,LIGOScientific:2021sio}.
}
The beyond-GR modifications in the modelling of these signals (including \PN theory, BH perturbation theory, effective-one-body approaches, and numerical relativity) are discussed in Sec.~\ref{sec:modling_beyond_GR}.

With \EMRIs one can constrain the multipolar structure of the primary object's
spacetime with exquisite precision, thus testing whether the primary object is
consistent with the Kerr \BH metric predicted in \GR
\edit{see, e.g., Ref.~\cite{Cardenas-Avendano:2024mqp} and references therein.}
The parametrization of the waveforms will depend on the type of test carried out. For instance, when testing the geometry of the dark objects inhabiting galactic centers assuming GR, the parameters would describe the deviations from Kerr, e.g., multipole moments, tidal parameters or post-Kerr parameters. On the other hand, when testing GR, the parameters would describe the modifications of GR, e.g., additional coupling constants, length scale of extra dimensions or higher-order corrections.

With both types of sources one can search for novel radiation channels
due to extra polarizations or additional charges present in beyond-\GR theories.

\subsubsection{Testing the Kerr-hypothesis: BH mimickers and echoes}
\paragraph{Description}

Exotic compact objects (ECOs) are horizonless objects which are predicted in some quantum gravity extensions of GR \cite{Nicolini:2005vd,Bena:2007kg,Giddings:2014ova,Koshelev:2017bxd,Abedi:2020ujo} and in the presence of exotic matter fields in the context of GR \cite{Liebling:2012fv,Giudice:2016zpa,Cardoso:2019rvt}.
The theoretical motivations for ECOs are the regularity of their inner structure and the
overcoming of semi-classical puzzles such as that of the information loss
\cite{Myers:1997qi,Das:2000su,Mathur:2009:1936-6612:133}. These ideas have inspired a plethora of
models including gravastars \cite{Mazur:2001fv,Mazur:2004fk}, boson stars
\cite{Feinblum:1968nwc,Kaup:1968zz,Ruffini:1969qy,Colpi:1986ye,Seidel:1991zh},
\edit{Proca stars~\cite{Brito:2015pxa},
}
wormholes \cite{Einstein:1935tc,Morris:1988cz,Damour:2007ap}, fuzzballs \cite{Mathur:2005zp,Mathur:2009:1936-6612:133} and others \cite{Bowers:1974tgi,Gimon:2007ur,Prescod-Weinstein:2009oap,Brustein:2016msz,Holdom:2016nek,Buoninfante:2019swn}.

ECOs are classified in terms of their compactness, reflectivity and possible extra degrees of freedom related to additional fields \cite{Cardoso:2019rvt,Wang:2018gin}. Two important categories are \cite{Cardoso:2017cqb}: \textit{ultracompact objects}, whose exterior spacetime has a photon sphere, and \textit{clean-photon-sphere objects} (ClePhOs), so compact that the round-trip time of the light between the photon sphere and the object's surface is longer than the instability timescale of photon orbits.
If the remnant of a merger is an ultracompact object, the ringdown signal differs from the BH ringdown at early stages and is dominated by the modified QNMs of the object \cite{Urbano:2018nrs,Maggio:2020jml}.
Conversely if the remnant of the merger is a ClePhO, the prompt ringdown is nearly indistinguishable from that of a BH because it is excited at the light ring \cite{Cardoso:2016rao}. The details of the object's interior appear at late times in the form of a modulated train of \GW echoes \cite{Cardoso:2016rao,Cardoso:2016oxy,Price:2017cjr,Correia:2018apm,Burgess:2018pmm,Huang:2019veb}. The time delay between echoes depends on the compactness of the object and/or the energy scale of new physics \cite{Oshita:2020dox}, whereas the amplitude is related to the reflectivity of the object~\cite{Cardoso:2019rvt}.

Many of these ECOs could be ruled out based on theoretical grounds. Horizonless compact objects are affected by an ergoregion instability when spinning sufficiently fast \cite{Friedman:1978ygc,1978RSPSA.364..211C,1996MNRAS.282..580Y,Kokkotas:2002sf}. The endpoint of the instability could be a slowly spinning ECO \cite{Cardoso:2014sna,Brito:2015oca} or dissipation within the object could lead to a stable remnant \cite{Maggio:2017ivp,Maggio:2018ivz}. Furthermore, ultracompact horizonless objects might be generically affected by a light-ring instability at the nonlinear level~\cite{Cardoso:2014sna,Cunha:2017qtt,Cunha:2022gde}.
Current and future \GW detectors will constrain models of ECOs in almost all the regions of their parameter space \cite{Cardoso:2007az,Fan:2017cfw,Barausse:2018vdb,Maggio:2020jml}.
In particular, searching for echoes in the post-merger signal of \MBHBs with LISA~\cite{Maggio:2019zyv} will provide a clean smoking gun of deviations from the standard, ``vacuum'', BH prediction. Furthermore, EMRIs could constrain the reflectivity of the primary object to unprecedented levels~\cite{Maggio:2021uge}.

\subsubsection{Expected source parameters}
The source parameters depend on the specific signal used to search for and constrain ECOs.
The echo signal depends on the parameters of the remnant (in particular mass and spin), on its compactness, and especially on its effective reflectivity, which is zero for a classical BH~\cite{Cardoso:2019rvt,Maggio:2021ans}. The reflectivity can be generically a complex function
of the frequency and of other remnant parameters.
In an inspiral, besides the standard binary parameters, ECOs are characterized by anomalous multipole moments and nonvanishing tidal deformability, sharing in this case properties similar to those of BHs in modified gravity theories.
In an EMRI, besides the different multipolar structure~\cite{Glampedakis:2017cgd,Raposo:2018xkf,Raposo:2020yjy,Bianchi:2020miz,Bianchi:2020bxa,Bena:2020see,Bena:2020uup,Herdeiro:2020kvf,Bah:2021jno,Fransen:2022jtw,Loutrel:2022ant,Vaglio:2022flq} and tidal deformability~\cite{Pani:2019cyc,Piovano:2022ojl} of the central object, a key parameter is again the (frequency-dependent) reflectivity~\cite{Maggio:2021uge}.



%

\newpage

\newpage

\section{Modelling requirements from data analysis}
\label{sec:modelling_requirements}

Ideally, waveform models should be infinitely accurate, evaluate instantly and be available in any format desired. In practice, none of these are achievable. The practical accuracy, efficiency and format requirements are set by the way LISA data is analyzed. This section starts with a brief overview of how LISA data analysis is expected to work. (For a more detailed description, see the LISA Data analysis whitepaper \cite{LDCwhitepaper}.) The remaining sections discuss how data analysis sets requirements on the accuracy, efficiency and formats of waveform models, providing the necessary framing for the discussion of waveform models in the rest of this whitepaper.

\subsection{Data analysis for LISA}
\label{sec:da_intro}

Contributor: Tyson Littenberg

The foundation of gravitational wave data analysis  is rooted in the conceptual simplicity of the measurement: observing relative changes in the separation between a collection of ``proof masses'' in free fall due to leading-order perturbations in the underlying spacetime metric, which propagate (effectively) uninhibited through the Universe. This is in stark contrast to, for example, electromagnetic observations, where the photons' propagation from source to detector is influenced by intervening material (e.g., dispersion, scattering, absorption, reprocessing), 
and then undergoes complicated interactions with the instruments themselves (e.g., focusing optics, filters, diffraction, absorption by the detector)  before registering as a signal. This is not to take away from the heroic effort and ingenuity required to develop the measurement system that is sensitive to the unfathomably small space-time perturbations themselves. However, given a detector that can achieve the necessary sensitivities, it is a tractable task to derive its response to incident gravitational waves from first principles.

As a result, \GW data analysis methods have primarily developed around a \emph{forward} problem, where the detector response is predicted from a hypothetical source and that predication is then tested against the data~\cite{Jaranowski:2005hz}.
There are notable exceptions, particularly in some searches for unmodelled \GW transients in ground-based interferometer data~\cite{Drago:2020kic}, which approach the analysis as an \emph{inverse} problem, starting with the observed data and working backward to solve for the input signal.  

Any analysis is only as good as the models that go into it. The phenomenal sensitivity, accuracy and precision of \GW observations is not achievable without highly accurate, coherent models for the gravitational waveforms themselves (the focus of this whitepaper), as well as the detector response and noise characteristics.

Under the assumption that the noise is Gaussian, the likelihood that the hypothetical model, parameterized by $\mathbf\theta$, would produce the observed data $\mathbf d$ is
\begin{equation}\label{eq:pedantic_likelihood}
	p(\mathbf d|\mathbf\theta) = \frac{1}{ \sqrt{\det( 2\pi\mathbf{C})}}e^{-\frac{1}{2} {\mathbf r}^\dagger  \mathbf C^{-1}  {\mathbf r} },
\end{equation}
where $\mathbf r = \mathbf d - \sum_i \mathbf h_i$ is a vector of all residual data samples after the discrete \GW signals $\mathbf h_i$ have been subtracted, $\mathbf{C}$ is the noise covariance matrix $C_{ij}\propto \langle n_i | n_j \rangle$, and $\mathbf n$ is noise such that, in the absence of any \GW signals, $\mathbf d=\mathbf n$.  The likelihood is testing whether the residual is consistent with the ansatz that, in the absence of discrete gravitational waves, the data is Gaussian characterized by $\mathbf{C}$.  Note the emphasis on \emph{discrete} gravitational waves -- a stochastic background of \GW signals appears in the data model as a ``noise'' term, which modifies $\mathbf C$ with particular covariances that set it apart from instrument noise. See \cite{Romano:2016dpx} for a detailed look at how the likelihood is derived, simplifications that produce slightly different forms of the equation in the literature, and a unified treatment of discrete and stochastic signals.

Also note that Eq.~(\ref{eq:pedantic_likelihood}) does not prescribe a representation of the data.  The game, as it were, is to represent the data in a way that minimizes the number of non-zero components of $\mathbf C^{-1}$ and thereby minimize the computational cost of evaluating the likelihood. It is the case under the assumption of stationary noise (i.e. $\mathbf{C}$ is constant over the observation time) that the discrete Fourier transform diagonalizes $\mathbf{C}$, which is why much of the \GW analysis literature is based in the Fourier domain. For LISA, due to the long duration signals expected, the assumption of stationary noise will be dubious at best, and so analysis methods may trend towards other representations for the data (e.g., time-frequency methods with short Fourier transforms, discrete wavelet transforms, etc.), but the fundamental likelihood function remains the same.  

The closest analog to the LISA analysis of individual sources is found in the analysis of ground-based interferometer data from the \LVK collaboration, which are heavily reliant on waveform models. 
There are two major differences between \LVK and LISA that limits the applicability of the analogy.  First: At current detector sensitivities, the rate of detectable sources is such that they are still sparsely distributed through the data, with typical signals present in the most sensitive frequency band of the detectors for $O(10\ {\rm s})$ at a rate of $O({\rm a\;few}/{\rm week})$.  LISA, on the other hand, will be signal-dominated, with tens of thousands of continuous Galactic sources overpowering the instrument noise below $O(3\ {\rm mHz})$ and, depending on the rate and mass distribution of massive black hole mergers, several extremely high SNR mergers in band for $O({\rm weeks})$ to $O({\rm months})$, overwhelming any other contributions to the data stream. Thus, whereas the \LVK searches are primarily ``data mining'' endeavors, sifting through a large volume of noisy data for rare and comparatively weak signals, LISA's primary challenges are twofold: Source confusion due to the large number of sources simultaneously detectable; and model accuracy due to both the large number of waveform cycles over which models must stay phase coherent, and to contend with such high SNRs so as to not contaminate lower-amplitude sources with residual power. 

The second key difference between the \LVK collaboration and LISA experiences is the volume of the parameter space itself.  Ground-based searches for compact mergers span a mass range for the components of O(1-100) M$_\odot$.  While challenging, this mass range is small enough that precomputed grids of template waveforms covering the parameter space can be used when searching for candidates (see, e.g., \cite{LIGOScientific:2016vbw,KAGRA:2021vkt}). The LISA parameter space for comparable-mass black hole mergers, for example, is several orders of magnitude larger, spanning $O(10^3-10^8\ {\rm M}_\odot)$, both eliminating the possibility of using grid-based methods and expanding the range of possible mass ratios encountered by the analysis.  To date, the most successful prototype LISA analyses have used stochastic sampling algorithms \cite{Katz:2021uax, Littenberg:2023xpl, Lackeos:2023eub}, still relying on template waveforms but using data-driven methods to concentrate waveform calculations in the high-probability regions of parameter space.  While \LVK analysis of compact mergers is hierarchical, with clear distinction between the ``search" and ``parameter estimation" steps, those two functions blur together for many prototype LISA pipelines. Stochastic sampling algorithms put more pressure on the computational efficiency of waveform calculations, since template generation is part of the analysis pipeline itself, as opposed to pre-computing and then reusing a (large) table of waveforms generated on a fixed grid. 

Note that while 3rd generation ground-based \GW detectors will trend towards higher event rates, signal durations and signal strengths, the LISA forecasted maximum signal strengths are uniquely in excess of signal-to-noise ratios $\sim 10^4$~\cite{Kaiser:2020tlg}. As a result, while the continued improvement of ground-based detectors and analysis methods naturally leads to evolution of the waveform models that directly benefit analysis of LISA data, the waveform development for 3rd generation detectors is necessary but not sufficient to fully achieve LISA's potential.  { LISA puts unique pressure on the accuracy, breadth of parameter space and computational efficiency of waveform models.} 

It is also worth considering that the signal-processing part of the LISA science ground segment is divided between two paradigms: the so called ``low latency'' and ``global'' analyses.  
The global analysis is the joint fit to all \GW signals in the data, and it is here where waveform accuracy is most important, in order to prevent mismatches with loud signals from contaminating weaker signals that are simultaneously present in the data stream.
The global analysis is computationally intensive, and having efficient waveform generation tools is necessary for it to be tractable.  However, the global-fit processing speed is more forgiving than that of the low-latency analyses, since the goal is to produce thorough source catalogs with a relatively relaxed release schedule on a $O({\rm monthly})$ to $O({\rm yearly})$ cadence.

For the low-latency analysis the trade-off is inverted, as computational speed and localization information are prioritized over all else.  The low-latency pipelines will run ${\sim}$daily, and will likely use a subset of the data (e.g.~by bootstrapping based on the most recently completed analysis). The primary goal is to provide actionable information for joint multimessenger observations of transients.  There are other functions of the low-latency pipelines, for example providing source-subtracted residuals to the instrument team for assessment of detector performance etc., but under the purview of waveform generation time is of the essence.  One thing to therefore consider at the architectural level of the waveform generation software is the need for tools that can be responsive to different demands on the speed/accuracy spectrum for processing sources, depending on the primary goal of the analysis.


\subsection{Accuracy requirements}
\label{sec:accuracy_requirements}

Coordinator: Deborah Ferguson and Maarten van de Meent \\
Contributors:  M.~Haney, R.~O'Shaughnessy

Inaccuracies in waveform models affect the analysis of LISA data in three main ways. First, if the modelled waveforms do not sufficiently resemble the signals produced by Nature, this can hamper our ability to detect and identify sources in the data. Second, errors in the model will introduce some level of bias in the estimation of the source parameters, and could potentially masquerade as beyond-\GR effects. Finally, if particularly loud sources are not perfectly subtracted from the data stream, their residuals can contaminate the searches for other sources. Below, we discuss the impact of these effects and how they lead to accuracy requirements for LISA waveform models.

\subsubsection{Detection and identification}

The impact of modeling errors on detection rates are fairly well understood in the case of a matched filter search for a single source in the data \citep{Flanagan:1997kp,Lindblom:2008cm,McWilliams:2010eq}. Suppose we have some waveform model, $h_{\rm model}(\vec{\lambda})$, depending on some set of parameters $\vec{\lambda}$, and we are looking for some true waveform $h_{\rm true}$ in the data, then the \emph{fitting factor}\edit{~\citep{Apostolatos:1995pj}} is defined by
\begin{equation}
\mathcal{F} = \max_{\vec{\lambda}}\frac{
	\nwip{h_{\rm true}}{h_{\rm model}(\vec{\lambda})}
}{
	\sqrt{ \nwip{h_{\rm true}}{h_{\rm true}}\nwip{h_{\rm model}(\vec{\lambda})}{h_{\rm model}(\vec{\lambda})} }
},
\end{equation}
where $\nwip{\cdot}{\cdot}$ is the noise weighted inner product \edit{from \eqref{eq:nwip}.}
 The fitting factor measures the effective loss in \SNR due to using an imperfect model. Consequently, if the used models have a fitting factor $\mathcal{F}$ for a particular source, the maximum range at which such a source can be detected is reduced by a factor $\mathcal{F}$. Whether this has any impact on the detection rate depends on the type of source. Some LISA \MBHB sources are so loud \edit{(in the higher mass seed scenario)} that even the earliest (and therefore furthest) such events would be easily detectable \cite{Barausse:2020mdt}, in which case the fitting factor of the used model has little to no effect on the detection rate.
For other, quieter sources (such as EMRIs, SOBHBs, or GBs) the detection rate is more range limited, and a poor fitting factor $\mathcal{F}$ could lead to a reduction of the number of detections by a factor $\mathcal{F}^3$ (assuming a uniform distribution of the source through a spatially flat universe). An appropriate norm for what degree of loss of sources is deemed admissible for achieving LISA's science goals needs to be established.  An additional consideration here is that increasing the number of unresolved sources could adversely affect the searches for others sources such as any cosmological \GW backgrounds~\cite{Pan:2019uyn}.

The above is valid for an idealized case, where the search is conducted using a continuous bank of templates. In practice, a search would use a discretized template bank, meaning that the effective fitting factor is increased due to the template spacing. Moreover, as noted in Section~\ref{sec:da_intro}, a fully coherent search of the LISA data seems infeasible. Instead, LISA pipelines will most likely employ semi-coherent \cite{Gair:2004iv,Chua:2017ujo} or stochastic \cite{Gair:2008zc,Cornish:2008zd} search strategies. In a semi-coherent search, template and data are both partitioned into short segments, and the template--data overlap for each segment is maximized over several extrinsic parameters. Waveforms used in such a search need only to stay phase-coherent over the shorter segments. For example, it was shown in \cite{Chua:2017ujo} that templates with $>97\%$ overlap accuracy over $\gtrsim10^6$ s will still be sufficient to detect $>50\%$ of EMRIs detectable with fully coherent, precise templates.

\subsubsection{Parameter estimation}
LISA parameter inference nominally involves a joint multi-source fit for all available sources in the data
\cite{Vallisneri:2008ye,MockLISADataChallengeTaskForce:2007iof,Babak:2008aa,MockLISADataChallengeTaskForce:2009wir}.
For the purposes of the discussion below, we approximate this process as independent parameter inference for individual sources, resolving the source from detector noise and the confusion noise of all other signals in the data.
Within that context,  the impact of waveform sytematics on parameter inference has been historically estimated using
well-understood analytic techniques; see, e.g., \cite{Vallisneri:2013rc,Cutler:2007mi}.   The most frequently
used ingredients in analytic waveform standards are the
match or \emph{faithfulness} ${\cal M}$, which is the overlap between a signal $h_{true}$ and template $h_{template}$ maximized over the coalescence
time and phase of the template; the \emph{Fisher matrix} $\left\langle
\partial_a h(\lambda)| \partial_b h(\lambda)\right \rangle$; and the inner product $\left \langle \partial_a h|\delta
h\right \rangle$  between derivatives of the waveform and residuals $\delta h$ between
two signal approximations.  
%
The simplest and most conservative waveform accuracy standards for parameter inference are expressed in terms of limits on the mismatch $1- {\cal  M}$, which nominally must be $\lesssim 1/\mathit{\SNR}^2$ to avoid introducing systematic
bias comparable to the statistical error \emph{for a single source} \cite{Flanagan:1997kp, Lindblom:2008cm, Chatziioannou:2017tdw}.  For a population of sources, in principle systematic biases could stack in hierarchical population inference, and the most conservative threshold would be $1/(N_s \mathit{\SNR}^2)$ where  $N_s$ is the typical number of sources \cite{Wysocki:2018mpo}.
However, \edit{these thresholds are likely to be much too conservative for many applications, see, e.g., \cite{Purrer:2019jcp}.} 


Because the impact of systematic biases depends strongly on the nature and scale of the bias relative to astrophysical features,  general conclusions about systematic bias cannot be drawn, and (barring negligible systematic error) must be assessed for each science goal individually by performing hierarchical population inference.
%
For example, most SOBHBs in quasicircular orbits are in wide, relatively slowly-evolving orbits in the LISA band.  For 90\% of LISA-relevant SOBHBs (in population models consistent with current \LVK observations), 2PN-accurate waveforms are sufficiently faithful for parameter estimation without biases \cite{Mangiagli:2018kpu}.
At the other extreme, MBHBs will have extremely high amplitudes, with minute statistical error~\cite{Lang:2007ge,Babak:2008aa,MockLISADataChallengeTaskForce:2009wir,MockLISADataChallengeTaskForce:2007iof}.
Nominally, the very conservative accuracy thresholds described
above would suggest a mismatch error target of order $1/(N_s \mathit{\SNR}^2) \simeq O(1/(10*(1000)^2)) \simeq 10^{-7}$, for statistical
errors to be small compared to statistical errors for the recovered population of MBHBs from \edit{$O(10)$ detections with $O(1000)$ \SNR as could occur in certain \MBHB formation scenarios \citep{Barausse:2020mdt}}.  This threshold may be
needed for applications that require joint inference on all massive MBHBs (e.g., tests of general relativity), but can
be dramatically relaxed for any astrophysical interpretation of the MBHB population.
Further studies will be necessary to establish specific requirements on waveform accuracy for each context in which parameter estimation is critical.

\subsubsection{Contamination}

Contamination effects occur when the residual from one source cannot be fully removed and impedes the interpretation of
other, subdominant sources in the data.
For such strong sources, systematic errors may be larger than statistical errors~\cite{Cutler:2007mi}.
Though measurement error may have little impact on the astrophysical
interpretation of the strong sources, the potentially substantial
residuals produced by inaccurate models of their gravitational
waves will introduce artifacts which contaminate
downstream data
analysis~\cite{Vallisneri:2008ye,MockLISADataChallengeTaskForce:2007iof,Babak:2008aa,MockLISADataChallengeTaskForce:2009wir,Porter:2014iva}.
For instance, for numerical relativity waveforms, insufficiently
resolved grids can cause significant residuals even for simulations
with precisely the same parameters as the observed
signal~\cite{Ferguson:2020xnm}.  The impact of errors arising from such imperfect modeling of strong sources can
only be properly assessed with full joint hierarchical inference,
using realistic models and contamination targets.
Because of the diversity of sources that could be contaminated by
residuals left by imperfect models for MBHB sources, much remains to
be done to comprehensively assess key questions: How do modelling
residuals from loud sources impact the detection and identification of
other sources?  Which sources are most affected?  And what are the
implications for the required modelling accuracy?

\subsection{Efficiency considerations}
\label{sec:efficiency_requirements}

Coordinator: Mark Hannam and Jonathan Thompson \\
Contributors: A.~Chua and M.~Katz



We shall discuss here the efficiency requirements imposed by search and inference for the two main classes of strong-field source that will be observable by LISA: MBHs and EMRIs. Relatively weaker-field sources such as GBs and SOBHs are a lesser concern; modeling of their signals (to the accuracy required for data analysis) is significantly easier, and computational cost is dominated by the need to attain a sampling resolution that is adequate for the typical duration, bandwidth and abundance of each source type. Thus the efficiency requirements for these sources may be addressed on the analysis end with suitable approximations, or at the modeling--analysis interface where techniques such as those in Section~\ref{sec:waveform_accel} may be applied.

\subsubsection{Extreme-mass-ratio inspirals}\label{sec:EMRI efficiency requirements}

EMRIs are the only LISA sources that combine the issue of strong-field complexity with that of long-lived signals \edit{(potentially staying in band for years)}, and thus they pose great challenges for both waveform modeling and data analysis. The main problem in EMRI search is \emph{information volume}: the space of LISA-observable EMRIs is gargantuan, requiring $10^{30}$--$10^{40}$ templates to cover in a naive grid-based search \cite{Gair:2004iv}.\footnote{As a point of reference, banks of at most $10^5$--$10^8$ templates are needed to detect \LVK BBH mergers \cite{Moore:2019pke}, \edit{and searches are often conducted with smaller banks of around $10^4$ templates in practice \cite{Sakon:2022ibh}.}} This is unfeasible regardless of waveform efficiency. One possible solution relies on semi-coherent filtering, which essentially returns less informative templates that cover larger regions in parameter space, allowing extensive searches to be performed in a viable amount of time. Assuming that computing performance in the 2030s lies around the $10^2$-teraflop regime (achievable at present with $\sim10^3$-node clusters), semi-coherent segments of $\lesssim10^6$ s would enable the analysis to be completed over the mission lifetime \cite{Gair:2004iv}.

In practice, EMRI search will probably use semi-coherent filtering within stochastic algorithms, rather than template banks. Thus the estimates in \cite{Gair:2004iv} are very conservative, as stochastic searches are generally far more efficient than grid-based searches at the same number of template evaluations, even with a large multiplier (e.g., repeated runs) to ensure proper coverage of the parameter space. Note, however, that a potentially large number of additional templates will be required in the assessment of detection significance for each candidate source, although it remains unclear whether such an assessment will eventually be performed in the search or inference stage.

\edit{The efficiency requirements for inference are more straightforward, in that LISA analysis algorithms will be built on similar tools to those used in \LVK parameter estimation, like Markov Chain Monte Carlo (MCMC)~\cite[e.g.]{Littenberg:2023xpl, Strub:2024kbe, Katz:2024oqg}. An estimated $10^6$--$10^9$ templates will be required per source posterior distribution. If we want to produce a posterior in less than 10 days we need waveforms that can be produced in $\lesssim 1 \mathrm{ms}$. LISA will detect $\sim10^4$ sources, so parallelization resources are inherently considered in our scaling estimate considering each individual source posterior as a single parallelized process taking $\lesssim10$ days. The estimated number of template evaluations per posterior also assumes that the prior regions for posterior sampling can be sufficiently localized to begin with; this is certainly possible in principle, but may require multiple search stages beforehand in a hierarchical approach.}

\subsubsection{Massive black hole sources}

Generally, MBH sources are expected to be detectable for $\sim$week to hours prior to the merger. Due to this detection time frame and large signal-to-noise ratio of MBH binaries, coherent analysis over the full waveform template will most likely be employed for both search and parameter estimation. Therefore, from a waveform production perspective, these two analyses are roughly the same even if search and parameter estimation employ different schemes for achieving their various goals \cite{Cornish:2020vtw}. One way a search algorithm may differ from a parameter estimation algorithm for MBHs, in terms of the waveform generation, is in the use of higher-order harmonics. Recent work \cite{Katz:2020hku,Marsat:2020rtl} has shown the importance of higher-order modes in parameter estimation for LISA. However, as detailed in \cite{Cornish:2020vtw}, searching over the dominant harmonic may be acceptable for initially and roughly locating sources throughout the high-dimensional parameter space. With this said, this is a matter of adding or removing modes (that we assume are available in a model), not altering the fundamental waveform generation method.  

Given this idea that the waveform generation is similar for both search and parameter estimation for MBHs, the following discusses how efficiency requirements relate to the stages of waveform creation. As discussed in Sec.\ \ref{sec:comp_techniques}, there are generally two main parts to waveform creation in the context of LISA: sparse, accurate waveform calculations and a scaling method to achieve a full waveform from the sparse information. MBH waveforms can be generated beginning with sparse calculations of the amplitude and phase for each harmonic mode with an accurate waveform model. Depending on the specific analysis type, these sparse calculations are then upsampled to the desired search or parameter estimation settings. Producing accurate and upsampled waveforms \textit{directly} from the accurate waveform generator is unnecessary and time-consuming: various methods allow for the upsampling of sparse, smooth functions in an accurate and much more efficient manner \edit{(e.g. see \cite{Garcia-Quiros:2020qlt,Katz:2021yft})}. This construction leads to two different waveform generation efficiency requirements. The first is the overall waveform (including upsampling). With similar requirements to EMRIs given in the section above, we expect to collect at least $\sim10^{6}$ MCMC samples for a given source posterior distribution. To accomplish this in $\sim10$ days, we need to generate waveforms at a rate of $\sim/1$s on a single \CPU core.

The other efficiency requirement deals with the separation of the accurate waveform generation from the scaling operation. Generally, the scaling operation is the bottleneck for LISA where waveforms can have up to $\sim10^6$--$10^8$ data points. Therefore, the requirement on the scaling part is that a waveform on a single \CPU core be scaled in $\sim1$ second or less. This condition, therefore, also sets the requirement for the accurate waveform portion assuming its sparsity prevents it from becoming the bottleneck. This requirement is that the sparsely sampled, accurate waveform must be of a similar or lower order of magnitude in timing when compared with the scaling operation.


\subsection{Interface and data format requirements}
\label{sec:interface_and_data_formats}
Contributors: Tyson Littenberg

An additional key consideration for optimally supporting LISA analyses is the need for a flexible interface between waveform generation software and analysis pipelines. It is important for verification and validation of pipelines, and as a means for cross-checking results, to have independently developed algorithms targeting the same sources. (There is also the added benefit of a constructive competition between development teams.) As per the previous discussion around Eq.~(\ref{eq:pedantic_likelihood}), different analysis pipelines will likely be built around different representations of the data.  Template generation is generally the computational bottleneck, and so it needs to be optimized for the application. At the same time, the benefits of having independent pipelines become liabilities if the analyses are not interfacing to the same waveform generation tools.  As a result, it is of paramount importance that the waveform and pipeline development teams are collaborating early and often to avoid unnecessary or redundant transformations of the waveforms by the analysis pipelines (consider, e.g., a waveform that is initially computed in the time domain but then output in the Fourier domain, being called by an analysis pipeline that uses a discrete wavelet domain representation of the data). 

The demand for flexibility of the waveform-analysis interface affects more than just the choice of a basis set used to represent the template. There have been promising developments in low-cost likelihood evaluations that use the instantaneous amplitude and phase of the template waveform, sampled on an adaptive grid, to concentrate computations to regions where the signal is changing most rapidly~\cite{Cornish:2020vtw,Marsat:2020rtl}. Such considerations are difficult to retroactively incorporate in established template generation algorithms, but present opportunities for increased efficiency if they are part of the original waveform algorithm designs.

\newpage

\section{Modelling approaches for compact binaries}
\label{sec:modelling_binaries}

Detecting and inferring the parameters of compact binaries systems requires waveform template which, in turn, necessitates solving the relativistic two-body problem.
Unlike in the Newtonian counterpart, it is not possible to only solve for the motion of the two bodies; one must solve for the dynamical evolution of the full spacetime.
In general there are no known closed-form solutions to the nonlinear Einstein field equations for radiating binaries and so a variety of techniques have been developed to compute solutions either numerically or via perturbative expansions.
The three main approaches that directly solve the Einstein field equations are numerical relativity (NR), post-Newtonian/Minkowskian (PN/PM) theory, and gravitational self-force (GSF).

\begin{figure}[t]
    \centering
    \includegraphics{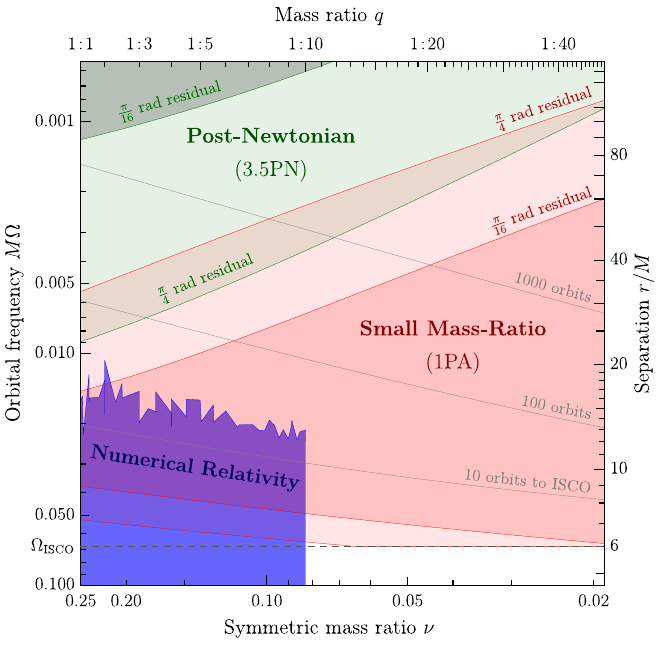}
    \caption{
	 Region of applicability of different approximation techniques for non-spinning quasi-circular binary \BH inspiral.
	 The 1PA region is a prediction derived from fitting NR data to an expansion of the form~\eqref{PA phase}.
	  The shaded regions indicate ranges within which the cumulative orbital phase-error is less than $\pi/4$ and $\pi/16$ radians, respectively.
	  Recent direct calculations of post-adiabatic (1PA) waveforms suggest the 1PA region shown is over-optimistic for $\nu > 0.1$, but they have borne out the prediction for $\nu < 0.1$ \cite{Albertini:2022rfe}.
	  The gray lines show the location of binaries with 10 (resp. 100 and 1000) orbits left before they reach the \ISCO.
	  This figure is reproduced from Ref.~\cite{vandeMeent:2020xgc}.}
    \label{fig:parameter_space}
\end{figure}

Each of these approaches has strengths and weaknesses that leads them to being best employed in different regions of the binary configuration parameter space.
Numerical Relativity directly solves the Einstein field equations to produce exact solutions up to numerical error.
Post-Newtonian theory analytically computes relativistic corrections to the binary's motion and \GW emission as a series expansion in powers of the orbital velocity as a fraction of the speed of light.
The closely related post-Minkowskian approach expands field equations around flat Minkowski spacetime in powers of the gravitational constant without any restriction of the velocity of the binary. Gravitational self-force expands the Einstein field equations in powers of the (small) mass ratio.
Fig.~\ref{fig:parameter_space} gives a quantitive description of the strengths and weakness of each approach in the orbital separation -- mass-ratio parameter space for non-spinning quasi-circular binaries.

In addition to the above approaches there are also effective frameworks that attempt to cover large portions of the parameter space.
The physically motivated Effective-One-Body model takes inspiration from the solution to the Newtonian problem and describes the binary as the motion of a test body in spacetime of a deformed single black hole.
Much progress can be made analytically with this approach by absorbing post-Newtonian and post-Minkowskian corrections, and further calibration can be applied using numerical results from the NR and GSF methods.
The Phenomenological (Phenom) waveform models do not attempt to solve the relativistic two-body problem.
Instead they aim to directly model the waveforms by building upon fast post-Newtonian models with further calibration from NR and GSF.
The EOB and Phenom models are heavily used in analysis of \GW data from current ground-based detectors.

In this section we outline the status of the above approaches and discuss the required development needed to reach the accuracy requirements outlined in Sec.~\ref{sec:accuracy_requirements}.
Concerns about the speed of waveform generation are addressed in Sec.~\ref{sec:waveform_accel}.
The status and requires for LISA of these approaches can be found in Sec.~\ref{sec:NR} for numerical relativity, Sec.~\ref{sec:weak_field} for weak-field post-Newtonian and post-Minkowskian expansions, Sec.~\ref{sec:GSF} GSF, Sec.~\ref{sec:EOB} for EOB, and Sec.~\ref{sec:phenom} for Phenomenological models.
These sections focus on modelling within \GR; see Sec.~\ref{sec:modling_beyond_GR} for a discussion on modelling in alternate theories of gravity.

\subsection{Numerical relativity}
\label{sec:NR}

Coordinators: Mekhi Dhesi, Deborah Ferguson and Deirdre Shoemaker \\
Contributors: Sebastiano Bernuzzi, Gabriele Bozzola,
Katy Clough,
Deborah Ferguson,
William Gabella,
Miguel Gracia,
Roland Haas,
Mark Hannam,
Eliu Huerta,
Sascha Husa,
Larry Kidder,
Pablo Laguna,
Carlos Lousto,
Geoffrey Lovelace,
David Neilsen,
Vasileios Paschalidis,
Harald Pfeiffer,
Geraint Pratten,
Hannes R\"uter,
Milton Ruiz,
Stuart Shapiro,
Jonathan Thompson,
Antonios Tsokaros,
Miguel Zilhao

\subsubsection{Description}

Numerical relativity (\NR) solves Einstein's field equations through
direct numerical integration on supercomputers (see e.g.\ the
books~\cite{AlcubierreBook2008, BonaPalenzuelaBona2009,
baumgarteShapiroBook, GourgoulhonBook2012, ShibataBook2015}), providing
the spacetime geometry and dynamics of the system in addition to the
gravitational radiation emitted.  Such solutions of the full nonlinear
field equations without approximations are critical to our
understanding of highly dynamic regimes without symmetries, such as
the late inspiral and merger of compact binaries.

The first steps at numerical solutions of Einstein's equations date back several decades~\cite{Hahn-Lindquist:1964,Smarr:1976qy} and tremendous
progress since then enabled the first simulations of inspiraling and merging binary black
holes~\cite{Pretorius:2005gq,Campanelli:2005dd,Baker:2005vv}.
These breakthroughs initiated an explosive growth of the field, as illustrated, e.g.\ in the reviews~\cite{Lehner:2001wq,Centrella:2010mx,Duez:2018jaf,Foucart:2022iwu}.
By now, NR has become a critical part of the modeling the
gravitational waves of the late inspiral and merger phase of
coalescing binaries (e.g.\ \cite{Mroue:2013xna,Husa:2015iqa,Boyle:2019kee,Hamilton:2023qkv}), which in turn underpins the analyses of all
observed gravitational waves from coalescing binaries.


In order to solve Einstein's field equations, modern NR codes
generally use the 3+1
approach~\cite{Arnowitt-Deser-Misner:1962,York:1979} in which the
four-dimensional spacetime is sliced into three-dimensional hypersurfaces, and Einstein's equations are reformulated as a
Cauchy problem with constraints.
Cauchy problem with constraints.
%
The first stage in numerically solving a compact binary inspiral is
the construction of initial data, for which Einstein's constraint equations are
reformulated as elliptic equations either in the context of the
conformal-thin-sandwich formulation~\cite{York1998,Pfeiffer:2002iy}
or the puncture approach~\cite{Brandt:1997tf}.
The resulting coupled nonlinear partial
differential equations are solved with custom-purpose elliptic solvers~\cite{Pfeiffer:2002wt,Ansorg:2004ds,2016ascl.soft08018G}.  Numerous improvements over the years increased the generality of the physical conditions that can be achieved and the numerical quality of the solution, e.g.~\cite{Lovelace:2008tw,Foucart:2008qt,Grandclement:2009ju,Ruchlin:2014zva,Ossokine:2015yla}.  Moreover, entirely new
codes were developed~\cite{Dietrich:2015pxa,
Rashti:2021ihv,
Assumpcao:2021fhq,
Vu:2021coj,
Papenfort:2021hod
}.
%
The time evolution is encoded in a set of coupled, hyperbolic partial differential equations with suitable gauge conditions and boundary conditions,
where three main approaches
have emerged:
the generalized-harmonic formulation~\cite{Pretorius:2004jg,Lindblom:2005qh,Szilagyi:2009qz},
the Baumgarte-Shapiro-Shibata-Nakamura formulation~\cite{Shibata:1995we,Baumgarte:1998te}
and different versions of the Z4 formulation of Einstein's equations~\cite{Bona:2003fj,Bernuzzi:2009ex,Ruiz:2010qj,Weyhausen:2011cg,Hilditch:2012fp,Alic:2011gg,Alic:2013xsa} (see Table~\ref{tab:NRcodes}).
%
Gravitational radiation is typically computed as gravitational strain~\cite{Reisswig:2010di} and must be propagated to future null infinity
(see review~\cite{Bishop:2016lgv}), which is often accomplished by
extrapolation of
\GW modes extracted at finite radius~\cite{Boyle:2009vi}.  Alternatively,
Cauchy-Characteristic
Extraction~\cite{Bishop:1996gt,Handmer:2014qha,Moxon:2021gbv} directly
yields the waveforms at infinity, with improved recovery of
non-oscillatory \GW modes and \GW memory effects~\cite{Mitman:2020pbt}.  Gauge conditions and transformations at future null infinity must be treated with care~\cite{Lehner:2007ip} to yield well-behaved numerical waveforms with
well-defined waveform modes~\cite{Mitman:2022kwt}.
Cauchy codes have difficulty resolving the memory effect~\cite{Boyle:2019kee,Mitman:2020pbt}.
Instead, most approaches to computing the memory effect use Cauchy NR waveforms (or NR calibrated waveform models) and the asymptotic Einstein equations to determine the unresolved memory effect in the Cauchy simulations that is required to satisfy Einstein's equations~\cite{Favata:2009ii,Talbot:2018sgr,Boersma:2020gxx,Khera:2020mcz, Mitman:2020bjf, Liu:2021zys}. This approximate approach agrees with the Cauchy-Characteristic extracted waveforms, though comparisons over a wider range of the BBH parameter space would be useful.

Once BBH simulations became possible, the
\NR community quickly achieved many firsts, including the first simulations of unequal-mass \BBH coalescences~\cite{Baker:2006vn,Herrmann:2006ks}, the first with spinning binaries~\cite{Campanelli:2006uy}, the first targeted eccentricity mergers~\cite{Hinder:2008kv} and first comparisons with \PN calculations ~\cite{Baker:2006ha,Buonanno:2006ui,Hannam:2007ik,Boyle:2007ft}.  \NR calculations of the merger revealed features of the non-linear regime, among them
that \BHs with spins positively aligned with the orbital angular momentum merge at higher frequency~\cite{Campanelli:2006uy}, spin hang-up configurations~\cite{Lousto:2011kp}, as well as the calculation of the recoil velocity imparted on the remnant \BH~\cite{Baker:2006vn,Gonzalez:2006md,Herrmann:2006ks};  in particular it was found that \BH spins oriented approximately parallel to the orbital plane can lead to recoil velocities of several 1000's km/s \cite{Campanelli:2007ew,Gonzalez:2007hi,Campanelli:2007cga}.

Analysis of \GW observations requires waveforms that cover the entire frequency band of the relevant detector at high accuracy. This has motivated
large efforts to improve the accuracy of NR simulations \cite{Boyle:2007ft,Husa:2007hp,Szilagyi:2009qz,Scheel:2014ina,Rosato:2021jsq,Etienne:2024ncu}, length of the inspiral~\cite{Mroue:2013xna,Szilagyi:2015rwa}, and  coverage of increasingly large portions of parameter space in increasing detail.  Parameter space exploration was initially performed through community-wide Numerical INJection Analysis (NINJA)~\cite{Aylott:2009tn,Ajith:2012az} and Numerical-Relativity-Analytical-Relativity (NRAR)~\cite{Hinder:2013oqa} collaborations, which also yielded important cross-checks between different codes~\cite{Hannam:2009hh,Hinder:2013oqa}.  Newer and more extensive parameter space surveys are listed in Table~\ref{tab:NRcodes}.  Waveform models
developed with \NR information are described in Sec.~\ref{sec:EOB} and~\ref{sec:phenom}.  NR surrogate models~\cite{Blackman:2017pcm,Varma:2019csw} are directly built on NR simulations, and are particularly important for the analysis of high-mass BBH systems like GW190521~\cite{LIGOScientific:2020iuh,LIGOScientific:2020ufj}.
NR simulations can also be directly used for \GW data-analysis, where they serve as synthetic signals~\cite{Schmidt:2017btt} for quantifying the response of \GW search and parameter-estimation pipelines~\cite{LIGOScientific:2014oec,LIGOScientific:2016ebw}, and to conduct indirect analyses of observations~\cite{LIGOScientific:2016kms,Lange:2017wki,Gayathri:2020coq}.


\NR also offers important information for the ringdown phase,  characterized by an exponential decay as the remnant \BH settles into a Kerr black hole.  \NR results determined the remnant parameters (mass, spin, recoil velocity)  \cite{Barausse:2009uz,Healy:2014yta,Hofmann:2016yih,Healy:2016lce,Healy:2018swt,Varma:2018aht,Varma:2019csw,Jimenez-Forteza:2016oae}.  Detailed NR calculations of the emitted gravitational waves during merger and ringdown yield the initial amplitudes and phases of quasi-normal ringdown modes and underpin theoretical studies of what information about nonlinear processes are accessible through the emitted gravitational waves~\cite{Mitman:2022qdl,Cheung:2022rbm}, and
how the ringdown phase can be used to test GR and to probe the no-hair and area theorems~\cite{LIGOScientific:2020tif,LIGOScientific:2021sio,Ghosh:2021mrv,Islam:2021pbd,Capano:2021etf,Finch:2022ynt,Isi:2020tac,Cotesta:2022pci}.
%
%

Due to our focus on the role of \NR in the LISA mission, this section primarily covers vacuum spacetimes  with a short discussion on environmental effects that include matter; the important work of \NR with neutron stars is not included.

\subsubsection{Suitable for what sources?}
\label{NR:sources}

The coalescence of   \BBHs is a primary source for LISA for a vast range of BH mass. Those \BBH systems
that merge in the LISA band require NR to produce
the waveforms during the late inspiral and merger.  In order to quantify the mass ranges that merge in the LISA band, we compute the masses for which  the frequency of ringdown and inspiral are both in the LISA band.
The dominant quasi-normal mode for a
non-spinning black hole has a frequency of $f=12.07 \mathrm{mHz}
\left(M/10^6M_\odot\right)^{-1}$ \cite{Berti:2005ys}, which will be
within the LISA band for BH masses of roughly $10^4 -10^8M_\odot$.  Turning to the inspiral, NR simulations of massive BBHs
typically cover a few tens of orbits before merger, starting at \GW
frequencies of $f\sim 1\mathrm{mHz} \left(M/10^6M_\odot\right)^{-1}$, placing this part of the inspiral into LISA band for $10^3 - 10^7
M_\odot$.  This estimate implies that  \BBHs of mass $10^4-10^8 M_\odot$  will have at least part of its waveform within
coverage of NR.  NR is capable of modeling these binaries for comparable mass ratios of less than 1:20, with attempts to reach  1:1000~\cite{Lousto:2022hoq}.

NR simulations become increasingly costly with increasing mass-ratio
and with very large spins, whereas there is only minor dependence of
computational cost on other BBH parameters like spin-direction and
orbital eccentricity.  Parameter space coverage is improving
over time.  The next section describes what parameters
NR currently covers and in Sec.~\ref{NR:challenges} we discuss where efforts are
needed to achieve complete coverage of the potential range of source
parameters.
\subsubsection{Status of the Field}\label{NR:status}
Since the breakthrough in \NR in 2005/2006~\cite{Pretorius:2005gq,Campanelli:2005dd,Baker:2005vv}
and the subsequent ``gold-rush'' to explore nonlinear phenomena
in black hole and neutron star mergers,
\NR has now matured into a reliable tool for accurately computing gravitational waveforms
needed to characterize \GW data.
In this section we briefly describe the landscape of current and future numerical
relativity codes, their cross-validation and the parameter space coverage of currently
available numerical waveforms.

The community has developed a number of successful \NR codes
that target a variety of goals for \BBH spacetimes including
covering increasing fractions of the parameter space
(so far mostly comparable mass ratios and moderate spins),
increasing number of \GW cycles, accuracy of waveforms and
interactions with matter.
Table~\ref{tab:NRcodes} provides a comprehensive list of currently available \NR codes,
some of which are briefly described in~Appendix \ref{app:NRcodes}.
Their capabilities have been sufficient for the detection and characterization of
\GW signals with current ground-based \GW detectors.
However, data collected with future space- or next generation ground-based instruments
will have a higher \SNR and its interpretation requires
much improved \NR waveforms.
\begin{table}[ht]
\centering
\renewcommand{\arraystretch}{1.05} 
\begin{tabular}{p{0.35\textwidth} c c c c c}
	\hline
  Code & Open & Public & Formulation & Hydro & Beyond \\
  & Source & catalog & & & GR \\
\hline

\texttt{AMSS-NCKU}  {\footnotesize \cite{Cao:2008wn, Cao:2013osa, Cao:2011fu, Hilditch:2012fp}} & Y & -- & BSSN/Z4c & -- & Y \\

\texttt{BAM}  {\footnotesize \cite{Bruegmann:2006ulg,Husa:2007hp,Thierfelder:2011yi,Dietrich:2015iva,Dietrich:2018phi,Gonzalez:2022mgo,Hamilton:2023qkv}} & -- & -- & BSSN/Z4c & Y & -- \\

\texttt{BAMPS}  {\footnotesize \cite{Bugner:2015gqa,Hilditch:2015aba,Renkhoff:2023nfw}} & -- & -- & GHG & Y & -- \\

\texttt{Dendro-GR}  {\footnotesize \cite{Fernando:2018mov,Fernando:2022php}} & Y & -- & BSSN/CCZ4 & -- & Y \\

\texttt{Einstein Toolkit}  {\footnotesize \cite{Loffler:2011ay,einsteintoolkit}} & Y & -- & BSSN/Z4c & Y & Y \\

$^{\ast}$\texttt{Canuda}  {\footnotesize \cite{Okawa:2014nda,Zilhao:2015tya,Witek:2018dmd}} & Y & -- & BSSN & -- & Y \\

$^{\ast}$\texttt{IllinoisGRMHD}  {\footnotesize \cite{Etienne:2015cea}} & Y & -- & BSSN & Y & -- \\

$^{\ast}$\texttt{LazEv}  {\footnotesize \cite{Campanelli:2005dd,Zlochower:2005bj,Healy:2017psd,Healy:2019jyf,Healy:2020vre,Healy:2022wdn}} & -- & & BSSN/CCZ4 & -- & -- \\

$^{\ast}$\texttt{Lean}  {\footnotesize \cite{Sperhake:2006cy,Berti:2013gfa}} & Partially & -- & BSSN & -- & Y \\

$^{\ast}$\texttt{MAYA}  {\footnotesize \cite{Jani:2016wkt}} & -- & & BSSN & -- & Y \\

$^{\ast}$\texttt{NRPy+}  {\footnotesize \cite{Ruchlin:2017com}} & Y & -- & BSSN & Y & -- \\

$^{\ast}$\texttt{SphericalNR}  {\footnotesize \cite{Mewes:2018szi,Mewes:2020vic}} & -- & -- & spherical BSSN & Y & -- \\

$^{\ast}$\texttt{Spritz}  {\footnotesize \cite{Cipolletta:2019geh,Spritzcode}} & Y & --& BSSN & Y & -- \\

$^{\ast}$\texttt{THC}  {\footnotesize \cite{Radice:2012cu, Radice:2013hxh, Radice:2013xpa,Dietrich:2018phi}} & Y & & BSSN/Z4c & Y & -- \\

$^{\ast}$\texttt{WhiskyMHD}  {\footnotesize \cite{Giacomazzo:2007ti,WhiskyWFs}} & -- & & BSSN & Y & -- \\

\texttt{ExaHyPE}  {\footnotesize \cite{Koppel:2017kiz}} & Y & -- & CCZ4 & Y & -- \\

\texttt{FIL}  {\footnotesize \cite{Most:2019kfe}} & -- & -- & BSSN/Z4c/CCZ4 & Y & -- \\

\texttt{GR-Athena++}  {\footnotesize \cite{Daszuta:2021ecf}} & Y & -- & Z4c & Y & -- \\

\texttt{GRChombo}  {\footnotesize \cite{Clough:2015sqa,GRChombo-website,Andrade:2021rbd}} & Y & -- & BSSN/CCZ4 & -- & Y \\

\texttt{HAD}  {\footnotesize \cite{Had,Liebling:2002qp,Lehner:2005vc}} & -- & -- & CCZ4 & Y & Y \\

\texttt{Illinois GRMHD}  {\footnotesize \cite{Etienne:2010ui,Sun:2022vri}} & -- & -- & BSSN & Y & -- \\

\texttt{MANGA/NRPy+}  {\footnotesize \cite{Chang:2020ktl}} & Partially & -- & BSSN & Y & -- \\

\texttt{BH@H/NRPy+}  {\footnotesize \cite{Ruchlin:2017com,BHcode}} & -- & -- & BSSN & -- & -- \\

\texttt{MHDuet}  {\footnotesize \cite{Palenzuela:2018sly,Liebling:2020jlq,MHDuetWeb}} & Y & -- & CCZ4 & Y & Y \\

\texttt{Nmesh}  {\footnotesize \cite{Tichy:2022hpa}} & -- & -- & GHG & Y & -- \\

\texttt{SACRA}  {\footnotesize \cite{Yamamoto:2008js,Kiuchi:2017pte,Kuan:2023trn,Shibata:2022gec,Shibata:2010wz,SACRA:catalog}} & -- & & BSSN/Z4c & Y & Y \\

\texttt{SACRA-SFS2D}  {\footnotesize \cite{Fujibayashi:2017puw,Shibata:2021xmo}} & -- & -- & BSSN/Z4c & Y & -- \\

\texttt{SGRID}  {\footnotesize \cite{Tichy:2006qn, Tichy:2009zr}} & Y & -- & BSSN & -- & -- \\

\texttt{SpEC}  {\footnotesize \cite{SpECsite,Boyle:2019kee,Mroue:2013xna,SXS:catalog}} & -- & & GHG & Y & Y \\

\texttt{SpECTRE}  {\footnotesize \cite{Kidder:2016hev,spectrecode}} & Y & -- & GHG & Y & -- \\

\texttt{SPHINCS}\_BSSN  {\footnotesize \cite{Rosswog:2020kwm}} & & -- & BSSN & SPH & -- \\
\hline
\end{tabular}
\caption{\label{tab:NRcodes}
List of numerical relativity evolution codes.
We indicate if a code is open-source, if it has been used to produce public gravitational waveform catalogs, the formulation of Einstein's equation used
(GHG: generalized harmonic,
BSSN: Baumgarte-Shapiro-Shibata-Nakamura,
CCZ4 / Z4c variants of the Z4 formulation,
GCFE: generalised conformal field equations),
if a code implements general relativistic hydrodynamics, and if it is capable of simulating compact objects beyond general relativity. SPH refers to smooth particle hydrodynamics. An asterisk indicates codes that are either (partially) based on the open-source Einstein Toolkit or are co-funded by its grant. Note this table was created jointly for this paper and Ref.~\cite{Foucart:2022iwu}.
}
\end{table}

In particular, LISA sources present a formidable challenge to \NR
(see Sec.~\ref{NR:challenges}),
both with regard to the characteristic features of the target \BBH systems
(e.g., high spins, large mass ratios, eccentric orbits, etc.)
and the quality of the \GW data due to, e.g., high signal-to-noise ratios and sources remaining in band for very long times.
Computational simulations that meet the accuracy requirements for these demanding
\BBH configurations, for sufficiently long times,
will require exascale computational resources.
It will also require continued research in \NR to develop new algorithms,
computational techniques and software to tackle these challenges faced by existing code
bases.
Some new codes are under development, and they are included in Table~\ref{tab:NRcodes}.

Given the complexity of \NR codes, cross-validation is an important component of code
verification and it is by no means trivial that results obtained with different code bases
agree.
They employ, for example,
different theoretical formulations of Einstein's equations for the evolution and initial
data and different methods for the \GW extraction.
They also employ different numerical techniques
(e.g, pseudo-spectral or high-order finite differences discretization)
and numerical implementations, e.g., for the grid structure.
First systematic code comparisons
and studies on the integration of \NR with data analysis and analytical approximations
have been conducted in Refs.~\cite{Baker:2007fb,Aylott:2009tn,LIGOScientific:2014oec,Hannam:2009hh,Hinder:2013oqa,Lovelace:2016uwp}.
For example,
the NINJA project~\cite{Aylott:2009tn,LIGOScientific:2014oec}
focused on building a \NR{} data analysis framework  and to develop first injection studies.
The initial study presented a sample of $23$ \NR waveforms of \BBH{s}
with moderate mass ratios $q\lesssim4$, considering only the dominant the $\ell=|m|=2$ \GW
mode.
The follow-up study increased to statistics to $60$ \NR waveforms.
While no cross-validation between the \NR waveforms were performed, NINJA was crucial to
identify technical and conceptual issues, that a hybridization with \PN or \EOB may be
needed and that more than only the dominant \GW mode may be necessary.
The Samurai project~\cite{Hannam:2009hh}
conducted a detailed cross-validation of gravitational waveforms obtained with five different \NR
codes. It focused on one \BBH system, namely an equal-mass, non-spinning, quasi-circular
(eccentricity $\lesssim 0.0016$)
binary completing about six orbits before their merger.
Focusing on the dominant $\ell=|m|=2$ multipole, it was found that the waveforms' amplitude
and phase agree within numerical error.
It was also found that these \NR waveforms were indistinguishable for \SNR$\leq14$ and
would yield mismatches of $\lesssim 10^{-3}$ for binaries with $M\sim60M_{\odot}$
(using the anticipated detector noise curves at the time).
The Numerical-Relativity--Analytical-Relativity (NRAR) collaboration~\cite{Hinder:2013oqa}
made important strides towards constructing more accurate inspiral-merger-ringdown
waveforms by combining analytical and \NR computations, and by defining new
accuracy standards that would be needed for \GW parameter estimation.
The collaboration pushed cross-validation of \NR codes to new frontiers by considering
quasi-circular \BBH{s} (with initial eccentricity $\leq 2\cdot 10^{-3}$)
with both unequal mass (up to $q\leq3$) and moderately spinning \BH{s} (up to
$\chi_{i}\sim\pm0.6$)
completing about $20$ \GW cycles before merger.
The study also included a binary of non-spinning \BH{s} with $q=10$, the highest mass ratio
for quasi-circular binaries at the time.
For the first time, the seven involved \NR codes performed $22$ targeted simulations to
meet the required accuracy standards. They employed the same analysis code to estimate the
uncertainties due to the numerical resolution and waveform extractions.
The simulations exhibited a relative amplitude error in the $\ell=|m|=2$ multipole of
$\lesssim 1\%$ and a cumulative phase error of $\lesssim0.25$rad.
Finally, Ref.~\cite{Lovelace:2016uwp} presented a comparison of targeted simulations for
the first \GW event, GW150914, concluding that the waveforms were sufficiently accurate
to effectively analyse LIGO data for comparable events.
A comparison between different time evolution formulations of Einstein's equations,
namely BSSN and Z4c, was performed in Refs.~\cite{Zlochower:2012fk,Hilditch:2012fp}.

The cross-validation studies between different \NR codes have been crucial
for building confidence in their results.
As summarized above, they have been conducted in limited regions
of parameter space (e.g., low mass ratios, some moderate spins, low eccentricity).
As the demand on \NR waveforms increases,
such as
covering a larger region of parameter space,
including eccentric or spin-precessing \BBH{s},
including higher multipoles (typically up to $\ell=8$) that are excited by asymmetric
systems
and further improving their accuracy,
extended validation studies become important.
With the technical and computational developments,
the numerical data becomes more and more sensitive to
mismatches, e.g., in the initial data, evolution or wave extraction.
Therefore, a careful study of the initial data and
initial parameters, of wave extraction techniques
(e.g., extrapolation of waveforms extracted at finite radii vs. Cauchy characteristic
extraction vs.~Cauchy characteristic matching) or the choice of asymptotic Bondi-van der Burg-Metzner-Sachs frame
will be needed~\cite{Mitman:2021xkq}.
The starting point for any simulation is the construction of constraint satisfying
initial data. After decades of work (see, e.g., Refs.~\cite{Cook:1994va, Brandt:1997tf,
Cook:2000vr, Ansorg:2004ds, Pfeiffer:2005zm, Gourgoulhon:2007tn, Lovelace:2008hd,Lovelace:2008tw,Ruchlin:2014zva,Tichy:2016vmv} and references therein),
multiple codes are now capable of generating \BBH initial
data where the BHs are on a quasi-circular orbit or where the orbit has non-zero eccentricity.

\begin{figure}[t]
    \centering
   \includegraphics[width=0.95\columnwidth,trim=30 20 200 20]{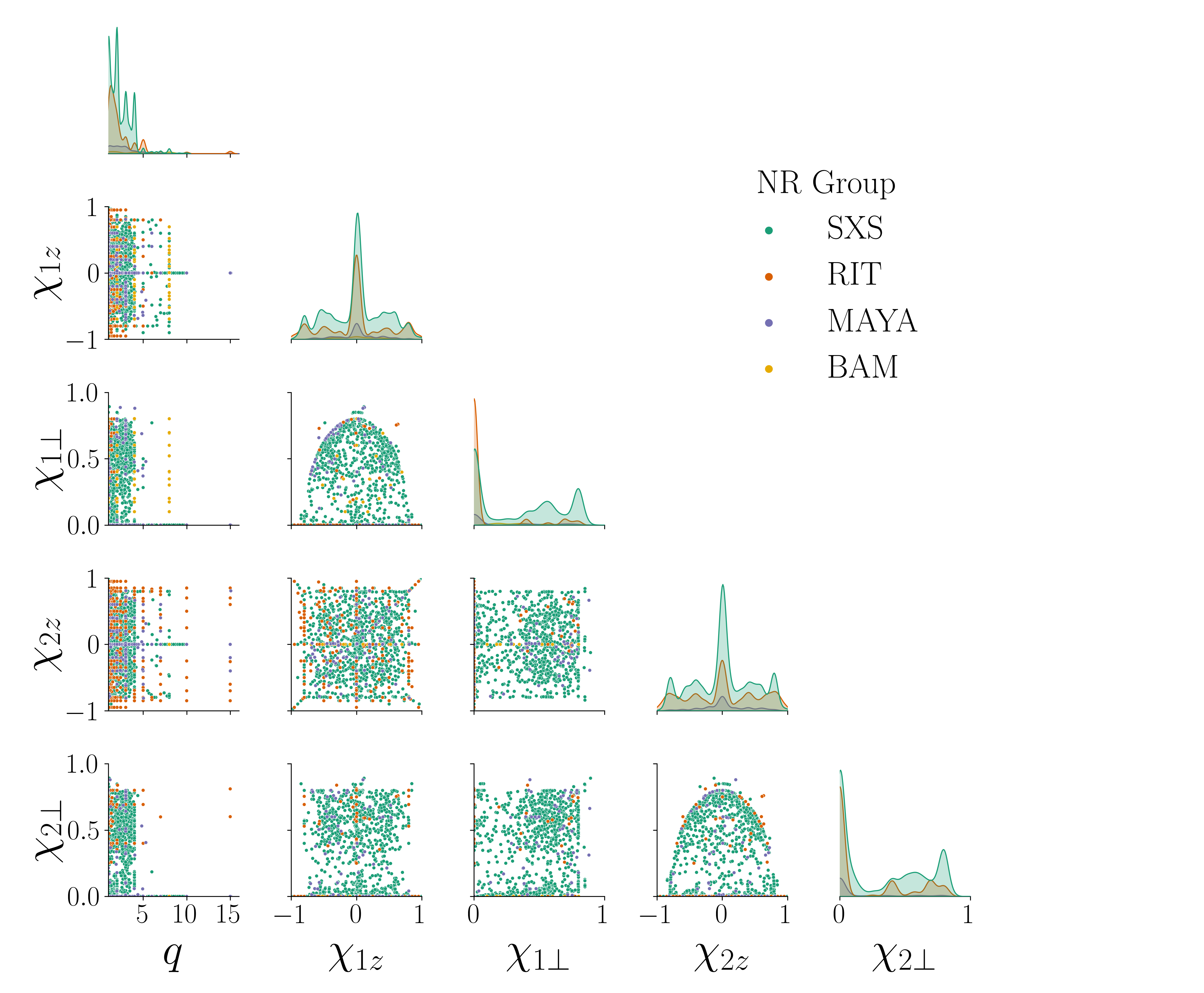}
   \caption{
     Parameter space coverage of public waveform catalogs in
     the quasi-circular limit.  Shown are mass-ratio $q$, projections of the BH spins onto the orbital angular momentum ($\chi_{1/2\,z}$), and magnitude of spin-components orthogonal to the orbital angular momentum $(\chi_{1/2\,\perp}$) for waveform catalogs of different NR groups.
     Data as of June 2023. 
   }
   \label{fig:coverage}
\end{figure}

Several groups worldwide have created public catalogs of \BBH merger simulations
with approximately $5,700$ waveforms available at the time of this
writing~\cite{Mroue:2013xna, Jani:2016wkt, Healy:2017psd, Boyle:2019kee, Healy:2019jyf,
Healy:2020vre,Healy:2022wdn,Hamilton:2023qkv,Ferguson:2023vta}; see also Table~\ref{tab:NRcodes}.
Nowadays, \NR waveforms survey several configurations in the extensive \BBH parameter
space spanned by their mass ratios, spin magnitudes and directions, and eccentricity.
Fig.~\ref{fig:coverage} illustrates the portion of the parameter space
covered by the publicly available waveforms at the time of this paper's publication.
Focussing on quasi-circular binaries, it shows that the parameter
space is best sampled for moderate mass ratios and spins.




Consequently, there are three dimensions in the \BBH parameter space --- spin, mass ratio
and eccentricity --- that require further waveform development.
For example, most runs with spinning \BH{s} concentrate on
spins $\chi\lesssim0.9$, because simulating \BH{s} with spin closer to the extremal Kerr limit
is technically challenging.
It requires a much higher numerical resolution and, more importantly,
the initial data is more difficult to construct.
There are some codes capable of constructing initial data for \BH{s} with spins above
$\chi=0.95$; see, e.g., Refs.~\cite{Lovelace:2008tw,Healy:2017vuz,Zlochower:2017bbg}.
Another challenging regime in \BBH simulations is that of unequal mass binaries.
Most simulations in the public catalogs cover the regime $q\lesssim8$
for spinning \BH{s}, and up to $q\lesssim18$ for nonspinning binaries, as can be seen in
Fig.~\ref{fig:coverage}.
However, simulations above $q=20$ are scarce:
the first high-mass ratio simulations up to $q=100$~\cite{Sperhake:2011ik} (see also~\cite{Fernando:2018mov})
and, more recently, up to $q=1024$~\cite{Lousto:2022hoq}
were achieved for head-on collisions.
The frontier for quasi-circular inspirals are mass ratios of
$q=128$~\cite{Lousto:2020tnb,Rosato:2021jsq}.
New techniques such as worldtube excision to model intermediate mass ratios of
$q=100$-$10^4$ are under development~\cite{Dhesi:2021yje,Wittek:2023nyi}.

In quasi-circular orbits, NR  \BBH simulations are well sampled in the regime of moderate mass ratio and spins, including precessing spins. As Fig.~\ref{fig:coverage} illustrates, the parameter space is best sampled over spins when the mass ratio is less than 1:4.    Increasing spins to the Kerr limit is challenging both as the resolution required for highly spinning black holes increases and, more importantly, the initial data are harder to construct.  Some codes, however, are capable of pushing  beyond $\chi=0.95$ ~\cite{Lovelace:2008tw,Healy:2017vuz}.


Most \NR simulations to-date have focused on quasi-circular inspirals because it is expected that any eccentricity present during a binary's formation will have been emitted through GW radiation by the time of merger~\cite{Peters:1964zz}; however, there are several astrophysics scenarios that predict eccentric \MBHBs~\cite{Armitage:2005xq,Bonetti:2018tpf},
produced eccentric numerical relativity waveforms~\cite{Hinder:2007qu,Huerta:2019oxn,Huerta:2016rwp,Ramos-Buades:2019uvh,Hinder:2017sxy} and other beyond circular evolution such as head on~\cite{Zilhao:2014zaa}, hyperbolic~\cite{Healy:2008js,Nelson:2019czq} and zoom-whirl~\cite{Healy:2009zm,Gold:2012tk,Gold:2009hr}.


Arguably the most challenging regime to simulate is that of highly unequal mass \BBHs.   While codes have been pushed to simulate up to 1:128 mass ratio~\cite{Lousto:2020tnb}, there are few simulations above 1:20  and fewer with generic spins and/or eccentricity.
The community has yet to achieve routine production of highly spinning ($\chi>0.9$) and highly unequal ($q>20$) configurations with or without eccentricity.
Higher order modes are excited with greater amplitudes as the binary parameters move away from non-spinning and equal mass.   Fortunately, NR codes routinely produce well resolved higher order modes up to and including about $\ell=8$.

In summary, the \BBH parameter space is currently covered as follows:
%
\begin{itemize}
\item Non-spinning \BBH mergers with mass ratio $q\leq18$~\cite{Jani:2016wkt,Healy:2017psd,Healy:2019jyf,Mroue:2013xna, Boyle:2019kee,Ajith:2012az,Gonzalez:2008bi,Hinder:2013oqa,Husa:2015iqa} --- see~\cite{Lousto:2010ut,Lousto:2020tnb} for binaries with  mass ratios $q=64,100$, and $128$.
\item \BBHs with moderate random spins and $q\leq
8$~\cite{Jani:2016wkt,Mroue:2013xna,Healy:2017psd,Healy:2019jyf,Hinder:2013oqa,Hamilton:2023qkv,Boyle:2019kee},
aligned-spin binaries of  $\chi\leq0.85$ with $q\leq18$~\cite{Husa:2015iqa,Healy:2020vre}, and a few with very high spins ($\chi\sim 0.99$)~\cite{Zlochower:2017bbg,Scheel:2014ina,Liu:2009al}.
\item Eccentric \BBHs with $q\leq 32$~\cite{Hinder:2007qu,Huerta:2019oxn,Ramos-Buades:2019uvh,Gayathri:2020coq,Healy:2020vre} with limited cycles.
\end{itemize}
Note that \BBH  simulations in \GR are scale-invariant with respect to the total mass of the system; therefore, a \BBH waveform can be re-scaled for both ground-based and space-based gravitational wave detectors without having to repeat the \NR computation. However, the mass scaling will determine the physical starting frequency of the signal, and therefore whether the \NR waveform covers the entire sensitivity band of the detector.




\subsubsection{Environmental Effects}


The NR community has made great strides in simulating binaries in  a matter-rich environment.  This Whitepaper focuses on ways the environment directly impacts the predicted gravitational waves; for EM signatures of BBH mergers, we refer to the Astrophysics Working Group White Paper~\cite{LISA:2022yao} and references therein.

The gravitational waves emitted during the merger of a BBH in an accretion disk are likely unaffected by the accreting matter~\cite{Bode:2009mt}, although Ref.~\cite{Fedrow:2017dpk} indicates that accretion disk densities greater than $10^6-10^7$ g cm$^{-3}$ 
would alter the coalescence dynamics enough to be relevant for ground-based \GW detectors.

In addition to EM counterparts, there are other possible sources for LISA modeled by NR.   One such class of signals arises from the magnetorotational collapse of a supermassive stars;
 see, e.g.,~\cite{2017A&G....58c3.22S} for a recent review and~\cite{Shibata:2006hr} for early \NR work. Recently the \GW signatures from such collapsing supermassive stars were calculated in~\cite{Shibata:2016vzw,Sun:2017voo}, where it was shown that LISA could observe such events out to redshift $z\simeq 3$.  Ref~\cite{Sun:2017voo} also computed possible EM counterparts and predicted that these systems could be sources of very long gamma-ray bursts.

Another potential source of gravitational waves detectable by LISA that is currently unmodeled comes from instabilities in accretion disks around \BHs. The particular case of the Papaloizou-Pringle instability~\cite{1984MNRAS.208..721P} of self-gravitating disks has been studied in~\cite{Kiuchi:2011re} without \BH spin, and in~\cite{Wessel:2020hvu} including spin. In~\cite{Wessel:2020hvu} it was shown that \BH spin could potentially increase the duration of the near monochromatic signal from the instability, and that
LISA could detect gravitational waves generated by the Papaloizou-Pringle instability around $10^5M_\odot$ \BHs out to redshift $z\simeq 1$.

Finally, for a detailed discussion of  environmental effects and the challenges they could present for precision tests of \GR,  such as dark matter environments, we refer the interested reader to the Fundamental Physics Whitepaper~\cite{LISA:2022kgy} and references therein.

\subsubsection{Challenges in NR}
\label{NR:challenges}
Improvements in hardware and numerical techniques continue
to speed up simulations, yet some corners of the binary parameter space
continue to demand challenging simulations that are costly in terms of
runtime and computational resources.
%
Three of the most pressing challenges facing NR are i) producing waveforms that cover the anticipated parameter space of  unequal mass, highly spinning, and eccentric binaries,
ii) producing long-lived waveforms with sufficient numbers of \GW cycles, and
iii) producing these waveforms
at standards of accuracy set by LISA's sensitivity.



\paragraph*{Parameter coverage:}  The \BBH parameter space for the LISA mission, as
 summarized in Table~(\ref{tab:SMBBHparams}) has mass ratios that span from 1:1 to 1:1000,  BH spins from 0 to 0.998, and a wide range of eccentricities.  As discussed in $\S$~\ref{NR:status}, almost all NR simulations to date are for $q\leq 8$, $\chi\leq 0.8$ and $e\approx 0$; and, therefore, there is significant work to be done to supply waveforms that cover the full potential parameter space for LISA.   We also discuss  LISA waveform catalogs.

Producing waveforms for systems with large mass ratios is a challenge for the broader waveform community with various approaches existing to bridge the gap between  \NR solutions and small mass ratio approximations. This is especially relevant for \IMRIs.
 High-mass ratios are demanding due to the need to resolve the smaller mass black hole and provide appropriate gauge conditions~\cite{Rosato:2021jsq}.  As discussed in $\S$~\ref{NR:status}, several  new codes are being developed by the NR community with the goal of having simulations with large mass ratios ($q>50$) be routinely possible.   In addition to pushing the NR capacity to larger mass ratios, new methods for modeling \IMRIs are currently being explored~\cite{Schutz2017, Dhesi:2021yje}, which combine black-hole perturbation theory and \NR techniques to significantly increase the numerical efficiency of simulations.
Coordination between the GSF and \NR communities as well as the construction of open
source platforms to share \NR ~\cite{Boyle:2019kee,Healy:2017psd,Jani:2016wkt} and GSF
waveforms, such as the \texttt{Einstein Toolkit}~\cite{Loffler:2011ay} and the
\texttt{Black Hole Perturbation Toolkit}~\cite{BHPToolkit}, will streamline and accelerate such endeavors.

A second challenge is producing many generic simulations with \BH spins that are very large in magnitude, i.e., approaching 1.   The simplest approach to solving the initial data problem is to use the Bowen and York method~\cite{Bowen:1980yu,Brandt:1997tf}; however, this method cannot construct \BHs with high spins, $\chi\gtrsim 0.93$~\cite{Cook:1989fb,Dain:2002ee,Lovelace:2010ne}.
In addition, resolving the region of spacetime near the horizons requires computationally expensive, very high resolution. This is exacerbated in evolution methods using excision to handle the singularities~\cite{Scheel:2014ina}.
As mentioned previously, \NR codes have successfully achieved some high spins, including several aligned-spin binaries with spins ($\chi\sim 0.99$)~\cite{Zlochower:2017bbg,Scheel:2014ina,Liu:2009al} by using new formulations that move beyond the conformally flat ansatz for puncture methods~\cite{Ruchlin:2014zva} and new techniques for handling the excision region~\cite{Scheel:2014ina}.

Finally, several astrophysical scenarios produce non-zero eccentricity in the LISA frequency band~\cite{Armitage:2005xq}.  \NR is capable of and has been producing eccentric runs~\cite{Hinder:2007qu,Huerta:2019oxn,Ramos-Buades:2019uvh,Gayathri:2020coq,Ramos-Buades:2022lgf,Joshi:2022ocr}; the challenge is simulations must start with the \BHs more widely separated than in quasi-circular configurations.  This is again a computational cost.

The source modeling community has continued to improve the modeling of these sources, developing
inspiral-merger-ringdown models that combine analytical approximations~\cite{Hinderer:2017jcs} with eccentric \NR waveforms~\cite{Hinder:2007qu,Huerta:2016rwp,Ramos-Buades:2019uvh,Hinder:2017sxy,Huerta:2017kez,Cao:2017ndf,Chiaramello:2020ehz,Liu:2019jpg,Gayathri:2020coq}. 

\paragraph*{Gravitational wave cycles:} The LISA frequency band may contain thousands of GW cycles, depending on the total mass of the massive binary, thus observing BBH signals for months or even years.  Given the accuracy of approximate waveform models calibrated with NR ~\cite{Hannam:2013oca,Bohe:2016gbl, Khan:2015jqa,Blackman:2017pcm,Husa:2015iqa}), and the recent developments in the self-force community~\cite{Dhesi:2021yje,Albertini:2022rfe}, it is likely  unrealistic and unnecessary to use \NR to describe the entire evolution of binary systems, from inspiral to ringdown.  We point the reader to the rest of section 4 in this paper for descriptions on the approaches for modeling the inspiral and inspiral-merger-ringdown waveforms.

\begin{figure}
  \includegraphics[width=0.95\textwidth,trim=80 10 100 10,clip=true]{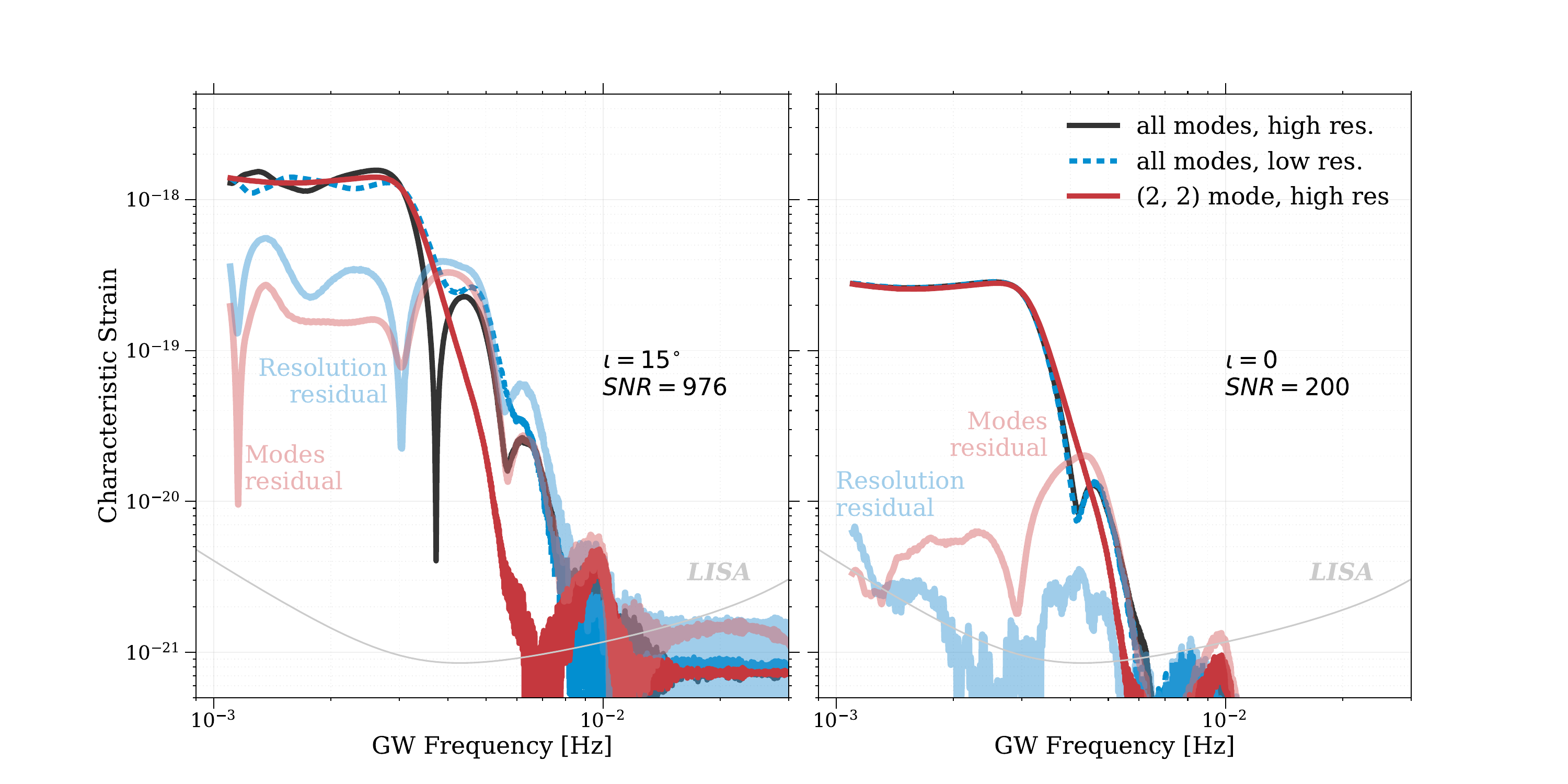}
  \caption{
  Illustration of how numerical errors in a template may
    leave a residual in high-SNR signals.  The left panel represents a
    BBH at high SNR (detector mass $M_{\rm det}=5\times 10^6M_\odot$,
    aligned spin of $\chi=0.2$ on the larger \BH at distance
    $D_L=30\,$Gpc and at inclination $\iota=15^\circ$).  The black
    line is the characteristic strain ($2 f \tilde{h}(f)$) obtained at
    high numerical resolution from all spherical harmonic modes.  The
    blue dashed line is the corresponding quantity obtained from a low
    resolution simulation, with the light blue line denoting the
    difference to the high-resolution waveform.  The solid red line is
    a high resolution template containing only the $(l,m) = (2,2)$
    mode, with the light red line showing the difference to the black
    line.  If this system were searched for with the low-resolution
    (or (2,2)-mode only) waveforms, then the shaded lines would remain
    as residuals after subtraction.  The SNR of the residual due to
    resolution in the left panel is 342, clearly indicating the need
    for higher accuracy NR simulations.  The right panel represents a
    more frequently expected system, where the SNR is 200, $D_L=158\,$Gpc and
    $\iota=0^\circ$.  Here, the residual due to resolution has an SNR
    of 2.4.  Figure adapted from~\cite{Ferguson:2020xnm}.  The LISA noise curve can be found at~\cite{Robson:2018ifk}. }
    \label{fig:why}
    \end{figure}

\paragraph*{Accuracy}
Massive \BBH mergers in LISA could have  high (up to $1\times 10^4$) signal-to-noise (SNR) ratios and \NR must model these loud signals accurately enough not only to infer the correct source properties, but also to subtract the signal from LISA data, revealing quieter, high-interest signals that may lay underneath. The full implications of these requirements for \NR waveform accuracy have yet to be assessed, but an initial study~\cite{Ferguson:2020xnm} suggests that \NR waveforms will require a substantial increase in accuracy, compared to today's most accurate waveforms, to be indistinguishable from a high-SNR observation with the same source properties.  A similar assessment for next-generation ground-based detectors concluded that \NR waveforms will need an order of magnitude more accuracy to avoid bias in the inferred source properties for the high signal-to-noise sources that these detectors will observe~\cite{Purrer:2019jcp}. The impact of using insufficiently accurate templates is highlighted in Figure~\ref{fig:why}.

These accuracy requirements are even more challenging for \NR if
the \BHs are precessing, if one or both \BHs are spinning nearly as
rapidly as possible, or if one \BH is much more massive than the
other.  Let us illustrate how quickly computational cost rises, by quantifying
computational cost increases relative to a fiducial baseline
simulation with mass-ratio $q_0$ and initial frequency $M\Omega_0$.  Cost
increases arise for multiple reasons.

First, the explicit evolution schemes used in \NR codes are subject to
the Courant-Friedrichs-Levy (CFL) condition, which limits the possible
time step to the dynamic time of the smaller \BH, $\Delta t\propto
m_2$.  As the mass ratio increases, the number of time-steps per unit
$M=m_1+m_2$ of evolution time will therefore increase as
\begin{equation}
  f_{\rm CFL}\sim\frac{q}{q_0}.
\end{equation}
Second, the total inspiral time of a BBH can be estimated by \PN
expansions~\cite{Blanchet:2013haa} as $T\propto 1/\nu
(M\Omega)^{-8/3}$, where $M\Omega$ is the initial dimensionless
frequency of a binary.  A change in mass ratio and initial frequency
will modify the computational cost in proportion to $T$, i.e. by a
factor
\begin{equation}
	f_{\Omega_{\rm low}}=\left(\frac{\nu}{\nu_0}\right)^{-1}\left(\frac{\Omega}{\Omega_0}\right)^{-8/3}.
\end{equation}
Thirdly, LISA has higher accuracy requirements than current \NR
simulations.  We denote by $f_{\tau}(A)$ the increase in computational
cost that is needed to reduce the truncation error $\tau$ by a factor
$A>1$, i.e. $\tau \rightarrow \tau/A$.

A fourth cause for increased computational cost arises from the desire
of achieving a certain overall phase error, which is important for \GW
waveform modeling.  For longer simulations, phase accuracy must be
preserved over more cycles.  From \PN theory, the orbital phase to
merger scales as $\Phi\propto \nu^{-1} (M\Omega_i)^{-5/3}$.  To achieve a
fixed overall phase-error (say, 1 radian), the relative phase error
must decrease $\propto 1/\Phi$, yielding an increase in computational cost by a factor
\begin{equation}
  f_\Phi\sim
  f_{\tau}\left(\left(\frac{\nu}{\nu_0}\right)^{-1}\left(\frac{\Omega}{\Omega_0}\right)^{-5/3}\right).
\end{equation}


The challenge now is that in order to achieve higher mass ratio simulations
starting at similar (or lower) initial frequency, and possibly also
achieving higher accuracy, each of the factors just
outlined increases very rapidly and that these factors multiply for an overall cost increase by a factor
\begin{equation}
C=f_{\rm CFL}\,f_{\Omega_{\rm low}}\,f_\tau\,f_\Phi.
\end{equation}

To obtain concrete numerical estimates, we assume that $f_\tau$ is a
power law, $f_\tau(A)=A^\alpha$.  This assumption is satisfied for
finite difference codes, where a convergence order of $k$ in $(3+1)$
dimension yields $\alpha=4/k$.
In terms of $\alpha$, the overall cost increase
is given by
\begin{equation}
C
\sim \left(\frac{q}{q_0}\right)^{2+\alpha}\,\left(\frac{\Omega_0}{\Omega}\right)^{8/3+5\alpha/3}\,A^\alpha,
\end{equation}
where we have also used $\nu\approx 1/q$, which is valid at large
$q$.

We can now make concrete estimates: Increasing the mass-ratio by a
factor of $2$, lowering the initial frequency by a factor of $1/2$,
and increasing accuracy by a factor of $2$, and assuming the \NR code
under consideration maintains perfect 8-th order convergence into the
regime of longer, more accurate simulations at higher mass-ratio
(i.e. $\alpha=1/2$), increases the computational cost two orders of
magnitude. More ambitiously, increasing the mass-ratio by a factor of
10, reducing initial frequency by a factor $1/4$ and increasing
accuracy by a factor $10$ increases the computational cost by five
orders of magnitude.  Codes based on spectral methods, e.g. SpEC, will
likely result in a smaller $f_\tau$; even so, increases in
computational cost are very challenging as one goes to longer
simulations at higher mass-ratio.

The NR codes that have been instrumental in providing public \BBH waveforms use either a spectral method or box-in-box mesh refinement.   These methods
scale well to the order of 1000 cores, but they cannot scale to hundreds of thousands or millions of cores.
For example, codes using high-order finite-difference stencils incur ever-increasing inter-core communication costs as the number of cores increases and this cost eventually becomes prohibitive.
These codes are  parallelization bound and do not
parallelize beyond current usage.  Since individual computer cores
are no longer dramatically increasing in speed (Moore's law has
4)  assessment of systematic bias in simulated LISA signals with currently available waveforms, and
5)  the development of \NR waveform for all potential parameters at the necessary accuracy.

\subsubsection{Future Codes and Catalogs}\label{sec:nr_figure_codes}

The LISA data analysis groups could have access to NR waveforms via a bespoke LISA catalog or as a public catalog. 
A bespoke catalog would have NR waveforms created specifically to meet those aforementioned requirements, potentially interfacing and merging existing catalogs and extending with its own dedicated entries
(similar to the LVK catalog \cite{Schmidt:2017btt}. 
If a bespoke LISA catalog is not created or deemed unnecessary due to a possible future abundance of NR public waveforms (current catalogs are indicated in Table \ref{tab:NRcodes}) we will need a software interface with existing NR waveform catalogs that is curated and cultivated with time. 
A full catalog of NR waveforms achieving all requirements will take time, possibly until the launch of LISA. We put forth the following priorities for the NR
waveforms: 1) an accuracy assessment of current and near-term achievable NR
waveforms, 2) the assessment of current and near achievable parameter space, 3) the
development of NR waveforms for the most likely LISA events at the required
accuracy, 4) assessment of systematic bias in simulated LISA signals with currently
available waveforms, and 5) the development of NR waveform for all potential
parameters at the necessary accuracy.





\newpage

\subsection{Weak field approximations (post-Newtonian/post-Minkowskian)}
\label{sec:weak_field}
Coordinator:  Laura Bernard and Chris Kavanagh\\
Contributors: A.~Antonelli, G.~Faye, J.~Garcia-Bellido, M.~Haney, F.~Larrouturou, A.~Le Tiec, O.~Long

\subsubsection{Description}\label{sec:PNdescript}

The \PN formalism is an approximation method in \GR that is well suited to describe the orbital motion of---and the \GW emission from---binary systems of compact objects, in a regime where the typical orbital velocities are small compared to the vacuum speed of light $c$ and the gravitational fields are weak. It has played and continues to play a central role in the construction of template banks for the detection and analysis of \GW signals generated by binary systems of black hole and/or neutron star, which are now routinely observed by the \LVK collaboration's detectors \cite{LIGOScientific:2018mvr,LIGOScientific:2020ibl,KAGRA:2021vkt}.
More precisely, PN results are at the core of several classes of \edit{waveform models}, including phenomenological waveforms and effective one-body waveforms (see Secs.~\ref{sec:EOB} and \ref{sec:phenom}).

In PN theory, the general relativistic corrections to the Newtonian motion and to the leading \GW emission (i.e. Einstein's quadrupole formulas) are computed in a systematic manner in powers of the small PN parameter $v^2/c^2 \sim G M / (c^2r)$, where $v$ and $r$ are the typical binary relative velocity and separation, $M$ is the sum of the component masses, and $v^2 \sim G M /r$ for bound orbits. Indeed, the most promising sources of gravitational waves for existing and planned interferometric detectors are bound systems of compact objects. By convention, a contribution of ``$n$PN'' order refers to equation-of-motion terms that are $O(1/c^{2n})$ smaller than the Newtonian acceleration, or, in the radiation field, smaller by that factor relative to the standard quadrupolar field.

The PN approximation dates back to the pioneering works of Lorentz and Droste \cite{LoDr.17}, as well as Einstein, Infeld and Hoffman \cite{Einstein:1938yz} \edit{and Fock~\cite{1939ZhETF...9..375F}}, who computed the leading 1PN corrections to the Newtonian equations of motion in a system of $N$ point masses. During the 1980s, those results were extended to 2.5PN to provide a rigorous basis for interpreting binary pulsar observations \cite{Damour:1981bh,Damour:1981ntn,Damour:1985,DaDe.86}. For binary systems of compact objects, the state of the art corresponds to gravitational waveforms that include all of the relativistic corrections up to 4PN order~\cite{Damour:2016abl, Marchand:2017pir,Foffa:2019rdf, Foffa:2019yfl}, in the simplest case of nonspinning bodies moving along a sequence of quasi-circular orbits.

Another complementary weak field approximation can be found by relaxing the small velocity assumption of the \PN approximation and demanding only that $G M / (c^2r)\ll 1$. This  scheme, known as the \PM approximation, has a long and venerable history, in the context of both unbound scattering where velocities can be arbitrarily large \cite{Westpfahl:1979gu,Westpfahl:1980mk} and far-zone--near-zone matching for bound orbits within certain \PN schemes (see e.g., Ref.~\cite{Blanchet:2013haa}).

As discussed below, much work has recently been dedicated to (i) push this accuracy to 4.5PN order, (ii) include the effects of the spins of the compact objects (see also Table~\ref{tab:PNresults}), (iii) generalize the results from circular to eccentric orbits, and (iv) explore the overlap and synergies with other approximation methods (\PM approximation, small-mass-ratio approximation, effective one-body model). The reader is referred to the review articles \cite{Futamase:2007zz,Blanchet:2009ggi,Schaefer:2009ehj,Blanchet:2013haa,Foffa:2013qca,Rothstein:2014sra,Porto:2016pyg,Schafer:2018jfw,Levi:2018nxp} and to the texbooks \cite{Maggiore:2007ulw,PoWiBook} for more information.

\begin{table}
\begin{tabular}{c||c|c|c|c||c|c|c|c}
 & \multicolumn{4}{c||}{Dynamics} & \multicolumn{4}{c}{Dissipative flux} \\
\hline
\hline
PN order & non-spinning & \multicolumn{3}{c||}{spinning} & non-spinning & \multicolumn{3}{c}{spinning} \\
\hline
 &  & SO & SS & higher spins &  & SO & SS & higher spins \\
\hline
\hline
0 & $\surd$ & - & - & - & - & - & - & - \\
\hline
1 & $\surd$ & - & - & - & - & - & - & - \\
\hline
1.5 & - & $\surd$ & - & - & - & - & - & - \\
\hline
2 & $\surd$ & - & $\surd$ & - & - & - & - & - \\
\hline
2.5 & $\surd$ & $\surd$ & - & - & $\surd$ & - & - & - \\
\hline
3 & $\surd$ & - & $\surd$ & - & - & - & - & - \\
\hline
3.5 & $\surd$ & $\surd$ & - & $\surd\ (S^3)$ & $\surd$ & - & - & - \\
\hline
4 & $\surd$ & - & $\surd$ & $\surd\ (S^4)$ & $\surd$ & $\surd$ & - & - \\
\hline
4.5 & $\ast$ & $\surd$  & - & $\surd\ (S^3)$ & $\surd$ & - & $\surd$ & - \\
\hline
5 & $\ast$ & - & $\surd$ & $\surd\ (S^4)$ & $\surd$ & $\surd$ & - & - \\
\hline
5.5 & $\ast$ &  &  & $\surd\ (S^5)$ & $\surd$ & $\surd$ & $\surd$ & - \\
\hline
6 &  &  &  & $\surd\ (S^6)$ & $\surd$ & $\surd$ & $\surd$ & $\surd\ (S^3)$ \\
\hline
6.5 &  &  &  & $\star$ & $\surd$ & $\surd$ & $\surd$ & \\
\hline
7 &  &  &  & $\star$ & $\surd$ &  &  & \\
\end{tabular}
\caption{State-of-the-art of known PN results for both the conservative and dissipative dynamics as well as for the gravitational flux. Contrary to the text, everything is stated as absolute order. For example, the 6PN absolute order for the non-spinning flux in the table corresponds to the 3.5PN relative order results as stated in the text. Hereditary effects \edit{enter at 2.5PN order and subsequent odd PN orders in the non-spinning dynamics and usually contains non-local contributions}. An instantaneous expression can be obtained by performing a low-eccentricity expansion. SO and SS refer to spin-orbit and spin-spin interactions respectively. \\ $\ast$ means that only a partial result is known at those orders. At 5 and 5.5PN order, the results were obtained by the combination of PN traditional techniques with scattering amplitudes and self-force. \\ $\star$ means that the dynamics is known at all leading order in the spin.} \label{tab:PNresults}
\end{table}

\subsubsection{Suitable for what sources?}

The \PN approximation is well suited to describe the dynamics and gravitational emission of compact binary systems in the early inspiral stage. While it can in principle be applied to model BH binaries of any mass ratio, it is expected to be more accurate for comparable to intermediate mass BHBs, as the gravitational field should remain small until the final cycles. For mass ratio above $q\gtrsim 100$, when the small body remains for long timescales close to the supermassive BH, the gravitational field is strong and the weak-field approximation will lose accuracy (see Fig.~\ref{fig:parameter_space}).

The \PM approximation is best suited to \CHE, which could be detected with LISA for a population of massive \PBH in dense clusters forming part of the halos of galaxies, as studied in Refs.~\cite{Garcia-Bellido:2017qal,Garcia-Bellido:2017knh} \edit{using a leading order PN approximation scheme}. It also provides valuable resummations of terms in the PN approximation and is used to improve bound compact binary coalescence models when combined with PN results.

PN waveforms are ideally suited to model binaries that are in an early stage of their inspiral while in the LISA band such as \SOBHBs and \GBs.

\subsubsection{Status}

\paragraph{Equations of motion and waveforms without spin.}

For an isolated, non-spinning compact binary system, the phase evolution and \GW modes during the inspiral stage are currently respectively known up to 4.5PN and 4PN~\cite{Marchand:2020fpt,Henry:2021cek,Larrouturou:2021dma,Larrouturou:2021gqo,Blanchet:2022vsm,Trestini:2022tot,Henry:2022ccf,Blanchet:2023bwj,Blanchet:2023sbv}. As for the LISA requirements, it was established in~\cite{Mangiagli:2018kpu} that the 2PN waveform will be enough for the parameter estimation of roughly 90\% of the detectable \SOBHBs, while the 3PN waveform phasing is needed for the systems that will merge within the mission lifetime. Nevertheless it will be quite valuable to push the current accuracy up to 4.5PN order for at least two main reasons. The first one is that it may be necessary for the parameter estimation of black hole binaries with large masses.
The second one is that this accuracy will be very beneficial for the calibration of numerical waveforms, and for testing the second-order self-force computations.

For quasi-circular orbits, the gravitational phase $\phi(t)$ is computed via the so-called energy \emph{flux-balance equation}, $\frac{\mathrm{d}E}{\mathrm{d}t} = - \mathcal{F}_\text{GW}$,
where $E$ is the conserved energy and $\mathcal{F}_\text{GW}$ is the flux of radiation emitted.
Both quantities are functions of the gravitational phase, which is thus obtained by a simple integration.
The amplitude and polarizations are obtained by the same radiative multipole moments that are used when computing the flux, as explained later.
For recent reviews, see~\cite{Buonanno:2014aza, Blanchet:2013haa, Goldberger:2007hy, Foffa:2013qca,Porto:2016pyg}.

At 4PN precision, three different techniques are used to compute the binding energy: the canonical Hamiltonian formulation of General Relativity~\cite{Jaranowski:2013lca, Jaranowski:2015lha, Damour:2014jta, Damour:2015isa, Damour:2016abl}, the Fokker Lagrangian approach~\cite{Bernard:2015njp, Bernard:2016wrg, Bernard:2017bvn, Marchand:2017pir, Bernard:2017ktp} and \EFT~\cite{Foffa:2012rn, Foffa:2011np, Galley:2015kus,Porto:2016pyg, Foffa:2019rdf, Foffa:2019yfl, Blumlein:2020pog}.
All of these methods derive the energy as a conserved quantity associated with the equations of motion for two point-like particles, and naturally give physically equivalent results.
The Hamiltonian approach has been pushed up to 6PN, yielding incomplete results~\cite{Bini:2020nsb}, see Sec.~\ref{subsub:weakfield_scattering}.
Exploiting synergies between traditional \PN methods and \EFT, the logarithmic contributions in the energy have also been computed up to 7PN~\cite{Blanchet:2019rjs}.

The flux of gravitational radiation can be expressed as a generalisation of the famous Einstein quadrupole formula~\cite{Einstein:1918btx,Landau:1975pou}, as~\cite{Thorne:1980ru}
\begin{align}
\mathcal{F}_\text{GW} &= \sum_{\ell \geq 2} \frac{G}{c^{2\ell+1}} \left[ \alpha_\ell \,U_L^{(1)} U_L^{(1)} + \frac{\beta_\ell}{c^2}\, V_L^{(1)} V_L^{(1)} \right]\nonumber\\
&= \sum_{\ell \geq 2} \frac{G}{c^{2\ell+1}} \left[ \alpha_\ell \,I_L^{(\ell+1)} I_L^{(\ell+1)} + \frac{\beta_\ell}{c^2}\, J_L^{(\ell+1)} J_L^{(\ell+1)} \right] + \mathcal{F}_\text{tails}\,,
\end{align}
where \edit{$L$ is a multi-index}, the $U_L$ (resp. $V_L$) are the mass (current) radiative multipole moments and the  $I_L$ (resp. $J_L$) are the mass (current) source multipole moments, the $\{\alpha_\ell,\beta_\ell\}$ are collections of numbers and the superscript $(n)$ denotes $n$ time derivatives.
To connect radiative moments to sources moments, their non-linear interactions (occurring during the propagation from the source to the detector) have been singled out in $\mathcal{F}_\text{tails}$.
The radiative moments are also directly involved in the computation of polarisations~\cite{Thorne:1980ru}, and thus their determination is crucial.

The contribution $\mathcal{F}_\text{tails}$ is known up to 4.5\PN order, using traditional PN methods~\cite{Marchand:2016vox}, and was confirmed by an independent PN re-expansion of resummed waveforms~\cite{Messina:2017yjg}.
The computation of the required moments is currently done using the Multipolar-post-Minkowskian-post-Newtonian (MPM-PN) algorithm~\cite{Blanchet:2013haa}.
The major piece of this work is the derivation of the mass quadrupole ($\ell = 2$) up to 4PN order: the main 
result has been obtained~\cite{Marchand:2020fpt,Larrouturou:2021dma,Larrouturou:2021gqo,Blanchet:2022vsm,Trestini:2022tot}, including the tail-of-memory contribution~\cite{Trestini:2023wwg}.
The other moments involved are the mass octupole and the current quadrupole, already known at the required order (3PN)~\cite{Faye:2014fra,Henry:2021cek}, and higher moments that are either known or trivial (i.e. needed at the Newtonian order only), see notably~\cite{Blanchet:2008je}.
Collecting all these results, the complete flux of gravitational radiation at infinity is
fully known at 4PN~\cite{Blanchet:2023sbv}
(and 4.5PN in the case of circular orbits~\cite{Marchand:2016vox}).
In the case of horizonless objects such as neutron stars, the flux-balance equation is used to compute the gravitational phase
for circular orbits at 4.5PN~\cite{Blanchet:2023bwj}.
However, if at least one object is a black hole, one must include an extra contribution due to the absorption of
gravitational waves by the black hole
horizon(s)~\cite{Poisson:1994yf,Tagoshi:1997jy,Alvi:2001mx,Porto:2007qi,Chatziioannou:2012gq}.
Note that two other methods are able to deal with high PN multipole moments, both currently developed up to 2\PN : the direct integration of the relaxed equations (DIRE)~\cite{Will:1996zj} and EFT~\cite{Leibovich:2019cxo}.
As half-\PN orders are often easier to compute than integer ones, one can probably push the results up to 4.5\PN order, including the amplitudes of the gravitational modes.
A first step in this program is to control the dissipative, radiation reaction effects in the equations of motion.
Table~\ref{tab:PNresults} summarizes the current knowledge regarding PN dynamics and flux without spin.

Finally, there have been some works trying to map the BMS (Bondi-Metzner-Sachs) flux-balance laws to equivalent PN results in the harmonic gauge~\cite{Compere:2019gft}. By transforming the metric from harmonic coordinates to radiative Newman-Unti (NU) coordinates, they were able to obtain the mass and angular momentum aspects and the Bondi shear as a function of the quadrupole moment~\cite{Blanchet:2020ngx}. In particular, they rederive the displacement memory effect (see also~\cite{Favata:2009ii}) and provide expressions for all Bondi aspects relevant to the study of leading and subleading memory effects~\cite{Blanchet:2023pce}.

\paragraph{Equations of motion and waveforms with spin.}
In the last decade, the effects of the proper rotation of the components of binary systems have
been included in the orbital equations of motion to 4PN order, extending previous works dating back to
the 70's~\cite{Barker:1975ae}. In the post-Newtonian terminology, the spin variable $S$ is rescaled as
as $S = c S_{\rm phys} = Gm^2 \chi$, with $\chi$ the dimensionless spin parameter comprised between $0$ and $1$. The leading spin interactions,  the so-called spin-orbit interaction (SO), couple the mass $m_1$
of a particle with the spin of the other, hence those contributions scale as
$\frac{G m_1}{c^2} \times \frac{S_2}{c} \sim \mathcal{O}(\frac{1}{c^3})$ and are regarded as
being of 1.5PN order. Meanwhile the first effect due to spin-spin interaction, scaling as
$\frac{G}{c^2}\times \frac{S_1}{c} \times \frac{S_2}{c} \sim \mathcal{O}(\frac{1}{c^4})$, is quadratic
in spin (SS) and arises at 2PN order. We refer to the successive subleading PN contributions to a given spin interaction (e.g. SO or SS) as the
next-to-leading (NL), next-to-next-to-leading (NNL) and so on.
Moving beyond the 2PN two-body dynamics including SO
interactions~\cite{Damour:1982}, it becomes increasingly difficult to resort to an explicit model,
where the bodies are described as small balls of fluid~\cite{Will:2005sn}. The most efficient
strategy consists in adopting an effective point of view and considering that the two objects are point
particles endowed with a classical spin.

The definition of spin in \GR was first introduced by
Mathisson~\cite{Mathisson:1937zz}, and rephrased later by Tulczyjew~\cite{Tulczyjew:1959}. The
covariant equations of evolution obeyed by a spinning test particle were obtained in their usual
form by Papapetrou~\cite{Papapetrou:1951pa} and later generalized by Dixon to bodies endowed with
higher-order multipole moments~\cite{Dixon:1974xoz}.
The equations of motion and precession served as a starting point to investigate the dynamics of
binary systems at next-to-leading (NL: 2.5PN)~\cite{Tagoshi:2000zg, Faye:2006gx} and
NNL (3.5PN) order linear in spin (SO interactions)~\cite{Marsat:2012fn}, assuming that all Dixon moments
other than the masses and the spins vanish. The metric is obtained by solving
iteratively Einstein's equations in harmonic coordinates~\cite{Bohe:2012mr} for the pole-dipole
stress-energy tensor. Later, to compute the SS corrections, \edit{first contributing} at NL (3PN) order~\cite{Bohe:2015ana}, the Dixon quadrupoles were added, so as to account for the self deformation of the
bodies produced by their own spins.

The direct computation of a generalized Lagrangian (i.e. depending on the accelerations) for two
spinning particles, using \EFT in harmonic coordinates, was performed in
stages from the mid-2000's on by integrating out the gravitational field entering the full `field
plus matter' Lagrangian with the help of standard Feymann diagram expansions~\cite{Porto:2006bt,
  Porto:2005ac, Levi:2015msa}. The NL $S_1$--$S_2$ (3PN) interactions~\cite{Porto:2008tb,Levi:2008nh} were computed first, before the NL $S_1$--$S_1$ (3PN)
contributions~\cite{Porto:2008jj} and NL SO (2.5PN) interactions~\cite{Levi:2010zu,Porto:2010tr} were considered. Equivalent results were obtained in parallel by means of
Hamiltonian methods based of the Arnowitt-Deser-Misner (ADM) formulation of \GR \cite{Steinhoff:2007mb,Steinhoff:2008ji,Steinhoff:2008zr}.

After the canonical treatment of the spin was better understood~\cite{Steinhoff:2008zr,Steinhoff:2009ei}, the
computations of the Hamiltonian were pushed up to NNL (3.5PN) order for the SO~\cite{Hartung:2011te,Hartung:2013dza}, to NNL (4PN) order for the $S_1$--$S_2$ interactions~\cite{Hartung:2011ea} and to NL (3PN) order for the $S^2$ interaction \cite{Hergt:2010pa}. A fully equivalent Lagrangian for the latter
effects was found using EFT ~\cite{Levi:2011eq}. Once the EFT degrees
of freedom and the gauge choice corresponding to the spin-supplementary condition were clearly
identified, the PN leading terms to all orders in spins were computed~\cite{Vines:2016qwa}, then \edit{beyond leading order corrections were added~\cite{Levi:2014gsa,Levi:2019kgk,Levi:2020lfn}} and the spin part of the harmonic Lagrangian was completed up to the NNNL (5PN) order~\cite{Kim:2022pou,Kim:2022bwv}.

Note that reaction forces, dissipative in essence, are
absent from the previous treatments, which exclusively describe the conservative dynamics.
Nonetheless, they can be computed from balance equations~\cite{Zeng:2007bq}, or more directly from the Papapetrou evolution equations, or from the
conservation of an appropriate stress-energy tensor~\cite{Wang:2007ntb}. One may even construct a Schwinger-Keldysh Lagrangian where
each degree of freedom is formally doubled. This was done for the NL SO and SS effects at 4PN and
4.5PN respectively in Refs.~\cite{Maia:2017gxn, Maia:2017yok}, and to all orders in spin in Ref.~\cite{Siemonsen:2017yux} (see also the approach developed for
ADM Hamiltonians~\cite{Wang:2011bt} at leading SO and SS orders).

Knowing the near-zone dynamics of two spinning particles, one can insert the corresponding PN metric
into the right-hand side of Einstein's equations in harmonic coordinates, which yields the
expression of the effective non-linear source entering the integrand of the multipole moments, as
defined in the multipolar post-Minkowskian (MPM) formalism~\cite{Blanchet:1998in}, or, equivalently,
in EFT~\cite{Ross:2012fc}; hence one gets the relevant mass, current and gauge moments in the
spinning case~\cite{Porto:2012as}. The expression for the radiative moments in terms of the former
quantities has been derived for general isolated systems~\cite{Faye:2014fra}, but the hereditary
terms therein are more delicate to evaluate than for mere point-mass particles on circular
orbits, since the orbital plane is now generically precessing. They have however been handled for
the leading~\cite{Blanchet:2011zv} and NL~\cite{Marsat:2013caa} SO interactions, which required the
full integration of the (approximately) conservative dynamics up to the considered order.

Finally concerning the gravitational radiation, the \GW flux has been obtained to NNL (4PN) order both for the SO and SS contributions~\cite{Bohe:2015ana,Cho:2022syn}.
The \GW phase, built from the radiative moments and the Noetherian energy, has been obtained within
the MPM formalism at the NNL (4PN) order for the SO contributions~\cite{Bohe:2013cla, Marsat:2013caa}
(in continuation of previous works at leading order (2PN)~\cite{Kidder:1992fr,Kidder:1995zr} and NL (3PN) orders~\cite{Blanchet:2006gy}), at the NL (3PN) order for the SS terms~\cite{Bohe:2015ana}, and at leading order for the spin cube contributions~\cite{Marsat:2014xea}.
Similar results were obtained with the EFT formalism~\cite{Porto:2010zg} for which state-of-the-art results are also the NNL (4PN) order~\cite{Cho:2022syn}. The waveform modes have been obtained to NNL (3.5PN) order~\cite{Henry:2022dzx} in the quasi-circular orbit approximations.
By contrast, the radiation amplitude has been computed up to the NL (2PN) order only~\cite{Buonanno:2012rv}, although it is currently provided in a ready-to-use form for precessing quasi-circular orbits to the even lower 1.5PN order~\cite{Arun:2008kb}.

See Table~\ref{tab:PNresults} for a summary of on the current knowledge about the PN spinning dynamics and flux.

\paragraph{Eccentric-orbit waveforms.}

The modeling of inspiral waveforms from compact binaries in eccentric orbits commonly relies on quasi-Keplerian parametrization (QKP) as a semi-analytic representation of the perturbative, slowly precessing post-Newtonian motion. The conservative motion as well as instantaneous and hereditary contributions to the secular (``orbital averaged'') evolution of the orbital elements are known to 3PN order, in both modified harmonic and ADM-type coordinates \cite{Damour:2004bz,Memmesheimer:2004cv,Arun:2007rg,Arun:2007sg,Arun:2009mc} and to 4PN order in the ADM-type coordinates only~\cite{Cho:2021oai}.

The complete 3PN-accurate \GW amplitudes from non-spinning eccentric binaries have been derived using the multipolar post-Minkowskian formalism, including all instantaneous, tail and non-linear contributions to the spherical harmonic modes \cite{Mishra:2015bqa, Boetzel:2019nfw, Ebersold:2019kdc}.
Going beyond the usual approximation of radiation reaction as an adiabatic process (and the associated ``orbital averaging'' in QKP), \cite{Damour:2004bz, Konigsdorffer:2006zt, Moore:2016qxz, Boetzel:2019nfw} provide post-adiabatic, oscillatory corrections to the secular evolution of \GW phase and amplitude.

For spinning eccentric binaries that have component spins aligned with the orbital angular momentum, the effects of spin-orbit and spin-spin couplings on the binary evolution and gravitational radiation have been worked out to leading order in QKP (i.e., up to 1.5PN and 2PN in the equations of motion) \cite{Klein:2010ti}. Subsequent work has aimed to extend the treatment of spins in QKP to higher PN orders \cite{Tessmer:2010hp, Tessmer:2012xr}. The waveforms modes to NL (3PN) order for aligned spins and eccentric orbits have been derived in ~\cite{Khalil:2021txt,Henry:2023tka}, including tail and memory contributions. While the instantaneous contributions were derived for generic motion, the hereditary contributions were computed first in a small-eccentricity expansion, then extended to larger eccentricities using a resummation. Regarding precessing eccentric systems, gravitational waveforms have been obtained to leading order in the precessing equation~\cite{Klein:2021jtd}. A fully analytical treatment has also been proposed up to 2PN order in the spin, including higher harmonics~\cite{Paul:2022xfy}.

In practice, the semi-analytic approach of QKP requires a numerical evolution of the orbit described by a coupled system of \ODEs and a root-finding method to solve the Kepler problem, and therefore lacks the computational efficiency required for most data analysis applications. The post-circular (PC) formalism \cite{Yunes:2009yz} provides a method to recast the time-domain response function $h(t)$ into a form that permits an approximate fully analytic Fourier transform in \SPA, under the assumption that the eccentricity is small, leading to non-spinning, eccentric Fourier-domain inspiral waveforms as a simple extension to the quasi-circular PN approximant TaylorF2. More recent work has extended the PC formalism to 3PN, with a bivariate expansion in eccentricity and the PN parameter \cite{Tanay:2016zog, Moore:2016qxz}, and has included previously unmodeled effects of periastron advance \cite{Tiwari:2019jtz}.

The parameter space coverage of Fourier-domain waveforms in the PC formalism is limited by the necessary expansion in small eccentricity. Newer models aim for validity in the range of moderate to high eccentricities, by utilizing numerical inversions in \SPA and resummations of hypergeometric functions to solve orbital dynamics \cite{Moore:2018kvz, Moore:2019xkm} or by
applying Pad\'e approximation on analytic PC schemes expanded into high orders in eccentricity.
A semi-analytic frequency-domain model for eccentric inspiral waveforms in the presence of spin-induced precession has been developed with the help of a shifted uniform asymptotics (SUA) technique to approximate the Fourier transform, a numerical treatment of the secular evolution coupled to the orbital-averaged spin-precession, and relying on a small eccentricity \edit{expansion~\cite{Klein:2018ybm}.}

\paragraph{Insight from scattering}
\label{subsub:weakfield_scattering}

One can gain surprising insight in the relativistic two-body problem by investigating \textit{unbound} orbits. Studying such systems implies analyzing the scattering of compact objects and the approximation to the two-body problem most naturally applicable to it, the \PM expansion.
Recently, there has been renewed interest in the subject, as it has been realized that \PM information from unbound systems can be transferred to bound ones, as done for instance via Hamiltonians \cite{Damour:2016gwp,Damour:2017zjx,Antonelli:2019ytb} or between gauge-invariant quantities \cite{Kalin:2019rwq,Kalin:2019inp, Bjerrum-Bohr:2019kec,Cho:2021arx,Saketh:2021sri}.
%
Moreover, \PM expansions can be independently obtained from scattering-amplitude calculations~\cite{Arkani-Hamed:2017jhn}, as done at 3PM order~\cite{Bern:2019crd,Bern:2019nnu,Kalin:2020fhe,Cheung:2020gyp,Kalin:2022hph} for the nonspinning sector (see also earlier results at 2\PM order \cite{Cheung:2018wkq,Bjerrum-Bohr:2018xdl,Cristofoli:2020uzm,Kosower:2018adc,Kalin:2020mvi} and \edit{preliminary results} at 4\PM \cite{Bern:2021dqo,Bern:2021yeh,Dlapa:2021npj,Dlapa:2021vgp,Dlapa:2022lmu}). 
Another method that has been successfully used is the worldline quantum field theory approach~\cite{Mogull:2020sak,Jakobsen:2021smu,Mougiakakos:2021ckm}.

The 3PM radiative contribution, which cancels a divergence in the 3PM conservative part, has been obtained in Refs.~\cite{Damour:2020tta,Bjerrum-Bohr:2021din,Herrmann:2021tct,Herrmann:2021lqe,DiVecchia:2021bdo,DiVecchia:2021ndb,DiVecchia:2020ymx}. The radiative effects in PM expansions have been further explored in~\cite{Brandhuber:2021eyq,Damgaard:2021ipf,Riva:2021vnj,Manohar:2022dea} and the radiative contributions to the scattering problem were obtained in a PN expansion~\cite{Bini:2020uiq,Bini:2021gat,Bini:2021qvf,Bini:2021jmj,Bini:2022yrk,Bini:2022xpp,Bini:2022enm}.

A Hamiltonian including spin-orbit coupling at the 2PM approximation has been
 obtained~\cite{Bini:2018ywr} by extracting the dynamical information from a scattering situation
 with the help of ``scattering holonomy''~\cite{Bini:2017xzy,Kalin:2019inp}. Likewise effects at `spin-squared' have been determined at the 2PM level using both amplitude \cite{Kosmopoulos:2021zoq} and the EFT formalism \cite{Liu:2021zxr,Chung:2020rrz,Bern:2020buy,Bautista:2021wfy,Jakobsen:2021lvp,Jakobsen:2021zvh}. To 3PM order, SO and SS contributions have been determined~\cite{Jakobsen:2022zsx}. Effects of arbitrary orders in spin in the scattering angle
 have also been studied by means of purely classical methods~\cite{Vines:2017hyw,Siemonsen:2017yux,Aoude:2022thd}. Those results have
 been recovered at 1PM order through the computation of quantum scattering amplitudes for
 minimally coupled massive spin-$n$ particles and gravitons in the classical limit, as $n$ goes to
 infinity~\cite{Maybee:2019jus}. Notably, the tree-level amplitude has been found to generate the full
 series of black-hole spin-induced multipole moments~\cite{Guevara:2019fsj}.
Tidal effects in PM expansions have also been investigated~\cite{Cheung:2020sdj,Bini:2020flp,Cheung:2020gbf,Bern:2020uwk,Kalin:2020lmz,Aoude:2020ygw,Haddad:2020que,Mougiakakos:2022sic}.

The aforementioned results have naturally prompted discussions and comparisons between the \GW astrophysics and scattering-amplitude communities, which is made possible by the use of pivotal gauge-invariant quantities for comparisons. The scattering angle of unbound compact-object interactions is one such pivotal quantity that, at least in a perturbative sense through the orders so-far considered, is thought to encapsulate the complete conservative dynamics.
Not only does it provide a common ground to exchange information between independent PM calculations \cite{Cristofoli:2019neg} or between \PM and \EOB schemes \cite{Damour:2016gwp,Damour:2017zjx,Vines:2017hyw,Vines:2018gqi,Antonelli:2019fmq,Damgaard:2021rnk,Khalil:2022ylj,Damour:2022ybd}, but it has also proven extremely useful to PN theory.

The \PM expansion of the scattering angle allows one to extract previously unknown \PN information through a synergistic combination of constraints from its simple mass-ratio dependence and gravitational self-force (GSF) results, as first realized at 5PN~\cite{Bini:2019nra,Bini:2020wpo}. Such construct has been exploited to partially calculate the 6PN dynamics~\cite{Bini:2020nsb,Bini:2020hmy}, allowing a nontrivial check at 3PM-6PN order of the 3PM result~\cite{Blumlein:2020znm,Cheung:2020gyp}, as well as to the generic spin-orbit and aligned bilinear-in-spin sectors at NNNL~\cite{Antonelli:2020aeb,Antonelli:2020ybz} (see also~\cite{Siemonsen:2019dsu} for more on the interface between PM and GSF theory).
GSF scattering observables can also provide a powerful handle on PM dynamics across all mass ratios. Calculations of the scattering angle to first-order (second-order) in the mass ratio fully determine the complete two-body Hamiltonian through 4PM (6PM) order~\cite{Damour:2019lcq}.

\paragraph{First laws and gauge invariant comparisons.}

The orbital dynamics of a binary system of compact objects exhibits a fundamental property, known as the first law of binary mechanics, that takes the form of a simple variational relation. This formula relates local properties of the individual bodies (e.g. their masses, spins, redshifts, spin precession frequencies), to global properties of the binary system, (e.g. gravitational binding energy, total angular momentum, radial action variable, fundamental frequencies). The first law was first established for binary systems of nonspinning compact objects moving along circular orbits \cite{LeTiec:2011ab}, as a particular case of a more general variational relation, valid for systems of black holes and extended matter sources \cite{Friedman:2001pf}. This first law was later extended to generic bound eccentric orbits \cite{LeTiec:2015kgg}, including the effect of the \GW tails that appear at the leading 4PN order \cite{Blanchet:2017rcn}, as well as to spinning compact binaries, for spin aligned or anti-aligned with the orbital angular momentum \cite{Blanchet:2012at}. In the context of the small mass-ratio approximation (see Sec. \ref{sec:GSF}), analogous relations were established for a test particle or a small self-gravitating body orbiting a Kerr black hole, by accounting for the conservative part of the first-order gravitational self-force (GSF) \cite{LeTiec:2013iux,Fujita:2016igj}. These various first laws of binary mechanics have proven useful for a broad variety of applications, including to:

\begin{itemize}
    \item Determine the numerical values of the ``ambiguity parameters'' that appeared in the derivations of the 4PN two-body equations of motion \cite{Jaranowski:2012eb,Jaranowski:2013lca,Jaranowski:2015lha,Damour:2014jta,Damour:2016abl,Bernard:2015njp,Bernard:2016wrg,Bernard:2017bvn};
    \item Compute the exact first-order conservative GSF contributions to the gravitational binding energy and angular momentum for circular-orbit nonspinning black hole binaries, allowing for a coordinate-invariant comparison to NR results \cite{LeTiec:2011dp};
    \item Calculate the GSF-induced correction to the frequency of the Schwarzschild \cite{LeTiec:2011dp,Akcay:2012ea} and Kerr \ISCO~\cite{Isoyama:2014mja,vandeMeent:2016hel};
    \item Calibrate the potentials that enter the EOB model for circular orbits \cite{Barausse:2011dq,Akcay:2012ea} and mildly eccentric orbits \cite{Akcay:2015pjz,Bini:2015bfb,Bini:2016qtx}, and spin-orbit couplings for spinning binaries~\cite{Bini:2015xua};
    \item Test the cosmic censorship conjecture in a scenario where a massive particle subject to first-order GSF falls into a nonspinning black hole along unbound orbits \cite{Colleoni:2015ena};
    \item Define the analogue of the redshift variable of a particle for black holes in NR simulations, this allowing further comparisons to the predictions of the PN and GSF approximations \cite{Zimmerman:2016ajr,LeTiec:2017ebm};
    \item Provide a benchmark for the calculations of the first-order GSF-induced frequency shift of the Schwarzschild innermost bound stable orbit \cite{Barack:2019agd} and the second-order GSF contribution to the gravitational binding energy \cite{Pound:2019lzj}.
\end{itemize}

The first law of binary mechanics for circular orbits has been extended to account for finite-size effects such as the rotationally-induced and tidally-induced quadrupole moments of the compact objects \cite{Ramond:2020gtn,Ramond:2022ctc}. Other interesting directions would be to extend the first law to generic precessing spinning compact binaries, and to establish it in the context of the PM approximation and for unbound orbits.

\subsubsection{Environmental effects}

The evolution of relativistic magneto-hydrodynamics, circumbinary disks around supermassive binary black holes have been studied in the weak-field approximation. This relies on the techniques of matched asymptotics to analytically describe a non-spinning binary system up to 2.5PN order~\cite{Johnson-McDaniel:2009tvj,Mundim:2013vca}, on which the magnetic field is then numerically evolved~\cite{Noble:2012xz,Zilhao:2014ida}. These results have later been extended to spinning, non-precessing binary black-holes~\cite{Gallouin:2012kb,Ireland:2015cjj} and to spinning, precessing black-hole spacetimes~\cite{Nakano:2016klh}.

Another astrophysical environmental effect concerns the presence of a third body around a massive binary back hole. Hierarchical triples may undergo several type of resonances~\cite{Kuntz:2021hhm}, that can result for example in \edit{von Zeipel-Kozai-Lidov oscillations~\cite{1910AN....183..345V,Kozai:1962zz,Lidov:1962wjn}}. This is an interchange between the eccentricity of the two-body inner orbit and its inclination relative to the plane of the third body studied. Such an effect is particularly interesting for LISA as it involves high eccentricity systems and can result in an enhancement of gravitational radiation. A general-relativistic treatment of triple hierarchical systems has been performed in a weak-field approximation up to 2.5PN order in the dynamics~\cite{Bonetti:2016eif}. The quadrupolar and octupolar waveforms have also been obtained by numerically integrating the three-body system trajectories from the PN expansion~\cite{Bonetti:2017hnb,Will:2017vjc,Chandramouli:2021kts}. Other studies on the \edit{von Zeipel-Kozai-Lidov} oscillations in \GR have employed an expansion in powers of the ratio between the two semi-major axis up the hexadecapole order~\cite{Bonetti:2017hnb} and later to second-order in the quadrupolar perturbation~\cite{Will:2020tri}. Finally, an effective two-body approach  to the hierarchical three-body problem has been proposed~\cite{Kuntz:2021ohi} and the mass quadrupole to 1PN order was derived under this formalism~\cite{Kuntz:2022onu}.

\subsubsection{Challenges}

Post-Newtonian information in data analysis enters primarily through the use of semi-analytical models (see Secs.~\ref{sec:EOB},\ref{sec:phenom}). To avoid parameter estimation biases due to waveform uncertainties, developments in PN theory will be needed to increase the accuracies of these models for the next generation of \GW detectors (see Sec.~\ref{sec:modelling_requirements} and \cite{Purrer:2019jcp} for a discussion in the context of third-generation ground-based detectors).

One of the most significant challenges in the PN formalism will be the completion of the 5PN order.
First, due to the so-called ``effacement principle''~\cite{Damour:1982wm}, the point-particle approximation breaks at this order.
Finite-size and tidal effects have to be taken into account, the latter being known up to 7.5PN~\cite{Flanagan:2007ix,Vines:2011ud,Damour:2012yf,Bini:2012gu,Henry:2020ski}.
In addition, due to the complexity of the computations involved, it seems highly unlikely to achieve the derivation of the complete gravitational phase at 5PN during the next decade, even if partial results are already obtained. In this regard, the interplay with \GSF and scattering amplitudes will be crucial to make significant progress in this direction.

Exploiting the links between asymptotic symmetries and hereditary effects in the PN formalism should improve the matching to \NR as memory effects are starting to be included in NR waveforms~\cite{Mitman:2021xkq} (see also Sec.~\ref{sec:NR}).

Regarding spinning binary systems, although the orbital dynamics are known up to the 4PN order, it might not
be sufficient for LISA data analysis.
In the near future one should primarily focus,
at lower orders, on the exterior gravitational field, outside the matter source, extending to
infinity. Indeed, new multipole moments will have to be computed to reach the same level of accuracy
for the \GW amplitude as for the flux, notably the 3PN SS contributions to the current quadrupole.
The next step will consist in moving on to 4PN order, to do as well as the current
equations of evolution. For this purpose, the NNL SS piece of the 4PN mass quadrupole will be
required.

Another important topic, in order to combine information coming from the \PN framework and numerical relativity,
will be  to connect the magnitude of the spin used in PN expressions to the rigorous
definitions employed in numerical simulations~\cite{Brown:1992br, Ashtekar:2001is}. The problem of
comparing the spin axis is even more intricate since, for now, its direction has not
been given a satisfactory unambiguous meaning in any of these schemes~\cite{Owen:2017yaj}.

Regarding eccentric waveforms, most current models have focused on a low eccentricity expansion and non-spinning or aligned spin systems.
It is crucial to make progress in both directions. First one has to go beyond the small eccentricity approximation, as some have started investigating~\cite{Moore:2018kvz,Moore:2019xkm}. Second, it will be important to have reliable fully precessing and eccentric waveforms in order to span all the parameter space expected for LISA sources.

Notwithstanding the amount of work and progress already made, studies of the scattering of compact objects still explore a relatively uncharted territory. Obvious extensions of the work reported above will involve pushing scattering calculations of both spinning and nonspinning systems to higher \PM orders in the conservative sector, both from a fully classical and amplitude approach. Such new information would help in obtaining new nontrivial PN information via the continued exploitation of the scattering angle's mass-ratio dependence, as well as continuing the fruitful exchange between the communities. The interconnectedness of the \PM, \PN and \GSF approximations should also be explored in the context of tidal effects (information from PM is available, e.g.,~\cite{Cheung:2020sdj}).
Scattering studies should also explore dissipative radiation-reaction effects. It is important to push these calculations, so that the accuracy gained in the dissipative dynamics goes on par with that of the conservative sector.


\newpage

\subsection{Small-mass-ratio approximation (gravitational self-force)}
\label{sec:GSF}


Coordinators: Marta Colleoni and Adam Pound \\
Contributors: S.~Akcay, E.~Barausse, B.~Bonga, R.~Brito, M.~Casals, G.~Compere, A.~Druart, L.~Durkan, A.~Heffernan, T.~Hinderer, S.~A.~Hughes, S.~Isoyama, C.~Kavanagh, L.~Kuchler, A.~Le Tiec, B.~Leather, G.~Lukes-Gerakopoulos, O.~Long, P.~Lynch, C.~Markakis,  A.~Maselli, J.~Mathews, C.~O'Toole,  Z.~Sam, A.~Spiers, S.~D.~Upton, M.~van de Meent, N.~Warburton, V.~Witzany

\newcommand{\eps}{\epsilon}

\subsubsection{Description}\label{Sec:GSF:Description}

When the secondary object in a binary is significantly smaller than the primary, we can treat the mass ratio $\eps = m_2/m_1 = 1/q$ as a small parameter and seek a perturbative solution for the spacetime metric, $g_{\alpha\beta}=g^{(0)}_{\alpha\beta}+\eps h^{(1)}_{\alpha\beta}+\eps^2 h^{(2)}_{\alpha\beta}+\ldots$ The background metric $g^{(0)}_{\alpha\beta}$ then describes the spacetime of the primary in isolation, typically taken to be a Kerr \BH. At zeroth order, the secondary behaves as a test mass in the background, moving on a geodesic of $g^{(0)}_{\alpha\beta}$. At subleading orders, it generates the perturbations $h^{(n)}_{\alpha\beta}$, which then exert a {\em self-force} back on it, accelerating it away from geodesic motion and driving the inspiral. These perturbations $h^{(n)}_{\alpha\beta}$ also encode the emitted waveforms.

The perturbative field equations and trajectory of the secondary are determined using {\em \GSF theory}, a collection of techniques for incorporating a small gravitating object into an external field~\cite{Poisson:2011nh,Harte:2014wya,Pound:2015tma,Barack:2018yvs,Pound:2021qin}. Using matched asymptotic expansions (or \EFT~\cite{Galley:2008ih}), the small object is reduced to a point particle, or more generally to a puncture in the spacetime geometry, equipped with the object's multipole moments. For a compact object, higher moments scale with higher powers of the small mass $m_2$, such that one additional multipole moment appears at each order in $\eps$. This multipolar particle (or puncture) is found to obey a generalized equivalence principle, behaving as a test body in a certain effective, smooth, vacuum metric $g^{(0)}_{\alpha\beta} + \eps h^{{\rm R}(1)}_{\alpha\beta}+\eps^2 h^{{\rm R}(2)}_{\alpha\beta}+\ldots$~\cite{Mino:1996nk,Quinn:1996am,Detweiler:2000gt,Detweiler:2002mi,Gralla:2008fg,Pound:2009sm, Gralla:2011zr,Harte:2011ku,Detweiler:2011tt,Pound:2012nt,Gralla:2012db,Harte:2014wya,Pound:2017psq,Upton:2021oxf}.

The {\em regular fields} $h^{{\rm R}(n)}_{\alpha\beta}$ that influence the secondary's motion can be calculated directly, using a puncture scheme in which they are the numerical variables~\cite{Barack:2007we,Vega:2007mc}. Alternatively, they can be extracted from the full fields $h^{(n)}_{\alpha\beta}$ using a mode-by-mode subtraction~\cite{Barack:1999wf,Barack:2002bt,Barack:2002mh,Barack:2001gx,Heffernan:2012su, Heffernan:2012vj, Heffernan:2021olv} or other methods~\cite{Casals:2009zh,Wardell:2014kea}, as reviewed in~\cite{Barack:2009ux,Wardell:2015ada,Pound:2021qin}. In either approach, the perturbative field equations can be solved using the methods of \BHPT~ \cite{Regge:1957td,Zerilli:1970wzz,Teukolsky:1973ha,Wald:1973,Chrzanowski:1975wv,Wald:1978vm,Kegeles:1979an,Sasaki:2003xr,Martel:2005ir,Barack:2005nr,BHPToolkit,Pound:2021qin}.

\begin{center}
\begin{figure}[t]
$\vcenter{\hbox{\includegraphics[width=0.4\textwidth]{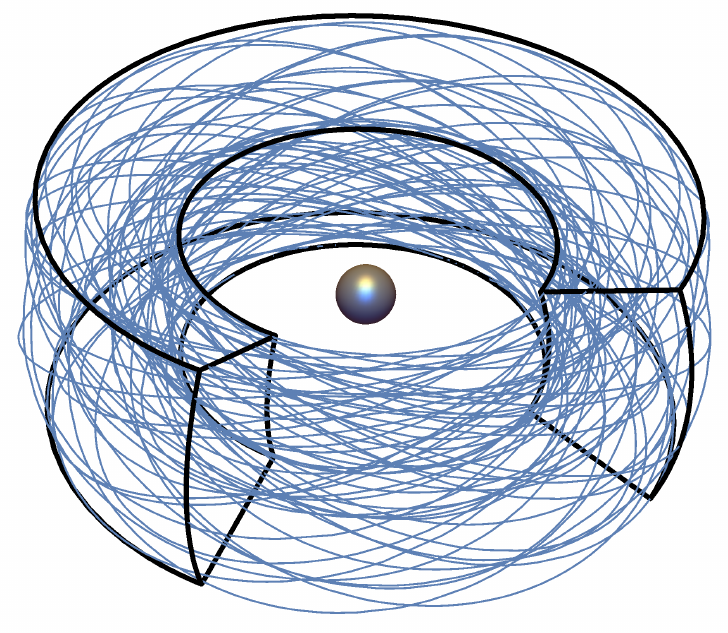}}}$
$\vcenter{\hbox{\includegraphics[width=0.4\textwidth]{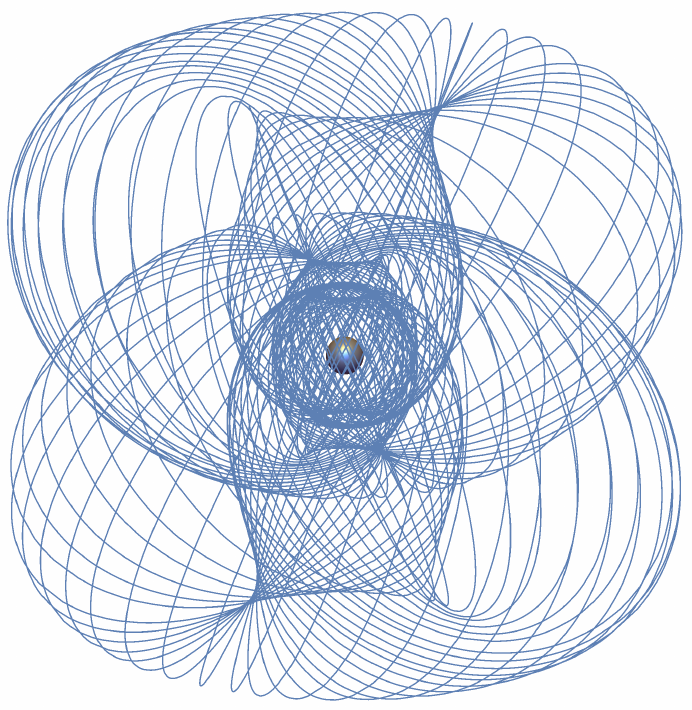}}}$
\caption{\label{Fig:Kerr geodesics} Geodesics of Kerr spacetime. Left: a non-resonant geodesic, which fills a toroidal shape. Right: a resonant geodesic. Images taken from~\cite{Barack:2018yvs}.}
\end{figure}
\end{center}

A key feature of this model is its clean separation of time scales: because the self-force is small, inspirals occur slowly, over $\sim 1/\eps$ orbits. On the time scale of a few orbits, the secondary's trajectory is approximately a bound geodesic of $g^{(0)}_{\alpha\beta}$ (illustrated in Fig.~\ref{Fig:Kerr geodesics}), which in Kerr spacetime is generically triperiodic, undergoing radial, polar, and azimuthal motion with frequencies $\Omega_r$, $\Omega_\theta$, and $\Omega_\phi$~\cite{Schmidt:2002qk,Mino:2003yg,Fujita:2009bp,Grossman:2011ps,Pound:2021qin}. Over the long inspiral, the frequencies slowly evolve due to dissipation. The field equations can therefore be solved using a two-timescale expansion~\cite{Mino:2008rr,Hinderer:2008dm,Pound:2010pj,Pound:2015wva,Miller:2020bft,two-timescale-0,Pound:2021qin}, $h^{(n)}_{\alpha\beta} = \sum_{k^A\in\mathbb{Z}}h^{(n,k^A)}_{\alpha\beta}(J^B,x^a)e^{ik^A \psi_A}$. Here $x^a$ are spatial coordinates, and all time dependence is encoded in the set of system parameters $J^A$ (the secondary's orbital energy, angular momentum, and Carter constant, the primary's mass and spin, etc.), and the set of orbital phases $\psi_A$. The parameters $J^A$ evolve slowly, on the inspiral time scale $t\sim 2\pi/(\eps\Omega)$, while the phases $\psi_A$ evolve on the orbital time scale $t\sim 2\pi/\Omega$, with evolution equations of the form~\cite{Hinderer:2008dm,VanDeMeent:2018cgn,Miller:2020bft,two-timescale-0,Pound:2021qin}
\begin{align}
\frac{dJ^A}{dt} = \eps G_{(1)}^A(J^B) + \eps^2 G^A_{(2)}(J^B) + O(\eps^3),\quad
\frac{d\psi_A}{dt} = \Omega^{(0)}_A(J^B)+\eps\Omega^{(1)}_A(J^B)+O(\eps^2).\label{GSFeq:two-timescale evolution}
\end{align}
Formulating the problem in this way enables practical methods of solving the Einstein equations~\cite{Miller:2020bft,Pound:2021qin} as well as facilitating rapid waveform generation~\cite{VanDeMeent:2018cgn,Pound:2021qin,Lynch:2021ogr}.

The two-timescale description will frame much of the discussion below. However, there are some special cases of interest that it does not apply to, such as scattering orbits~\cite{Hopper:2017iyq,Hopper:2017qus,Long:2021ufh,Barack:2022pde,Long:2022sdq}. Even in an inspiral, there are regions of parameter space in which the approximation breaks down: at the end of the inspiral, when the secondary transitions into a final plunge into the primary~\cite{Buonanno:2000ef,Ori:2000zn,Apte:2019txp,Burke:2019yek,Compere:2019cqe,Compere:2021zfj}, and during resonances, which occur when at least two of the frequencies $\Omega_A$ have a rational ratio, causing a linear combination of phases, $k^A \psi_A$, to become approximately stationary~\cite{Tanaka:2005ue,Flanagan:2010cd,Lukes-Gerakopoulos:2021ybx}. Such resonances can arise from a variety of physical causes, outlined below, with significant observational consequences.

Many of the core tools in \GSF modelling, as well as advanced codes described in the sections below, have been consolidated in the open-source Black Hole Perturbation Toolkit~\cite{BHPToolkit}, which provides a hub for \GSF code development, as well as in the Black Hole Perturbation Club~\cite{BHPClub}. In particular, the \FEW package~\cite{Katz:2021yft}, which exploits \GSF models' multiscale structure, has provided a flexible framework for rapid waveform generation.

\subsubsection{Suitable for what sources?}\label{Sec:GSF:Sources}

Historically, the \GSF approximation has been motivated by \EMRIs, with mass ratios in the interval $10^{-7} \lesssim \eps \lesssim 10^{-3}$ as far as LISA sources are concerned (see Table \ref{Tab:EMRIparameters}). The \GSF model's accuracy for a typical \EMRI can be estimated from the evolution equations~\eqref{GSFeq:two-timescale evolution}. On the inspiral time scale, the solutions  for the orbital phases, and therefore the \GW phase, take the form
\begin{equation}\label{PA phase}
\psi_A = \frac{1}{\eps}\left[\psi^{(0)}_A + \eps \psi^{(1)}_A + O(\eps^2)\right]\!.
\end{equation}
Following Ref.~\cite{Hinderer:2008dm}, the leading term in this expansion is referred to as {\em adiabatic} order (0PA), and the $n$th subleading term as $n$th {\em post-adiabatic} order ($n$PA). An adiabatic approximation, which has large phase errors $\sim \eps^0$ over an inspiral, is expected to suffice for detection of \EMRIs using the semi-coherent searches discussed in Sec.~\ref{sec:EMRI efficiency requirements}. A 1PA approximation, which will have phase errors $\sim \eps\ll 1$ rad over the final year of inspiral, should suffice for parameter extraction even for loud \EMRIs with \SNR $> 50$. 

A large and growing body of evidence \cite{Fitchett:1984qn,Anninos:1995vf,Favata:2004wz,LeTiec:2011bk,LeTiec:2011dp,Sperhake:2011ik,LeTiec:2013uey,Nagar:2013sga,vandeMeent:2016hel,Zimmerman:2016ajr,LeTiec:2017ebm,Rifat:2019ltp,vandeMeent:2020xgc,Warburton:2021kwk,Wardell:2021fyy,Albertini:2022rfe,Ramos-Buades:2022lgf} suggests that the \GSF approximation can also accurately model \IMRIs (see Table \ref{Tab:IMRIparameters}). It can even provide a useful model of \CO binaries with \textit{comparable} masses when it is re-expressed as an expansion in powers of the symmetric mass ratio $\nu \equiv m_1 m_2 / (m_1 + m_2)^2 = \eps + O(\eps^2)$. We refer to Ref.~\cite{LeTiec:2014oez} for an early review 
and Refs.~\cite{Albertini:2022rfe,Ramos-Buades:2022lgf} for recent analyses of the accuracy of 1PA models in the $1\lesssim q\lesssim 100$ regime. Further study is required to assess whether 1PA models meet expected accuracy requirements for specific classes of astrophysical sources in this range of mass ratios. 

Besides compact binary inspirals, the \GSF approximation can be used for other classes of sources.
A leading-order \GSF model, comprising a point source on a geodesic trajectory, should be suitable for \XMRIs \cite{Amaro-Seoane:2019umn,Gourgoulhon:2019iyu} and fly-by burst signals~\cite{Hopper:2017iyq,Hopper:2017qus,Long:2021ufh,Barack:2022pde,Long:2022sdq}. \GSF theory may also be relevant for modelling gravitational waves from cosmic strings~\cite{Blanco-Pillado:2018ael,Chernoff:2018evo,Blanco-Pillado:2019nto}. Inspirals of less compact bodies into \MBHs (e.g., white or brown dwarfs or main-sequence stars prior to tidal disruption), and three-body ``binary \EMRIs'', can be modelled by including sufficiently high multipole moments in the particle or puncture~\cite{Steinhoff:2012rw,Chen:2018axp,Rahman:2021eay}. 





\subsubsection{Status}\label{Sec:GSF:Status}

\begin{center}
\begin{figure}[t]
\includegraphics[scale=0.679]{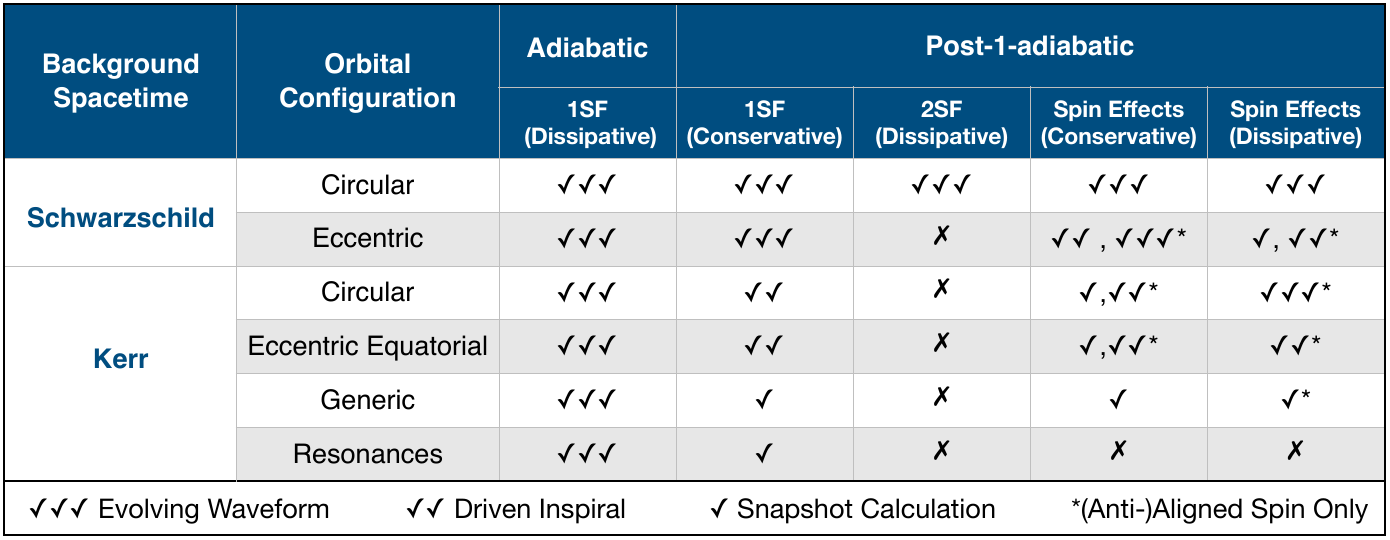}
\caption{\label{Fig:GSF progress} Progress in modelling EMRIs using \GSF methods. `1SF' and `2SF' indicate calculations involving the first or second-order self-force, respectively, and `spin effects' indicates calculations that take into account the secondary's spin (the spin of the primary is accounted for in all `Kerr' calculations. `Snapshot calculations' (single tick) are ones in which self-force effects are calculated on fixed geodesic orbits.  }
\end{figure}
\end{center}

The current status of \GSF models is summarized in Fig.~\ref{Fig:GSF progress}, which also summarizes the necessary inputs for a waveform model at 0PA and 1PA order. 

\paragraph{Adiabatic approximation}


At leading order in the two-timescale expansion, only the geodesic frequencies $\Omega^{(0)}_A$ and time-averaged dissipative piece $G_{(1)}^A$ of the first-order self-force are required to drive the evolution~\cite{Mino:2003yg,Mino:2005an}.  The secondary's motion in this approximation can be regarded as an adiabatic inspiral through a sequence of geodesic orbits
~\cite{Hinderer:2008dm,Hughes:2005qb}. The waveform can then be built from a corresponding  sequence of ``snapshot'' Fourier mode amplitudes together with the leading phases in Eq.~\eqref{PA phase}. Alternatively, outside the two-timescale scheme, a time-domain field equation can be solved with an adiabatically inspiraling source particle~\cite{Sundararajan:2008zm,Barausse:2011kb,Taracchini:2014zpa,Harms:2014dqa,Rifat:2019ltp}.

$G_{(1)}^A$ is most conveniently computed using ``flux-balance'' formulas~\cite{Galtsov:1982hwm,Drasco:2005is,Sago:2005gd,Isoyama:2013yor,Isoyama:2018sib}, in which the ``work'' done by the dissipative self-force balances the fluxes of gravitational waves to infinity and down the horizon (or flux-like quantities in the case of the Carter constant).  Methods of computing fluxes in \GSF theory are well developed~\cite{Press:1973zz,Teukolsky:1974yv,Nakamura:1987zz,Martel:2003jj,Poisson:2004cw,Sundararajan:2007jg,Sundararajan:2008zm,Zenginoglu:2011zz,Harms:2014dqa}, and have been numerically implemented for generic (inclined and eccentric) geodesics about a Kerr \BH ~\cite{Hughes:1999bq,Glampedakis:2002ya,Drasco:2005kz,Fujita:2009us}. The fluxes have also been calculated analytically (by expanding $h^{(1)}_{\alpha\beta}$ in a \PN series~\cite{Mano:1996vt,Mino:1997bx,Sasaki:2003xr}) to high \PN order for circular~\cite{Fujita:2012cm, Fujita:2014eta, Shah:2014tka}, eccentric~\cite{Tagoshi:1995sh,Munna:2020juq,Munna:2020som}, and generic orbital configurations~\cite{Sago:2005fn,Ganz:2007rf,Sago:2015rpa}.


After decades of progress~\cite{Cutler:1994pb,Finn:2000sy,Hughes:2001jr,Fujita:2010xj,Pan:2010hz,Gralla:2016qfw,Burke:2019yek, Chua:2020stf,Fujita:2020zxe}, adiabatic inspirals and waveforms are now being computed for generic orbits in Kerr, both numerically~\cite{Hughes:2021exa} and within the analytical BHPT-\PN framework mentioned above~\cite{Isoyama:2021jjd}.

Once fluxes and waveform mode amplitudes have been computed across the parameter space, adiabatic waveforms can be efficiently generated using a combination of neural network and reduced-order techniques \cite{Chua:2020stf,Katz:2021yft}, a method that lends itself to \GPU acceleration techniques; see Sec.~\ref{sec:hardware_accelerators}. This approach should ultimately meet the efficiency requirements described in Sec.~\ref{sec:EMRI efficiency requirements}. However, due to the high-dimensional parameter space, significant work remains to populate the parameter space. As a consequence, data analysis development has so far relied on semirelativistic “kludge” models that seek to capture all the qualitative features of EMRI waveforms while sacrificing accuracy~\cite{Barack:2003fp,Babak:2006uv,Chua:2015mua,Chua:2017ujo,EKS}. These models typically combine (i) post-Newtonian information about the orbital dynamics, (ii) qualitative (and sometimes quantitative) features of the fully relativistic 0PA orbital evolution, and (iii) a post-Newtonian (e.g., Peters-Mathews~\cite{Peters:1963ux}) multipolar formula for the emitted waveform. The most advanced of such models, the Augmented Analytical Kludge~\cite{Chua:2017ujo}, is available in the the \FEW package~\cite{Katz:2021yft}.

\paragraph{Post-1-adiabatic self-force effects: 1SF}

The 1PA terms in the two-timescale expansion ($\Omega^{(1)}_A$ and $G^A_{(2)}$) take as input the conservative piece of the first-order self-force (1SF, $\propto h^{(1)}_{\alpha\beta}$) and the dissipative piece of the second-order self-force (2SF, $\propto h^{(2)}_{\alpha\beta}$). Once these ingredients have been computed, a 1PA waveform-generation scheme takes the same form as a 0PA one~\cite{Miller:2020bft,Pound:2021qin}.

Numerical computations of the full first-order self-force have made tremendous progress, evolving from Lorenz-gauge calculations \cite{Barack:2005nr,Barack:2007tm,Barack:2010tm,Dolan:2012jg,Akcay:2013wfa,Osburn:2014hoa,Isoyama:2014mja,Wardell:2015ada} to more efficient methods relying on radiation gauges \cite{Keidl:2010pm,Shah:2010bi,Pound:2013faa,Merlin:2014qda,vandeMeent:2015lxa,vandeMeent:2016pee,Merlin:2016boc,vanDeMeent:2017oet}. These developments culminated in 1SF calculations along generic bound geodesics in Kerr spacetime \cite{vandeMeent:2017bcc}, and work is still ongoing to further improve 1SF methods~\cite{Toomani:2021jlo,Dolan:2021ijg,PanossoMacedo:2022fdi}. 

High-precision numerics and BHPT-\PN methods have also made possible high-order \PN expansions of numerous conservative invariants, such as the Detweiler redshift~\cite{Shah:2012gu,Shah:2013uya,Bini:2014nfa,Hopper:2015icj,vandeMeent:2015lxa,Johnson-McDaniel:2015vva,Kavanagh:2016idg, Bini:2018zde,Bini:2019lcd}, 
the periastron advance \cite{vandeMeent:2016hel,Bini:2019zjj}, and other invariants~\cite{Nolan:2015vpa,Shah:2015nva,Bini:2015kja,Bini:2018ylh,Bini:2019lkm,Munna:2022xts}, which have played a key role in the synergies with \PN theory and \EOB described in Secs.~\ref{sec:weak_field} and~\ref{sec:EOB}.

Modern 1SF calculations mostly rely on frequency-domain methods and mode-by-mode subtraction or puncture schemes to calculate $h^{\rm R(1)}_{\alpha\beta}$. However, there have also been advances in \TD calculations based on either finite-difference schemes~\cite{Barack:2017oir, Hughes:2005qb,Long:2021ufh,DaSilva:2023xif} or spectral methods~\cite{Canizares:2008dp,Canizares:2010hu,Canizares:2011fd,Field:2009kk,Diener:2011cc,Markakis:2014nja,DaSilva:2023xif,Vishal:2023fye}, along with improved time-stepping methods~\cite{OBoyle:2022yhp,Markakis:2023pfh,DaSilva:2023xif,OBoyle:2023jqo}. There is also ongoing work to directly obtain the retarded Green function \cite{Casals:2013mpa, Wardell:2014kea, Casals:2019heg, Yang:2013shb,OToole:2020ejc}, which would then allow direct evaluation of the self-force~\cite{Mino:1996nk, Quinn:1996am}.

As at 0PA, calculations of the self-force and of \GW amplitudes across the parameter space have been used to simulate self-forced inspirals~\cite{Warburton:2011fk,Osburn:2015duj,VanDeMeent:2018cgn,Lynch:2021ogr,Lynch:2023gpu} and generate waveforms~\cite{Lackeos:2012de,Osburn:2015duj}. To date, this has only been done for equatorial orbits and quasi-spherical orbits~\cite{Lynch:2023gpu}, due to the computational expense of current 1SF calculations for generic orbits. 

\begin{center}
\begin{figure}[t]
\includegraphics[width=\textwidth]{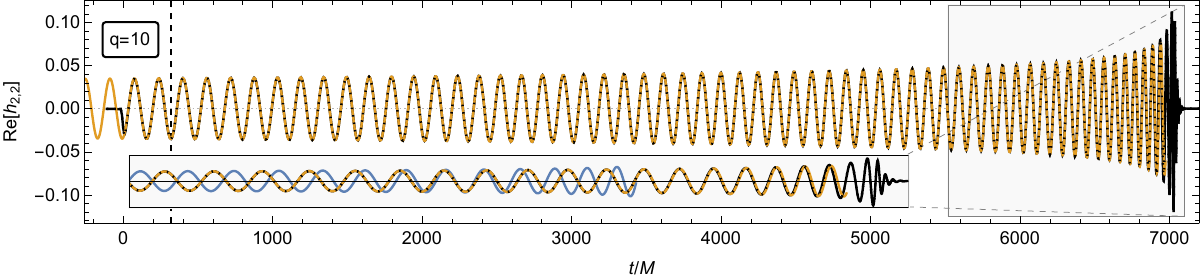}
\caption{\label{Fig:1PA waveform}1PA \GSF waveform for a quasicircular, nonspinning binary with mass ratio $q=10$ (orange). The inset shows a zoomed portion of the waveform near the merger. Also included for comparison are the 0PA \GSF waveform (blue, inset only) and the waveform for the same binary produced using an NR simulation in the SXS catalog~\cite{Boyle:2019kee} (SXS:BBH:1107, in black). The three waveforms are aligned in time and phase at $t=320M$, when the orbital separation is $\approx 13.83M$. Image reproduced from Ref.~\cite{Wardell:2021fyy}.}
\end{figure}
\end{center}

\paragraph{Post-1-adiabatic self-force effects: 2SF} The dissipative piece of the second-order self-force ($G_{(2)}^A$) contributes to the \GW phasing at the same 1PA order as the first-order conservative self-force, but calculations of it are less mature. After years of development of the governing formalism~\cite{Rosenthal:2006iy,Detweiler:2011tt,Pound:2012nt,Gralla:2012db,Pound:2012dk,Pound:2015fma,Pound:2017psq,Upton:2021oxf} and of practical implementation methods~\cite{Pound:2014xva,Warburton:2013lea,Wardell:2015ada,Pound:2015wva,Miller:2016hjv,Miller:2017coe,Pound:2021qin,Durkan:2022fvm,Spiers:2023mor}, Refs.~\cite{Pound:2019lzj,Warburton:2021kwk,Wardell:2021fyy} recently carried out the first concrete calculations of physical second-order quantitities in the restricted case of quasicircular orbits around a Schwarzschild \BH. These calculations culminated in the first complete 1PA waveforms in Ref.~\cite{Wardell:2021fyy}. Figure \ref{Fig:1PA waveform} shows a comparison between one of these 1PA waveforms and an NR waveform for $q=10$. As alluded to in Sec.~\ref{Sec:GSF:Sources}, the 1PA approximation agrees well with NR even at this moderate mass ratio.

These 1PA waveforms have also more recently been extended to include a slowly spinning primary~\cite{Mathews:Capra2022}. Sec.~\ref{Sec:GSF:Challenges} discusses the next major barrier in \GSF calculations: 2SF calculations with a rapidly spinning primary, eccentricity, and inclination.

\paragraph{Transient resonances}

Resonances are a ubiquitous feature of the strong-field dynamics around \BHs
in GR~\cite{Brink:2013nna,Brink:2015roa}, and their observational imprints on waveforms can be significant.
Transient self-force resonances, between the orbital frequencies $\Omega_r$ and $\Omega_\theta$, will have strong observational consequences for EMRIs~\cite{Mino:2005an,Tanaka:2005ue,Levin:2008mq,
Apostolatos:2009vu,Flanagan:2010cd}.
Essentially \textit{all} LISA-type \EMRIs pass through at least one dynamically significant resonance,
leading to a large, $O(1/\sqrt{\eps})$ (i.e.,\ 0.5PA) contribution to the waveform phase~\cite{Ruangsri:2013hra,Berry:2016bit,Mihaylov:2017qwn}.
 The magnitude of this effect depends sensitively on the orbital phase at the resonance. As a result, modelling it requires 1PA accuracy prior to resonance. There are ongoing efforts to understand the impact of these resonances and include them in evolutions~\cite{Gair:2011mr,Flanagan:2012kg,vandeMeent:2013sza,Isoyama:2013yor,Lewis:2016lgx,Isoyama:2018sib,Isoyama:2021jjd,Lukes-Gerakopoulos:2021ybx,Nasipak:2021qfu,Nasipak:2022xjh,Gupta:2022fbe}.

There also occur resonances between the $r$ and $\varphi$ frequencies and between the $\theta$ and $\varphi$ frequencies. These do not change the intrinsic inspiral dynamics, but they can lead to a strong net emission of linear momentum that results in a `kick' to the system's center of mass~\cite{Hirata:2010xn,vandeMeent:2014raa}.
The maximum kick velocity can reach $\sim 30,000 \times \eps^{3/2}$~km/s (depending on the primary spin and on the orbital eccentricity): for IMRI systems, these values could be comparable to the escape velocities from globular
clusters, which are typically of the order of a few tens of km/s~\cite{Merritt:2004xa,Antonini:2016gqe}.

\paragraph{Merger and ringdown}

The two-timescale expansion breaks down as the secondary object approaches the separatrix between bound and plunging orbits. There, the secondary enters a gradual transition across the separatrix~\cite{Buonanno:2000ef,Ori:2000zn,Burke:2019yek,Compere:2019cqe} followed by an approximately geodesic plunge. The transition motion has been mostly studied within the \EOB framework using resummed \PN expansions~\cite{Buonanno:2000ef,Buonanno:2005xu,Nagar:2006xv,Damour:2007xr,Damour:2009kr,Bernuzzi:2010ty,Bernuzzi:2010xj,Bernuzzi:2011aj,Pan:2013rra} or within the \GSF approach using simplifying approximations~\cite{Ori:2000zn,Sundararajan:2008bw,Kesden:2011ma,Taracchini:2014zpa,Apte:2019txp,Compere:2019cqe,Burke:2019yek}. Merger-ringdown waveforms have been generated from the final plunge at first order~\cite{Folacci:2018cic,Rom:2022uvv}, and complete inspiral-merger-ringdown waveforms have been generated by solving time-domain equations for first-order perturbations sourced by the combined inspiral-transition-plunge motion~\cite{Barausse:2011kb,Taracchini:2014zpa,Harms:2014dqa,Rifat:2019ltp,Islam:2022laz}. Only recently, Refs.~\cite{Compere:2021iwh,Compere:2021zfj,Kuchler:2024esj} took a first step toward 1PA/2SF inspiral-merger-ringdown waveforms by developing a systematic expansion of the transition motion and metric perturbation that matches to the two-timescale expansion for quasicircular, equatorial inspirals into a Kerr \BH. 

\paragraph{Spin and finite-size effects}
When its multipole structure is accounted for, the secondary obeys the Mathisson-Papapetrou-Dixon equations of a test body~\cite{Mathisson:1937zz,Papapetrou:1951pa,Dixon:1970zza,Dixon:1970zz,Dixon:1974xoz} in the effective metric; such a result is expected to hold to all perturbative orders for both \BHs and material bodies~\cite{Thorne:1984mz,Harte:2011ku,Harte:2014wya}. 

The spin contributes first-order conservative and second-order dissipative forces in the equations of motion, leading to 1PA, $\mathcal{O}(\eps^0)$ contributions to the final inspiral phase~\cite{Mathews:2021rod}. 
Spin also generally breaks the integrability of Kerr geodesic motion \cite{Suzuki:1996gm}; the resulting potential phenomena, like prolonged resonances, can leave a significant imprint on the gravitational waves \cite{1983RSPSA.385..229R,Suzuki:1999si,Ruangsri:2015cvg,Witzany:2019nml,Zelenka:2019nyp} and may contribute $\mathcal{O}(1)$ to the \GW phase. Away from these resonances, the spinning-particle motion is perturbatively separable to linear order in spin~\cite{Witzany:2019nml} and easily incorporated into a two-timescale approximation~\cite{WitzanyPoundBarack,Mathews:2021rod}.


For dissipative effects, a spin-flux balance law has been established~\cite{Akcay:2019bvk}, and \GW fluxes were computed for a variety of binary configurations~\cite{Mino:1995fm,Tanaka:1996ht,Han:2010tp,Harms:2015ixa,Harms:2016ctx,Lukes-Gerakopoulos:2017vkj,Nagar:2019wrt,Akcay:2019bvk,Zelenka:2019nyp,Piovano:2020zin,Skoupy:2021asz}. Waveforms from generic inspirals into a Schwarzschild \BH have also been computed including 1SF and first-order conservative spin effects~\cite{Warburton:2017sxk} but excluding dissipative spin effects. Waveforms including {\em all} 1PA spin effects from circular equatorial inspirals into Schwarzschild ~\cite{Mathews:2021rod} and Kerr \BHs~\cite{Piovano:2021iwv} have been produced, and the spin's complete 1PA contribution to the \GW phase has been calculated for eccentric equatorial inspirals into a Kerr \BH~\cite{Skoupy:2022adh}. A formulation of the first-order conservative spin-forced motion for generic orbits about a Kerr \BH~\cite{Drummond:2022xej,Drummond:2022efc} paved the way for calculations of the spin's complete impact on fully generic 1PA waveforms, beginning with recent snapshot computations of energy and angular momentum fluxes~\cite{Skoupy:2023lih}.

The secondary's higher moments are unlikely to contribute at 1PA order. The tidal quadrupole's contribution to the acceleration scales with $\eps^4$~\cite{Binnington:2009bb} and can therefore only become relevant for non-compact objects. The spin also induces a quadrupole deformation of the secondary, which creates an acceleration $\sim\eps^2$~\cite{Steinhoff:2012rw,Rahman:2021eay} that is unlikely important for EMRIs (up to resonances) but might be relevant for IMRIs~\cite{Rahman:2021eay}.

\subsubsection{Environmental effects}\label{Sec:GSF:BeyondGR}



In the preceding sections, we have focused on the simplest \GSF model: an isolated vacuum \BH orbited by a single body. This will not fully describe astrophysical small-mass-ratio binaries.  Accreting matter will be present at some level, and there are likely to be additional bodies nearby. Accounting for these effects will be necessary to achieve the stringent accuracy requirements that \EMRIs, in particular, place on \GSF models. 

There are three ways we can modify \GSF models to include such effects: (i) perturbatively, by adding ``small''  matter fields or metric perturbations to the spacetime, which then exert small forces on the secondary; (ii) non-perturbatively, by modifying the background spacetime and introducing ``large'' matter fields; (iii) by changing boundary conditions, imagining (for example) modifications very near the \BH horizon or non-asymptotically flat perturbations due to external matter. 
Sec.~\ref{sec:modling_beyond_GR} discusses how these classes of modifications can also be used to incorporate beyond-GR effects.


Although important aspects of the astrophysical environment of \EMRIs are uncertain, it is expected that there will be other stellar-mass bodies nearby. These bodies create an additional metric perturbation, which should be sufficiently small to be treated as a linear perturbation. The influence of the perturbation is generally negligible, except at moments when two or more of the orbital frequencies characterising the perturbed \EMRI system become commensurate.  At those times, an \EMRI experiences a tidal resonance~\cite{Yang:2017aht}, an orbital resonance akin to the mean motion resonance known from planetary dynamics~\cite{Yang:2019iqa}. Calculations across the parameter space showed that a single resonance can dephase the waveform
by several radians over the inspiral~\cite{Bonga:2019ycj,Gupta:2021cno,Gupta:2022fbe,Gupta:2022jdt}, and that most \EMRIs will cross multiple tidal resonances before plunge. 
If LISA can reliably measure these resonances, they can be used to learn about the tidal environment of \EMRI systems.

Accretion and gas interactions will also be present at some level in any astrophysical \EMRI system. 
Order-of-magnitude estimates suggest 
they will be negligible  unless the primary \BH powers an AGN (in which case it would be surrounded by an 
accretion disk with which the secondary is likely to interact)~\cite{Barausse:2014tra,Barausse:2014pra,Barausse:2007dy,Sukova:2021thm}. 
Since AGNs are believed to make up 1--10\% of local galaxies, only a comparable fraction of \EMRIs are expected to be significantly 
affected by accretion and gas interactions. For those systems, accretion onto the primary and secondary, as well as 
dynamical friction from the disk and planetary-like migration within it, are expected to cause secular effects 
comparable to those of GW fluxes~\cite{Barausse:2014tra,Barausse:2014pra,Barausse:2007dy}. Recently work has begun to incorporate the disk and other potential nonvacuum effects into \GSF models by adding torques to 0PA Kerr models~\cite{Speri:2022upm} or by working perturbatively on an exact nonvacuum background, assuming a spherically symmetric dark matter distribution~\cite{Cardoso:2022whc,Destounis:2022obl}.  

The direct gravitational pull from AGN disks is likely to be negligible~\cite{Barausse:2014tra}, but if more dense disks/rings exist in nature, the effect may be significant~\cite{Barausse:2006vt}. Exact spacetime solutions describing thin disks or rings around \BHs ~\cite{Lemos:1993qp,Basovnik:2016awa,Semerak:2020kbp} show that orbits in such situations are non-integrable, exhibiting characteristic phenomena like chaos and prolonged resonances \cite{Semerak:2015dya}. 
If these effects are significant in realistic scenarios, \GSF models should be able to account for them through perturbative corrections on top of a Kerr background~\cite{Yunes:2005ve,LeTiec:2020bos,Gupta:2022fbe,Polcar:2022bwv}.

\subsubsection{Challenges}\label{Sec:GSF:Challenges}

The principal goal of \GSF waveform modelling is to develop complete 1PA models for generic orbital configurations around a spinning, Kerr \BH. To be complete, these models must include the spin of the companion, transitions across resonances, and the final plunge. They must also be sufficiently modular to incorporate beyond-GR and environmental effects, and there are strong motivations to explore other regions of the parameter space, such as scatter orbits and comparable masses. We summarize here the main challenges in developing and implementing such models.

\paragraph{Post-1-adiabatic calculations}

1PA models are currently missing two ingredients: the effects of the companion's spin for generic binary configurations, and dissipative 2SF effects; see Fig.~\ref{Fig:GSF progress}. Both must be incorporated into a unified two-timescale expansion of the field equations, the orbital motion, and the spin evolution. 

While substantial work remains to calculate 1PA spin effects for generic binary configurations, these calculations can leverage existing methods for point-particle sources. Therefore, 2SF calculations represent the overriding obstacle to 1PA accuracy. Practical 2SF calculations have, thus far,  been restricted to quasicircular orbits in a Schwarzschild background~\cite{Pound:2019lzj,Warburton:2021kwk,Wardell:2021fyy}. As discussed in Sec.~\ref{sec:sources_EMRIs}, the majority of \EMRIs are expected to have significant eccentricity and may have highly precessing orbital planes when they enter the LISA band, meaning 2SF techniques need to be extended to cover generic orbital configurations. They must also be extended to the realistic case of a Kerr background. 

Eccentricity and inclination bring multiple challenges. Most recent self-force calculations have utilized  decompositions into Fourier modes (the natural setting of the two-timescale expansion) and angular harmonics. Eccentric, inclined orbits can require $\sim 10^5$ modes, all of which will couple to one another in the second-order source. The sum of Fourier modes can also suffer from poor convergence: in the existing 2SF puncture scheme, the source for $h^{\rm R(2)}_{\alpha\beta}$ has finite differentiability on the puncture's worldline, 
leading to  slow power-law convergence (the Gibbs phenomenon). 
At first order, similar problems were overcome using methods of extended solutions~\cite{Barack:2008ms, Warburton:2011hp,Hopper:2012ty,Akcay:2013wfa,Osburn:2014hoa,vandeMeent:2015lxa,vandeMeent:2017bcc} that restore exponential convergence. These have inspired a new scheme, applicable at second order, known as the method of extended effective sources~\cite{Leather:2023dzj}, which has been demonstrated in the case of a scalar-field toy model for eccentric orbits about a Schwarzschild \BH. Work now remains to apply it to gravitational perturbations, both at first and second order, and to orbits in Kerr spacetime. 


The extension to Kerr spacetime brings more challenges. Second-order calculations in Schwarzschild have so far relied on directly solving the perturbative Einstein equations, which are not separable in Kerr spacetime. At first order, this problem was overcome using radiation-gauge methods, in which the metric perturbation is reconstructed from a solution to the (fully separable) Teukolsky equation~\cite{Teukolsky:1973ha,Chrzanowski:1975wv, Kegeles:1979an}. One path to 2SF calculations is to extend that method to second order~\cite{Campanelli:1998jv,Green:2019nam,Toomani:2021jlo,Spiers:2023cip}. The standard method of metric reconstruction~\cite{Chrzanowski:1975wv, Kegeles:1979an} fails beyond linear order, but recent work found an extension to all orders~\cite{Green:2019nam,Toomani:2021jlo}. 
Recent work on second-order flux-balance laws~\cite{Grant:Capra2023,Sam:Capra2024} also suggests that 1PA rates of change of energy and angular momentum (though perhaps not the Carter constant) can be computed directly from a solution to the second-order Teukolsky equation, without the need for metric reconstruction. 

An additional obstacle is that in radiation-gauge implementations, the first-order metric perturbation has gauge singularities  that extend away from the particle. There is a rigorous procedure to extract physical 1SF quantities despite these singularities~\cite{Pound:2013faa,Merlin:2016boc,vanDeMeent:2017oet}, but the singularities become ill defined in the second-order field equations~\cite{Toomani:2021jlo}. Several avenues are being explored to resolve this problem~\cite{Toomani:2021jlo,Dolan:2021ijg,Osburn:2022bby,Dolan:2023enf}.

There are also challenges common to all these 2SF calculations: at second order, the field equations have a noncompact source that falls off slowly at large distances~\cite{Pound:2015wva} and is burdensome to compute due to the strong nonlinear singularity at the particle~\cite{Miller:2016hjv}. Recent work~\cite{Upton:2021oxf,Spiers:2023cip} has shown that both problems might be mitigated by using gauges adapted to the lightcone structure of the perturbed spacetime. 
However, additional work will be required to implement these gauge choices in a practical numerical scheme.


\paragraph{Covering the parameter space} Even once numerical implementations of all necessary ingredients are available, spanning the full \EMRI parameter space remains a considerable challenge at both 0PA and 1PA orders. This is due to the high dimensionality of the parameter space and the high computational burden of self-force calculations, particularly at second order. Covering the \EMRI parameter space will likely involve  a combination of (i) using analytic results to reduce the region of the parameter space where high-precision interpolation of numerical data is required, (ii) better interpolation methods, and (iii) improvements in computational efficiency of current numerical calculations. We address each of these three below in the context of 0PA inspirals, 1PA inspirals, and waveform calculations.

Computing adiabatic inspirals requires interpolating the rates of change ($G_{(1)}^A$) of the orbital energy, angular momentum, and Carter constant across the four-dimensional parameter space of the primary spin and three orbital elements. 
For an adiabatic model sufficiently accurate to build 1PA corrections upon, we need to interpolate $G_{(1)}^A$ to better than a relative accuracy of $\eps^{-1}$. The central challenge is then constructing such a high-accuracy interpolant over the 4D parameter space. Recent work has demonstrated advantages of Chebyshev interpolation for this purpose~\cite{Lynch:2021ogr}. The region of the parameter space, and the accuracy of the interpolation of the numerical data, can also be reduced by constructing global fits informed by analytic results. 

At 1PA order we must also compute the change in the mass of the primary during the inspiral~\cite{Miller:2020bft}, and so the parameter space grows to five dimensions. 
There are also corrections due the spin of the secondary, but fortunately, these can be added on separately. 
At 1PA order the accuracy requirements of the contributions are much lower, at $\sim10^{-2}-10^{-3}$ relative \cite{Osburn:2015duj}. This should allow analytic results to assist in reducing the parameter space that numerical results need to cover. Such analytical results could be obtained by extending BHPT-\PN calculations. Alternatively, they could be obtained from \EOB dynamics, following the programme in Refs.~\cite{Han:2011qz,Han:2014ana,Han:2016zee,Han:2017evx,Zhang:2020rxy,Zhang:2021fgy,Shen:2023pje}. Regions of high eccentricity may also be more easily covered using advanced time-domain codes~\cite{Canizares:2010yx,Field:2009kk,Diener:2011cc,Markakis:2014nja,Barack:2017oir,Long:2021ufh,OBoyle:2022yhp,DaSilva:2023xif,Vishal:2023fye}.

\paragraph{Extending the parameter space} 

The challenges above are centered on ``vanilla'' regions of the parameter space: the inspiral phase away from resonances, which is amenable to a two-timescale expansion. However, it is also critical to include accurate transitions across resonances, particularly in the \EMRI regime. In addition to the dominant $r$-$\theta$ resonances that occur in a self-forced inspiral, prolonged resonances may occur due to spin, external matter, or a non-Kerr central object~\cite{Apostolatos:2009vu,Destounis:2021mqv,Lukes-Gerakopoulos:2021ybx}. 
Similarly, the transition from inspiral to plunge can also be important, particularly for more comparable masses~\cite{Rifat:2019ltp,vandeMeent:2020xgc,Albertini:2022rfe}. Both resonances and the plunge will require matching the two-timescale expansion to specialized approximations in those parameter regions~\cite{vandeMeent:2013sza,Berry:2016bit,Pound:2021qin,Lukes-Gerakopoulos:2021ybx,Compere:2021iwh,Compere:2021zfj,Gupta:2022fbe}.

There are also reasons to compute self-force effects on scattering orbits, whether to model hyperbolic, burst sources or to inform \PM and \PN dynamics, as described in \ref{subsub:weakfield_scattering} (and possibly to infer properties of  bound, self-forced orbits~\cite{Kalin:2019inp,Gonzo:2023goe}). 
Explorations of scatter orbits have only recently begun~\cite{Barack:2022pde,Long:2022sdq,Barack:2023oqp,Whittall:2023xjp}. While bound-orbit self-force calculations can utilise the orbit's discrete Fourier spectrum, scatter orbits have a continuous spectrum~\cite{Hopper:2017qus,Hopper:2017iyq}, suggesting some clear advantages to simulations in the time domain~\cite{Barack:2019agd,Barack:2022pde,Long:2022sdq}. 

In addition to including more of the two-body GR parameter space, considerable work must be done to include possible beyond-GR and environmental effects in \GSF models. Fortunately, the \GSF model is relatively modular. This means most additional effects can be added separately, as described in Sec.~\ref{Sec:GSF beyond GR}, and are readily incorporated into frameworks such as~\FEW~\cite{Katz:2021yft}. However, modelling these additional effects will be particularly challenging if they are not amenable to a two-timescale treatment, or if they are too large to be treated perturbatively.


\newpage

\subsection{Perturbation theory for post-merger waveforms (quasi-normal modes)}
\label{sec:QNMs}

Coordinators: Stephen R.~Green and Laura Sberna  \\
Contributors: E.~Berti, R.~P.~Macedo, P.~Mourier, N.~Oshita, M.~van de Meent


\subsubsection{Description}\label{Sec:QNM:Description}

Following a compact binary merger, the remnant object settles into a
stationary state through a process known as the ``ringdown''. The
ringdown signal is interesting because it involves a collection of
discrete modes that encode information about the final object. For a
BH merger, the no-hair theorems of general relativity predict that
the final state is itself a Kerr BH (see,
e.g.,~\cite{Chrusciel:2012jk}), yielding precise predictions for the ringdown frequencies. We consider this case here.


\begin{center}
  \begin{figure}[t]
    \centering
    \includegraphics[width=0.458\textwidth]{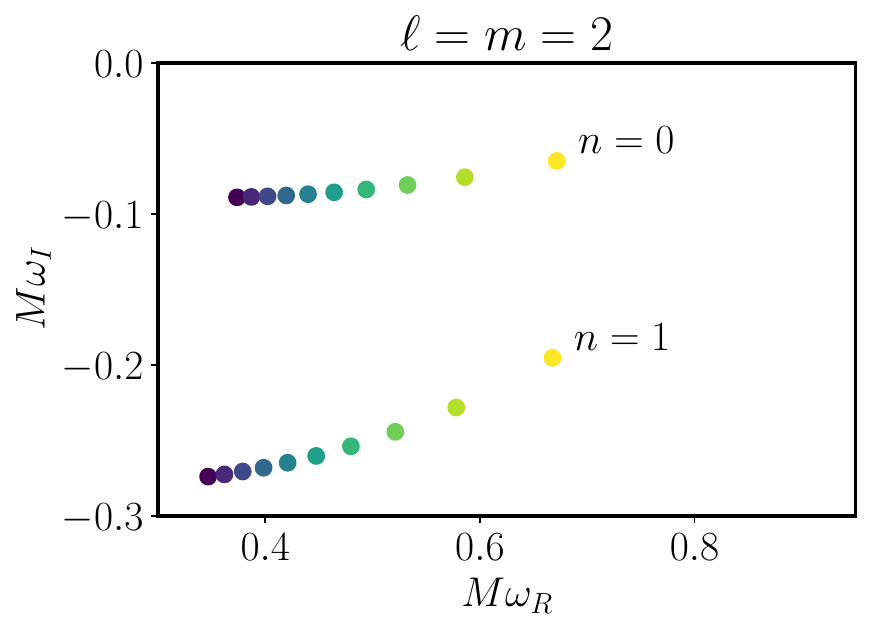}
    \includegraphics[width=0.522\textwidth]{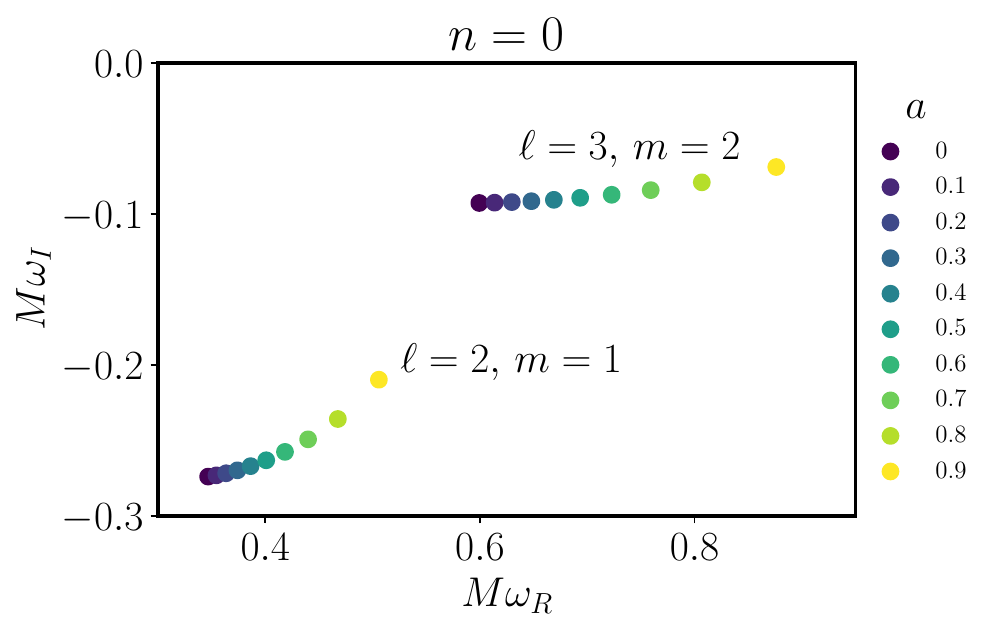}
    \caption{\label{Fig:QNMs} Real and imaginary parts of the Kerr quasinormal mode frequencies, coloured by the BH spin (from $a=0$ in dark blue to $a=0.9$ in yellow). Left: $\ell=m=2$ fundamental and first overtone modes. Right: modes with different angular numbers $(\ell,m)$.}\label{fig:qnms}
  \end{figure}
\end{center}

At late times, the ringdown is best described using \BHPT. This involves expanding the spacetime metric $g_{\alpha\beta}=g^{(0)}_{\alpha\beta}+\epsilon g^{(1)}_{\alpha\beta}+\epsilon^2 g^{(2)}_{\alpha\beta}+...$~\cite{Regge:1957td,Zerilli:1970se,Zerilli:1970wzz} and solving order-by-order in the Einstein equation. At zeroth order, the background metric $g^{(0)}_{\alpha\beta}$ describes the remnant Kerr BH. At first order, $g^{(1)}_{ab}$ satisfies the Einstein equation linearized about Kerr. Rather than work directly with the metric perturbation, it is convenient to use instead the Newman-Penrose formalism, expressing the perturbation in terms of the (complex) Weyl curvature scalars. Equations for the $s=\pm 2$ linearized Weyl scalars $\psi_0$ and $\psi_4$ famously decouple yielding a separable wave-like equation discovered by Teukolsky~\cite{Teukolsky:1972my,Teukolsky:1973ha,Press:1973zz,Teukolsky:1974yv}. Both $\psi_0$ and $\psi_4$ uniquely describe $g^{(1)}_{ab}$ up to gauge and perturbations to other Kerr spacetimes~\cite{Wald:1973}, and in particular describe the \GW degrees of freedom, with $\psi_4$ most relevant for gravitational radiation at infinity. Similarly, spin-weight $s=\pm 1$ and $s=0$ Teukolsky equations describe electromagnetic and scalar-field perturbations, respectively.

Homogeneous solutions to the Teukolsky equation can be obtained by imposing boundary conditions corresponding to the physical requirement that radiation is \edit{outgoing} at the BH horizon and outgoing at infinity. For each spin weight, this eigenvalue problem admits a countably infinite, discrete spectrum of complex-frequency \emph{quasinormal modes (QNMs)}~\cite{Press:1971wr,Chandrasekhar:1975zza}, see Fig.~\ref{fig:qnms}. Each QNM is labelled by three integers: two spin-weighted spheroidal-harmonic quantum numbers used to separate the angular dependence ($\ell\ge2$, $-\ell\le m\le \ell$) and an ``overtone number'' $n\ge0$, which sorts the frequencies in ascending order of
the absolute value of their imaginary part. Negative (real) frequency modes are related to modes with negative $m$ by the symmetry $-\omega^*_{n\ell m}=\omega_{n\ell-m}$.
Kerr quasinormal frequencies (QNFs) have a strictly negative imaginary part, hence the modes decay with time, due to absorption by the BH horizon and radiation to infinity. These frequencies are moreover uniquely determined by the Kerr mass and spin in agreement with no-hair theorems. For detailed reviews on QNMs, see, e.g.,~\cite{Kokkotas:1999bd,Nollert:1999ji,Berti:2009kk,Berti:2018vdi}.

In contrast to normal modes, QNMs of Kerr  do not form a complete basis, and therefore cannot fully describe the ringdown for all times, even at linear order. At \emph{early} times, the linear response to a perturbation is dominated by a ``direct'' emission of radiation (also known as the prompt response)~\cite{Jensen:1985in,Leaver1986,Nollert:1999ji}. At \emph{late} times, the perturbation is dominated by a power-law ``tail'' (associated with a branch cut in the frequency plane of the Green's function of the Teukolsky equation), which arises due to the asymptotic properties of the wave-equation potential at large distances from the horizon~\cite{Leaver1986,Price:1971fb}. QNMs are most important at \emph{intermediate} times, and for realistic inspirals they tend to dominate the majority of the post-merger signal. They have therefore proven extremely relevant for waveform modeling, either to analyze the ringdown in isolation or for combined inspiral-merger-ringdown waveforms. Indeed, a single QNM can be used to infer the mass and spin of the remnant, whereas additional modes or inspiral information enable consistency checks on \GR~\cite{Detweiler:1980gk,Dreyer:2003bv,Berti:2005ys,Isi:2019aib,LIGOScientific:2021sio}.   


\subsubsection{Suitable for what sources?}\label{Sec:QNM:Sources}

LISA is expected to detect the ringdown of MBHB and IMRI sources for which the merger occurs during the LISA observational time, and for which remnant QNFs fall within the LISA frequency band~\cite{Berti:2005ys,Berti:2007zu,Berti:2016lat}. For EMRIs, the ringdown is likely to be too weak to resolve, a potential exception being those with near-extremal primaries~\cite{Compere:2017hsi,Oshita:2022yry}.


One can easily estimate the mass range of remnants with ringdowns
within the LISA band. Indeed, the longest lived (fundamental, $\ell=m=2, n=0$)
QNM has frequency
$f_0 \simeq 0.4/(2 \pi  M_{\rm {final}})$, with the exact
value depending on the remnant's spin.
For this to lie within the
LISA band, $[10^{-4},10^{-1}]$~Hz, the mass of the remnant must therefore
fall within the range $[10^{8},10^{5}]~\text{M}_{\odot}$. Ringdown
modelling will be particularly important for MBHBs at the higher
end of the mass spectrum, since their inspiral will take place outside
the LISA band and contribute little to the SNR~\cite{Baibhav:2020tma}.


\subsubsection{Status}\label{Sec:QNM:Status}
At linear order in \BHPT and for most of the duration of the signal, the ringdown is well described by a superposition of QNMs. For a given BH mass and spin, the Kerr QNFs were first computed numerically in \cite{Detweiler:1980gk}. The most accurate and reliable numerical method is Leaver's continued fraction approach~\cite{Leaver:1985ax,Nollert:1993zz,Onozawa:1996ux} (see Mathematica notebook and data at \cite{Berti_website}) and in particular its spectral refinement~\cite{Cook:2014cta,Hughes:1999bq} (implemented in the Python package \texttt{qnm}~\cite{Stein:2019mop}).
Kerr QNFs can also be estimated analytically using WKB (Wentzel–Kramers–Brillouin) or WKB-inspired approximations~\cite{Blome:1981azp,Schutz:1985km,Seidel:1989bp,Kokkotas:1991vz,Dolan:2009nk,Dolan:2010wr}.
QNM modeling should be performed in a certain preferred BMS
frame~\cite{MaganaZertuche:2021syq, Mitman:2022kwt}.  Afterwards it
can be transformed to a different desired reference frame,
e.g., a frame co-precessing with the
binary~\cite{Hamilton:2021pkf,Hamilton:2023znn}, a
frame set prior to a merger kick~\cite{Gerosa:2016vip},
or the post-Newtonian preferred frame appropriate to modeling the early
inspiral~\cite{Mitman:2022kwt}.


In addition to the frequencies, the QNM component of the ringdown signal also comprises the amplitude and phase of each mode. Unlike the frequencies, these quantities are not uniquely determined by the remnant, but rather depend on the particular initial conditions that led to its formation.
In principle, the amplitudes and phases can be computed from the spacetime right after the formation of a common BH horizon~\cite{Leaver1986,Berti:2006wq}, and hence from the binary parameters~\cite{Kamaretsos:2011um,Kamaretsos:2012bs}. However, this data is hard to predict analytically, and moreover challenging to extract from NR simulations. Common practice is therefore to calibrate the desired mode amplitudes and phases by fitting against NR simulations~\cite{Berti:2007fi,Kamaretsos:2011um,London:2014cma,Baibhav:2017jhs,London:2018gaq,MaganaZertuche:2021syq,Forteza:2022tgq,Baibhav:2023clw}. Alternatively, initial data can be estimated at the level of the Teukolsky equation using an Ori-Thorne procedure~\cite{Taracchini:2014zpa,Lim:2019xrb,Hughes:2019zmt,Apte:2019txp,Lim:2022veo}.


For binary BH mergers of comparable masses, the linear ringdown model---a superposition of a finite number of QNMs with arbitrary amplitudes and phases---has been extensively compared against NR simulations. At late times ($\gtrsim 10 M$ after the peak of the waveform), the general consensus is that the linear model provides a good, stable, and consistent fit to the numerical waveform. At these times, the waveform is well described by a small number of overtones (dominated by the $\ell=m=2$ multipole and overtones with $n\lesssim 2$), and fitting for the frequencies and decay rates allows for accurate inference of the remnant mass and spin~\cite{Buonanno:2006ui,Baibhav:2017jhs,Giesler:2019uxc,Cook:2020otn,JimenezForteza:2020cve,Ota:2021ypb,Li:2021wgz,MaganaZertuche:2021syq,Baibhav:2023clw}. More surprisingly, several studies indicate that the linear model applies even at much earlier times, provided higher $(\ell, m, n)$ modes are included in the model~\cite{Giesler:2019uxc,Cook:2020otn,JimenezForteza:2020cve,MaganaZertuche:2021syq}. 
However, the relevance of the linear model at these times ($\lesssim 10 M$ after the peak of the waveform) and, in particular, the role of spherical-spheroidal mode mixing, nonlinear modes and higher overtones ($n>2$), is still actively debated~\cite{Bhagwat:2019dtm,MaganaZertuche:2021syq,Forteza:2021wfq,Mitman:2022qdl,Ma:2022wpv,Ma:2023vvr,Ma:2023cwe,Baibhav:2023clw,Nee:2023osy,Zhu:2023mzv}.

A complete linear-order model of the post-merger signal should also include back-scattering of radiation against the background potential. This is well approximated by a power-law tail  $h\sim t^{-n_{\rm tail}}$ at late times, where $n_{\rm tail}=7$ for generic gravitational perturbations of Kerr~\cite{Price:1971fb,Barack:1999st,Tiglio:2007jp}. So far the tail contribution has not been confidently identified in NR simulations of binary BH mergers, so it is usually neglected in waveform modeling. 


At very early times (around the peak amplitude) we expect higher order \BHPT to become relevant. Beyond-linear-order perturbations satisfy the same equations as at linear order, but now with source terms made up of lower-order perturbations~\cite{Gleiser:1995gx,Campanelli:1998jv,Brizuela:2009qd,Nakano:2007cj,Pazos:2010xf,Loutrel:2020wbw,Spiers:2023cip}. It is therefore natural to expect the signal to deviate from a simple superposition of linear QNMs. Such deviations could include new driven frequencies that are combinations of the first-order QNFs~\cite{Lagos:2022otp,Bucciotti:2023ets}, or corrections to the first order modes, in terms of the mode amplitudes, phases, and frequency spectrum~\cite{Sberna:2021eui,Ripley:2020xby}. Indeed, a driven second-order mode (which in quasicircular mergers appears in the $\ell=m=4$ multipole, sourced by the square of the $\ell=m=2$ fundamental mode) was recently identified in binary BH simulations~\cite{London:2014cma,Ma:2022wpv,Cheung:2022rbm,Mitman:2022qdl,Khera:2023oyf}.
Notably, the amplitude of the second-order mode is comparable to the
amplitude from linear order, raising questions about perturbative convergence.

Detailed second-order calculations remain challenging, and are complicated by aspects of metric reconstruction and regularization of singularities. Nevertheless, second-order Kerr perturbations in the ringdown context were recently calculated numerically~\cite{Ripley:2020xby}. Additional analytic progress in the higher-order ringdown has included the development of a bilinear form under which Kerr QNMs are orthogonal \cite{Green:2022htq,Cannizzaro:2023jle} (see also~\cite{London:2020uva}), new methods to reconstruct the metric in the presence of a first-order source \cite{Green:2019nam}, and the use of analytic approximations~\cite{Bucciotti:2023ets}. These calculations have considerable overlap with the self-force problem, where significant progress has recently been made at second order, see Sec.~\ref{Sec:GSF:Status}.

\subsubsection{Challenges}\label{Sec:QNM:Challenges}
The main challenges in post-merger signal modeling using \BHPT concern the inclusion of higher modes, the instability and quick damping of overtones, non-QNM ringdown components (e.g., from tails or the prompt response), and nonlinear effects.

Higher $(\ell, m, n)$ modes are predicted by \BHPT to arise in generic binary mergers. However, there is no consensus on whether it is consistent to include more than a few overtones ($n\ge 1$) in a purely linear model \cite{Giesler:2019uxc,Baibhav:2023clw,Nee:2023osy}. The challenge lies in the fact that higher overtones decay rapidly, and therefore are relevant at early times, when nonlinearities are also expected to become more significant. 
%
Despite providing good fits to the signal, higher overtones
might simply play the role of fitting noise, nonlinearities, or non-QNM components (such as tails or the prompt response) in the signal. 
For asymmetric binaries (with precession, eccentricity, or unequal mass ratios) angular modes beyond $\ell=m=2$ may also have significant amplitudes (see e.g.~\cite{Capano:2021etf,MaganaZertuche:2021syq,Li:2021wgz,Oshita:2022pkc}), and their fundamental ($n=0$) modes have lifetimes similar to the fundamental $\ell=m=2$ mode.

Recent theoretical studies also indicate that the QNM spectrum itself could be unstable~\cite{Nollert:1996rf,Aguirregabiria:1996zy,Vishveshwara:1996jgz,Jaramillo:2020tuu,Jaramillo:2021tmt,Cheung:2021bol,Konoplya:2023owh}, with overtones particularly sensitive to nonlinear perturbations, small environmental effects, and deviations from vacuum general relativity. However the experimental implications of this instability are not fully understood.

The size of power-law tail contributions due to back-scattering has yet to be estimated for binary BHs in full GR. Reference~\cite{Baibhav:2023clw} cautioned that, at the linear level, the tail contribution can be comparable to that of high-overtone QNMs with $n\simeq 5$. \edit{The tail might play an even larger role in eccentric binary mergers, as suggested by perturbative solutions \cite{Albanesi:2023bgi} and some NR simulations \cite{Carullo:2023tff}.} This should motivate a more careful study of the contribution of back-scattering effects in the post-merger signal.

No study has yet quantified which (if any) LISA-band ringdown sources will require higher-order perturbation theory. Ideally, starting from a first-order perturbation, a nonlinear model would predict detailed corrections including the amplitude of driven modes, shifts in amplitudes and phases for first-order modes, and any frequency drifts. Some initial progress towards such a model includes agnostic fits of numerical relativity waveform catalogues~\cite{Baibhav:2023clw}, numerical studies~\cite{Ripley:2020xby}, and developments in BH perturbation theory~\cite{Green:2022htq,Spiers:2023cip}.

Some studies have speculated that nonlinearities could be stronger for near-extreme remnants, whose spectrum contains long-lived modes with commensurate frequencies. In this limit, nonlinear effects could lead to gravitational turbulence~\cite{Yang:2014tla} and connect with the Aretakis instability of extremal BHs~\cite{Aretakis:2011gz}. To assess these possibilities, it will be necessary to extend calculations beyond scalar-field toy models~\cite{Yang:2014tla} to full general relativity, see for example Ref.~\cite{Redondo-Yuste:2023seq}. 



\newpage

\subsection{Effective-one-body waveform models}
\label{sec:EOB}
Coordinators: Tanja Hinderer, Geraint Pratten\\
Contributors: S.~Akcay, A.~Antonelli,  S.~Bernuzzi, A.~Buonanno, J.~Garcia-Bellido, A.~Nagar, L.~Pompili


\subsubsection{General description}
\label{subsec:generalEOB}
\begin{figure}
    \centering
    \includegraphics[width=0.98\textwidth,trim=30 70 30 100,clip=true]{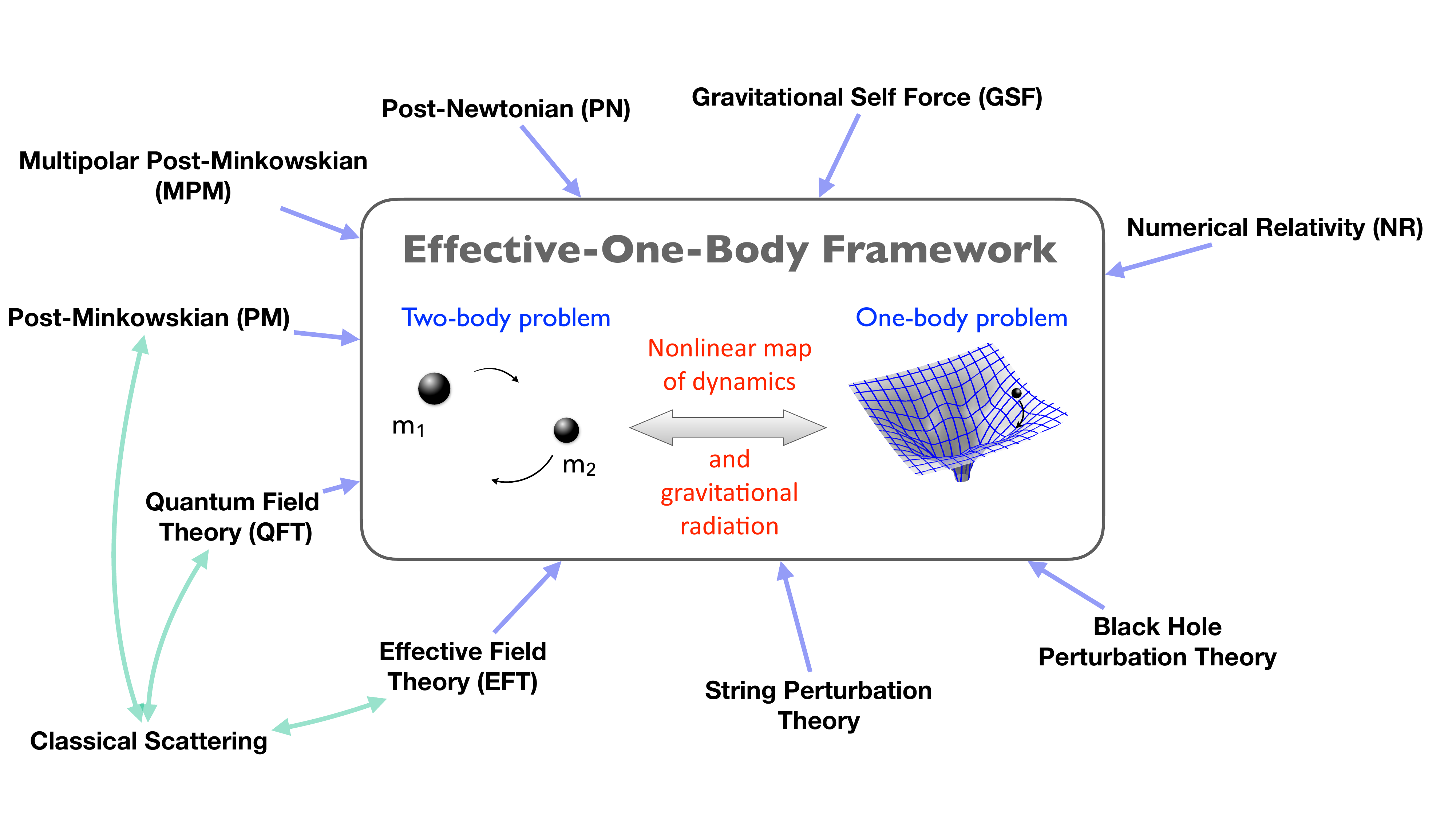}
    \caption{Theoretical inputs to the EOB framework. The EOB theory draws on a variety of perturbative results from numerous approaches in different regimes, as well as from NR simulations, as illustrated here.}
\label{fig:EOBmodels}
\end{figure}

The effective-one-body (EOB) approach was originally introduced in Refs.~\cite{Buonanno:1998gg,Buonanno:2000ef,Damour:2000we,Damour:2001tu,Buonanno:2005xu}
with the aim of providing \GW detectors with semianalytic waveform models for the entire coalescence of compact-object binaries  (i.e., the inspiral, plunge, merger and
ringdown), resumming PN information around the strong-field test-body limit. Since the breakthrough in NR in 2005~\cite{Pretorius:2005gq,Campanelli:2005dd,Baker:2005vv}, the
EOB framework has incorporated information from the NR simulations, thus producing highly-accurate waveform models
for \GW observations (e.g., see the review articles\edit{~\cite{Damour:2008yg,Buonanno:2014aza,Damour:2016bks}} and discussion below). Over the years, the EOB framework
has been extended to the scattering problem, and has incorporated analytical information from other methods, such as the PM approach and GSF theory, as illustrated in Fig.~\ref{fig:EOBmodels}; \edit{see e.g.~\cite{Berkovits:2022ivl,Buonanno:2022pgc} for overviews of the role of string perturbation theory, QFT, and worldline EFT}.

The EOB waveform models consist of three main building blocks: 1) the Hamiltonian, which describes the conservative
dynamics, 2) the radiation-reaction (RR) force, which accounts for the
energy and angular momentum losses due to \GW emission, and 3) the
inspiral-merger-ringdown waveform modes, built upon improved PN resummations
for the inspiral part, and functional forms calibrated to NR waveforms
for the merger-ringdown signal. We now briefly review these three key ingredients.

A fundamental pillar of the EOB approach is the map of the real two-body dynamics into that of an effective test mass or test spin in a deformed Schwarzschild or Kerr background, with the deformation parameter being the symmetric mass ratio $\nu = \mu/M$, where $\mu=m_1m_2/M$ is the reduced mass and $M = m_1+m_2$ is the total mass. More specifically, the real or EOB Hamiltonian, $H_{\rm EOB}$,
is related to the effective Hamiltonian, $H_{\rm eff}$,  by~\cite{Buonanno:1998gg}
\begin{equation}
\label{EOBmap}
H_{\rm EOB}=M \sqrt{1+2\nu \left(\frac{H_{\rm eff}}{\mu}-1\right)}.
\end{equation}
Interestingly, such a relation agrees with calculations in quantum electrodynamics~\cite{Brezin:1970zr} aimed at deriving an approximate
binding energy for charged particles with comparable masses in the eikonal approximation; it has been shown to hold exactly at 1PM order~\cite{Damour:2016gwp}, and it has been extensively used in scattering-theory computations~\cite{Damour:2016gwp,Vines:2017hyw,Vines:2018gqi,Damour:2017zjx}. The above energy-map~\eqref{EOBmap} achieves a concise resummation of PN information into the Hamiltonian via a small number of terms. In the center-of-mass frame, the EOB equations of motion read:
\begin{equation}
\label{EOBEOM}
\frac{d\boldsymbol{r}}{d t}=\frac{\partial H_{\rm EOB}}{\partial \boldsymbol{p}}\,, \quad \quad
\frac{d\boldsymbol{p}}{d t}=-\frac{\partial H_{\rm EOB}}{\partial \boldsymbol{r}}
    +\bm{\mathcal{F}}\,, \qquad    \frac{d\boldsymbol{S}_{1,2}}{d t}=\frac{\partial H_{\rm EOB}}{\partial \boldsymbol{S}_{1,2}}\times \boldsymbol{S}_{1,2},
\end{equation}
where $\boldsymbol{r}$ and $\boldsymbol{p}$ are the canonical variables, notably the relative position and momentum, respectively, $\bm{\mathcal{F}}$ is the RR force, and $\boldsymbol{S}_i$ with $i = 1,2$ are the spins of the compact objects. For example, in the nonspinning limit, where the dynamical variables reduce to $(r,\phi,p_r,p_\phi)$, the effective Hamiltonian in the gauge
of Refs.~\cite{Buonanno:1998gg,Damour:2000we} reads
\begin{equation}
H_\text{eff} = \mu\,\sqrt{A_\nu(r) \left[\mu^2 + A_\nu(r) \bar{D}_\nu(r)\, p_r^2 + \frac{p_\phi^2}{r^2} + Q_\nu(r,p_r)\right]}\,,
\end{equation}
where the potentials $A_\nu(r)$ and 
\edit{$\bar{D}_\nu(r)$}
differ from the Schwarzschild ones due to PN corrections depending on $\nu$. The higher-order (yet) unknown PN corrections in $A_\nu(r)$ might be informed to NR simulations. The potential $Q_\nu(r)$ is a non-geodesic term that needs to be introduced at 3PN order~\cite{Damour:2000we} to preserve the mapping (\ref{EOBmap}), and reduces to zero in the test-mass limit.


Precessing-spin EOB waveforms for the inspiral-merger-ringdown were first built in Ref.~\cite{Buonanno:2005xu} using the nonspinning EOB Hamiltonian~\cite{Buonanno:1998gg,Buonanno:2000ef, Damour:2000we} augmented with a spinning PN Hamiltonian~\cite{Blanchet:2013haa}. The EOB Hamiltonian for spinning objects was first developed in Ref.~\cite{Damour:2001tu}, and then in Refs~\cite{Damour:2008qf,Nagar:2011fx,Damour:2014sva,Balmelli:2015zsa}.
These papers follow the structure of the Hamiltonian advocated in Ref.~\cite{Damour:2001tu}
that in the large mass-ratio limit reduces to the one of a (nonspinning)
test mass on a Kerr background.
Furthermore, another
line of research, which started in Refs.~\cite{Barausse:2009aa,Barausse:2009xi},
built EOB Hamiltonians for spinning objects such that, in the large mass-ratio limit, they reduce to the one of a test spin on a Kerr background~\cite{Barausse:2009aa,Barausse:2011ys}.
The two different spinning Hamiltonians were comprehensively compared in Refs.~\cite{Rettegno:2019tzh,Khalil:2020mmr}.

In the original EOB model~\cite{Buonanno:2000ef}, the radiation-reaction force of Eq.~(\ref{EOBEOM})
was given by a suitable Pad\'e resummation of the PN-expanded energy flux at 2.5PN order,
following the seminal work of Ref.~\cite{Damour:1997ub}.
Subsequently, the (quasi-circular) radiation-reaction force, that is the
flux of angular momentum, has been expressed as the sum of factorized and resummed multipoles according
to the procedure introduced in Refs.~\cite{Damour:2007xr,Damour:2008gu}.
This approach was then first extended to spinning bodies in Ref.~\cite{Pan:2010hz} and
then
improved in Refs.~\cite{Taracchini:2012ig,Taracchini:2013rva,Nagar:2016ayt,Messina:2018ghh,Nagar:2019wrt,Cotesta:2018fcv} by
means of additional factorizations and resummation of the orbital and spin parts,
and the inclusion of higher-order PN terms.

The factorization of each waveform multipole proposed in Ref.~\cite{Damour:2008gu} reads
\begin{equation}
	\label{hlm}
	h_{\ell m}^{\text{insp-plunge}}(t)=h_{\ell m}^{(N,\epsilon)}\,S_\text{eff}^{(\epsilon)}\, T_{\ell m}\, e^{i\delta_{\ell m}}\,
	(\rho_{\ell m})^\ell\,{h}^{\rm NQC}_{\ell m}\,,
\end{equation}
where $h_{\ell m}^{(N,\epsilon)}$ is the Newtonian contribution and
$\epsilon$ denotes the parity of the mode. The factor $T_{\ell m}$
resums an infinite number of leading-order logarithms arising from tail effects~\cite{Damour:2007xr}
(see also Ref.~\cite{Faye:2014fra}), the term $e^{i\delta_{\ell m}}$
is a residual phase correction due to sub-leading order logarithms in hereditary contributions,
and the functions $\rho_{\ell m}$ are the residual amplitude corrections. These functions
were originally obtained as PN-expansions~\cite{Damour:2008gu}, but in some versions of the EOB models, it
was found useful
to further resum them using Pad\'e approximants for improved strong-field robustness~\cite{Nagar:2016ayt,Messina:2018ghh,Nagar:2019wrt}. Another approach
consists of calibrating effective high-order PN parameters entering the $\rho_{\ell m}$'s
to improve their accuracy, doing so either using NR data~\cite{Taracchini:2012ig,Taracchini:2013rva,Cotesta:2018fcv,Pompili:2023tna} or
second-order GSF results~\cite{vandeMeent:2023ols}.

The factor ${h}^{\rm NQC}_{\ell m}$ is the phenomenological next-to-quasi-circular (NQC) correction~\cite{Damour:2002vi} to the waveform that is informed by NR simulations, and it is designed to correctly shape the
waveform during the late plunge up to merger, where the motion is not quasi-circular and the resummed quasi-circular waveform lacks information.
The complete EOB waveform is constructed by attaching the merger-ringdown mode, $h_{\ell m}^{\rm merger-RD}(t)$, to the inspiral-plunge one, $h_{\ell m}^{\rm insp-plunge}(t)$, at a suitable matching time $t=t_{\rm match}$, around the peak of the EOB orbital frequency (which approximately corresponds to merger time), that is~\cite{Buonanno:2000ef}:
\begin{equation}
    h_{\ell m}(t) = h_{\ell m}^{\rm insp-plunge-merger}(t) \, \Theta \left( t_{\rm match}^{\ell m}-t \right)
    + h_{\ell m}^{\rm ringdown}(t) \, \Theta \left( t-t_{\rm match}^{\ell m} \right) \,,
\end{equation}
where $\Theta(t)$ is the Heaviside step function.
Inspired by results for the infall of a
test mass into a BH~\cite{Davis:1972dm} and the close-limit approximation~\cite{Price:1994pm}, several EOB waveform models used a
superposition of QNMs for the dominant mode
~\cite{Buonanno:2000ef,Damour:2001tu,Buonanno:2006ui,Buonanno:2007pf,Damour:2009kr,Pan:2011gk,Barausse:2011kb,Taracchini:2012ig,Damour:2012ky,Pan:2013rra,Taracchini:2013rva,Babak:2016tgq}. More recently, an NR-informed fit of the amplitude and phase has become standard ~\cite{Damour:2014yha}. This framework is similar in spirit to the rotating source approximation developed in Ref.~\cite{Baker:2008mj}, but with significant technical differences.
An important input into the merger-ringdown part of the EOB model is the mapping
between progenitor binary parameters and the mass and spin of the
final remnant. This is needed to determine the frequency of QNMs of the remnant, and
it is obtained from NR results~\cite{Jimenez-Forteza:2016oae, Hofmann:2016yih}.

There is considerable freedom in modeling and resumming the EOB Hamiltonian, radiation-reaction force and \GW modes, and  spin effects. Those different choices, together with variations in the gauge adopted and deformations of the potentials in the Kerr spacetime, have led to
two main EOB families: ~\texttt{SEOBNR} (e.g., see Refs.~\cite{Bohe:2016gbl,Cotesta:2018fcv,Ossokine:2020kjp,Ramos-Buades:2023ehm}) and \texttt{TEOBResumS} (e.g., see Refs.~\cite{Damour:2014sva,Nagar:2018zoe,Nagar:2020pcj,Gamba:2021ydi}),
which we discuss in detail in Secs.~\ref{subsec:SEOBNR} and~\ref{subsec:TEOBResum} below.
The EOB waveform program advances along two interrelated directions.
The first comprises theoretical advances on the underlying structure
and mapping of analytical results from PN, PM and GSF approaches.
The second involves testing and refining the
models using NR information and rendering them available for use in
data analysis. The EOB waveform models are intrinsically in the time domain
(TD), which provides a more direct relation between source physics
and asymptotic radiation than frequency-domain models.
Modeling both the dynamics and radiation is useful for gaining deeper insights
into strong-field nonlinearities
and it enables more thorough tests of the robustness of different aspects
of the waveform based on examining the behavior of various characteristic
quantities under changes in the binary's parameters, and on comparing different
gauge-invariant quantities between EOB, NR, or perturbative
results --- for example the binding energy and fluxes~\cite{Boyle:2008ge,Damour:2011fu,LeTiec:2011dp,Bernuzzi:2014owa,Ossokine:2017dge,Antonelli:2019ytb}, periastron advance~\cite{LeTiec:2011bk,LeTiec:2013uey,Hinderer:2013uwa}, scattering angles~\cite{Damour:2014afa,Hopper:2022rwo,Khalil:2022ylj,Damour:2022ybd}
or spin-precession invariant~\cite{Hinderer:2013uwa} and redshift\edit{~\cite{LeTiec:2011dp, Barausse:2011dq,Bini:2016cje,Akcay:2012ea}}. At the same time, this also makes EOB models
computationally more expensive than
IMR waveform models, which describe only the asymptotic \GW
signals in the \FD, as discussed in Sec.~\ref{sec:phenom}.
However, the computational cost is not an insurmountable problem. 
Much recent work has significantly improved the
computational efficiency of EOB models through the development of optimized codes~\cite{Devine:2016ovp,Knowles:2018hqq,Nagar:2018gnk,Rettegno:2019tzh,Gamba:2020ljo},
 by building \ROMs~\cite{Field:2013cfa,Purrer:2014fza,Purrer:2015tud,Lackey:2016krb,
Lackey:2018zvw,Cotesta:2020qhw,Gadre:2022sed,Tissino:2022thn,Khan:2020fso,Thomas:2022rmc,Pompili:2023tna}, implementing the \edit{post}-adiabatic
approximation~\cite{Nagar:2018gnk,Rettegno:2019tzh,Gamba:2020ljo,Mihaylov:2021bpf}, and making use of machine-learning methods\edit{~\cite{Schmidt:2020yuu,Dax:2022pxd, Tissino:2022thn,Thomas:2022rmc,Khan:2020fso}}. Close Hyperbolic Encounters could also be searched for with both ground- and space-based interferometers using machine learning \cite{Morras:2021atg}.

\EOB waveform models have been constructed for quasi-circular non-spinning\edit{~\cite{Buonanno:2000ef,Buonanno:2006ui,Buonanno:2007pf,Damour:2007yf,Damour:2007vq,Damour:2008te,Buonanno:2009qa,Pan:2011gk,Damour:2012ky,Nagar:2019wds}} and 
spinning\edit{~\cite{Buonanno:2005xu,Pan:2009wj,Nagar:2011fx,Damour:2014sva,
Taracchini:2012ig,Taracchini:2013rva,Bohe:2016gbl,Cotesta:2018fcv,Pan:2013rra,Babak:2016tgq,Ossokine:2020kjp,Nagar:2018plt,Nagar:2018zoe,Akcay:2020qrj,Gamba:2021ydi,Pompili:2023tna,Ramos-Buades:2023ehm,Nagar:2020pcj,Nagar:2021xnh}} binaries\edit{, building also on further developments of EOB Hamiltonians~\cite{Damour:2015isa,Damour:2000we,Damour:2001tu,Damour:2008qf,Balmelli:2015lva,Balmelli:2015zsa,Barausse:2009xi,Barausse:2011ys,Khalil:2020mmr} and resummations of the radiative sector~\cite{Damour:2008gu}}. Furthermore, orbital eccentricity~\cite{Bini:2012ji,Hinderer:2017jcs,Chiaramello:2020ehz,Nagar:2021gss,Khalil:2021txt,Ramos-Buades:2021adz,Albanesi:2022xge} \edit{(see also~\cite{Cao:2017ndf,Liu:2021pkr})} and matter effects~\cite{Bernuzzi:2014owa,Hinderer:2016eia,Steinhoff:2016rfi,Akcay:2018yyh,Steinhoff:2021dsn,Matas:2020wab,Gonzalez:2022prs} (see Sec.~\ref{sec:EOBbeyondGRandBSM}) , as well as information from PM~\cite{Damour:2016gwp,Damour:2017zjx,Antonelli:2019ytb,Damgaard:2021rnk,Khalil:2022ylj,Damour:2022ybd}
and conservative GSF information~\cite{Damour:2009sm,Yunes:2009ef,Yunes:2010zj,Barausse:2011dq,Akcay:2012ea,
Antonelli:2019fmq,Nagar:2022fep,Albertini:2022rfe,Albertini:2022dmc,vandeMeent:2023ols}
have been also incorporated in EOB models. 
The required effort to advance the EOB modeling involves: 1) obtaining new analytical information and mapping it into the EOB framework; 
2) resumming the results into full waveform models with additional flexibility for calibrations to NR; 
and 3) testing and comparing models to assess their performance and accuracy. We describe each of the two main EOB
families in more detail \edit{in Secs.~\ref{subsec:SEOBNR} and~\ref{subsec:TEOBResum} below}.

\begin{table}[t]
    \scriptsize{
    \begin{tblr}{
                hline{1-26}={solid},
                hline{28}={solid},
                vlines,
                colspec={Q[l,m]Q[l,m]Q[c,m]Q[c,m]Q[c,m]Q[c,m]},
                rows = {abovesep=1pt,belowsep=1pt},
                columns = {rightsep=3pt,leftsep=3pt},
                }
    family &  waveform model & spins & ecc./hyper. & NR \& Teukolsky~calib.\ region & CP-frame modes \edit{$(\ell,|m|)$}\\
    $\;\;1^{\rm st}$
    &{$^*$\texttt{EOBNRv1}\\ \cite{Buonanno:2007pf} }& \SetCell[r=2]{c} & \SetCell[r=18]{c} & $q\leq 4 $ & \SetCell[r=3]{c} $(2,2)$\\
    \SetCell[r=4]{l} $\;\;2^{\rm nd}$
    & {$\;\;$\texttt{EOBNRv2} \cite{Buonanno:2009qa}} & & & $q\leq 3$ & $(2,2)$ \\
    & {$\;\;$\texttt{SEOBNRv0} \cite{Pan:2009wj}} & &  & $q=1, \chi_{1,2}=\pm 0.4$ & $(2,2)$\\
    &{$^*$\texttt{EOBNRv2HM} \cite{Pan:2011gk}} & \SetCell[r=2]{c} & & \SetCell[r=2]{c} $q\leq 6$ & \SetCell[r=2]{c} {$(2,2),(2,1),(3,3),$\\$(4,4),(5,5)$}\\
    &{$^*$\texttt{EOBNRv2HM\_ROM} \cite{Marsat:2020rtl}} & & & $q\leq 6$ & {$(2,2),(2,1),(3,3),$\\$(4,4),(5,5)$}\\
    \SetCell[r=6]{l}$\;\;3^{\rm rd}$ &{$^*$\texttt{SEOBNRv1} \cite{Taracchini:2012ig}} & \SetCell[r=4]{c} $\checkmark$ &  & \SetCell[r=2]{c}  $q\leq 6, \chi_{1,2}=\pm 0.4$ & \SetCell[r=4]{c} $(2,2)$ \\
    &{$^*$\texttt{SEOBNRv1\_ROM} \cite{Purrer:2014fza}} &$\checkmark$ &  &  &\\
    &{$^*$\texttt{SEOBNRv2} \cite{Taracchini:2013rva}} &$\checkmark$ & & \SetCell[r=4]{c} {NR: $q\leq 8, |\chi_{1,2}|\leq 0.98$ \\ Teuk.: $q=1000, |\chi_{1,2}|\leq 0.99$} \\
    &{$^*$\texttt{SEOBNRv2\_ROM} \cite{Purrer:2015tud}} &$\checkmark$ &  &  &\\
    & {$\;\;$\texttt{SEOBNRv2P} \cite{Pan:2013rra}} & \SetCell[r=2]{c}  $\checkmark \checkmark$ &  & & \SetCell[r=2]{c} $(2,2),(2,1)$  \\
    & {$^*$\texttt{SEOBNRv3P} \cite{Taracchini:2013rva,Babak:2016tgq}} & $\checkmark \checkmark$ & & &  \\
    \SetCell[r=9]{l} $\;\;4^{\rm th}$
    &{$^*$\texttt{SEOBNRv4} \cite{Bohe:2016gbl}} & \SetCell[r=2]{c}  $\checkmark$ & & \SetCell[r=9]{c} {NR: $q\leq 8, |\chi_{1,2}|\leq 0.995$ \\ Teuk.: $q=1000, |\chi_{1,2}|\leq 0.99$} & \SetCell[r=2]{c} $(2,2)$ \\
    &{$^*$\texttt{SEOBNRv4\_ROM} \cite{Bohe:2016gbl}} & $\checkmark$ & & & $(2,2)$ \\
    &{$^*$\texttt{SEOBNRv4P} \cite{Ossokine:2020kjp}} & $\checkmark \checkmark$  & &  & $(2,2), (2,1)$ \\
    &{$^*$\texttt{SEOBNRv4HM} \cite{Cotesta:2018fcv}} & \SetCell[r=2]{c} $\checkmark$ & & & \SetCell[r=5]{c} {$(2,2),(2,1),(3,3),$\\$(4,4),(5,5)$} \\
    &{$^*$\texttt{SEOBNRv4HM\_ROM} \cite{Cotesta:2020qhw}} & $\checkmark$ &  & & \\
    &{$^*$\texttt{SEOBNRv4PHM} \cite{Ossokine:2020kjp}} & \SetCell[r=2]{c} $\checkmark\checkmark$ &  &   & \\
    &{$^*$\texttt{SEOBNRv4PHM\_surr} \cite{Gadre:2022sed}} &$\checkmark\checkmark$ &  &  &  \\
    &{$^*$\texttt{SEOBNRv4EHM} \cite{Ramos-Buades:2021adz}} &$\checkmark$ & $\checkmark$ & &    \\
    &{$^*$\texttt{SEOBNRv4PHM\_4dq2} \cite{Thomas:2022rmc}} &$\checkmark\checkmark$ & \SetCell[r=5]{c}  &  & $(2,2),(2,1),(3,3),(4,4)$ \\
    \SetCell[r=4]{l} $\;\;5^{\rm th}$
    &{$^*$\texttt{SEOBNRv5} \cite{Pompili:2023tna}} & \SetCell[r=3]{c} $\checkmark$ & & \SetCell[r=4]{c} {NR: $q\leq 20, |\chi_{1,2}|\leq 0.998$ \\ Teuk.: $q=1000, |\chi_{1,2}|\leq 0.99$} & \SetCell[r=2]{c} $(2,2)$ \\
    &{$^*$\texttt{SEOBNRv5\_ROM} \cite{Pompili:2023tna}} & $\checkmark$ & & & \\
    &{$^*$\texttt{SEOBNRv5HM} \cite{Pompili:2023tna}} & $\checkmark$ & & & \SetCell[r=2]{c} {$(2,2),(2,1),(3,2),(3,3)$,\\$(4,3),(4,4),(5,5)$} \\
    &{$^*$\texttt{SEOBNRv5PHM} \cite{Ramos-Buades:2023ehm}} & $\checkmark\checkmark$ & & & \\
    %
    \SetCell[c=6]{l}$\checkmark$ aligned spin, $\checkmark\checkmark$ arbitrary spin orientations.\\
    \SetCell[c=6]{l}{$^*$ Implemented in \texttt{LALsuite}, and for the latest version (v5) in the open-source Python package  \\ \phantom{$^*$ } \texttt{pySEOBNR}~\cite{Mihaylov:2023bkc}.}
    \end{tblr}}
    \caption{Progress in the development of ${\tt SEOBNR}$ models. The NR-calibration region refers only to calibration against aligned-spin NR waveforms. We indicate the coprecessing modes as CP modes. The surrogate models ${\tt SEOBNRv4PHM\_surr}$ and $\tt {SEOBNRv4PHM\_4dq2}$, and the eccentric/hyperbolic model \texttt{SEOBNRv4EHM} are implemented in LALSuite, but they have not been reviewed, yet, and they are not publicly available at the moment. The surrogate models are limited in length and binary's parameter space (see Refs.~\cite{Gadre:2022sed,Thomas:2022rmc} for details). }
    \label{tab:SEOBNR}
\end{table}

\subsubsection{Suitable for what sources?}
 Inspiral-merger-ringdown signals from massive black-hole binaries (MBHBs) will be within the LISA band (see Sec.~\ref{sec:sources_MBHBs}), 
 making EOB models naturally applicable for these sources. The bulk of MBHB events observable by LISA are expected to have mass 
 ratios $q\lesssim 10$, but some systems may have mass ratios of up to several hundreds. Most tests and calibrations of EOB models 
 have been for $q\lesssim 20$, but the fact that EOB models interpolate between this regime and the test-body limit with information 
 from BH perturbation theory makes them structurally well-suited for larger-$q$ systems. Spin precession and eccentricity effects, 
 both simultaneously relevant for MBHBs and IMRIs, have \edit{mainly} been separately included in EOB models, \edit{and there has been recent 
 progress on going beyond this, e.g.~\cite{Gamba:2024cvy,Liu:2023ldr}}. These considerations apply not only for MBHBs but
also  for describing intermediate mass-ratio binaries (IMRIs) discussed in Sec.~\ref{sec:sources_IMRIs}.
Importantly, EOB models have been either validated~\cite{Albertini:2022rfe,Albertini:2022dmc} or improved using 
perturbation-theory and GSF information~\cite{Yunes:2009ef,Yunes:2010zj,Barausse:2011dq,Antonelli:2019fmq,Nagar:2022fep,vandeMeent:2023ols},
which is relevant for IMRIs and also EMRIs.
EOB models are also suitable to stellar-origin binaries Sec.~\ref{sec:sources_SOBHBs}, 
whose early-inspiral signals will be within the LISA band. For inspirals, EOB models provide a more accurate though 
less efficient description than pure PN-based models.
Using EOB waveforms will further be important for connecting the LISA portion of SOBHB signals with the corresponding merger signals measurable in future ground-based detectors.
The EOB approach also allows for the inclusion of additional physical effects, for instance, from gravity theories beyond GR and environmental effects as explained in Sec.~\ref{sec:EOBbeyondGRandBSM}.

As discussed in Sec.~\ref{ssec:EOBChallenges}, while EOB models are naturally suited to the above sources, significant further advances in the modeling accuracy, complexity of physical effects, robustness over a wide range of parameter space, and efficiency
will be required for their use in LISA data analysis.

\subsubsection{The \texttt{SEOBNR} waveform models}
\label{subsec:SEOBNR}

The \texttt{EOBNR} family of waveform models~\footnote{The generic name \texttt{SEOBNRvnEPHM} indicates that the version \texttt{vn} of the \EOB model  is calibrated to NR simulations (NR), includes spin (S) and precessional (P) effects, eccentricity (E)  and higher modes (HM).} has been developed with two main goals: i) make use of the most accurate analytical information for the two-body dynamics and gravitational radiation (\PN, \PM, \GSF) and results from NR and Teukolsky-code simulations to build physical, highly-accurate waveform models of compact-object binaries, and ii) make them available to \GW detectors for searches and inference studies. The history, since 2007, and the current status of \texttt{EOBNR} models are illustrated in Table~\ref{tab:SEOBNR}. These models have been implemented into the LIGO Algorithms Library (\texttt{LALsuite}) \cite{lalsuite}, and more recently in the open-source Python package  \texttt{pySEOBNR}~\cite{Mihaylov:2023bkc}. These codes were then reviewed by the \LVK Collaboration. Here we summarize the main milestones, focusing on the binary BHs, while leaving details of the modeling improvements (two-body inspiraling dynamics, transition merger to ringdown, RR effects, \GW modes, resummation of EOB potentials, etc.) to the corresponding publications.

Building on the original work on the EOB framework~\cite{Buonanno:1998gg,Buonanno:2000ef,Damour:2000we,Buonanno:2005xu}, the first \texttt{EOBNR} model calibrated to nonspinning NR waveforms was developed in Ref.~\cite{Buonanno:2007pf}, following initial comparisons to NR equal-mass nonspinning binaries in Ref.~\cite{Buonanno:2006ui},
in which the importance of including overtones in the EOB description of the merger-ringdown signal was pointed out.
\texttt{EOBNRv1} was employed by Initial and Enhanced  LIGO, and Virgo for the first searches of coalescing binary BHs~\cite{Ochsner:2010hea,LIGOScientific:2011hqo,LIGOScientific:2011jth,LIGOScientific:2012vij}. The calibration to spinning binaries with equal masses was considered in Ref.~\cite{Pan:2009wj}, employing an EOB Hamiltonian for a test mass in a deformed Kerr spacetime~\cite{Damour:2008qf}. The nonspinning waveforms with higher modes were first modeled in Ref.~\cite{Pan:2011gk} with the improved  factorized waveforms~\cite{Damour:2007xr,Damour:2008gu,Pan:2010hz}, thus marking the second generation of \texttt{EOBNR} models.

The third-generation of \texttt{EOBNR} models encompassed significant advances in including spin effects in the two-body dynamics and radiation, resumming perturbative information,  and calibration to NR simulations. An EOB Hamiltonian for a test spin in a deformed Kerr spacetime was derived in Refs.~\cite{Barausse:2009aa, Barausse:2009xi, Barausse:2011ys}. It included all PN corrections in the test-body limit, at linear order in the test-body's spin. To enforce the presence of a photon orbit (and peak of the orbital frequency) for aligned-spin  binary BHs, it used a logarithmic (instead of Pad\'e) resummation of the EOB potentials. Those improvements led to the aligned-spin model \texttt{SEOBNRv1}~\cite{Taracchini:2012ig}, and the  first spin-precessing  model, \texttt{SEOBNRv2P}~\cite{Pan:2013rra}, which adopted the co-precessing frame description of Refs.~\cite{Buonanno:2002fy,Schmidt:2010it} to efficiently handle precessional effects. Those models also included information for the merger-ringdown
waveforms from the test-body limit, notably from time-domain Teukolsky waveforms obtained in Refs.~\cite{Barausse:2011kb}.  
With the availability of a larger set of accurate SXS NR waveforms, and further analytical results from PN theory, the model 
was upgraded to \texttt{SEOBNRv2}~\cite{Taracchini:2013rva}, with its precessing version known as \texttt{SEOBNRv3P}~\cite{Babak:2016tgq}\edit{, 
which is often
also referred to simply as~\texttt{SEOBNRv3}}.
The \edit{aligned-spin versions of these} third-generation \texttt{SEOBNR} models were used in the template bank of the modeled searches 
\edit{in the \texttt{gstLAL} and \texttt{pyCBC} pipelines}, and for parameter-estimation of \GW signals detected during the first 
observing run (O1) of Advanced LIGO\edit{~\cite{LIGOScientific:2016aoc,LIGOScientific:2018mvr,LIGOScientific:2016dsl}}.

The fourth generation of models included substantial and more efficient (via Markov-chain--Monte-Carlo methods) re-calibrations of
the aligned-spin baseline models, such as \texttt{SEOBNRv4}~\cite{Bohe:2016gbl}, and phenomenological ansatzes for 
the merger-ringdown based on Ref.~\cite{Damour:2014yha}. The \texttt{SEOBNRv4} model was completed by including 
higher modes~\cite{Cotesta:2018fcv}, arbitrary spin orientations~\cite{Ossokine:2020kjp}, and  extended to eccentric and hyperbolic orbits 
for aligned spins in Ref.~\cite{Ramos-Buades:2021adz}, using \GW modes recast in factorized form in Ref.~\cite{Khalil:2021txt} 
(see also Refs.~\cite{Cao:2017ndf,Liu:2019jpg,Liu:2021pkr,Hinderer:2017jcs} for other eccentric \texttt{EOBNR} models). 
These updated models for quasi-circular orbits were used for the template banks of modeled searches, and extensively 
for inference studies  during the second (O2) and third (O3) runs of Advanced LIGO and 
Virgo\edit{~\cite{LIGOScientific:2020ibl,KAGRA:2021vkt,LIGOScientific:2020ufj,LIGOScientific:2020stg,LIGOScientific:2020zkf,LIGOScientific:2018mvr}}.

Among the highlights of the fifth generation of models, they were calibrated to NR waveforms with larger mass-ratios and spins using 
a catalog of 442 SXS simulations~\cite{Pompili:2023tna}, and they incorporate for the first-time information from second-order GSF 
in the nonspinning modes and radiation-reaction force~\cite{vandeMeent:2023ols}. The models include several  higher modes, notably 
the $|m|=\ell$ modes for $\ell=2\mbox{--}5$ \edit{and} the \edit{$(2,\pm 1)$, $(3,\pm 2)$ and $(4,\pm 3)$} modes. Differently from the \edit{\texttt{SEOBNRv3P}} and \texttt{SEOBNRv4} Hamiltonians, 
the \texttt{SEOBNRv5} reduces in the test-body limit to the one of a test mass in the Kerr spacetime. 
The accurate precessing-spin dynamics of \texttt{SEOBNRv5PHM}~\cite{Ramos-Buades:2023ehm} was obtained including partial 
precessing-spin information in the EOB Hamiltonian in the co-precessing frame by orbit averaging the in-plane spin components of the 
full precessing Hamiltonian at 4PN order~\cite{Khalil:2023kep}. Furthermore, the evolution equations for the spin and angular momentum 
vectors were  computed in a PN-expanded, orbit-averaged form for quasi-circular orbits, similarly to what was done in
Refs.\edit{~\cite{Estelles:2020osj,Ackley:2020atn,Akcay:2020qrj}}, but with important differences due to the distinct gauge and 
spin supplementary conditions~\cite{Khalil:2023kep}, and the inclusion of orbit-average in-plane spin effects. Thanks to a 
simpler (but more approximated) precessing-spin inspiraling dynamics, which allows for the use of the post-adiabatic 
approximation~\cite{Rettegno:2019tzh,Mihaylov:2021bpf},  and the new high-performance Python 
infrastructure \texttt{pySEOBNR}~\cite{Mihaylov:2023bkc}, the computational efficiency of the \texttt{SEOBNRv5PHM} models has
improved significantly, generally $\sim 8 \mbox{--} 20$ times faster than the previous generation \texttt{SEOBNRv4PHM}. This is
particularly important in view of the use of these waveform models with next-generation detectors on the ground and with LISA,
which have a much broader bandwidth.

Furthermore, \texttt{SEOBNR} waveform models have also been used to extend the NR surrogate models 
(see Sec.~\ref{sec:comp_techniques}) to lower frequencies, by hybridizing them~\cite{Varma:2018mmi,Yoo:2022erv,Yoo:2023spi}, and
to build IMR phenomenological waveform models (see Sec.~\ref{sec:phenom}). For the latter, \texttt{SEOBNR} models 
have been employed to construct time-domain hybrid NR waveforms, and also to calibrate the phenomenological models
in regions of the parameter space where NR waveforms are not available~\cite{Khan:2015jqa,Pratten:2020fqn,Estelles:2020twz}.

Quite importantly for LISA and next-generation detectors on the ground, 
the computational efficiency of the time-domain \texttt{SEOBNR} models can also be improved by building surrogate and ROM
versions~\cite{Purrer:2014fza,Purrer:2015tud,Bohe:2016gbl,Gadre:2022sed,Pompili:2023tna}. Lastly, the 
efficient and flexible \texttt{pySEOBNR} infrastructure~\cite{Mihaylov:2023bkc}, which uses Bayesian algorithms
to calibrate waveforms against NR simulations, will allow to swiftly include and test new analytical information (PN,PM,GSF)
(and their possible resummations) as soon as they become available. This is  crucial to improve waveform models by at least two orders of
magnitude to match the expected waveform accuracy requirements.

Extensions of the \texttt{SEOBNR} models to extreme mass-ratio inspirals were obtained 
in Refs.~\cite{Yunes:2009ef,Yunes:2010zj,Taracchini:2013rva,Taracchini:2014zpa}, and 
applications to non-vacuum binary systems and gravity theories beyond GR are discussed in Sec.~\ref{sec:EOBbeyondGRandBSM}.

\subsubsection{The \texttt{TEOBResumS} waveform models}
\label{subsec:TEOBResum}

\begin{table}[t]
    \centering
    \scriptsize{\begin{tabular}{|cc|cc|}
    \hline
    \multicolumn{2}{|c|}{Physical content} & \texttt{GIOTTO} &   \texttt{DALI} \\[0.3em]
    \hline 
    \multirow{5}{*}{Analytic information} & $A(r)$ & Pseudo $5$PN, resummed & Pseudo $5$PN, resummed \\
                                          & $D(r)$ & $3$PN, resummed & $5$PN, resummed \\
                                          & $Q(r, p_{r_*})$ & $3$PN  & $5$PN (local) \\
                                          & $G_S, G_{S_*}$ & $3.5$PN &  $3.5$PN\\
                                          & $r_c$ & NLO &  NLO \\[0.25em]
    \hline
    \multirow{5}{*}{NR information}       & $a^6_c$         &  \multicolumn{2}{c|}{Effective parameter in $A(r)$ } \\
                                          & $c_{\rm N3LO}$  &  \multicolumn{2}{c|}{Effective parameter in $G_S, G_{S_*}$ } \\
                                          & NQCs            &  \multicolumn{2}{c|}{Ensure correct transition between plunge and merger} \\
                                          & Ringdown        &  \multicolumn{2}{c|}{Phenomenological model, quasi-circular} \\
                                          & \hspace*{-2em}BBH NR Validation region$\!\!$ & \multicolumn{2}{c|}{ \edit{$q\leq10$} and test-mass with $|\chi_{1,2}|\geq 0.99$; $10<q<128$ no-spins$\!\!$} \\[0.25em]
    \hline
    \multirow{2}{*}{Spins}          & Aligned    & $\checkmark$ & $\checkmark$ \\
                                    & Precessing & $\checkmark$ & - \\[0.25em]
    \hline
    Orbital dynamics  & & Circular & Generic (bound \& open)\\[0.25em]
    \hline
    \multirow{2}{*}{\!\! CP modes} 	& Inspiral to merger & \multicolumn{2}{c|}{$(\ell, |m|) \leq 8$}\\
                                       		& Merger/ringdown    & \multicolumn{2}{c|}{$(\ell, |m|) = (2,2), (2,1), (3,2), (3,3), (4,2), (4,4), (5,5)$}\\[0.25em]
 \hline
    \end{tabular}}
\caption{Current/default physics incorporated in \texttt{TEOBResumS}.
\texttt{TEOBResumS} is developed open source and publicly available at \url{https://bitbucket.org/eob_ihes/teobresums/}. 
Symbols are defined in the text. Historical milestones in the model developments and the associated references \edit{as well as} robustness tests
and the detailed parameter space coverage can be found on the Wiki page and are continuously updated. \texttt{TEOBResumS} can also be
installed via {\tt pip install teobresums}. The code is interfaced to state-of-art \GW data-analysis pipelines, including \edit{Bilby} and 
\edit{PyCBC}. The code uses semantic versioning since the deployment of the \texttt{GIOTTO} version. Earlier versions of \texttt{TEOBResumS} are
implemented in \edit{LALSuite~\cite{lalsuite}} and reviewed by LVK.}
\label{tab:TEOBResumS}
\end{table}

The \texttt{TEOBResumS} model 
builds upon early EOB developments at the Institut des Hautes Etudes Scientifiques (IHES)
that define the basics structure of the model. 
The name suggests the key features: arbitrary compact binaries with tidal (T) and generic spins (S) interactions are modeled by making systematic use of analytical resummations. This name has first appeared in the tidal model of Ref.~\cite{Bernuzzi:2014owa} and in the tidal and spinning model of Ref.~\cite{Nagar:2018zoe}.
The latter include the factorized waveform described above~\cite{Damour:2008gu,Damour:2009kr},
the systematic resummation of EOB potentials via Pad\'e approximants~\cite{Damour:2002qh,Damour:2008qf,Damour:2009kr,Nagar:2011fx,Damour:2012ky},
NQC corrections~\cite{Nagar:2006xv,Damour:2007xr,Damour:2007vq,Damour:2008te,Damour:2012ky},
attachment of the ringdown around the peak of the orbital frequency (i.e., light-ring crossing)~\cite{Damour:2007xr,Damour:2012ky}, the concept of centrifugal radius for incorporating
spin-spin interactions~\cite{Damour:2014sva}, resummed gyro-gravitomagnetic functions~\cite{Damour:2008qf,Nagar:2011fx,Damour:2014sva} and the factorized
NR-informed ringdown waveform~\cite{Damour:2014yha}.
These developments made heavy use of results in the test-mass limit obtained by
means of a new approach to black hole perturbation theory
\cite{Nagar:2006xv,Damour:2007xr,Bernuzzi:2010xj,Bernuzzi:2011aj,Harms:2014dqa} to understand
each physical element entering the structure of the waveform and striving for physical completeness, simplicity, accuracy and efficiency.
Currently, \texttt{TEOBResumS-GIOTTO} is a unified framework to generate \edit{inspiral waveforms for any type of quasi-circular compact binary
and complete inspiral-merger-postmerger waveforms
for BBH and BHNS binaries~\cite{Riemenschneider:2021ppj,Gonzalez:2022prs}}. 

\texttt{TEOBResumS-Dalí} is the model's extension to generic orbits and describes bound orbits with arbitrary eccentricity \cite{Chiaramello:2020ehz,Nagar:2021gss,Nagar:2021xnh,Bonino:2022hkj}, hyperbolic orbits and scattering~\cite{Nagar:2020xsk,Gamba:2021gap,Hopper:2022rwo}. The key features of the models are summarized in Table~\ref{tab:TEOBResumS} and described in the following.
An extension of the model that relies on GSF-informed potentials~\cite{Akcay:2015pjz,Nagar:2022fep} so as to generate
extreme-mass-ratio-inspirals (EMRIs) is available~\cite{Albertini:2022dmc,Albertini:2022rfe},
and work on scalar-tensor gravity (building upon Refs.~\cite{Julie:2017pkb,Julie:2017ucp})
is also currently in progress~\cite{Jain:2022nxs,Jain:2023fvt,Jain:2023vlf}.

The structure of the spin-aligned effective \texttt{TEOBResumS} Hamiltonian is
\begin{equation}
	H_\text{eff}=H_\text{eff}^\text{orbital}+H^\text{SO}_\text{eff},
\end{equation}
where $H_\text{eff}^\text{orbital}$ incorporates even-in-spin contributions
(spin-spin couplings), while $H^\text{SO}_\text{eff}$ incorporates odd-in-spin ones
(spin-orbit couplings). The spin-orbit contribution reads
\begin{equation}
	H^\text{SO}_\text{eff}=G_S S + G_{S_*}S_*,
\end{equation}
where $(G_S,G_{S_*})$ are the gyro-gravitomagnetic functions and $S\equiv S_1+S_2$,
while $S_*=m_2/m_1 S_1+m_1/m_2 S_2$ are useful combinations of the spins~\cite{Damour:2007nc,Damour:2008qf,Nagar:2011fx}.
In the large-mass-ratio limit, $S$ becomes the spin of the largest black hole, while $S_*$ \edit{reduces to $q$ times} the spin of the small body.
The orbital contribution $H_\text{eff}^\text{orbital}$ incorporates the three EOB potentials $(A,D,Q)$. The $A$ function employs the full 4PN information~\cite{Damour:2015isa} augmented by an effective 5PN term, parametrized by the coefficient $a_6^c$, which is informed by NR simulations. In \texttt{GIOTTO}, both $(D,Q)$ are kept at 3PN accuracy.
In the latest versions of \texttt{Dalí}, the full 5PN information is being currently explored~\cite{Nagar:2023zxh}.
The spin-orbit sector includes next-to-next-to-leading-order (NNLO)
analytical information~\cite{Nagar:2011fx} in the (resummed) gyro-gravitomagnetic
functions~\cite{Damour:2014sva} augmented by an effective $N^3LO$ coefficient $c_3$
that is informed by NR simulations. Spin-spin couplings are incorporated using
the concept of centrifugal radius~\cite{Damour:2014sva}, $r_c$,  and starting from NLO accuracy.
\texttt{TEOBResumS} can deal with generically oriented spins, thus incorporating
the precession of the orbital plane and the related modulation of the waveform.
Spin-precession in the quasicircular case is incorporated with an efficient yet accurate hybrid EOB-post-Newtonian scheme based on the common ``twist approach"~\cite{Akcay:2020qrj,Gamba:2021ydi,Gonzalez:2022prs}.

In \texttt{GIOTTO}, the radiation reaction and multipolar waveform (up to $\ell=m=8$) are implemented as an upgraded version of the factorized and resummed procedure introduced in Ref.~\cite{Damour:2008gu}; see Refs.~\cite{Nagar:2020pcj,Messina:2018ghh}. In addition, NR information is used to determine NQC waveform corrections (multipole by multipole) so as to correctly shape the waveform around merger. The latter is then attached to a NR-informed, phenomenological description of the ringdown based on Ref.~\cite{Nagar:2020pcj}. Importantly, NR information is included in such a way to maintain  consistency between waveform and fluxes \cite{Riemenschneider:2021ppj}.
The leading-order horizon absorption terms are implemented in the model since its early version~\cite{Damour:2014sva}.
In \texttt{Dalí}, the radiation reaction on generic orbits is obtained by incorporating generic Newtonian prefactors in the factorized quasicircular EOB waveform~\cite{Chiaramello:2020ehz,Nagar:2021gss}. This approach, which is extended to include resummed 2PN terms, has been verified against exact Teukolsky fluxes for highly eccentric and hyperbolic geodesics and currently provides the best available representation of the radiation reaction \cite{Placidi:2021rkh,Albanesi:2021rby,Albanesi:2022ywx}.
A new paradigm to incorporate PN results into EOB, which promises a further boost in performance, has been recently proposed in Ref.~\cite{Albanesi:2022xge}.
%
The inclusion of matter effects and application to probing the nature of compact objects is discussed in Sec.~\ref{sec:EOBbeyondGRandBSM}.

On the algorithmic side, the development of \texttt{TEOBResumS} has introduced two key analytical acceleration techniques, namely the post-adiabatic (PA) approximation of the EOB dynamics~\cite{Nagar:2018gnk} and the EOBSPA~\cite{Gamba:2020ljo}. The PA approximation is an iterative procedure to solve the circularized EOB dynamics at given radii (or frequency) thus bypassing the need to numerically solve the system of ODEs. Performances are reported in e.g. Refs~\cite{Nagar:2018gnk,Nagar:2018plt,Riemenschneider:2021ppj,Tissino:2022thn} and show that \texttt{TEOBResumS}'s waveform generation times are competitive with some of the fastest waveforms and machine learning approximants for long signals (i.e. BNS inspirals), while retaining the original EOB accuracy.
The EOBSPA is a technique to generate multipolar \FD EOB waveforms using the stationary phase approximation. Ref.~\cite{Gamba:2020ljo} illustrates that the EOBSPA is computationally competitive with current phenomenological and surrogate models, and can generate (virtually) arbitrarily long and faithful waveforms up to merger for any BNS. Currently, the EOBSPA is also the only alternative to PN methods for efficiently generating intermediate-mass binary black hole inspiral waveforms for LISA~\cite{Gamba:2020ljo}.

\subsubsection{Environmental effects}

The EOB approach provides considerable freedom for incorporating yet
unknown effects, such as beyond-\GR gravity (see
Sec.~\ref{ssec:phenom_beyond_GR}), exotic BH-like objects, or
environmental effects.  Analyzing the presence of such effects
critically relies on accurate \GR waveforms for comparison and
benefits from the techniques developed for \GR. While the modeling of astrophysical
environments has not yet been explored directly within the
EOB waveforms program, there is in principle no obstruction to
incorporating them. For example, perturbative and numerical studies of
environmental effects are already available (as discussed in Secs.~\ref{sec:weak_field}, \ref{sec:GSF} and \ref{sec:NR}) and have identified key
distinctive features, which could be mapped into EOB models similarly
to what has already been accomplished for other physical
effects. Likewise, existing capabilities of using EOB-baseline
models for parameterized tests of \GR could be extended to include
environmental effects in a parameterized, phenomenological way at the
level of waveforms. These two approaches could be useful for different
purposes.

\subsubsection{Challenges}
\label{ssec:EOBChallenges}

Employing EOB waveforms for \GW signals of LISA sources
requires a significant continued effort to improve the accuracy  and
include all physical effects. Importantly, the waveform models must
incorporate the simultaneous effects of high spins at generic
orientations, eccentricity, and large mass ratios $q$, all of which also
require many more higher modes. Furthermore, we
require additional flexibility to discover new physics from beyond GR
and the Standard Model of particle physics, as discussed in
Sec.~\ref{sec:modling_beyond_GR}. In terms of accuracy, the models
must describe high signal-to-noise-ratio events of MBHBs, at more than
an order of magnitude higher than the calibration standards achieved
to date, and they have to be tested for the expected LISA detector
response.

The effort to meet these requirements involves substantial further
model developments on the theoretical and computational fronts. On the
theoretical side, advances on the structure of the EOB models to
incorporate more information from perturbative schemes are
essential. This effort relies on inputs from higher-order calculations
with more realistic physics from PN, PM, and GSF to develop robust EOB
models for a wide range of the parameter space. While several
proof-of-principle studies have been carried out on including PM and
GSF information, more work is needed to incorporate them in full
state-of-the-art models in a way that can readily be updated as more
information becomes available. \edit{This also applies to the inclusion of memory effects}.
Methods for including precessing-spins and
moderate eccentricity effects (where the systems circularize before
merger) are in place but only separately. In principle,
the EOB Hamiltonian is fully generic but other components of the model
are not, and require further structural developments to
incorporate larger eccentricities, as well as the simultaneous
effect of precessing spins. Further considerations will also be required on
the sets of dynamical variables used to evolve the inspirals, for
example, action-angle variables versus the canonical Cartesian coordinates,
or a mixture of them. The higher complexity when including both
precession and eccentricity effects is challenging, and
requires due care to ensure gauge-invariant comparisons. As for
existing models, insights from the small-mass-ratio limit will likely
prove useful to address this challenge.

On the practical side, significant further work on testing,
calibrating, and optimizing the EOB models against NR results is
crucial to attain the accuracy required for LISA sources discussed in
Sec.~\ref{sec:accuracy_requirements}. This in turn relies on the
availability of accurate NR waveforms over a wider parameter space, as
detailed in Sec.~\ref{sec:NR}. Current EOB models are only calibrated
in the aligned-spin sector. However, this would need to be changed,
and the calibration be extended to precessing spins to achieve the
much higher accuracy requirements of LISA sources.

Furthermore, accurate waveforms for LISA that include eccentricity, precessing spins, large mass ratio, higher modes, and means to test for new physics have a highly complex structure characterized by a large number of different
frequencies. Accurately capturing these features significantly
slows down the computations of waveforms and enlarges the dimensionality of the parameter space. While EOB waveforms are a priori less efficient than closed-form models, there exist a number of approaches to overcome these shortcomings, as discussed above in Secs.~\ref{subsec:generalEOB}, ~\ref{subsec:SEOBNR} and ~\ref{subsec:TEOBResum}.

\newpage

\subsection{Phenomenological waveform models}
\label{sec:phenom}

Coordinators: Sascha Husa, Maria Haney \\
Contributors: M.~Colleoni, M.~Hannam, A.~Heffernan, J.~Thompson

\subsubsection{Description}

After the binary black hole breakthrough of 2005, a pragmatic approach to developing waveform models for compact binary coalescence was required. Such models needed to describe the waveform from inspiral to ringdown, be
tuned to \NR, and suit a wide range of \GW data analysis applications. 
Key requirements for broad applicability were (and still are) computational efficiency and broad coverage of the parameter space.
This has led to several generations of frequency-domain and more recently also time-domain models, which have been implemented in the open source {\tt LALSuite} framework \cite{lalsuite} 
under the name of IMRPhenom, and which are constructed in terms of piecewise closed form expressions for the amplitude and phase of spherical harmonic modes.
A careful choice of the closed-form expressions allows the maximal compression of information about the waveform into a small number of coefficients that vary across the parameter space. 
More recently, fast ODE integration techniques have also been developed to 
model the evolution of the component spins in precessing binaries \cite{Estelles:2021gvs}.
The principal objective of the phenomenological waveform program has been to deliver waveform models that keep up with requirements of data analysis applications and are updated to become 
increasingly accurate as detectors become more sensitive. 

\begin{figure} 
    \centering
    \includegraphics[width=0.495\textwidth,trim=30 50 70 80,clip=true]{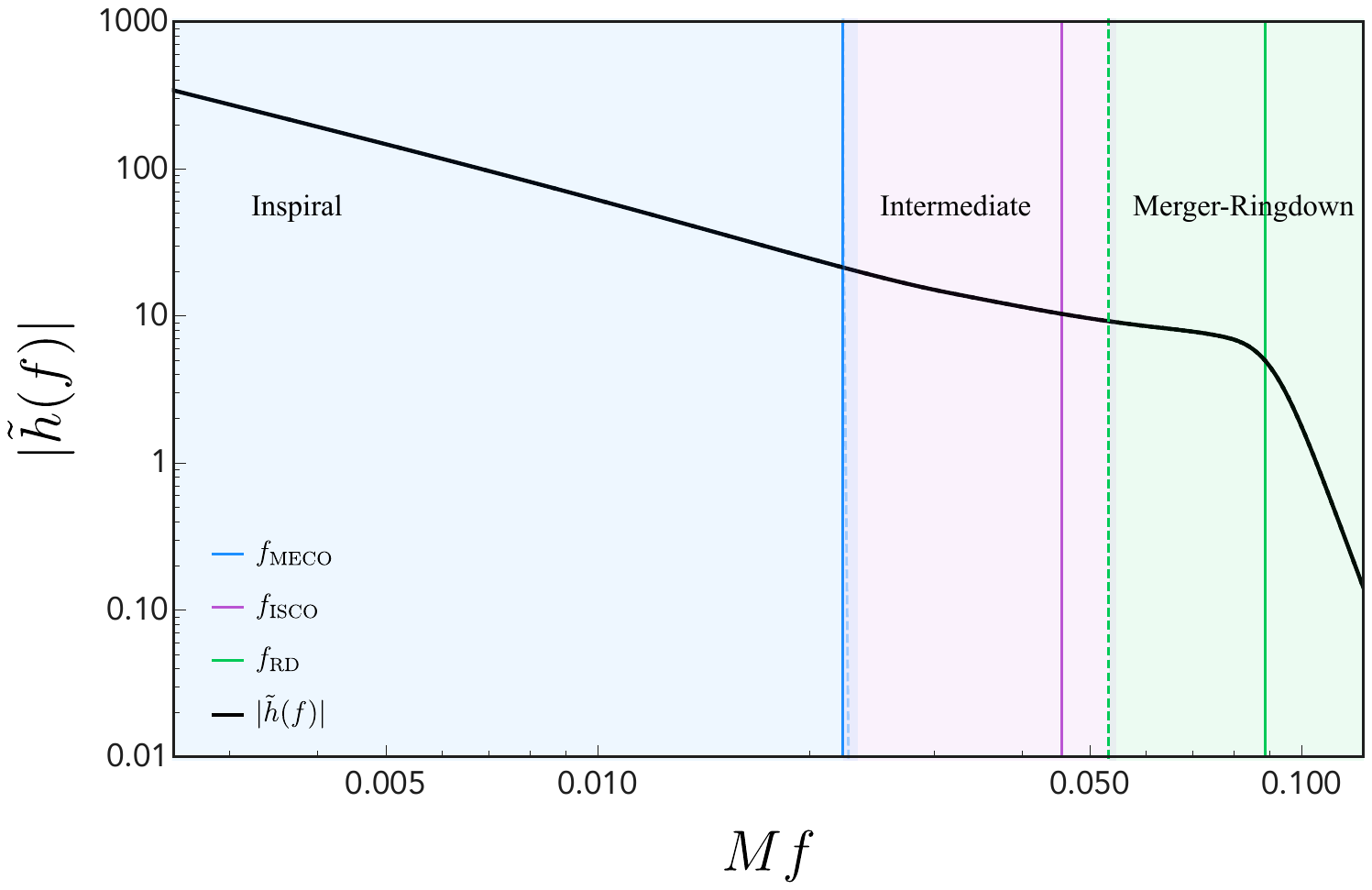}
    \includegraphics[width=0.495\textwidth,trim=30 50 70 80,clip=true]{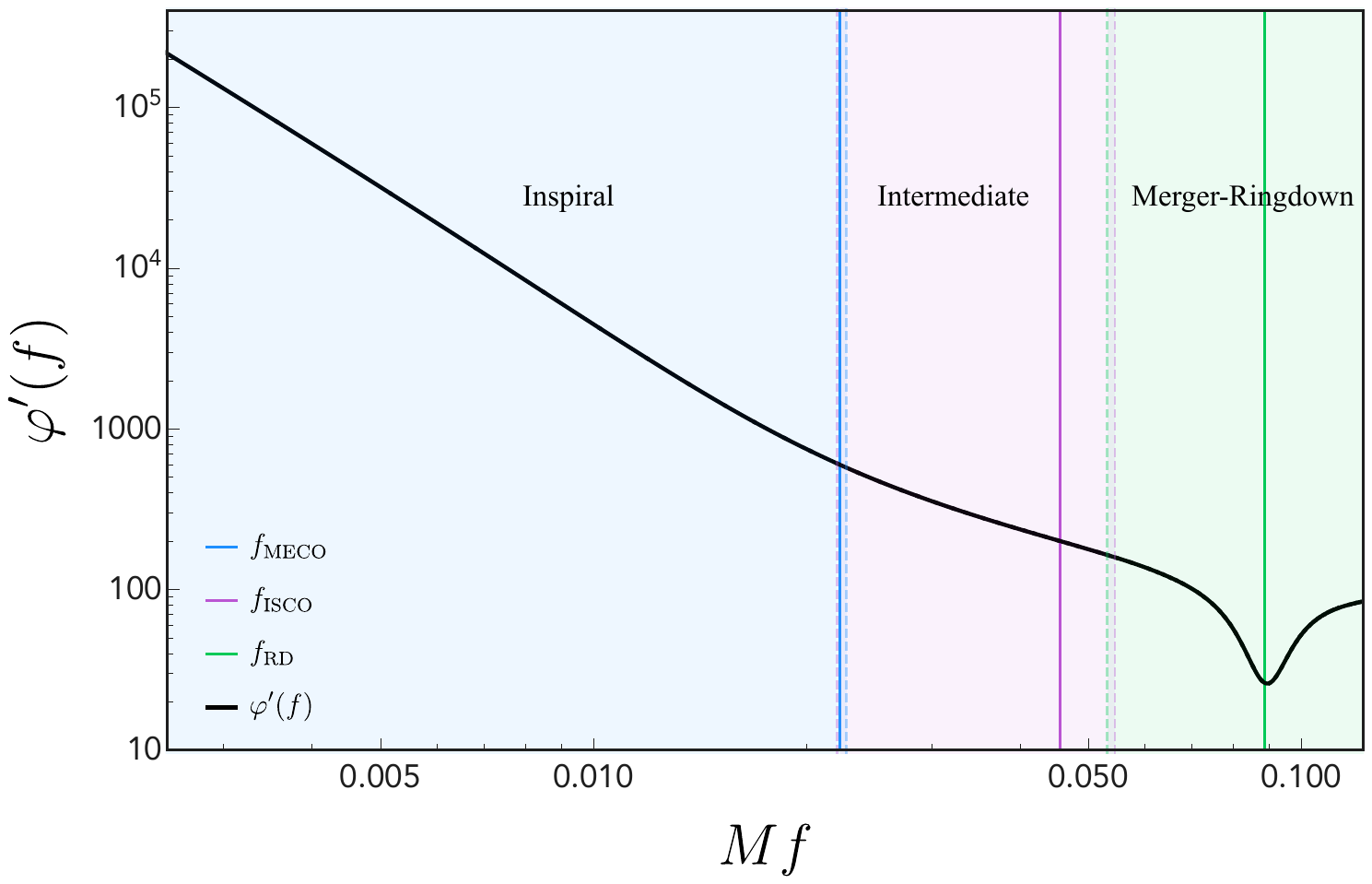}
    \caption{Illustration of the regions used in Phenom models exemplified by the IMRPhenomX model. 
The plots show the Fourier domain amplitude (left) and phase derivative (right) of the $\ell=2, m=-2$ spherical harmonic mode.  Vertical lines
mark the frequencies of the \ISCO for the remnant black hole,
the MECO (minimum energy circular orbit as defined in \cite{Cabero:2016ayq}, see also \cite{Blanchet:2001id}) and the ringdown frequency of the fundamental mode.  Figure adapted from \cite{Pratten:2020fqn}.}
    \label{fig:PhenomRegions}
\end{figure}

To construct the models, one proceeds in three stages: First, an appropriate piecewise ansatz is developed for simple functions, e.g. the amplitude and phase of spherical or spheroidal 
harmonics. This ansatz is split into regions, for example corresponding to the inspiral, merger, and ringdown, where physical insight about the different regions can be exploited.
Fig.~\ref{fig:PhenomRegions} shows the three regions used in current models, where in each such region closed-form expressions are developed to approximate a discrete data set of 
calibration waveforms.
 Future upgrades may increase the number of regions to increase accuracy.
The analytical ansatz attempts to incorporate physical insight, e.g. regarding perturbative information concerning the inspiral and ringdown.
For the inspiral, an ansatz is typically constructed as a deformation of a post-Newtonian description. In the ringdown, black hole perturbation theory can be used to link features 
in the waveform to the quasinormal frequencies determined by the final mass and spin of a black hole. The ansatz is then fitted to each waveform in a calibration data set, 
resulting in a set of generalized coefficients for each waveform. This stage is usually referred to as the ``direct fit''.
The input data consist of NR data and perturbative descriptions at low frequency, such as post-Newtonian expansions or EOB models. Most typically, these types of input data are used 
in the form of hybrid waveforms, which are constructed by gluing inspiral approximants to shorter \NR data. 
For the PhenomD, PhenomX and PhenomT models, waveforms from the \texttt{SEOBNR} family have been employed to construct such hybrids~\cite{Khan:2015jqa,Pratten:2020fqn,Estelles:2020twz}.
Finally, the coefficients are interpolated across the parameter space, which we will refer to as the ``parameter space fit''. In both stages of fits it is essential to avoid both 
overfitting and underfitting.

Waveform model development has been increasingly ``data driven'', adapting phenomenological descriptions to the available data sets, while also using physical insight and perturbative 
information. The inspiral descriptions, for example, extend post-Newtonian expansions, and the black hole ringdown is formulated in terms of exponentially damped oscillations with complex 
frequencies that depend on the dimensionless spin of the remnant.
In addition, approximate maps have been used, e.g. to map the non-precessing waveforms to precessing ones. Such maps: \vspace{-0.5em}
\begin{itemize}
\item Allow to increase the region of the parameter space where the models 
can be employed before sufficiently extensive catalogues of NR waveforms are available. 
\item Give guidance for future more sophisticated models that are calibrated to NR.
\end{itemize}

The ``phenom'' approach sketched above facilitates the development of simple and robust codes and very rapid evaluation of the waveforms, and is also well suited to benefit from 
parallelisation, e.g., through \GPUs  \cite{Katz:2020hku} or similar hardware.
Phenom models have already been used in computationally efficient parameter estimation studies of MBHB sources for LISA \cite{Katz:2020hku}.
Initially, only frequency domain models were developed \cite{Ajith:2007qp,Ajith:2007kx,Ajith:2009bn,Santamaria:2010yb,Hannam:2013oca,Husa:2015iqa,Khan:2015jqa,Bohe:T1500602,London:2017bcn,Khan:2018fmp,Khan:2019kot,Dietrich:2018uni,Dietrich:2019kaq,Thompson:2020nei,Pratten:2020fqn,Garcia-Quiros:2020qpx,Garcia-Quiros:2020qlt,Pratten:2020ceb,Colleoni:2023czp}, 
since they are naturally adapted to matched filter techniques carried out in the frequency domain. More recently, time domain models  have also been developed 
\cite{Estelles:2020osj,Estelles:2020twz,Estelles:2021gvs}, which can simplify modelling complex phenomena such as precession or eccentricity.

Four generations of such models have been constructed to date in the frequency domain, progressing from the $\ell=\vert m\vert =2$ mode for quasi-circular non-spinning binaries to 
multi-mode precessing waveforms, as will be described in Sec.~\ref{sec:phenom:status}. 
The flexibility of the phenomenological approach is also well suited to parameterize unknown information, e.g., beyond-GR effects (see Sec.~\ref{ssec:phenom_beyond_GR}).

\subsubsection{Suitable for what sources?}

Massive black hole binaries  have already been modelled with the phenom approach, i.e., comparable-mass black-hole binaries, described in 
Sec.~\ref{sec:sources_MBHBs}.  The bulk of such binaries detected by LISA is expected to have mass ratios $q=m_1/m_2$ up to $\sim10$ (see Sec.~\ref{ssec:MBHB_expected_source_parameters}), 
where phenom models have already been calibrated to \NR. The main challenge to model such binaries for LISA is to significantly increase the accuracy, 
corresponding to the much higher expected values for the signal-to-noise-ratio ({\SNR}) with LISA. However, a tail up to mass ratios of a few times $10^{3}$ also has to 
be expected (Sec.~\ref{ssec:MBHB_expected_source_parameters}). The latest generation of phenom models has indeed already been calibrated to perturbative numerical waveforms that 
correspond to solutions of the Teukolsky equation \cite{Pratten:2020fqn,Garcia-Quiros:2020qpx} at mass ratios of
 $\approx 10^{3}$. Significant progress will still be required regarding both the input calibration waveforms and the models to increase the accuracy, in particular for larger spins 
 and  the inspiral. Developments and challenges toward computing large mass ratio NR waveforms are discussed in Sec.~\ref{sec:NR}, including synergies between traditional numerical 
 methods, analytical and self-force methods.
In principle, the phenom approach may be well suited to some EMRI systems as well. 
Extension to EMRIs seems most feasible for non-precessing quasi-circular systems, but may not be practical 
for precessing or eccentric systems, where the signal can be extremely complex. 

Stellar-origin binaries (Sec.~\ref{sec:sources_SOBHBs}) and Galactic binaries (Sec.~\ref{sec:sources_GBs}) are types of sources for which the phenomenological approach is likely to 
provide a well suited framework. It can be adapted to produce computationally efficient models that can be tuned to data analysis requirements. However, to our knowledge, no work 
has yet been carried out in this direction. 
Finally, one of the strengths of the phenom approach is precisely that it allows for the easy inclusion of approximate phenomenological models of physical effects. In this sense, 
parameterised modifications to GR or models of environmental effects would in principle be straightforward to include.

\subsubsection{Status}\label{sec:phenom:status}

To date four generations of \FD models have been developed, as well as a first generation of time domain models, which correspond to the most recent \FD models in terms of parameter 
space coverage and accuracy. These models have been implemented in the LIGO Algorithms library (LALsuite) \cite{lalsuite} and reviewed by the LIGO-Virgo-KAGRA collaboration.
With the exception of \cite{Hamilton:2021pkf,Thompson:2023ase}, these models  
have only been calibrated to non-precessing quasi-circular numerical waveforms, and precession has been implemented via the ``twisting'' 
approximation \cite{Hannam:2013oca,Bohe:T1500602,London:2017bcn,Khan:2018fmp,Khan:2019kot,Pratten:2020ceb,Estelles:2020osj}, which is based on the fact that in the inspiral 
the precession timescale is much smaller than the orbital timescale. Consequently, the precessing motion contributes little to the \GW luminosity, and the inspiral rate is dominated 
by the non-precessing dynamics. One can thus approximately map non-precessing to precessing waveforms, and in a further step one can use fast \PN approximants to describe 
the Euler angles that rotate the orbital plane and waveform. An effective analytical single-spin description for the Euler angles including next-to-next-to-leading order (NNLO) 
orbit-averaged \PN effects has been developed in \cite{Marsat:2013wwa}, and a double spin approximation based on a multiple-scale analysis (MSA) has been developed 
in \cite{Chatziioannou:2017tdw}.
A key shortcoming of implementations of the twisting-up approach in the frequency domain \cite{Hannam:2013oca,Bohe:T1500602,London:2017bcn,Khan:2018fmp,Khan:2019kot,Pratten:2020ceb} 
has been the use of the \SPA, which is not appropriate for the late inspiral, merger and ringdown. Possible strategies for going beyond the \SPA have been 
discussed in \cite{Marsat:2018oam}, but will require further study and testing. Further shortcomings of the twisting-up approach have recently been discussed in \cite{Ramos-Buades:2020noq}.

Before discussing the current generation of models we briefly summarize the historical development of the first three generations:
\begin{enumerate}
\item The first generation consists of the PhenomA model for the  $l=\vert m\vert =2$ modes of non-spinning binaries \cite{Ajith:2007qp,Ajith:2007kx}.
\item Second generation models PhenomB \cite{Ajith:2009bn} and PhenomC \cite{Santamaria:2010yb} included non-precessing spins, where the two spin degrees of freedom are described 
by a single effective spin. Also, the precessing PhenomP \cite{Hannam:2013oca} model was constructed from PhenomC using an approximate ``twisting-up'' map between precessing and 
non-precessing waveforms.
\item The PhenomD \cite{Husa:2015iqa,Khan:2015jqa} model for the  $l=|m|=2$-mode significantly improved the fidelity of the phenomenological ansatz and the quality of the 
\NR calibration. Several models have been derived from IMRPhenomD: 
\begin{itemize}
\item PhenomPv2 \cite{Bohe:T1500602} updates the PhenomP model for precession.
\item PhenomHM \cite{London:2017bcn} adds subdominant harmonics based on an approximate map from the $l=|m|=2$ mode to subdominant harmonics. 
\item PhenomPv3 \cite{Khan:2018fmp} upgrades the NNLO effective single-spin description of 
precession that had been used for the first two versions of PhenomP to a double-spin description based on the MSA approximation \cite{Chatziioannou:2017tdw}.
\item PhenomPv3HM \cite{Khan:2019kot} includes subdominant harmonics in the twisting-up construction. 
\item Matter effects have been included in terms of 
tidal phase corrections \cite{Dietrich:2018uni,Dietrich:2019kaq} and amplitude corrections of NS-BH systems \cite{Thompson:2020nei}.
\end{itemize}

The PhenomD model has been implemented for use in the LISA data challenge infrastructure \cite{LISA-LCST-SGS-MAN-001}, and has been used in the ``Radler'' edition of the 
LISA data challenge. The PhenomD and PhenomPv2 models have been used for parameter estimation for observed \GW events since the first detection of gravitational waves, 
the GW150914 event \cite{LIGOScientific:2016vlm}. Its derivatives including subdominant harmonics
have also been used to analyse events during the third observing run.
A detailed study of waveform systematics \cite{LIGOScientific:2016ebw} found these models to be accurate enough for the GW150914 event, but also highlighted the need for 
further improvements. This is consistent with a more recent study concerning the need for further improvements for upcoming upgrades to ground-based \GW detectors \cite{Purrer:2019jcp}.
\end{enumerate}

The PhenomX \cite{Pratten:2020fqn,Garcia-Quiros:2020qpx,Garcia-Quiros:2020qlt,Pratten:2020ceb} family provides a thorough upgrade of third generation phenom models. PhenomXAS  \cite{Pratten:2020fqn} replaces PhenomD for the dominant mode, increasing the accuracy by roughly two orders of magnitude in terms of mismatch: \vspace{-0.5em}
\begin{itemize}
\item The number of NR waveforms used for calibration is increased from 20 to $\sim400$.
\item Teukolsky waveforms are included in the data set in order to extend the model to EMRI waveforms.
\item Final spin and mass include EMRI information derived from circular geodesics. 
\item Both spin degrees of freedom are calibrated to numerical data. 
\item Heuristic parameter space fits have been replaced by the systematic hierarchical fitting approach described in \cite{Jimenez-Forteza:2016oae}. An ansatz is selected among classes of polynomial and rational functions by minimizing not only the RMS errors, but also information criteria that approximate a full Bayesian approach, which avoids overfitting and underfitting. This method has, however, not yet been applied to 
dimensions larger than 3 (i.e. precession or spinning eccentric binaries). 
\end{itemize}
PhenomXHM \cite{Garcia-Quiros:2020qpx} extends PhenomXAS to the sub-dominant modes, and PhenomXPHM \cite{Pratten:2020ceb} adds precession via the twisting-up procedure, and allows 
to switch between the NNLO and MSA descriptions for the Euler angles. A first calibration of Phenom models to precessing NR simulations has recently been 
presented in the frequency domain \cite{Hamilton:2021pkf,Hamilton:2023znn}, paving the way for further increasing the accuracy of precessing and eventually generic models in the future. Another important step in improving precessing models has been the incorporation of the mode asymmetry associated with large recoils \cite{Bruegmann:2007bri,Ramos-Buades:2020noq,Kalaghatgi:2020gsq} in the frequency domain \cite{Ghosh:2023mhc,Thompson:2023ase}.  Extensions to BNS and BHNS systems following the approaches of third generation models are in progress \cite{Colleoni:2023czp,Abac:2023ujg}.

The fourth generation PhenomX family demonstrates that a significant increase in the accuracy of phenom models can be paired with a significant decrease in computational cost. 
Consequently, PhenomX provides the fastest Inspiral-Merger-Ringdown (IMR) waveforms currently available for data analysis without \GPU acceleration \cite{Garcia-Quiros:2020qlt,Pratten:2020ceb}. 
Parameter estimation application with GPU acceleration for PhenomHM has been presented in  \cite{Katz:2020hku}. Part of the improvement in efficiency is due to 
the ``multibanding method'', which is based on \cite{Vinciguerra:2017ngf}: analytical error estimates determine the grid spacing for interpolation from a coarse grid, 
and a standard iterative scheme is used to rapidly evaluate the complex exponentials required to compute the waveform from the amplitude, phase, and the Euler angles 
used in  ``twisting up''.

Most recently, the time-domain PhenomT model family \cite{Estelles:2020osj,Estelles:2020twz,Estelles:2021gvs} has been constructed to mirror the features of PhenomX in the time domain. 
A key motivation for the development of time domain models is that they do not require an analytical approximation to the Fourier transform in order to obtain explicit expressions 
for a ``twisted-up'' precessing waveform. This also simplifies the incorporation of analytical results, e.g., concerning the precessing ringdown frequencies; similar simplifications are hoped for when modelling eccentricity. The time domain model has also recently been extended to include the $\ell=2, m=0$ spherical harmonic, which
contains the leading contribution to the gravitational wave memory effect \cite{Rossello-Sastre:2024zlr}.   For  future developments, it is foreseen that both frequency and time domain models will be upgraded in parallel, 
and that each of the two ``branches'' will benefit from progress with the other.

The historical development and current status of phenom models is sketched in Table~\ref{tab:Phenom}.

\begin{table}[t]
    \centering
    \scriptsize{
    \begin{tblr}{
                hline{1-19}={solid},
                hline{21}={solid},
                vlines,
                colspec={Q[l]Q[c]Q[l]Q[c]Q[c]Q[c]Q[c]},
                rows = {abovesep=1pt,belowsep=1pt},
                columns = {rightsep=2pt,leftsep=2pt},
                }
    Family & Domain & Waveform Model & Spins & CP-Frame Modes $(\ell,|m|)$ & Eccentricity & Calibration Region\\
    $\;\;1^{\rm st}$ gen.
    & \SetCell[r=13]{c} {FD} & {\texttt{IMRPhenomA}\\ \cite{Ajith:2007qp,Ajith:2007kx}} & $\times$ & \SetCell[r=7]{c} {$(2,2)$} & \SetCell[r=9]{c} {no} & $q\leq 4$\\
    \SetCell[r=3]{l} {$\;\;2^{\rm nd}$ gen.}
    & &{\texttt{IMRPhenomB}\\ \cite{Ajith:2009bn}} & $\checkmark$ &  &  & \SetCell[r=3]{c} {$q\leq 4, |\chi_{1,2}|\leq 0.75$ \\ $|\chi_{1,2}|\leq 0.85$ (for $q=1$)}\\
    & & {\texttt{IMRPhenomC}\\ \cite{Santamaria:2010yb}} & $\checkmark$ &  &  & \\
    & & {\texttt{IMRPhenomP}\\ \cite{Hannam:2013oca}} & $\checkmark \checkmark$ &  &  & \\
    \SetCell[r=5]{l} {$\;\;3^{\rm rd}$ gen.}
    & & {\texttt{IMRPhenomD}\\ \cite{Husa:2015iqa}\\ \cite{Khan:2015jqa}} & $\checkmark$ &  &  & \SetCell[r=5]{c} {$q\leq 18, |\chi_{1,2}|\leq 0.85 $ \\ $-0.95 \leq |\chi_{1,2}| \leq 0.98$ \\ (for $q=1$)}\\
    & & {\texttt{IMRPhenomPv2}\\ \cite{Bohe:T1500602}} & $\checkmark \checkmark$ &  &  & \\
    & & {\texttt{IMRPhenomPv3}\\ \cite{Khan:2018fmp}} & $\checkmark \checkmark$ &  &  & \\
    & & {\texttt{IMRPhenomHM}\\ \cite{London:2017bcn}} & $\checkmark$ & \SetCell[r=2]{c} {$(2,2),(2,1),(3,3),$\\ $(4,3),(4,4)$} &  & \\
    & & {\texttt{IMRPhenomPv3HM}\\ \cite{Khan:2019kot}} & $\checkmark \checkmark$ &  &  & \\
    \SetCell[r=8]{l} {$\;\;4^{\rm th}$ gen.}
    & & {\texttt{IMRPhenomXAS}\\ \cite{Pratten:2020fqn}} & $\checkmark$ & \SetCell[r=2]{c} {$(2,2)$} & \SetCell[r=4]{c} {in dev.} & \SetCell[r=8]{c} {NR: \\ $q \leq 18, |\chi_{1,2}| \leq 0.99$ \\ Teukolsky: \\ $q \leq 1000$}\\
    & & {\texttt{IMRPhenomP}\\ \cite{Pratten:2020ceb}} & $\checkmark \checkmark$ &  &  & \\
    & & {\texttt{IMRPhenomXHM}\\ \cite{Garcia-Quiros:2020qpx}} & $\checkmark$ & \SetCell[r=2]{c} {$(2,2),(2,1),(3,2),$\\ $(3,3),(4,4)$} &  & \\
    & & {\texttt{IMRPhenomPHM}\\ \cite{Pratten:2020ceb}} & $\checkmark \checkmark$ &  &  & \\
    & \SetCell[r=4]{c} {TD} & {\texttt{IMRPhenomT}\\ \cite{Estelles:2020osj}} & $\checkmark$ & \SetCell[r=2]{c} {$(2,2)$} & \SetCell[r=4]{c} {in dev.} & \\
    & & {\texttt{IMRPhenomTP}\\ \cite{Estelles:2020osj}} & $\checkmark \checkmark$ &  &  & \\
    & & {\texttt{IMRPhenomTHM}\\ \cite{Estelles:2020twz}} & $\checkmark$ & \SetCell[r=2]{c} {$(2,2),(2,1),(3,3),$\\$(4,4),(5,5)$} &  & \\
    & & {\texttt{IMRPhenomTPHM}\\ \cite{Estelles:2021gvs}} & $\checkmark \checkmark$ &  &  & \\
     \SetCell[c=7]{c} {$\times$ no spins $\checkmark$ spins aligned with orbital angular momentum $\checkmark \checkmark$ precessing spins}\\
     \SetCell[c=7]{c} {CP: mode content in co-precessing frame}\\
    \end{tblr}
    }
    \caption{Progress in the development of phenomenological waveform models in frequency domain (\FD) and time domain (\TD). \texttt{IMRPhenom} models are implemented in \texttt{LALSuite} \cite{lalsuite} and reviewed by the LIGO-Virgo-KAGRA collaboration.}
    \label{tab:Phenom}
\end{table}

\subsubsection{Environmental effects}\label{sec:phenom:beyondGR}

The phenom approach provides significant freedom to incorporate additional features, such as poorly known subdominant \GR effects,
beyond-\GR effects (see Sec.~\ref{ssec:phenom_beyond_GR}), and environmental effects. 
We note that analysing the presence of such effects also requires accurate \GR waveforms for comparison and will benefit from the techniques developed for \GR. Work toward tests of \GR or the presence of environmental effects may call for improving the accuracy or other features of \GR models. While the modelling of environmental effects has not yet been explored in the phenomenological waveform program, the phenomenological models are primed for their addition in two ways: \vspace{-0.5em}
\begin{enumerate}
\item The modular structure of phenomenological waveforms should facilitate the incorporation of known environmental effects, e.g. by augmenting \PN information about the inspiral with information about environmental effects. 
\item Environmental effects for which a quantitative model is not yet available 
could be incorporated into the phenom ansatz in a parameterised way, similar to existing \GW tests for theory-agnostic deviations from GR. 
\end{enumerate}
In both cases it will be useful to have both frequency and time domain phenomenological waveform models available, allowing one to choose the natural domain for a given effect.

\subsubsection{Challenges}\label{sec:phenom:challenges}

Waveform modelling for comparable mass binaries in general faces four main challenges: \vspace{-0.5em}
\begin{enumerate}
\item The availability of a sufficient number of high-accuracy numerical relativity waveforms throughout the parameter space, including precession and eccentricity. (This is not specific for  phenom models.)
\item The lack of accurate analytical descriptions, in particular for precession and eccentricity. Here, specific challenges arise for the phenom approach, which uses ``deformed'' \PN expressions as the basis for constructing a computationally efficient inspiral ansatz. This will benefit in particular from further analytical developments in precession and eccentricity.
\item The development of modelling techniques that can produce efficient models from high-dimensional data sets. For the phenom approach, it may turn out to be difficult to develop closed form expressions that accurately represent the full morphology of precessing and eccentric binaries, and to accurately interpolate the full parameter space without compromising computational speed.
\item Development of an overall data analysis strategy and concrete code framework. The flexibility and modularity of the phenom approach could be exploited to develop variants of 
models with different tradeoffs between accuracy and speed, or accuracy and broad coverage of parameter space, which can then be utilised to optimise computational efficiency. 
\end{enumerate}
Significant coupling between these challenges is foreseen, e.g. the number and quality of NR waveforms required will also depend on the progress with analytical results. In addition, there will be more sophisticated data analysis approaches, which could exploit the advantages and disadvantages of different models, or tunable parameters, which could allow one to trade accuracy for speed. These could eventually give guidance for the development of waveform models that are designed as an integral part of the data analysis strategy.
These challenges are not specific to LISA, but apply in general to further model improvements. LISA, however, poses particular challenges due to the extreme accuracies required for the loudest events, and also due to the more complicated nature of the detector response (see e.g.~\cite{Marsat:2018oam}). Strategies of how to best employ waveform models in data analysis applications,
resolving tradeoffs of accuracy versus computational cost, can not be decoupled from considerations regarding the detector response, and new research will be required to exploit the simplicity and flexibility of phenomenological waveform models in order to accelerate the evaluation of the waveform together with the detector response.

The non-precessing quasi-circular sector will provide important initial guidance on how to connect with EMRI descriptions (or to which degree this is possible). This will provide an arena for 
toy-model explorations of highly accurate models in limited regions of parameter space, determining the limits of accuracy and computational efficiency that can be achieved, 
as well as studying issues of accuracy and efficiency related to the LISA detector response. It is therefore important to continue pushing to higher accuracies with these models.

The development of accurate phenomenological models of precessing systems is expected in both the time and frequency domains. Time-domain models will build on \cite{Estelles:2020osj}, which is foreseen to serve as a testing ground to further extend the calibration to \NR, and will facilitate to use time domain implementations of the LISA response function.  Further progress in the precessing inspiral will also crucially depend on progress with approximations that reduce the need for NR calibration for the inspiral across the large parameter space of precessing binaries, such as those based on the MSA \cite{Chatziioannou:2017tdw} or dynamical renormalisation group (DRG) \cite{Galley:2016zee,Yang:2019oqm}.
In order to accurately describe the late inspiral, merger and ringdown of precessing systems in the frequency domain, further tests and implementations of the strategies discussed in \cite{Marsat:2018oam} will be required, such as analytical treatments that go beyond the \SPA.

Eccentric waveform modes are again likely to require cross-pollination between frequency and time domain models (motivating in part the development of PhenomT in parallel to PhenomX), and significant advances in the development of analytical descriptions for the inspiral phase of spinning and eccentric binaries.
This is particularly true where binary systems show both precession and eccentricity. It will be important to understand whether simple approaches twisting up eccentric waveforms will be sufficient for moderate \SNR. Further investigations will determine if these prescriptions are adequate to determine binary parameters with sufficient accuracy, such that only a moderate number of NR waveforms is required to develop an accurate local model for high \SNR events.

A key challenge is to develop models that are accurate across a large region of parameter space, describing well the changes of waveform morphology, e.g., as the mass ratio increases, or as spin alignment migrates between aligned and anti-aligned with the orbital angular momentum. For current ground-based detectors, parameter estimation posteriors are typically rather broad, so  refined models in a smaller region of the parameter space would have very limited use. For LISA, however, it is foreseen that a hierarchy of models will be developed: those that are sufficiently accurate to identify signals across large portions of parameter space, 
and others that further increase the accuracy for smaller regions of parameter space, which would be used for the most accurate parameter estimation of the loudest signals. Models with parameters allowing one to trade accuracy for computational efficiency would be appropriate for hierarchical approaches to data analysis. This approach could also be extended further by mixing descriptions in terms of closed-form expressions with equation-solving, e.g. to improve the accuracy of describing eccentricity or precession.
An important open question is how to incorporate results from the self-force program (see Sec.~\ref{sec:GSF}) into phenomenological models, and how to use ideas 
developed for the phenomenological waveforms program to accelerate self-force waveforms.

\newpage

\section{Waveform generation acceleration}
\label{sec:waveform_accel}

Coordinator: Alvin Chua, Michael Katz\\
Contributors: S.~Field\\

The stringent efficiency requirements described in Sec.~\ref{sec:efficiency_requirements} naturally necessitate the acceleration of waveform models, in order to enable the LISA science analysis. The two main strategies to achieve this are the development of improved computational techniques, and hardware acceleration.

\subsection{Computational techniques}\label{sec:comp_techniques}

\begin{figure}
    \centering
    \includegraphics[width=0.34\textwidth]{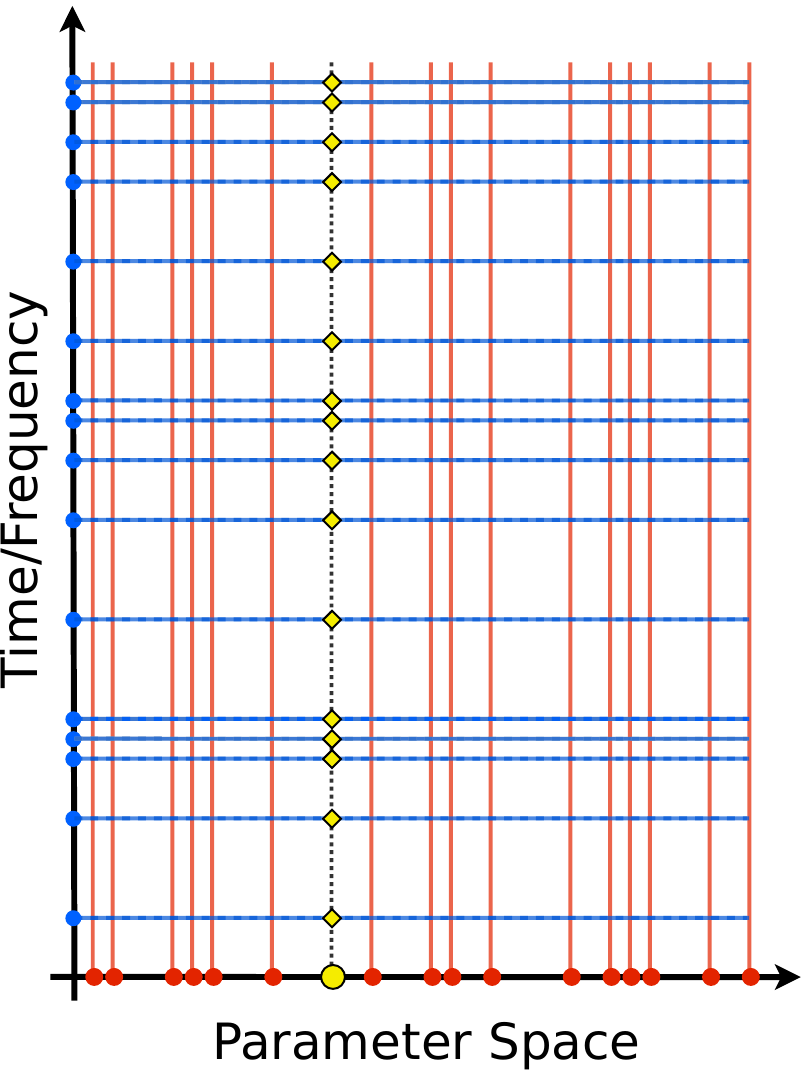}
    \caption{
    Waveform acceleration can be achieved by building a surrogate model through a sequence of three steps.
    (Step 1) The red solid lines show the waveform training data evaluated at specific parameter values.
    (Step 2) Reduced-order modeling is used to compress the waveform data through a dimensional-reduction step, and the blue dots represent the data that needs to be modeled after compression.
    (Step 3) The blue lines indicate fits for the compressed waveform data across the 
    parameter space.
    The yellow dot shows a generic parameter value, say $\pmb{\lambda}_0$, that is not
    in the training set. To compute the waveform for this value,
    each fit is evaluated at $\pmb{\lambda}_0$ (the yellow diamonds), and then the
    reduced-order representation is used to reconstruct the decompressed waveform (the dotted black line). }
    \label{fig:Surrogate}
\end{figure}

To maximize the science returns from highly relativistic LISA sources with exacting modeling requirements, we require waveform-acceleration techniques that do not compromise waveform accuracy. The general solution to this problem is to construct interpolants or fits for high-fidelity waveform data, where the data comes from an accurate underlying model that is too slow to use directly in data analysis studies (e.g., as it involves solving \PDEs \edit{ or coupled ODEs}). This data can describe either the full time-/frequency-series representation of the waveform, or specific waveform components that are more computationally expensive. While many waveform acceleration techniques have appeared in the literature, most rely on the three steps summarized below \edit{and illustrated in Fig.~\ref{fig:Surrogate}.}

\subsubsection{Step 1: Data representation}

A gravitational waveform's parametric dependence is generally too complicated to model directly with an acceptable degree of accuracy. Instead, the waveform is typically decomposed into waveform data pieces that are simpler, more slowly varying functions of parameters and time. If we were to take a machine-learning viewpoint of the problem, data pieces are features of the data, and discovering what problem-specific features to extract is a feature engineering task. How should we define our features?

For non-precessing \BBHs in quasi-circular orbits, an amplitude and phase decomposition for each $(\ell,m)$ angular-harmonic mode is an obvious choice for our data pieces.
For precessing systems, the problem becomes significantly more complicated as the modes have a rich signal morphology. Here, the general approach has been to transform the modes to a co-rotating frame where the system has significantly reduced dynamical 
features~\cite{Blackman:2017dfb,Blackman:2017pcm,Varma:2019csw}. It is often advantageous to further decompose the data into symmetric and asymmetric mode combinations~\cite{Blackman:2017dfb,Blackman:2017pcm,Varma:2019csw}. The price of this simplification is that the frame dynamics must also be modeled, so that the co-rotating modes can be transformed back into the inertial frame of the detector.
For eccentric BBH systems, which will be important for LISA science, the optimal data representation is currently unknown.

The highly disparate masses and large separation of timescales in \EMRI systems allow their inspirals to be modeled within the two-timescale framework \cite{Miller:2020bft}.
This allows the trajectory of the inspiral through the orbital configuration phase space to be computed in milliseconds \cite{Katz:2021yft, Wardell:2021fyy}.
The waveform is then computed by sampling the metric amplitudes along this inspiral typically by evaluating a precomputed interpolant of the amplitudes.
Most often these amplitudes are computed in the frequency domain so the waveform is decomposed into harmonics of not just angular dependence, but also the fundamental frequencies in the system \cite{Hughes:2001jr,Drasco:2005kz}.
Generically precessing and eccentric \EMRI waveforms are extremely complex, but the above description effectively breaks the waveform down into slowly evolving sinusoids. 
The only drawback is that there are far more modes to compute and sum for a typical waveform ($\gtrsim10^5$), though in practice many of these do not carry much power \cite{Katz:2021yft}.

\subsubsection{Step 2: Reduced-order modeling}

The waveform quantities of interest often reside in a representation space of extremely high dimensionality $D$, and must generally be interpolated or fitted over a parameter space of modest dimensionality $d$ as well. For example, the full waveform itself for strong-field LISA sources typically has $D\gtrsim10^4$ (signals of $\gtrsim10^5\,\mathrm{s}$ sampled at $\gtrsim0.1\,\mathrm{Hz}$) and $d\sim10$. These computational difficulties can often be mitigated through the use of dimensional-reduction tools \cite{Prudhomme2001,barrault2004empirical, Binev10convergencerates2,Devore2012,Maday:2002,Quarteroni:2011, Cannon:2012gq,Cannon:2010qh,Cannon:2011xk,Cannon:2011rj,Field:2011mf,Herrmann:2012xpx}, the collection of which is broadly referred to as  \ROMs.

\ROMs aim to remove redundancies in large data sets, making them more amenable to analysis, and identifying relevant features for further approximation. It typically begins with a greedy algorithm or singular value decomposition that identifies the set of $n$ ``most important'' values in parameter space, whose associated waveform quantities span a \textit{reduced-basis} space within which the model is well approximated \cite{Field:2011mf,Herrmann:2012xpx,Field:2013cfa,Purrer:2014fza}. This serves as both a dimensional-reduction step as well as a parametric-sampling step.

While invaluable for reducing computational cost on the representation end ($n\ll D$), the compression provided by \ROMs does not actually lessen the intrinsic complexity of the model. Indeed, the successful application of \ROMs relies on sufficient coverage of the parameter space with training data in the first place---or, more realistically, on truncation of the parameter domain to compensate. \ROM themselves also do not address the issue of interpolating or fitting the reduced representation at high accuracy, which can be challenging even in parameter-spaces with $d\gtrsim3$.


\subsubsection{Step 3: Interpolation and fitting}

The last step is to interpolate or fit each compressed data piece (viewed as scalar- or vector-field data) over $d>3$ dimensions, and to do so both accurately ($\lesssim10^{-6}$ error) and efficiently ($\lesssim1$ seconds). In the context of sources and models for ground-based observing, a variety of choices have appeared, including polynomials~\cite{Field:2013cfa}, splines~\cite{Purrer:2014fza}, Gaussian process regression~\cite{Doctor:2017csx}, deep neural networks \cite{Chua:2018woh,Khan:2020fso,Thomas:2022rmc,Khan:2021czv}, and forward stepwise greedy regression using a custom basis~\cite{Blackman:2017pcm}. A recent review article provides informative comparisons between some methods~\cite{Setyawati:2019xzw}. The accuracy and efficiency goals of models for ground-based observing are more modest, however, and it is unclear if these methods will continue to work for LISA \MBHs~\cite{Purrer:2019jcp,Ferguson:2020xnm}, or for \EMRIs in full generality~\cite{Chua:2020stf}. As such, other interpolation and regression methods should continue to be explored.

\subsubsection{Status of techniques}

In the \GW literature, the term surrogate modeling is sometimes used as a generalization of \ROMs (for the full waveform) to include additional steps beyond dimensional reduction. 
\ROM surrogates have been built for many different waveform families, regions of parameter space, and signal durations. 
Early models focused on closed-form waveform families largely as a proof-of-principle exercise \cite{Herrmann:2012xpx,Field:2011mf,Caudill:2011kv,Cannon:2011rj, Cannon:2011xk,Cannon:2010qh}. 
Since then, ROM techniques have been further developed and refined for various \LVK (hence comparable-mass) applications: EOB \cite{Field:2013cfa,Purrer:2014fza, Purrer:2015tud,Cotesta:2020qhw,Gadre:2022sed,Khan:2020fso,Thomas:2022rmc}, NR \cite{Blackman:2015pia, Blackman:2017dfb,Blackman:2017pcm,Varma:2019csw,Varma:2018mmi}, multiple subdominant modes \cite{Field:2013cfa,Cotesta:2020qhw,Blackman:2015pia,Blackman:2017dfb,Blackman:2017pcm,Varma:2019csw,Varma:2018mmi}, eccentric binaries \cite{Islam:2021mha}, frequency-domain waveforms \cite{Purrer:2014fza,Purrer:2015tud,Cotesta:2020qhw}, time-domain waveforms \cite{Field:2013cfa,Blackman:2015pia, Blackman:2017dfb,Blackman:2017pcm,Varma:2019csw, Varma:2018mmi}, generically precessing systems \cite{Blackman:2015pia,Blackman:2017dfb,Blackman:2017pcm,Varma:2019csw}, neutron star inspirals with tidal effects \cite{Lackey:2016krb,Lackey:2018zvw,Barkett:2019tus,Matas:2020wab,Abac:2023ujg}, and waveforms with $\sim10^4$ cycles \cite{Purrer:2014fza,Purrer:2015tud,Cotesta:2020qhw,Varma:2018mmi,Yoo:2023spi}.

When applied to computationally expensive models such as \EOB or \NR, \ROM surrogates can accelerate the generation of a single waveform by factors of $10^2$ (\EOB models; described by \ODEs) to $10^8$ (\NR models; described by partial differential equations), while being nearly indistinguishable to the underlying model. \EOB surrogates are extensively used as part of \LVK parameter-estimation efforts as well as template-bank detection \cite{LIGOScientific:2018mvr,LIGOScientific:2016vlm,LIGOScientific:2021usb,KAGRA:2021vkt,LIGOScientific:2020ibl}. NR surrogates have been used in numerous targeted follow-up studies on specific \BBH events \cite{LIGOScientific:2020stg,LIGOScientific:2020iuh,Kumar:2018hml,Islam:2023zzj}. As such, \ROM techniques have been essential to the widespread use of both \EOB and \NR waveforms for realistic data analysis efforts with \LVK data. 

The present surrogate framework is ill-suited to \EMRI waveform modeling, where the increased duration and mode complexity combine to increase the information content of the model by at least 20 orders of magnitude \cite{Gair:2004iv,Chua:2017ujo,Moore:2019pke}. Surrogates might be useful in special cases of the \IMRI regime, where these difficulties are lessened. There has been recent work on the first models \cite{Rifat:2019ltp,Islam:2022laz} for quasi-circular, non-spinning \IMRI systems, based on solutions to the time-domain Teukolsky equation rescaled to fit NR results.
The latest surrogate waveform model in this series covers the last $30,000$M of the inspiral-merger-ringdown signal, with mass ratios ranging from 2.5 to $10^4$. To put this in context, for a million-Solar-mass system, this is equivalent to approximately 2 days of signal, which is inadequate for the expected observable duration of months to years for an \EMRI of that mass. However, it should suffice to describe an \IMRI signal of similar total mass and a mass ratio of $\sim10^3$.

In the angular- and frequency-based decomposition of an \EMRI model, the problem separates naturally into two main parts: that of generating the inspiral trajectory (from which the waveform phasing is straightforwardly derived), and that of generating the waveform amplitude for each of the $\sim10^5$ constituent modes. Fast frameworks for \EMRI trajectory generation already exit \cite{VanDeMeent:2018cgn,Miller:2020bft}. They present no computational difficulties at adiabatic order (for search); however, their extension to post-adiabatic order (for inference) will require interpolating the first-order fluxes to a relative precision of the mass ratio ($\sim\epsilon$) \cite{Osburn:2015duj}. Trajectory-level interpolation with \ROM is unlikely to be of practical use, due to issues of accuracy and parameter-space extensiveness. The accuracy requirements on the waveform amplitudes are far less stringent, though, and it is here that ROM compression and neural-network interpolation has been applied to good effect (in the mode direction, rather than the time/frequency direction) \cite{Chua:2020stf}.

\subsubsection{Future challenges}

LISA data analysis has three distinctive features that will require new ideas in waveform acceleration. First, the waveforms are significantly longer in duration, which will stress both computational resources and current \ROM techniques.
Second, as compared to \LVK waveform modeling, LISA accuracy requirements are significantly more demanding. 
The fitting schemes will thus need to deliver higher accuracies while also contending with increased dimensionality. 
Finally, the orbital motion of LISA introduces a time-dependency in the response of the detector, which also needs to be modeled efficiently for data analysis applications; see, e.g., \cite{Marsat:2018oam}.

In terms of source-specific challenges, eccentricity is expected to play a more important role in the modeling and interpretation of LISA \MBHs.
This will necessarily increase the problem from 7 to 9 parametric dimensions. 
While this should not pose significant difficulty for the dimensional-reduction step \cite{Herrmann:2012xpx,Blackman:2014maa}, the parameter-fitting step will be vastly more difficult, and it is unlikely that any of the existing techniques will continue to perform well. 
For EMRIs, the main difficulty is ensuring that the waveform-amplitude fit retains its accuracy when extended from eccentric Schwarzschild orbits at present \cite{Chua:2020stf} to generic Kerr orbits with a much larger parameter and representation space.

\subsection{Hardware accelerators / configurations}\label{sec:hardware_accelerators}

In addition to computational techniques, generating waveforms with a variety of hardware accelerators can vastly improve analysis time. 
Hardware acceleration, in a general sense, means hardware designed for a specific type of optimization when compared to general-purpose \CPUs. 
Some examples of commonly used hardware accelerators include \GPUs, Field-Programmable Gate Arrays (FPGAs) \cite{vanderbauwhede2013high}, Tensor Processing Units (TPUs) \cite{paper2021state}, and Artificial Intelligence (AI)-accelerators \cite{mishra2023artificial}.

\GPUs are designed specifically for parallel computations. 
These devices leverage a large number of compute cores ($\sim1000s$) and specialized memory structure to make a large number of independent calculations simultaneously. 
From an algorithmic standpoint, many of the methods discussed in Sec.~\ref{sec:comp_techniques} can be implemented on \GPUs to achieve considerable gains in efficiency. 
These include, but are not limited to, artificial neural networks (matrix multiplication), interpolation, parallel sorting, and basis transformation. 
Additionally, some accurate waveform calculations can be implemented in parallel, allowing for all or parts of waveforms to be implemented entirely for \GPUs. 
Programming for \GPUs is generally performed in \texttt{C/C++} or, as of more recently, \texttt{Python}. 

While \GPUs take advantage of their parallel architecture, they are still programmed using specific software for a fixed hardware unit. 
FPGAs, on the other hand, represent ``programmable hardware.'' 
These devices are programmed at the device level to implement customized hardware flow designed for specific tasks. 
Their most common usage is in neural networks and signal processing. 
Programming on FPGAs is done in \texttt{Verilog} or \texttt{VHDL} \cite{chu2011fpga}, although some wrappers in \texttt{C/C++} are starting to become available.

\subsubsection{Status}

Recently FPGAs have started to be used by the GW community \cite{Que:2021cqo,Martins:2024rsj} for data analysis but, to the best of our knowledge, they have not been used for waveform generation.
In recent years, \GPUs have seen increased interest from the community for use in waveform creation, as well as \GW-related analysis in general. 
\GPUs have been used for precessing waveforms, as well as population analysis related to ground-based observations from the \LVK observing network \cite{Talbot:2019okv,Edwards:2023sak}. In this work, the waveform and the detector response were calculated directly on the \GPU hardware. To limit the cost of Bayesian inference in ground-based observing, \GPUs were recently used to greatly accelerate the creation of surrogate waveform models using techniques such as artificial neural networks \cite{Khan:2020fso}. The creation of \MBHB waveforms has also been implemented for LISA, including both the source-frame waveforms and the LISA response function for use in LISA parameter estimation studies \cite{Katz:2020hku} ($\sim$500x acceleration). The generation scheme for this work calculated, in parallel, accurate waveforms and LISA response information on sparse grids. Then, to scale to the full data stream, cubic spline interpolation was leveraged. This technique is generally well-suited for parallelization.  

For \EMRIs, \GPUs have been used to improve the efficiency of black hole perturbation theory calculations within the \GSF community \cite{McKennon:2012iq,Khanna:2010jv,Field:2020rjr}. In this work, the many harmonics of \EMRI metric perturbation are computed in parallel as the orbit evolves. In terms of data-analysis related waveform creation for \EMRIs, the \AAK \cite{Chua:2017ujo} waveform from the \texttt{EMRI Kludge Suite} was accelerated with \GPUs (\texttt{gpuAAK}) for use in the LISA Data Challenges ($\sim$1000x). Additionally, the first fully relativistic fast and accurate \EMRI waveform that goes beyond the kludge approximations was implemented on \GPUs resulting in a considerable increase in efficiency ($\sim$2000x) \cite{Chua:2020stf,Katz:2021yft}. Both the \texttt{gpuAAK} and the fast and accurate \EMRI waveforms generate sparse orbital and phase trajectories in serial on a \CPU. These sparse trajectories are then scaled with spline interpolation to the full waveform sampling rate. This scaling is performed for each separate harmonic mode so that they can be properly and accurately combined into the final waveform. It is actually this summation over $\sim10^6-10^7$ time points, each with a combination of $\sim$1000s of harmonic modes, that most requires the \GPU architecture for optimal results. 

\subsubsection{Challenges}

A central challenge when working with \GPUs, FPGAs, or any other specialized hardware is the trade-off between efficiency and streamlined code creation and maintenance. For optimal efficiency, it is generally necessary to implement code specific to each type of hardware. This could mean implementation in an entirely different language with, for example, an FPGA, or an adjustment for different hardware or software requirements for each application. An example of this is the use of shared memory in \GPUs. Shared memory is a small amount of memory available on the \GPU chip, which is much faster to access in comparison to the larger global memory located off of the chip. Leveraging shared memory is essential to achieving optimal efficiency for many applications. There is no use of this form of shared memory on \CPUs. Therefore, it may be desirable to implement a fast \GPU version and a mirror \CPU version that effectively uses the same exact code as the \GPU version with slight adjustments. On the other hand, a \CPU version optimized for a \CPU may be needed, creating two separate codes. In general, it is helpful to minimize the amount of code duplication as much as possible, but that may be governed by efficiency requirements. 

Many applications require some parallelization over \CPUs in addition to specific hardware considerations. Understanding how to best leverage all resources is key to achieving maximal efficiency. With that said, maximal efficiency should be achieved while maintaining a constant user interface. This allows for the application of a waveform within a desired program while using a variety of hardware. 

Another general concern about the use of specialized hardware is just that: it is specialized. Purchasing specialized hardware in place of \CPUs will penalize the breadth at which the hardware can be used. This has generally hurt the availability of accelerators at large. Similarly, accelerators come at a variety of prices. Academic-grade \GPUs (indicating they have strong performance in double precision) are generally expensive. However, the newest generations of accelerators combined with modern algorithms within the \GW field have proven that efficiency is not only highly desirable, but cost efficient when compared to the equivalent number of \CPUs (up to $\sim$ 1000s). Fortunately, researchers now often have access to clusters of CPUs, GPUs, TPUs, and AI-accelerators through local, national, and international scale computing centres. Ultimately, the combination of a variety of hardware will be necessary to accomplish the scientific goals of the LISA mission. 


\newpage

\section{Modelling for beyond GR, beyond Standard Model, and cosmic string sources}
\label{sec:modling_beyond_GR}
Coordinators: Richard Brito and Daniele Vernieri \\
Contributors:
J.~Aurrekoetxea, L.~Bernard, G.~Bozzola, A.~Cárdenas-Avendaño, D.~Chernoff, K.~Clough,
T.~Helfer, T.~Hinderer, E.~Lim, G.~Lukes-Gerakopoulos, E.~Maggio, A.~Maselli, D.~Nichols,
J.~Nov\'{a}k, M.~Okounkova, P.~Pani, G.~Pappas, V.~Paschalidis, M.~Ruiz, A.~Toubiana,
A.~Tsokaros, J.~Wachter, B.~Wardell, H.~Witek\\

As explained in Sec.~\ref{sec:sources_beyond_GR}, the observation of gravitational waves with LISA has an enormous potential to probe physics beyond GR and the Standard Model of particle physics, in regimes that have so far not been explored~\cite{Barausse:2020rsu,LISA:2022kgy}. In order to fulfill this potential, we need to construct accurate waveforms that take into account signatures of new physics. As we will discuss in this Section, this is in fact a very challenging task that for most cases is still at its early stages of development.

In this section we provide a brief summary of how beyond GR theories and the effect of beyond Standard Model physics can be modelled using the techniques presented in Sec.~\ref{sec:modelling_binaries}. The discussion will be mostly concerned with modelling the \emph{generation} of gravitational waves by LISA sources in the presence of new physics. Beyond-GR effects can also affect the way gravitational waves \emph{propagate} on cosmological scales, see e.g.~\cite{Mirshekari:2011yq,Saltas:2014dha,Belgacem:2017ihm,LISACosmologyWorkingGroup:2019mwx,Bonilla:2019mbm,DAgostino:2019hvh,Allahyari:2021enz,DAgostino:2022tdk,LISACosmologyWorkingGroup:2022jok}. In general, separation of scales allows to add modifications to the \GW propagation on top of any waveform model for which the generation problem is understood and therefore we do not discuss this possibility here. More details on this problem can be found in Refs.~\cite{LISACosmologyWorkingGroup:2019mwx,LISACosmologyWorkingGroup:2022jok}. We will also not discuss in detail the potential implications that LISA observations will have for fundamental physics, since a more detailed review on this issue can be found in LISA's Fundamental Physics Working Group white paper~\cite{LISA:2022kgy}.

This section follows a similar structure to Sec.~\ref{sec:modelling_binaries} where various modelling approaches for compact binaries are discussed. However, the modelling of gravitational waves emitted by cosmic strings is discussed in a separate subsection~\ref{ssec:cosmic_strings}, mainly because these sources require modelling approaches that are specific to cosmic strings.


\subsection{Numerical relativity}

\subsubsection{Beyond GR}\label{sec:NR_beyondGR}

NR simulations of BBH mergers in beyond-GR theories are only at their infancy and have only been performed in a handful of theories, e.g.~\cite{Healy:2011ef,Berti:2013gfa,Cao:2013osa,Okounkova:2017yby,Hirschmann:2017psw,Witek:2018dmd,Okounkova:2019dfo,Okounkova:2019zjf,East:2020hgw,Figueras:2021abd,East:2021bqk,AresteSalo:2022hua,Ripley:2022cdh,Cayuso2023}. Even though most studies have so far focused on proof-of-principle simulations, in recent years there has been significant improvements in this topic.

The earliest \NR simulations of BBH mergers in beyond GR theories focused on a particular
class of scalar-tensor theories that are known to possess a well-posed initial value
problem~\cite{Salgado:2008xh,Healy:2011ef,Berti:2013gfa,Cao:2013osa}.
In these scalar-tensor theories, \BHs satisfy no-hair
theorems~\cite{Hawking:1972qk,Sotiriou:2011dz},
and BBH mergers are indistinguishable from GR.
However, in more complex extensions of \GR, or for scalar matter with nontrivial boundary
conditions
(e.g., cosmological boundary conditions or large scalar-field gradients),
\BHs can carry a scalar charge, and these solutions need not be unique.

One of the main difficulties when trying to do NR simulations in more complex theories is the fact that the well-posedness of the initial value problem
(see e.g.~\cite{baumgarteShapiroBook,Sarbach:2012pr,Hilditch:2013sba})
in alternative theories is for most cases particularly challenging to establish
~\cite{Delsate:2014hba,Papallo:2017qvl,Papallo:2017ddx,Cayuso:2017iqc,Sarbach:2019yso,Ripley:2019hxt,Ripley:2019irj,Kovacs:2019jqj,Kovacs:2020pns,Kovacs:2020ywu,Witek:2020uzz,Julie:2020vov,Bezares:2020wkn,terHaar:2020xxb,Silva:2020omi,AresteSalo:2022hua,Ripley:2022cdh,Barausse:2022rvg,deRham:2023ngf}.

To circumvent these problems, two treatments have been proposed in which alternative theories are considered as effective field theories of gravity. One approach uses an
{\textit{order-by-order expansion}}~\cite{Benkel:2016rlz,Benkel:2016kcq,Okounkova:2017yby,Witek:2018dmd,Okounkova:2019dfo,Okounkova:2019zjf,Doneva:2022byd},
in which the spacetime metric and extra fields are expanded around GR,
order-by-order in the coupling constant~\footnote{Sometimes this is also called
``order-reduction scheme.'' This is not to be confused with order-reduction schemes in
which the field equations themselves are used to replace higher-order curvature terms
that are then kept only to a given (reduced) order in the coupling constant.},
 to guarantee well-posedness.
The second approach, the so-called \textit{equation-fixing} method, proposes to use a
scheme inspired by the Israel-Stewart treatment of viscous relativistic hydrodynamics, in
which an effective damping controls higher frequency modes while preserving the physics
of the low frequency modes (nearly)
untouched~\cite{Cayuso:2017iqc,Allwright:2018rut,Cayuso:2020lca,Franchini:2022ukz,Cayuso2023}.
In this second approach the alternative theory equations of motion are viewed as a
low-energy effective deviation from the Einstein equations, such that the theory can be
captured by the low-energy degrees of freedom. Although these methods are quite generic,
they have so far been mainly employed in theories with higher derivative terms in the
metric~\cite{Benkel:2016rlz,Benkel:2016kcq,Okounkova:2017yby,Witek:2018dmd,Okounkova:2019dfo,Okounkova:2019zjf,Okounkova:2020rqw,Cayuso:2020lca,Silva:2020omi,Doneva:2022byd,Elley:2022ept,Cayuso2023},
such as quadratic theories of gravity with an extra scalar field (for which the
order-by-order expansion was used), namely dynamical Chern-Simons gravity (dCS)~\cite{Alexander:2009tp, Green:1984sg, Taveras:2008yf, Mercuri:2009zt}, Einstein dilaton Gauss-Bonnet gravity (EdGB)~\cite{Kanti:1995vq, Cayuso2023, Moura:2006pz, Berti:2015itd} or more generic Einstein scalar Gauss-Bonnet theories (ESGB)~\cite{Doneva:2017bvd,Silva:2017uqg,Antoniou:2017acq}. Unlike the simplest scalar-tensor theories, in dCS and EdGB BHs naturally possess scalar ``hair'', and therefore deviations from GR naturally occur. Some ESGB theories are prone to BH scalarization, i.e., a process in which GR BHs become unstable against the spontaneous development of scalar hair for large enough spacetime curvatures~\cite{Doneva:2017bvd,Silva:2017uqg,Antoniou:2017acq,Doneva:2022ewd} (see also~\cite{Dima:2020yac,Herdeiro:2020wei,Hod:2020jjy,Berti:2020kgk,Doneva:2021dqn,Doneva:2021dcc,Elley:2022ept} for a similar process but where scalarization is controlled by the magnitude of the BH spin).

In particular, using the order-by-order expansion mentioned above, BBH simulations were
done in dCS~\cite{Okounkova:2019dfo,Okounkova:2019zjf} and EdGB~\cite{Okounkova:2020rqw,
Witek:2018dmd}. Importantly, this scheme also simplifies the problem of covering the
parameter space needed to build IMR waveforms for parameter estimation purposes. Since NR
is \textit{scale-invariant} and the dCS and EdGB coupling parameters are scaled out of
the order-by-order expansion scheme~\cite{Okounkova:2019dfo}, a simulation for one set of BBH parameters (masses and spins) can be applied to any total mass, and any valid value of the coupling constants. With enough NR simulations, one can thus hope to cover the the full BBH parameter space, build a surrogate model for parameter estimation~\cite{Varma:2019csw}, and perform model-dependent tests of GR with gravitational waves. Work towards model-dependent tests using full IMR signals obtained through this method has been done in Ref.~\cite{Okounkova:2022grv}, where parameter estimation was performed on full BBH waveforms in dCS.

Despite their promises and advantages, the schemes mentioned above also suffer from some
important problems. The main problem with the order-by-order expansion scheme
is the fact that solutions may be plagued by secularly growing solutions signaling the breakdown of
perturbation theory on long timescales. As proposed in Ref.~\cite{GalvezGhersi:2021sxs},
such problem could potentially be solved using renormalization-group techniques that can
be used to build solutions valid over the secular timescale.
For the equation-fixing method on the other hand, there are first studies in spherical
symmetry~\cite{Cayuso:2020lca,Thaalba:2023fmq,Franchini:2022ukz,Bezares:2021yek}
and for BBH mergers in some beyond-GR theories~\cite{Cayuso2023}.
A more detailed quantitative analysis is needed
to identify how much a given ``equation-fixing'' choice
affects low-energy modes in scenarios of interest, and to fully characterize
the accuracy of these approximate schemes~\cite{Coates:2023swo}.

These problems can be avoided for cases where a given theory in its full form can be shown to have a well posed initial value problem. Important work in this direction has recently been done.
For example, nonlinear evolutions of BBHs  in quadratic gravity were presented in
Ref.~\cite{Held:2023aap}.
Furthermore, BBHs simulations have been performed in cubic Horndeski
theories~\cite{Figueras:2021abd} which are known to be well-posed in the standard gauges
used in numerical GR~\cite{Kovacs:2019jqj}. For more generic theories, it has been
recently shown that, at sufficiently weak coupling, the equations of motion for Horndeski
gravity theories possess a well posed initial value problem in a modified version of the
generalized harmonic gauge formulation~\cite{Kovacs:2020pns,Kovacs:2020ywu}. Quite
remarkably, using this formulation it could in principle be possible to perform
simulations in all Horndeski theories of gravity without having to resort to the two
schemes mentioned above~\cite{Ripley:2022cdh}. In particular, Ref.~\cite{East:2020hgw}
used this modified generalized harmonic (MGH) formulation to perform the first numerical
simulations of BBH mergers in shift-symmetric ESGB gravity, without approximations, which
were also compared against PN results in Ref.~\cite{Corman:2022xqg}. Simulations of head-on BH collisions in ESGB theories that exhibit spontaneous BH
scalarization~\cite{East:2021bqk} and simulations of binary neutron stars in
shift-symmetric Einstein-scalar-Gauss-Bonnet~\cite{East:2022rqi} have also been done.
Based on the MGH formulation, a modified version of the CCZ4 formulation of Einstein's
equations has also been proposed, which has been used to prove the well-posedness of the
most general scalar-tensor theory of gravity with up to four derivatives, in singularity
avoiding coordinates~\cite{AresteSalo:2022hua}. This was then used to perform (circular
and equal-mass) BBH merger simulations in such theories~\cite{AresteSalo:2022hua,Doneva:2023oww}. 
More recently, using shift-symmetric ESGB gravity as a benchmark theory, Ref.~\cite{Corman:2024cdr} compared the results obtained with the MGH formulation against results obtained using the order-by-order expansion and equation-fixing methods. This work confirms that the order-by-order approach cannot faithfully track the solutions when the corrections to GR are non-negligible, whereas the equation-fixing method is able provide consistent solutions.

Finally, NR simulations have also been done in theories in which BHs possess an electric
or magnetic charge~\cite{Zilhao:2012gp,Zilhao:2013nda,Hirschmann:2017psw,Bozzola:2019aaw,Bozzola:2020mjx,Luna:2022udb}, possibly
coming from mini-charged dark matter~\cite{Cardoso:2016olt}),  primordial magnetic
monopoles~\cite{Preskill:1984by}), or in specific classes of scalar-tensor-vector
modified theories of gravity~\cite{Moffat:2005si}. In all these cases the field equations
can be cast in a form that is mathematically equivalent to Einstein-Maxwell theory (or to
Einstein-Maxwell with an extra scalar field for some
theories~\cite{Hirschmann:2017psw,Fernandes:2019rez}), which is known to possess a
well-posed initial boundary value problem~\cite{Alcubierre:2009ij}.

\subsubsection{Boson clouds}
\label{sec:NR_bosonclouds}

Simulations of BHs surrounded by massive bosonic fields have been widely studied, focusing mostly on the evolution of bosonic fields around isolated BHs (see
e.g.~\cite{Witek:2012tr,Barranco:2013rua,Okawa:2014sxa,Okawa:2014nda,Okawa:2015fsa,Zilhao:2015tya,Sanchis-Gual:2015sxa,Sanchis-Gual:2016jst,Clough:2019jpm,Bamber:2020bpu,Traykova:2021dua,Wang:2022hra,Clough:2022ygm,East:2022ppo}). This has led to the first successful nonlinear evolution of the
superradiant instability of minimally-coupled Proca fields around a spinning BH in~\cite{East:2017ovw,East:2018glu}. These works have laid down the possibility to study
the impact of minimally-coupled bosonic fields in BBH systems within
NR~\cite{Bernard:2019nkv,Cardoso:2020hca,Ikeda:2020xvt,Choudhary:2020pxy,Zhang:2022rex,Bamber:2022pbs,Cheng2023InPrep} but such studies are still in their infancy.

\subsubsection{Exotic compact objects}\label{sec:NR_ECOs}

ECOs come in many different flavors, and the state of numerical simulations of ECO mergers
vary widely, depending on the source.
For example, simulations of string-theoretical models including fuzzballs, wormholes,
firewalls or gravastars have not yet been performed because the theoretical foundations ---
the field content, equations of motion, equations of state (where applicable), solution
space --- are under active investigation.
Formulations suitable for numerical relativity are not yet available.

In contrast, binary boson stars that are compact objects composed of real or complex
bosonic (scalar or vector) fields, with and without self interaction potentials, have
received a lot of attention~\cite{Liebling:2012fv,Palenzuela:2006wp, Palenzuela:2007dm,
Choptuik:2009ww,Brito:2015yga,Brito:2015yfh, Cardoso:2016oxy,Bezares:2017mzk,
Palenzuela:2017kcg, Bezares:2018qwa,Helfer:2018vtq,
Widdicombe:2019woy,Sanchis-Gual:2018oui,CalderonBustillo:2020fyi,Bezares:2022obu,CalderonBustillo:2022cja,Sanchis-Gual:2022mkk,Sanchis-Gual:2022zsr,Jaramillo:2022zwg,Freitas:2022xvg,Croft:2022bxq,Evstafyeva:2022bpr,Siemonsen:2023hko,Siemonsen:2024snb}.
These boson stars may be treated as potential dark matter components (see Sec.~\ref{sec:sources_beyond_GR}),
or simply as toy model proxies for unknown exotic objects.
Some works have also considered mixed collisions of BHs or neutron stars with these
types of bosonic ECOs~\cite{Dietrich:2018bvi, Clough:2018exo} as well as collisions of fermion-boson stars~\cite{Bezares:2019jcb}.
Besides these, simulations of extremely compact perfect fluid stars have also been performed in~\cite{Tsokaros:2019lnx,Tsokaros:2019mlz}.
Most binary boson star simulations have explored new phenomena that arise during their
merger or the qualitative structure of the gravitational radiation emitted.
For model specific tests of gravity more accurate waveforms will be needed which, in turn,
requires improved initial data and more accurate numerical evolutions.

In summary, despite the progress in the last few years, most of the studies mentioned
above have only provided a proof of principle for the stable evolution of some families
of ECOs and of fundamental fields around BHs in NR, and have primarily aimed at
identifying significant qualitative differences between their signals and those of
traditional binary candidates. Key areas for improvements include the numerical accuracy
of such simulations and the quality of the initial
data~\cite{Helfer:2021brt,Aurrekoetxea:2022mpw,Bamber:2022pbs,Evstafyeva:2022bpr,Cheng2023InPrep}. For example for boson stars, a formalism is currently lacking to ensure that the superimposed boosted objects are not in an excited state, the risk being that such artifacts in the initial data are wrongly attributed as features of the ECO signal. Less challenging, but equally important refinements are reducing the eccentricity of the initial orbits (most simulations thus far have been head-on mergers, or only achieve a few orbits before the plunge), and initial errors in the Hamiltonian and momentum constraints (many simulations rely on constraint damping which reduces control over the initial data parameters).
Efforts to solve some of these problems are currently underway, in particular see~\cite{Aurrekoetxea:2022mpw} where a novel method to solve the initial data constraints has been formulated and that could be particularly useful when the sources involve fundamental fields. See also Ref.~\cite{Siemonsen:2023hko} where numerical simulations of binary boson stars were performed using constraint satisfying initial data.

Whilst feasible given these improvements, a significant effort would be required to create a template bank of waveforms of a similar quality to those used for BH and neutron star mergers. This is true even if the class of objects were restricted to a simple model, such as minimally coupled, non self-interacting, non spinning complex scalar boson stars.
Refining the NR tools with which we study such objects is work that is merited in the run up to the mission. In particular, key goals are (1) unambiguously identifying characteristic deviations in the merger signals of ECOs and of BHs surrounded by fundamental fields (e.g. to be used to motivate parameterised deviations from vacuum BBH waveforms) and (2) expanding the classes of objects for which simulations of mergers have been performed to qualitative examples in all sufficiently predictive ECO cases.
\subsection{Weak-field approximations}
\subsubsection{Beyond GR}\label{sssec:weak_field_beyond GR}

Applying weak field approximation techniques to theories beyond GR raises several difficulties
which have prevented obtaining high-order PN results, with the exception of a small set of alternative theories of gravity.
For most cases only the leading non-GR correction is known (see e.g.~\cite{Yunes:2016jcc,Tahura:2018zuq}). The most advanced results concern scalar-tensor theories, for which the equations of motion are known up to 3PN order~\cite{Mirshekari:2013vb,Bernard:2018hta,Bernard:2018ivi}, including finite-size effects~\cite{Bernard:2019yfz,vanGemeren:2023rhh}. The gravitational flux and waveform are known up to 2PN order~\cite{Lang:2013fna,Sennett:2016klh} while the scalar energy flux is known to 1.5PN order beyond the quadrupolar formula~\cite{Lang:2014osa,Bernard:2022noq}. The scalar-tensor equations of motion have been used to obtain similar results in other theories such as Einstein-Maxwell-scalar~\cite{Julie:2017rpw,Julie:2018lfp}
and ESGB theories~\cite{Julie:2019sab,Shiralilou:2020gah,Shiralilou:2021mfl,vanGemeren:2023rhh}.
The main difficulty encountered in more involved theories, such as the Horndeski family~\cite{Kobayashi:2019hrl}, is related to the need for a mechanism to screen the fifth force in the solar system. As an example, the Vainshtein mechanism requires to keep non-linearities in the description of the dynamics even in the weak field limit, rendering the definition of a perturbative method challenging~\cite{Chu:2012kz,deRham:2012fg,deRham:2012fw,Barausse:2015wia,McManus:2017itv,Dar:2018dra,Kuntz:2019zef,Brax:2020ujo}.

Given the many different alternative theories of gravity proposed over the years and the difficulty in constructing accurate waveforms in many of these theories, the most common approach to build beyond GR waveforms for data analysis purposes has been to use theory-agnostic phenomenological frameworks,
such as the parametrised post-Einsteinian (ppE) formalism~\cite{Yunes:2009ke},
the TIGER pipeline~\cite{Agathos:2013upa,Meidam:2017dgf} or the flexible theory-independent (FTI) approach~\cite{Mehta:2022pcn}.
In these frameworks, deviations from GR are assumed to be small and are treated as perturbative corrections to the signal predicted by GR.
For example, in the ppE formalism the inspiral phase of the frequency domain waveform is given by:
\begin{equation}\label{ppE}
 \tilde{h}(f)=\tilde{h}_{GR}(f) [1+\alpha_i (\pi \mathcal{M} f)^{a_i}] e^{i\beta_j (\pi \mathcal{M} f)^{b_j}}\,,
\end{equation}
where $\tilde{h}_{GR}(f)$ is the frequency domain waveform of GR, $\mathcal{M}$ is the chirp mass of the binary, while the deviations from GR are described by the dimensionless parameters $\alpha_i$,$\beta_j$, $a_i$ and $b_j$. The index $i$ and $j$ indicate the PN order at which the modification enters and are also introduced as a reminder that modifications could enter at different PN orders, in which case we should sum over all of them. Different values for the generic beyond GR parameters can be mapped to distinct physical effects and gravitational theories (see e.g.~\cite{Barausse:2016eii,Yagi:2011xp,Yagi:2015oca,Yunes:2009bv,Yagi:2011yu,Yunes:2011we,Mirshekari:2011yq,Carson:2019kkh,Perkins:2020tra}), allowing to use this formalism to place constraints on such theories.
Extensions of the ppE formalism to include higher-harmonic \GW modes have also been recently formulated~\cite{Mezzasoma:2022pjb}, as well as an extension for precessing binaries~\cite{Loutrel:2022xok} (the TIGER framework already considers precessing BBH waveforms, however in this case the modifications to the GW phase are only applied in the coprecessing frame~\cite{Meidam:2017dgf}). The latter was based on the recent computation of time- and frequency-domain analytical waveforms emitted for quasicircular precessing BH binaries in dCS gravity~\cite{Loutrel:2022tbk} and in binaries composed of deformed compact objects with generic mass quadrupole moments~\cite{Loutrel:2022ant}.

As final remarks, we should note that while the ppE, TIGER and FTI approaches are very useful and
powerful methods to perform null tests of GR, a qualitative and quantitative
interpretation of a constraint on the beyond GR parameters (or measurements of a deviation from
GR) only makes sense when accessed with respect to specific theories~\cite{Chua:2020oxn};
see e.g. Refs.~\cite{Nair:2019iur,Perkins:2021mhb}. 
Theory-agnostic approaches should only be seen as complementary, and not as alternatives, to explicit computations of waveforms in specific alternative theories. Moreover, the mapping from the beyond GR parameters to specific theories also comes with important caveats. In particular, typically, only the leading-order modification to the PN coefficients is considered in the mapping (however see Ref.~\cite{Perkins:2022fhr}, where it was shown that, for a wide class of theories, the main effect of including higher-order terms is to mildly strengthen constraints when compared to constraints obtained using only the leading-order modification). Furthermore, these approaches break down in the high-frequency part of the signal, where the PN expansion is not applicable. In both the TIGER and the FTI framework, the beyond GR modifications to the phase are tapered off at some transition frequency, while also assuring that the waveform remains smooth at this transition point, to enforce that these modifications are only applied in the inspiral regime (in the TIGER approach the beyond GR coefficients are simply set to zero at the transition frequency between the inspiral and the ``intermediate'' regime, whereas the FTI method applies a tapering function over a given frequency window that ensures that the beyond GR coefficients smoothly transition to zero at high frequencies). There is no guarantee that these approaches provide a good description of specific beyond GR theories at high-frequencies. These issues motivate even further the need for computations of full waveforms in specific alternative theories.

\subsubsection{Dark matter and boson clouds}\label{sssec:weak_field_DM}

Weak field approximations have been widely used to study the impact of dark matter environments on binary systems~\cite{Eda:2013gg,Barausse:2014tra,Cardoso:2019rou}. Such an environment can affect the binary dynamics through dynamical friction~\cite{Chandrasekhar:1943ys}, accretion and the gravitational pull of the environment itself~\cite{Eda:2013gg,Barausse:2014tra,Cardoso:2019rou}. These effects typically modify the waveform at negative PN orders and can be captured within the ppE formalism presented above~\cite{Barausse:2014tra,Cardoso:2019rou}.

In this context, Newtonian approximations have also been used to study binary systems moving inside bosonic structures, mainly motivated by the fact that ultralight bosons are promising candidates to describe dark matter~\cite{Hui:2016ltb}. For example, Refs.~\cite{Annulli:2020ilw,Annulli:2020lyc} studied how compact binaries travelling through these structures would be affected by dynamical friction and emission of scalar radiation. In particular, it was found that, for sources in the LISA band, scalar radiation affects the gravitational waveform at leading -6 PN order with respect to the dominant quadrupolar term~\cite{Annulli:2020ilw,Annulli:2020lyc}. In addition, BHs surrounded by boson clouds have been studied within a weak-field approximation~\cite{Baumann:2018vus,Baumann:2019eav,Baumann:2019ztm,Zhang:2018kib,Zhang:2019eid,Berti:2019wnn,Cardoso:2020hca,Takahashi:2021eso,Takahashi:2021yhy,Tong:2022bbl} showing that a plethora of signatures, including tidally induced resonances~\cite{Baumann:2018vus,Baumann:2019ztm}, floating and kicking orbits~\cite{Cardoso:2011xi,Zhang:2018kib,Baumann:2019ztm,Ferreira:2017pth} and ionization of the cloud~\cite{Takahashi:2021yhy,Baumann:2021fkf,Baumann:2022pkl,Tomaselli:2023ysb} can occur.
Finally, it was recently shown that sufficiently light bosonic fields can also be bound to and engulf a binary BH system as a whole, showing a rich phenomenology~\cite{Wong:2019kru,Wong:2020qom,Ikeda:2020xvt}, which could potentially lead to additional signatures besides the ones we discussed above.
A main problem with those studies is that they have mostly been performed within a weak-field approximation, neglecting high PN corrections and nonlinear effects close to merger. It would be especially important to extend such studies beyond the weak-field regime and further explore the physics of ultralight fields around binary BHs with full NR simulations (see Sec.~\ref{sec:NR_bosonclouds} for recent attempts).

\subsubsection{Exotic compact objects}

In the inspiral phase, the nature of the coalescing bodies can be studied through (i) their multipolar structure; (ii) (the absence of) tidal heating; and (iii) their tidal deformability. In particular, the multipole moments of an ECO will be different from those of their Kerr counterpart~\cite{Glampedakis:2017cgd,Raposo:2018xkf,Raposo:2020yjy,Bianchi:2020miz,Bianchi:2020bxa,Bena:2020see,Bena:2020uup,Herdeiro:2020kvf,Bah:2021jno,Fransen:2022jtw,Loutrel:2022ant,Vaglio:2022flq}, which at leading order enters as a 2PN correction to the \GW phase due to the object's quadrupole moment~\cite{Poisson:1997ha,Loutrel:2022ant}. We note that in many cases the multipolar structure of the ECO is only expected to be different from the Kerr metric if the ECO is spinning (as is the case for the simplest boson star models~\cite{Ryan:1996nk,Vaglio:2022flq}), although examples of non-spinning (i.e. static) ECOs with a different multipolar structure from the Schwarzschild metric also exist~\cite{Raposo:2018xkf,Herdeiro:2020kvf,Herdeiro:2023wqf}.
In addition, tidal interactions during the coalescence also provide unique signatures able to disentangle BHs from ECOs in the form of tidal heating. For BH binaries a small fraction of the emitted radiation is lost through the horizon~\cite{Hughes:2001jr,Alvi:2001mx,Poisson:2009di,Cardoso:2012zn,Poisson:2018qqd}. Absorption at the horizon introduces a 2.5PN (4PN) $\times \log v$ correction to the \GW phase of spinning (nonspinning) binaries, relative to the leading term.
On the other hand, exotic matter is expected to weakly interact with gravitational waves, leading to a smaller absorption during the inspiral, and therefore to a suppressed contribution to the  accumulated \GW phase from tidal heating~\cite{Maggio:2017ivp,Maselli:2017cmm,Oshita:2019sat,Datta:2019epe,Datta:2020rvo}.
Finally, tidal deformations can be strong enough, especially during the late stages of the inspiral, to modify the binary's orbital evolution leaving an imprint on the emitted waveform encoded in the ECO's tidal Love numbers~\cite{Pani:2015tga,Cardoso:2017cfl}, which at leading order introduces a 5PN correction to the \GW phase~\cite{Flanagan:2007ix}.
Overall more work is needed to construct fully consistent inspiral waveforms for ECOs that incorporate all these ingredients. For example, for the inspiral of boson stars with quartic interactions, such a waveform was only recently constructed in Refs.~\cite{Pacilio:2020jza,Vaglio:2023lrd}. It would be important to extend this to other types of ECOs.
\subsection{Perturbation theory for post-merger waveforms}

\subsubsection{Beyond GR}

The ringdown phase of a \GW signal emitted by a BBH merger event can be described using BH perturbation theory and is dominated by the \QNMs of the remnant~\cite{Kokkotas:1999bd,Berti:2009kk}. Owing to the very large SNR with which LISA will be able to detect the ringdown phase of supermassive BHs~\cite{Berti:2005ys,Kamaretsos:2011um}, we expect to be able to measure several QNMs for a single event up to very large distances~\cite{Dreyer:2003bv,Berti:2005ys,Berti:2016lat,Baibhav:2018rfk}. This will allow to perform precise consistency tests of the QNMs and test the hypothesis that the remnant of a BBH merger is well described by the Kerr metric in GR~\cite{Dreyer:2003bv,Berti:2005ys,Kamaretsos:2011um,Gossan:2011ha,Meidam:2014jpa,Brito:2018rfr,Carullo:2018sfu,Carullo:2019flw,Giesler:2019uxc,Isi:2019aib,Maselli:2019mjd,Bhagwat:2019dtm,Bao:2019kgt,Ota:2019bzl,Hughes:2019zmt,JimenezForteza:2020cve,Ghosh:2021mrv,Bhagwat:2021kwv,Forteza:2022tgq,Lim:2022veo}. Due to these prospects, significant work has been done in recent years in order to accurately model the ringdown phase of BBH mergers within GR~\cite{Damour:2014yha,London:2014cma,London:2018gaq,London:2018nxs,Giesler:2019uxc,Cook:2020otn,Dhani:2020nik,Finch:2021iip,MaganaZertuche:2021syq,Li:2021wgz,Lim:2022veo,Mitman:2022qdl,Cheung:2022rbm}.

For beyond GR theories, however, there has been significantly less progress, even though
QNM measurements can, in principle, be powerful probes of beyond GR physics. In
particular, for beyond GR theories, the ringdown of a BBH merger is typically expected to
differ from the ringdown in GR (although not necessarily for all non-GR theories).
Computations of QNMs in beyond GR theories are very challenging, especially for rotating
BHs for which the separability of the perturbative equations present in
GR~\cite{Teukolsky:1972my} is in general absent.
Then, one may employ numerical methods
to compute QNMs for non-separable equations, see e.g.
Refs.~\cite{Dias:2015wqa,Dias:2021yju,Dias:2022oqm,Li:2022pcy,Chung:2023zdq,Chung:2023wkd,Chung:2024vaf,Chung:2024ira}).
In addition, most rotating BH solutions in alternative theories, for the cases where they
differ from Kerr, are only known either perturbatively, through a small-spin expansion
around non-spinning
backgrounds~\cite{Pani:2011gy,Ayzenberg:2014aka,Maselli:2015tta,Cardoso:2018ptl,Cano:2023qqm},
or they are given in the form of numerical solutions of the field equations
(e.g.~\cite{Kleihaus:2011tg,Herdeiro:2014goa,Stein:2014xba,Herdeiro:2016tmi,Cunha:2019dwb,Sullivan:2019vyi,Sullivan:2020zpf}),
complicating the problem even further.
Furthermore, BH spacetimes may change their character as compared to GR. For example,
it was shown that BHs in ESGB and dCS gravity are of Petrov-type I~\cite{Owen:2021eez}
 (instead of the more symmetric type-D spacetimes like the Kerr metric).
Therefore, even though QNMs have been computed for a handful of theories and BH solutions (see e.g.~\cite{Barausse:2008xv,Molina:2010fb,Barausse:2014tra,Blazquez-Salcedo:2016enn,Glampedakis:2017cgd,Tattersall:2017erk,Franciolini:2018uyq,Cardoso:2018ptl,McManus:2019ulj,Glampedakis:2019dqh,Cardoso:2019mqo,Silva:2019scu,Pierini:2021jxd,Pierini:2022eim}), up to very recently the vast majority of these results either assumed non-rotating or slowly-spinning BH backgrounds or relied on approximations such as the eikonal/geometric optics approximation. Methods to circumvent some of these problems were formulated in Refs.~\cite{Cano:2020cao,Li:2022pcy,Hussain:2022ins,Ghosh:2023etd,Wagle:2023fwl,Cano:2023tmv,Cano:2023jbk,Cano:2024jkd,Cano:2024ezp} where modified Teukolsky equations governing the perturbations of non-Kerr spinning BHs arising in theories beyond GR, were derived in the case where the deviations from the Kerr geometry are small. Another approach would be to rely on fits of time-domain waveforms computed from numerical simulations in selected classes of theories beyond GR. However, this is quite a challenging task, even in GR~\cite{London:2014cma,Baibhav:2017jhs,London:2018gaq,London:2018nxs,Giesler:2019uxc,Cook:2020otn,Dhani:2020nik,Finch:2021iip,Forteza:2021wfq,MaganaZertuche:2021syq,Li:2021wgz,Mitman:2022qdl,Cheung:2022rbm}, besides the fact that such simulations are still only possible for a handful of cases, has emphasized in Sec.~\ref{sec:NR_beyondGR}.

\subsubsection{Exotic compact objects}

The vibration spectra of ECOs have been computed for a wide class of models, although mostly for spherically symmetric configurations, including: boson stars \cite{Berti:2006qt,Macedo:2013jja}, gravastars \cite{Chirenti:2007mk,Pani:2009ss,Chirenti:2016hzd,Volkel:2017ofl}, wormholes \cite{Cardoso:2016rao,Konoplya:2016hmd,Bueno:2017hyj}, quantum structures \cite{Cardoso:2005gj,Eperon:2016cdd,Cardoso:2016oxy,Barcelo:2017lnx,Brustein:2017koc,Wang:2019rcf,Chakraborty:2022zlq}, and in a model-indepedent fashion using the membrane paradigm~\cite{Maggio:2020jml}. Typically, the QNMs of ECOs differ from the BH QNMs due to the presence of a surface instead of an event horizon~\cite{Cardoso:2019rvt} and the excitation of the internal oscillation modes of the objects~\cite{Ferrari:2000sr,Pani:2018flj,Glampedakis:2017cgd}.
In addition, the isospectrality of axial and polar modes of spherically symmetric BHs in GR~\cite{Chandrasekhar:1975zza} is broken in ECOs, which are instead expected to emit a characteristic ``mode doublet''. The detection of such doublets would be a irrevocable signature of new physics~\cite{Berti:2009kk,Brustein:2017koc,Cardoso:2019mqo,Maggio:2020jml}.
The formation of an exotic ultracompact object may also lead to the emission of \GW echoes in the post-merger signal~\cite{Cardoso:2016rao, Cardoso:2017cqb} (see also Sec.~\ref{sec:sources_beyond_GR}). Several phenomenological \GW templates for echoes have been developed based on standard GR ringdown templates with additional parameters~\cite{Abedi:2016hgu,Nakano:2017fvh,Wang:2018gin}, the superposition of sine-Gaussians~\cite{Maselli:2017tfq}, hybrid methods that put together information from perturbation theory and the pre-merger orbital dynamics~\cite{LongoMicchi:2020cwm,Ma:2022xmp,Zhong:2022wnb}, using the close-limit approximation~\cite{Annulli:2021ccn}, and analytical models that explicitly depend on the physical parameters of the ECO~\cite{Mark:2017dnq,Testa:2018bzd,Maggio:2019zyv,Xin:2021zir}.

Finally, we note that although most of these works have so far relied on perturbative methods, fully nonlinear simulations of head-on collisions of boson stars in the large-mass-ratio regime, resulting in spinning horizonless remnants with stable light rings, have recently been performed~\cite{Siemonsen:2024snb}. Those simulations confirm the picture that the prompt post-merger signal approaches that of Kerr black holes in the large-compactness limit with the subsequent emission containing periodically appearing bursts akin to \GW echoes.

\subsection{Small mass-ratio approximation}\label{Sec:GSF beyond GR}

\subsubsection{Beyond GR} \label{beyondGR}

Although there have been substantial advances in GSF models in GR (see Sec.~\ref{sec:GSF}), self-force calculations for theories beyond GR are in their infancy. The most detailed studies in this vein have so far investigated changes to the emitted \GW flux for scalar-tensor and higher-order curvature theories of gravity~\cite{Sopuerta:2009iy,Pani:2011xj,Cardoso:2011xi,Yunes:2011aa,Canizares:2012is,Blazquez-Salcedo:2016enn,Fujita:2016yav}, and for theories of massive gravity when assuming a Schwarzschild BH background~\cite{Cardoso:2018zhm}; there are only a few examples of formulations of a full self-force theory beyond GR~\cite{Zimmerman:2015hua,Spiers:2023cva}. But recently it has been argued that for a vast class of theories, no-hair theorems or separation of scales lead metric and scalar perturbations to decouple~\cite{Maselli:2020zgv}.
In particular, in the large class of higher-derivative gravity models, the modification to
GR scales with the curvature, i.e., with the inverse of the mass.
Consider, for example, the Kretschmann curvature scalar $\mathcal{K}=R_{\mu\nu\rho\sigma}
R^{\mu\nu\rho\sigma}$.
On the horizon of a Schwarzschild BH of mass $M$ it scales as $\mathcal{K}\sim M^{-4}$.
Consequently, higher-curvature modifications to a supermassive BH are negligible.
This allows the background spacetime to be treated as the Kerr solution, and changes in
the binary's evolution to be fully controlled by the scalar field charge of the small BH~\cite{Maselli:2020zgv,Maselli:2021men,Barsanti:2022ana,Zhang:2022rfr,Guo:2022euk}. In this regard, the study of small-mass-ratio binaries in generalized scalar-tensor theories of gravity can benefit from efforts already devoted to investigating self-forces on scalar charges, and their orbital evolution around BHs~\cite{Barack:2000zq,Detweiler:2002gi,Diaz-Rivera:2004nim,Warburton:2010eq,Warburton:2011hp,Warburton:2014bya,Gralla:2015rpa,Castillo:2018ibo,Nasipak:2019hxh,Compere:2019wfw,Nasipak:2021qfu,Heffernan:2021olv,Barack:2022pde,Nasipak:2022xjh}.

\subsubsection{Dark matter and boson clouds}\label{sssec:SF_DM}

For EMRI and IMRI systems, the largest effect of dark matter is to produce a significant drag force through dynamical friction~\cite{Chandrasekhar:1943ys}, which increases the rate of inspiral of the small compact object~\cite{Macedo:2013qea,Barausse:2014tra,Eda:2014kra,Cardoso:2019rou,Vicente:2022ivh} (accretion and the usual gravitational pull of the dark matter can be significant, but are in general subleading effects~\cite{Eda:2013gg,Barausse:2014tra,Yue:2017iwc,Cardoso:2019rou}). This typically requires very high densities of dark matter surrounding the intermediate or supermassive BH~\cite{Barausse:2014tra,Eda:2014kra}. The required densities can be achieved when the IMBH or SMBH adiabatically grows in a dark-matter halo, and the distribution of dark matter near the BH gets compressed into a ``spike'' with significantly higher densities than the surrounding halo~\cite{Quinlan:1994ed,Gondolo:1999ef,Ullio:2001fb,Kim:2022mdj}.
Sufficiently large densities may also be achieved for clouds of ultralight bosons formed around MBHs, as discussed in Sec.~\ref{sec:sources_beyond_GR}. If they exist, these clouds would introduce additional metric perturbations that perturb the companion's orbit, and they would respond to the gravity of the companion. If the cloud contributes a sizable fraction of the BH's mass, its gravity would necessitate a change in the background geometry (see Refs.~\cite{Collodel:2021jwi,Delgado:2023wnj} for such an example), dramatically complicating the GSF model.
Building accurate IMRI/EMRI waveforms evolving in a dark matter spike or boson cloud is also crucial to distinguish these environments from other astrophysical environments, such as accretion disks. In fact, recent studies strongly suggest that these different types of environments leave distinguishable signatures in IMRI or EMRI waveforms~\cite{Hannuksela:2018izj,Hannuksela:2019vip,Coogan:2021uqv,Speri:2022upm,Cole:2022yzw,Becker:2022wlo,Kim:2022mdj}. Therefore the detection of such systems could also be used to identify and learn about the environment surrounding supermassive BHs.

To date, studies of the gravitational effects of dark matter spikes and boson clouds on binary systems, such as EMRIs and IMRIs, have mostly relied on Newtonian or post-Newtonian approximations for the orbit and the matter distribution~\cite{Eda:2013gg,Eda:2014kra,Yue:2017iwc,Yue:2018vtk,Hannuksela:2018izj,Hannuksela:2019vip,Ferreira:2017pth,Berti:2019wnn,Baumann:2018vus,Zhang:2018kib,Zhang:2019eid,Kavanagh:2020cfn,Baumann:2021fkf,Baumann:2022pkl,Speeney:2022ryg,Kim:2022mdj,Tomaselli:2023ysb}. Generalizing some of these calculations to the fully relativistic case is in principle feasible. Indeed, first steps in this direction were recently done in Refs.~\cite{Cardoso:2021wlq,Destounis:2022obl,Cardoso:2022whc} where a generic and fully-relativistic formalism to study EMRI systems in spherically-symmetric, non-vacuum BH spacetimes was developed. For boson clouds, a fully relativistic framework that assumes that the impact of the cloud on the BH geometry can be treated perturbatively, was proposed in Ref.~\cite{Brito:2023pyl}.

However, there are important open problems to be solved. First, generalizing the formulation of Refs.~\cite{Cardoso:2021wlq,Destounis:2022obl,Cardoso:2022whc,Brito:2023pyl} to non-spherically-symmetric backgrounds is a non-trivial task, and secondly, for dynamic matter distributions~\cite{Kavanagh:2020cfn}, evolving the matter environment coupled to the binary system at first post-adiabatic order will likely be a complex problem.

%

Finally, it is worth noting that boson clouds themselves can also be strong sources of nearly-monochromatic gravitational waves potentially detectable by LISA~\cite{Arvanitaki:2009fg,Arvanitaki:2010sy,Yoshino:2013ofa,Arvanitaki:2014wva,Baryakhtar:2017ngi,Brito:2017zvb,Isi:2018pzk,Siemonsen:2019ebd,Brito:2020lup,Zhu:2020tht,Siemonsen:2022yyf}. The gravitational waves emitted by these sources can be computed using the same techniques typically used in the small mass-ratio approximation: one can consider the cloud as being a small perturber of a Kerr background geometry and use the Teukolsky formalism to compute \GW emission from a boson cloud~\cite{Yoshino:2013ofa,Brito:2017zvb,Siemonsen:2019ebd,Brito:2020lup,Siemonsen:2022yyf}.

\subsubsection{Beyond GR BHs and exotic compact objects}
A key goal of the LISA mission, is to determine whether the dark central objects in galactic cores are genuine BHs, and if so, whether they are accurately described by GR BHs. With EMRIs we expect to be able to constrain fractional deviations from the quadrupole moment of the Kerr solution at a level smaller than $10^{-4}$~\cite{Barack:2006pq,Barausse:2020rsu}, which would allow to impose stringent constraints on ECOs and non-GR BHs. Work to include these modifications in GSF models has been done by considering the large central object as a ``bumpy'' BH. Bumpy BHs are spacetimes that include the Kerr limit, but that differ in a parameterized way, typically by modifying the spacetime's multipole moment structure~\cite{Manko:1992ra,Ryan:1995wh,Collins:2004ex,Vigeland:2011ji,Moore:2017lxy,Xin:2018urr,Fransen:2022jtw}. The ``bumps'' can be introduced directly into the background metric, or (if they are sufficiently small) they can be treated as additional metric perturbations.
A benefit of the bumpy BH framework is that it is agnostic about the mechanism which produces the BH's bumps. They could describe non-GR physics (to exactly recover BH solutions in specific alternative theories~\cite{Jackiw:2003pm,Yunes:2009hc,Vigeland:2011ji}, for example), or if the central object is an ECO, they could be tidally induced by the companion, encoding the ECO's tidal Love numbers which are potentially detectable with EMRIs~\cite{Pani:2019cyc,Datta:2021hvm,Piovano:2022ojl}. Because bumpy BHs break the symmetries of Kerr that yield integrable motion~\cite{Carter:1968rr}, orbits in these spacetimes are generically chaotic~\cite{Gair:2007kr,Lukes-Gerakopoulos:2017jub,Destounis:2020kss,Destounis:2021rko} and subject to prolonged resonances~\cite{Lukes-Gerakopoulos:2010ipp,Destounis:2020kss,Destounis:2021mqv,Destounis:2021rko}.
Most of the work considering EMRIs around spacetimes with a multipolar structure different from Kerr, constructed approximate ``kludge'' waveforms~\cite{Xin:2018urr,Chua:2018yng,Fransen:2022jtw} generated considering geodesics in a perturbed Kerr spacetime with parameters evolved using post-Newtonian equations~\cite{Glampedakis:2005cf,Barack:2006pq,Gair:2007kr,Apostolatos:2009vu,Moore:2017lxy}. While these approaches are believed to reproduce the main features of the orbit, they will not be enough to get to the precision required to determine that the inspiral is indeed an inspiral into a Kerr BH or not~\cite{Sasaki:2003xr,Gair:2007kr}, and much more work is needed to build such waveforms. The additional parameters in such models will also typically increase the possibility of degeneracies, making it important to fully understand possible discernible features, such as characteristic variations of the amplitude and the energy emission rate~\cite{Suzuki:1999si} or the appearance of prolonged resonances~\cite{Apostolatos:2009vu,Destounis:2021mqv,Destounis:2021rko}.
If the central object is an ECO, in addition to nonintegrable motion and geodesic resonances~\cite{Destounis:2023khj},
it will also lack a true event horizon. This causes a change in boundary conditions: in a BH spacetime, fields are purely ingoing on the horizon, while in an ECO spacetime, fields obey a boundary condition of (partial) reflection at the effective surface.  The effect of this reflection has been taken into account for the dissipative first-order self-force in~\cite{Li:2007qu,Cardoso:2019nis,Datta:2019epe,Sago:2021iku,Maggio:2021uge,Sago:2022bbj}. In addition, the modes of the ECO could also be excited during the inspiral of the small compact object~\cite{Pani:2010em,Macedo:2013jja,Cardoso:2019nis,Asali:2020wup,Fransen:2020prl,Sago:2021iku,Maggio:2021uge,Cardoso:2022fbq} (see Ref.~\cite{Cardoso:2022fbq} where the conditions for such modes to be effectively excited during the inspiral were studied in more detail). Depending on the object's properties, this could cause the rate of the inspiral to change significantly in the vicinity of each resonance. For the dissipative first-order self-force, the effect on the emitted \GW flux has been modelled for gravastars~\cite{Pani:2010em}, boson stars~\cite{Macedo:2013jja} and an exotic ultracompact object with an exterior Schwarzschild or Kerr geometry but with a hard surface close to the would-be horizon~\cite{Cardoso:2019nis,Sago:2021iku,Maggio:2021uge}. An object plunging in the interior of an ECO would also be subject to additional forces that can dominate the dynamics, such as dynamical friction~\cite{Chandrasekhar:1943ys} and accretion. The impact of these effects on the orbit dynamics and \GW waveforms has been modeled for small compact objects plunging onto boson stars using a Newtonian approximation~\cite{Macedo:2013qea} (see also~\cite{Kesden:2004qx} for earlier work on the same subject where dynamical friction and accretion had been neglected).

\subsection{Effective-one-body waveform models}
\label{sec:EOBbeyondGRandBSM}
EOB waveforms with generic deviations from GR have been developed based on a
parameterized form for performing theory-agnostic tests. These include deformations to
the \GW phase (see Eq.~\eqref{ppE}) which are added as corrections to frequency-domain EOB
waveforms in
GR~\cite{LIGOScientific:2018dkp,Sennett:2019bpc,Mehta:2022pcn,LIGOScientific:2020tif,LIGOScientific:2021sio},
or parameterized deformations of the QNMs in time-domain EOB waveforms, which have been
used to perform tests of the no-hair theorem in GR and constrain higher-order curvature
theories of
gravity~\cite{Brito:2018rfr,Ghosh:2021mrv,Silva:2022srr,LIGOScientific:2020tif,LIGOScientific:2021sio,Maggio:2022hre}.

Recent work has also computed the EOB dynamics for non-spinning binaries in various theories beyond GR as a foundational step towards enabling model-dependent tests. Results are available for scalar-tensor theories of gravity~\cite{Julie:2017pkb,Julie:2017ucp,Jain:2022nxs,Julie:2022qux,Jain:2023fvt,Jain:2023vlf}, Einstein-Maxwell-dilaton theories~\cite{Khalil:2018aaj} and ESGB gravity~\cite{Julie:2022qux}.
The EOB approach has also been used to compute the two-body potential energy of point-like particles in theories exhibiting a Vainshtein screening~\cite{Kuntz:2019plo}. 
Developing these results into full EOB waveform models and including more realistic physics remains a challenging task, however important steps in this direction were recently done in Ref.~\cite{Julie:2024fwy} where the first example of a full IMR waveform model for a beyond-GR theory, namely ESGB, was built. This waveform builds on top of SEOBNRv5PHM~\cite{Ramos-Buades:2023ehm}, the state-of-the-art EOB multipolar waveform model for
spin-precessing binary BHs, and includes corrections from ESGB gravity in both the inspiral and ringdown while also adding nuisance parameters to marginalize over the uncertainty in the merger morphology.

For non-vacuum binaries involving e.g. ECOs or boson clouds around BHs, a number of generic matter effects during the inspiral change the \GW signals away from those of a BH binary (see discussion in previous subsections). Several of the dominant matter effects for neutron star binaries are included in EOB models that could also have broader applicability to ECOs or non-vacuum BHs. These effects are described in a parameterized form, where the nature of the object is encoded in characteristic coefficients such as the bodies' multipole moments, tidal and rotational Love numbers, and QNM frequencies.  Specifically, the effects of rotational deformation and adiabatic tides are currently included in the TEOBResum family of models discussed in Sec.~\ref{sec:EOB} in the way described in Refs.~\cite{Damour:2009wj,Bini:2012gu,Bernuzzi:2012ci,Bini:2014zxa,Bernuzzi:2014owa,Nagar:2018zoe,Nagar:2018gnk,Akcay:2018yyh,Nagar:2018plt,Gonzalez:2022prs}, with the state-of-the-art models incorporating the known adiabatic gravitoelectric and -magnetic tidal terms up to $\ell=8$ and $\ell=3$~\cite{Gamba:2023mww}, respectively, in different resummation schemes~\cite{Bernuzzi:2012ci,Bini:2014zxa,Bernuzzi:2014kca,Akcay:2018yyh,Gamba:2023mww}, dynamical tides from the fundamental modes~\cite{Gamba:2022mgx,Steinhoff:2016rfi,Hinderer:2016eia} as well as nonlinear-in-spin effects dependent on the nature and interior structure of the object~\cite{Nagar:2018zoe,Nagar:2018plt}.  The SEOBNR models include tidal effects using the dynamical tidal models for the fundamental modes from~\cite{Steinhoff:2016rfi,Hinderer:2016eia}, which have recently been further developed~\cite{Steinhoff:2021dsn,Mandal:2023lgy,Gupta:2020lnv}. Matter effects have also been incorporated in the ROMs of SEOBNR waveforms in the frequency domain by augmenting the ROMs developed for BBHs with analytical closed-form expressions correcting the \GW phase to include tidal and spin-induced multipole effects based on tidal models that are specific to calibrations to NR simulations of neutron star binaries, which could in principle be replaced by more general frequency-domain tidal models, and PN calculations for the spin-induced multipole effects~\cite{Dietrich:2017aum,Dietrich:2018uni,Dietrich:2019kaq,Abac:2023ujg}.

However, significant work remains. Besides the improvement of current EOB models already discussed in Sec.~\ref{sec:EOB}, other improvements that are specific to ECOs or non-vacuum BHs include, for example, incorporating more effects from spins and relativistic phenomena in the matter sector (which are likely to be more important for compact ECOs than for neutron stars); incorporating parameterized absorption coefficients; and the inclusion of effects specific to dark matter spikes and boson clouds already discussed in Secs.~\ref{sssec:weak_field_DM} and~\ref{sssec:SF_DM}. In addition, a remaining challenge is to develop full IMR waveforms. Possible avenues for constructing full waveforms for binaries with matter have been explored, e.g., in the context of binary neutron stars~\cite{Breschi:2019srl}, by using NR-informed analytical models that build on quasiuniversal features found in these systems~\cite{Bernuzzi:2015rla,Breschi:2019srl,Breschi:2022ens}, and for the tidal disruption in BH-neutron-star binaries~\cite{Zappa:2019ntl,Matas:2020wab,Gonzalez:2022prs}. To achieve similar complete waveforms for ECOs, it is worth mentioning that a phenomenological model was proposed in Ref.~\cite{Toubiana:2020lzd}, which used EOB waveforms with tidal effects to describe the inspiral phase and a toy model inspired by Ref.~\cite{Takami:2014tva} to model the post-merger dynamics of the system. Whether this model is an accurate enough description for known ECOs, e.g. boson stars, remains to be fully studied.

\subsection{Phenomenological waveform models}\label{ssec:phenom_beyond_GR}

The phenomenological waveform models of Sec.~\ref{sec:phenom} already provide standard tools for tests of GR with the current generation of ground-based detectors~\cite{LIGOScientific:2016lio,LIGOScientific:2018dkp,LIGOScientific:2019fpa,LIGOScientific:2020tif,LIGOScientific:2021sio}. In particular, parameterized tests (in the spirit of the ppE formalism, see Sec.~\ref{sssec:weak_field_beyond GR}) have been employed to probe theory-agnostic deviations from GR in the Phenom ansatz describing the phase evolution of an observed signal, namely using the TIGER pipeline~\cite{Agathos:2013upa,Meidam:2017dgf} although, as shown in Ref.~\cite{Mehta:2022pcn}, the FTI framework can also be applied to Phenom waveform models (note that in all \LVK papers so far the FTI framework has only been applied on EOB waveforms, whereas parameterized tests with Phenom waveforms typically use the TIGER approach~\cite{LIGOScientific:2018dkp,LIGOScientific:2019fpa,LIGOScientific:2020tif,LIGOScientific:2021sio}). Moreover, Phenom models have also been used to test the BH nature of compact binaries from waveform signatures of spin-induced quadrupole moments~\cite{Krishnendu:2019tjp}. At a more fundamental level, such tests depend on the parameterization of the waveform model, and face a number of practical and conceptual problems, e.g., varying the phenomenological coefficients in such models by amounts that are too large, pathologies may arise in the waveform (see e.g. Ref.~\cite{Johnson-McDaniel:2021yge}). Some improvements can be expected from adding time domain models~\cite{Estelles:2020osj,Estelles:2020twz,Estelles:2021gvs} to the existing frequency domain models, thus providing very different waveform parameterizations to be varied. Going beyond theory agnostic tests, the flexibility of phenomenological models also provides a natural ground to develop models for beyond GR theories or ``exotic'' physics, such as boson stars and boson clouds around BHs, possibly starting from recent work on BNS and BHNS models \cite{Colleoni:2023czp,Abac:2023ujg}. A key problem is the very large parameter space for such theories, and significant technical and conceptual progress will be required to develop mature approaches. Natural first steps for the near future would be, for example, to calibrate accurate models to small ``toy model'' regions of some beyond GR theories, and to develop simple toy models that do not require any calibration to numerical waveforms.

\subsection{Modeling cosmic strings}\label{ssec:cosmic_strings}

Waveform models for \GW emission from cosmic strings are needed for LISA
for three types of anticipated searches: for the stochastic
background, for individual bursts and for individual, coherent,
nonlinear oscillatory waveforms. There are significant uncertainties
in the properties of the underlying string elements (the number of
objects, the presence of cusps, kinks and small scale structure, the
size distribution of the objects, the spatial and velocity
distribution of the objects). Here, however, we concentrate on the
modeling of the {\it waveforms} generated by one or more loops while
also mentioning some of the incompletely known loop properties that
will influence the waveforms.

\subsubsection{Nambu--Goto, smooth loops with cusp/kink features}

We begin with minimally coupled string networks with loops that are
smooth on large scales and include cusps or kinks but no other small
scale structure.

Early research on this topic focused on computing the individual loop waveforms to lowest order in the string tension $\mu_S$, by considering relatively simple loops and using standard weak-field approximations for GW generation~\cite{Garfinkle:1988yi,Vachaspati:1987sq,Burden:1985md,Garfinkle:1987yw,Allen:1991bk,Allen:1994ev,Allen:1994iq,Casper:1995ub,Allen:2000ia}. 
Both the burst waveforms and the relevant statistical averages for the background implicitly follow. FFT techniques work well to estimate the
power at low to moderate harmonics \cite{Allen:1991bk,Blanco-Pillado:2017oxo}
and additional numerical techniques have been devised specifically to handle
higher harmonics \cite{Suresh:2023hkz}.

In the case of \LVK, small scale features dominate emission at high frequencies.
A similar situation pertains to LISA, which will be sensitive to the high
frequencies of the astrophysically relevant low tension strings.
The development of an asymptotic approach
Refs.~\cite{Damour:2000wa,Damour:2001bk,Damour:2004kw} was an
important quantitative and qualitative advance. High frequency
emission sourced by cusps and kinks is approximately independent of
the large scale loop structure. The important features of cusp
waveforms in this limit are that (1) the cusp power spectrum falls off
at high frequencies as $f^{-4/3}$; and (2) the \GWs
from the cusp are strongly beamed in the direction of cusp motion
and exponentially suppressed at frequencies 
$f \gtrsim 1/(\theta^3 T)$
for angle $\theta$ with respect to that direction (with $T$ the periodicity of the string).
The waveform's time dependence is
$\propto |t-t_c|^{1/3}$, where $t_c$ is the time at
which the cusp formation event is noted by an
observer situated in the beam of emission ($\theta=0$), and so the
cusp waveform is called \emph{sharp} or \emph{spiky}. This sharpness
is softened for $0<\theta\ll 1$. Similar analyses are applicable to
kinks.

The genericity of the high frequency predictions for cusps and kinks,
the independence with respect to the large scale loop structure and
the simplicity of the templates are all important, and the results have
been exploited in astrophysically-motivated searches
\cite{Siemens:2006vk,ShapiroKey:2008ckh,LIGOScientific:2009bje,LIGOScientific:2019ppi}.

The outlook and need for waveforms is as follows:

(1) The existing asymptotic treatments of kink and cusp emission are
likely to be sufficient for modeling an unresolved stochastic wave
background. That is because the energy flux derived from averaging
over time, space and angle, which is implicit in such a calculation,
depends upon low-order moments of the loop's full beam pattern
that can be derived from the asymptotic results.

(2) More refined models of the burst waveform that do not, strictly
speaking, lie in the asymptotic regime are needed. Such waveforms
can account for geometric effects when the observer does not lie
directly in the cusp direction and for anisotropies in the beam shape.
These can be found by extending the asymptotic treatment.

(3) Existing calculational methodologies for full loops are
accurate and efficient at low to moderate modes.  LISA and pulsar
timing arrays will probe both low and high order modes.
Hybridization methods giving models of the beam that span the full range of
modes will be required for any loop sources that stand out above the
stochastic background \cite{Suresh:2023hkz}.

\subsubsection{Nambu--Goto, small-scale structure}

Each time a long horizon-crossing string intercommutes it gains two
kinks. Small scale structure builds up on the string
\cite{Austin:1993rg,Polchinski:2007rg,Martins:2014kda,Vieira:2016vht,Martins:2020jbq}.
Gravitational backreaction probably smoothens the string at small
scales \cite{Bennett:1989ak,Quashnock:1990wv,Siemens:2001dx,Siemens:2002dj}.
The wigglyness that remains has been
evaluated \cite{Austin:1993rg,Polchinski:2007rg} but never determined
directly by simulation.  Ref.~\cite{Siemens:2003ra} found that
small-scale structure near the cusp rounded off the sharpness of the
cusp waveform but affected only observers very near to the direction
of cusp motion. Alternatively, Ref.~\cite{Polchinski:2007rg} found that
the small scale structure enhanced intercommutation on a newly formed
loop and excised the cusp when it first begins to form. It will
be important to investigate the effect of small structure on
loop dynamics and the resultant waveforms. This can be done
parametrically as the amount and character of the wigglyness
increases on otherwise simple loops.

\subsubsection{Nambu--Goto for Pseudocusps}

As a last note, in addition to cusp waveforms, some authors~\cite{Stott:2016loe} have studied the bursts due to ``pseudocusps'', i.e., string trajectories which very closely approach, but do not strictly reach, the cusp configuration. These pseudocusp burst waveforms are of lower amplitude than true cusps, and follow the same power spectrum decay of $f^{-4/3}$ at high frequencies. Accounting for them in a search can lead to an enhancement in the number of expected burst events.

\subsubsection{Beyond Nambu--Goto for cusps}

Gauge theory strings with energy scale $\mu_S^{1/2}$ have a
characteristic core width $\sim \mu_S^{-1/2}$ whereas classical
superstrings are always well-represented by the zero thickness
Nambu--Goto limit.  When a cusp forms the string instantaneously
doubles back upon itself.  Ref.~\cite{Blanco-Pillado:1998tyu} predicts
that a region of string around the cusp, of the order of the string
thickness, will annihilate into particles due to overlap. For strings
with large tensions (near that of the GUT scale) this modification to
the cusp shape is at a small enough scale that it does not
meaningfully affect the cusp waveform. However, as tension decreases
the physical scale increases.  It may prove useful to compare the
waveform of the gauge theory cusp to the superstring cusp. This can be
done by parametrically varying the core width in Abelian-Higgs
simulations.

A final assumption about the standard cusp waveform model is that it
relies only on properties universal to all strings. However,
superstrings possess additional features that could modify their cusp
waveforms. Ref.~\cite{Damour:2004kw} suggested modifications to the
amplitude based on the dimensionless loop-length parameter
$\alpha$. For example, cusps very near to Y-junctions will have their
burst amplitude modified by a correction factor that depends on the
nearby string geometry~\cite{Binetruy:2009vt,Binetruy:2010bq,Binetruy:2010cc}, while the extra degrees
of freedom on superstrings (extra dimensions) can lead to a strong
suppression of both cusp rate and burst
amplitude~\cite{OCallaghan:2010mtk}.

\subsubsection{Cosmic strings and \GW production in numerical relativity}

While in most cases, perturbative methods (e.g., Nambu--Goto strings coupled to perturbative gravity \cite{Quashnock:1990wv}) are sufficient, situations may arise when perturbative methods are insufficient and gravitational backreaction becomes of ${\cal O}(1)$, even if the string tension $G\mu_S$ is small. For example, a string with a sharp kink can easily lead to a locally strong field situation \cite{Jenkins:2020ctp}. In such strong field limits, the full machinery of NR must be employed. The construction of full NR solutions for cosmic strings was recently achieved by \cite{Helfer:2018qgv,Aurrekoetxea:2020tuw}, where stable dynamical solutions for an Abelian-Higgs string minimally coupled to gravity was found. This opens the door to not just the detailed exploration of the string dynamics, but also the characterization of the \GW signals that can then be searched for.

The Abelian-Higgs model minimally coupled to gravity can be recast into the standard BSSN/CCZ4 formalism, and solved using finite differences (we refer the reader to  \cite{Helfer:2018qgv} for details). Presently, solutions for infinite static strings, planar loops \cite{Helfer:2018qgv} and single traveling wave solutions \cite{Garfinkle:1990jq} have been constructed. Once a string configuration is constructed and successfully evolved, the work required to extract useful \GW signals from them follows very similarly to that of any standard NR simulations, see Sec.~\ref{sec:NR}.

As an example, in \cite{Aurrekoetxea:2020tuw}, the gravitational waveform emitted during the BH formation of a collapsing circular cosmic string was computed, and it was shown that the waveform is dominated by the BH formation and ringdown phase, with the primary contribution being the $\ell=2$, $m=0$ mode (as opposed to quasi-circular compact binary merger signals which are dominated by the $\ell=2$, $m=2$ mode). Intriguingly, due to the large asymmetric ejection of material in the formation process, a large \GW memory was also seen.

The future prospects for using NR to compute \GW signals from such strong gravity events is very promising. For example, a full general relativistic treatment of the \GW emission from cosmic string cusps and kinks is long overdue, but is now within reach. Presently, the  main obstacle to explore more complicated string configurations is the need for the construction of accurate initial conditions -- a usual problem in any NR endeavor, here made more difficult by the presence of fundamental matter fields. Nevertheless, this area is ripe for further exploration.

\newpage

\section{Afterword}
\label{sec:discussion}
LISA will usher in an era of millihertz \GW astronomy that will open a window to new classes of sources and offer unprecedented opportunities to probe our understanding of the universe. Integral to this scientific vision are predictions of the source waveforms, which are necessary for the detection and interpretation of \GW events.

This white paper has analyzed the question of waveform preparedness in the LISA era. It has reviewed the modelling requirements from astrophysical and data analysis perspectives, and has analyzed the status of the main approaches to modelling waveforms from compact binaries both in vacuum \GR and in the presence of environmental influences or effects from new physics beyond \GR and the Standard Model.
This white paper also provides guidance on new developments that are needed
for the different approaches, both in terms of modelling accuracy and the covered
parameter space, so that gravitational wave models will be ready for the LISA
era.

The experience with ground-based \GW detectors has to some extent prepared the community. 
For example, several approaches that predict the waveforms for comparable mass black hole binaries can be applied directly to massive black hole binaries due to the simple mass scaling of \GR.
However, these models are not currently sufficient to fully realize LISA's science goals.
In particular they insufficiently cover the parameters of eccentricity, highly asymmetric mass-ratios, and high spin magnitudes. 
In addition, the possibility of very strong massive black hole binary signals places additional demands on the accuracy of these models, well beyond what can currently be achieved.

For extreme-mass-ratio inspirals, leading-order (adiabatic) models that are sufficient to enable some LISA science are becoming available, and these models can be evaluted fast enough for use in LISA data analysis.
In order to enable the full spectrum of LISA science with EMRIs post-adiabatic models are needed.
These have recently been developed for the simplest orbital configurations (non-spinning, quasi-circular).
It also looks likely that these models can cover much of the intermediate mass ratio parameter space.
Significant work remains to extend these post-adiabatic models to cover the full parameter space of precessing and eccentric parameter binaries where both components are spinning.

The modelling of sources in theories beyond GR is still in its infancy with significant effort needed. 
For example, the majority of computations have focused on higher derivative gravity, that are most interesting in the high curvature regime, or on Horndeski models.
Moreover, simulations of compact binaries in extensions of \GR are, at the time of writing this white paper, proof-of-concept. 
That is, they have been done for nearly equal-mass binaries, typically of non spinning black holes, and at relatively low resolution.
This is comparable to the state of numerical relativity in \GR about $15$ years ago (i.e., few years after the breakthrough in 2005).
Likewise, perturbative or PN calculations have only started to go beyond the leading order in the correction.
To perform theory-specific tests of gravity, more work is needed,
both in terms of modelling accuracy and in covering (parts of) the
parameter space of black hole masses, spins and additional theory-specific parameters.

The good news is that in all approaches there are clear ideas of what needs to be done in order to reach the modelling requirements for LISA. Doing so will require a concerted effort over the next decade. 
Given sufficient resources, we are positive that all goals can be achieved.

We particularly recommend that the community conduct more investigations into the waveform standards and waveform interface.  While lessons have been learned from ground-based detectors, LISA will offer unique challenges that will need to be addressed by the waveform and data analysis communities working together.



\newpage
\appendix
\section{Descriptions of NR codes}
\label{app:NRcodes}


Description of some numerical relativity codes that focus, in particular,
on modelling LISA sources. A more extensive list is given in Table~\ref{tab:NRcodes}.


\begin{description}
\item[BAM]
BAM is a modular code framework initially developed at the University
of Jena and is now maintained and further developed by the CoRe
(Computational Relativity)
Collaboration~\cite{Bruegmann:2006ulg,Husa:2007hp,Thierfelder:2011yi,Dietrich:2015iva}.
BAM employs finite differences for the discretization of spacetime fields,
and high-resolution shock-capturing methods for the evolution of general
relativistic hydrodynamics variables.
It features automated code generation with Mathematica, adaptive mesh
refinement (AMR) in space and time, and parallelization with MPI and
OpenMP for large scale HPC, recently at 70\% on up to 50k cores.
BAM is employed for the simulation of compact binary mergers,
e.g.,~\cite{Hamilton:2023qkv,Bernuzzi:2012ci,Dietrich:2015pxa,Kolsch:2021lub}.
A major focus is on binary black hole, binary neutron star, and black hole -
neutron star systems~\cite{Chaurasia:2021zgt},
plus more exotic compact objects or merger
scenarios~\cite{Dietrich:2018bvi}.
Recent updates allow the study of radiation
hydrodynamics~\cite{Gieg:2022mut}, magnetohydrodynamics~\cite{Neuweiler:2024jae}, and the usage
of an entropy-viscosity limiter for the flux
computation~\cite{Doulis:2022vkx}.
BAM has contributed numerical data for the development and validation
of e.g.\ Phenom-type and EOB waveform models. Several hundred binary
neutron star merger simulations are available as part of the existing CoRe
database~\cite{Dietrich:2018phi, Gonzalez:2022mgo}.

\item[BAMPS]
The Jena-Lisbon collaboration is developing the code bamps, a new and
highly scalable NR code for exascale applications using
pseudo-spectral and DG methods~\cite{Hilditch:2015aba,Bugner:2015gqa}.
A key feature is hp-refinement for spectral element
methods~\cite{Renkhoff:2023nfw}, which allows highly scalable and
accurate simulations, currently with a focus on smooth fields.
To date, bamps has already been employed for the
study of GW collapse~\cite{Hilditch:2015aba,Fernandez:2022hyx},
simulations with scalar fields for collapse and for boson stars are in
development.
Further applications include simple neutron star
spacetimes~\cite{Bugner:2015gqa}, and studies of new techniques
regarding the dual
foliation framework~\cite{Bhattacharyya:2021dti}.
\item[Dendro-GR]
Dendro-GR \cite{Fernando:2018mov,Fernando:2019xyz,Fernando:2017abc} uses an unstructured grid with a localized, wavelet-based refinement scheme to evolve the BSSN formulation of the Einstein equations using moving punctures.  In test simulations of a \BBH with a mass ratio of $10:1$, Dendro-GR shows good scaling to over $10^5$ cores.

\item[Einstein Toolkit]
 The Einstein Toolkit~\cite{Loffler:2011ay,einsteintoolkit} is an open-source cyberinfrastructure for computational astrophysics, with more than $300$ registered users worldwide.
It implements Einstein's equations, or extensions thereof, in the BSSN or CCZ4 formulation together with the moving puncture gauge.
  It employs the method of lines, in which spatial derivatives are implemented as up to
$8^{\rm th}$ order finite differences
together with a collection of direct time integration techniques
(e.g., fourth order Runge-Kutta integrator).
  The Einstein Toolkit is based on the Cactus computational toolkit~\cite{Cactuscode:web}
  and the Carpet boxes-in-boxes AMR package~\cite{Schnetter:2003rb,CarpetCode:web}.
It also provides a multipatch infrastructure
consisting of Cartesian boxes-in-boxes meshes in the region where the binaries evolve
(typical size $\sim 50 \dots 100M$), and a spherical outer region in the wave zone
(typical radius $\sim 1000 \dots 2000M$) provided by the Llama code~\cite{LlamaWeb}.
  Kreiss-Oliger dissipation is employed to reduce high frequency noise at refinement boundaries.
 The toolkit uses hybrid MPI/OpenMP parallelization.
  The Einstein Toolkit has been a broad community project with a multitude of software
  that is based on its infrastructure or that has been developed within its framework.
  The additional software is indicated by asterisks in Table~\ref{tab:NRcodes}.


\item [Einstein Toolkit -- CarpetX]
The Einstein Toolkit consortium is developing a new AMR driver, CarpetX~\cite{roland_haas_2022_7245853}, for the Einstein Toolkit.
CarpetX leverages the AMReX framework~\cite{AMReX_JOSS}
to provide scalable adaptive mesh refinement for physics modules.
CarpetX will provide: efficient support for both CPUs and GPUs;
parallelization via MPI and OpenMP; SIMD vectorization;
scalable I/O based on the ADIOS2 file format~\cite{GODOY2020100561}
and the openPMD metadata standard;
and more.
CarpetX offers block-structured 
AMR based
on local error estimates, exact conservation across mesh interfaces, higher order
prolongation operators, and scalable elliptic solvers based on PETSc~\cite{petsc-web-page}.
CarpetX is available as open source and was used in production in
Ref.~\cite{Shankar:2022ful}.

\item[GR-Athena++]
Another example is GR-Athena++\cite{Daszuta:2021ecf} based on the astrophysical (radiation) magnetohydrodynamics code Athena++~\cite{2020ApJS..249....4S}. GR-Athena++ leverages on Athena++'s oct-tree AMR and exploits hybrid parallelism at different levels. The standard distributed approach is augmented by a dynamical tasklist for the procedures in each mesh block that overlaps calculation and communication (thus mitigating some of the overhead associated with high order finite differencing). In-core vectorization and other optimization techniques of the basic kernels \cite{jlpea8020015} are also employed. This approach leads to a high parallel efficiency, with strong scaling efficiencies above 95\% for up to $10^4\,$CPUs and excellent weak scaling is up to $10^5\,$CPUs measured in a production BBH setup with AMR.
\item[GRChombo]
  GRChombo \cite{Clough:2015sqa, Andrade:2021rbd, GRChombo-website} is an open-source NR code built using an optimized version of the publicly available library Chombo \cite{Chombo} developed at LBNL. It is based on established methods of solving the Einstein equations, but the highly flexible AMR capability and templating over physics classes makes it suited to strong gravity problems in beyond-GR scenarios (e.g. Horndeski gravity \cite{Figueras:2020dzx}, cosmic strings \cite{Aurrekoetxea:2020tuw}), which have been key targets of the code. GRChombo is written entirely in \texttt{C++14}, using hybrid MPI/OpenMP parallelism and vector intrinsics (in particular, the evolution calculations are explicitly vectorised) to achieve improved performance on the latest architectures. It has good strong scaling up to around 4,000 cores for typical binary BH problems, at which point it becomes limited by problem size as with most traditional codes.

The next release of Chombo is designed to be performance portable to different heterogeneous architectures (including GPUs), which GRChombo should be able to leverage to improve scaling further.

\item[Illinois GRMHD]
The Illinois GRMHD evolution code~\cite{Duez:2005sf,Etienne:2010ui,Sun:2022vri}
 solves the Einstein field equations via the BSSN scheme with
puncture gauge conditions, utilizes a conservative high-resolution, shock-capturing (HRSC) scheme for the matter
and adapts the Carpet infrastructure to implement AMR.
The code solves the magnetic induction equation by introducing a
magnetic vector potential, which guarantees that the B-field remains divergence-free for any
interpolation scheme used on refinement level boundaries.
This formulation reduces to a standard constrained transport (CT) scheme on uniform grids.
An improved Lorenz gauge condition for evolving the vector potential in AMR
without the appearance of spurious B-fields on refinement level boundaries is implemented.
This gauge exhibits no zero-speed modes, enabling spurious magnetic effects to propagate off the grid quickly.
The GRMHD code recently added a radiation module~\cite{Ruiz:2023hit} built to handle transport (photons or neutrinos)
via the M1 moment formalism.
A version of the Illinois GRMHD code has been released as an open-source module that has been
ported to the Einstein Toolkit~\cite{Etienne:2015cea}, where it has been documented and rewritten to make it
more user-friendly, modular and efficient.
The code has been used to simulate numerous scenarios related to LISA sources
(e.g., BHBH mergers, BHBH mergers in gaseous clouds and magnetized disks, etc.)
and their associated gravitational waveforms.

\item[MHDuet]
   The publicly available code {\sc MHDuet}~\cite{MHDuetWeb,Palenzuela:2018sly,Liebling:2020jlq,Vigano:2020ouc,Bezares:2021dma} is generated by the open-source platform {\sc Simflowny} ~\cite{2021CoPhC.25907675P} to run under the {\sc SAMRAI}~\cite{SamraiWeb} infrastructure, which provides the parallelization and the adaptive mesh refinement. There are different versions of the code, either to study alternative gravity theories or to include large-eddy-simulations and neutrino transport.
\item[SACRA]
The SimulAtor for Compact objects in Relativistic Astrophysics (SACRA) code solves Einstein's equation with the BSSN-puncture formulation together with a Z4c constraint propagation prescription and the relativistic hydrodynamics equation with the HRSC scheme~\cite{Yamamoto:2008js,Kiuchi:2017pte}. It implements the box-in-box conservative AMR with
parallelization by MPI and OpenMP. The code has been used to simulate compact binary mergers such as NSNS mergers, BHNS mergers, and BHBH mergers. The NSNS waveform catalog SACRA Gravitational Wave Data Bank is available at \cite{SACRA:catalog}.
The latest version of SACRA has a couple of branches, such as SACRA-TD, to simulate the tidal disruption of an ordinary star by the SMBH~\cite{Lam:2022yeg}, or SACRA-MG to simulate a compact object in scalar-Gauss-Bonnet theory~\cite{Kuan:2023trn} and scalar-tensor theory~\cite{Shibata:2022gec}. SACRA also has a version to solve high-dimensional Einstein’s equation~\cite{Shibata:2010wz}.
In addition, it features modules for neutrino radiation transport~\cite{Sekiguchi:2012uc}, MHD~\cite{Shibata:2005gp,Kiuchi:2022ubj} and viscous hydrodynamics~\cite{Shibata:2017jyf}.

\item [SACRA-SFS2D] is a viscous-radiation hydrodynamics~\cite{Fujibayashi:2017puw} in axisymmetry and
radiation magnetohydrodynamics code~\cite{Shibata:2021xmo} in full general relativity.
These codes can be applied to the collapse of massive stars and supermassive stars to a
stellar-mass and supermassive black hole~\cite{Uchida:2017qwn}, to the post-merger evolution
of the neutron-star
binaries~\cite{Fujibayashi:2022ftg}, and to collapsar modeling~\cite{Fujibayashi:2023oyt}.

\item[SpEC] The Spectral Einstein Code
  (\texttt{SpEC})~\cite{SpECsite} is a pseudo-spectral GR code
  developed by the Simulating eXtreme Spacetimes collaboration.
  \texttt{SpEC} is capable of evolving BBH and binaries with neutron
  stars.  It incorporates an elliptic solver to construct initial
  data~\cite{Pfeiffer:2002wt,Foucart:2008qt,Ossokine:2015yla},
  eccentricity reduction~\cite{Pfeiffer:2007yz, Buonanno:2010yk,
    Mroue:2012kv}, constraint-preserving and non-reflective outer
  boundary conditions~\cite{Lindblom:2005qh, Rinne:2006vv,
    Rinne:2007ui}, and implements hp-AMR~\cite{Lovelace:2010ne,
    Szilagyi:2014fna}.  Extracted gravitational waves are extrapolated
  to future null
  infinity~\cite{Boyle:2009vi,Taylor:2013zia,Boyle:2019kee} and
  corrected for center-of-mass drift~\cite{Woodford:2019tlo}.
  \texttt{SpEC} is a highly accurate NR code for BBH, and has
  computed large waveform catalogs of BBH inspirals encompassing
  several tens of orbits~\cite{Mroue:2013xna,Boyle:2019kee}.
  \texttt{SpEC} has evolved BBH with spins as large as
  0.998~\cite{Chatziioannou:2018wqx}, inspirals as long as 170
  orbits~\cite{Szilagyi:2015rwa} as well as eccentric
  binaries~\cite{Ramos-Buades:2022lgf}.

\item[SpECTRE]
The Simulating eXtreme Spacetimes collaboration is developing the
next-generation code SpECTRE~\cite{spectrecode, Kidder:2016hev}, which
uses a discontinuous-Galerkin-finite-difference hybrid
method~\cite{Deppe:2021ada, Deppe:2021bhi, Legred:2023zet} for accurate and
robust neutron star simulations combined with task-based parallelism, showing
good scaling to over $600,000$ cores~\cite{Kidder:2016hev}. Using
Cauchy-Characteristic extraction, SpECTRE is able to resolve gravitational wave
memory~\cite{Moxon:2021gbv, Mitman:2020pbt} and improve hybridization with
post-Newtonian waveforms by ensuring numerical relativity waveforms are in the
same BMS frame as the post-Newtonian
waveforms~\cite{Mitman:2022kwt}.
Improvements for efficiently simulating binary
black holes with mass ratios $q>10$ aim to reduce the computational cost of such
simulations by a factor of $q$~\cite{Wittek:2023nyi}.
\texttt{SpECTRE} also incorporates a flexible elliptic solver~\cite{Fischer:2021voj,Vu:2021mhw,Vu:2021coj}.

\item[Spritz]
\texttt{Spritz} is an open-source code that solves the equations of general relativistic magnetohydrodynamics and that can take into account finite temperature nuclear equations of state and neutrino emission~\cite{Cipolletta:2019geh,Cipolletta:2020kgq,Kalinani:2021ofv}. The code is based on the Einstein Toolkit framework and implements also high-order methods for the hydrodynamic equations.

\item[WhiskyMHD]
\texttt{WhiskyMHD} is a fully general relativistic magnetohydrodynamic code based on the Einstein Toolkit~\cite{Giacomazzo:2007ti}. The code has been used mainly to perform simulations of binary neutron star mergers, but it has also been used to perform the first simulations of magnetized plasma around merging supermassive black holes in the ideal magnetohydrodynamic approximation~\cite{Giacomazzo:2012iv}.

\end{description}

%

\section*{Acknowledgements}

We would like to thank our internal reviewers, Leor Barack, Christopher Berry, and Nelson Christensen for extremely helpful feedback.
We are also grateful to two anonymous reviewers for their detailed comments which resulted in many improvements to this white paper.
Sarp Ak\c{c}ay acknowledges University College Dublin's Ad Astra Fund.
Pau Amaro Seoane acknowledges the funds from the ``European Union NextGenerationEU/PRTR'', Programa de Planes Complementarios I+D+I (ref. ASFAE/2022/014).
Enrico Barausse acknowledges support from the European Union’s H2020 ERC Consolidator Grant “GRavity from Astrophysical to Microscopic Scales” (Grant No.~GRAMS-815673) and the EU Horizon 2020 Research and Innovation Programme under the Marie Sklodowska-Curie Grant Agreement No.~101007855.
Emanuele Berti is supported by NSF Grants No.~AST-2006538, PHY-2207502, PHY-090003 and PHY-20043; by NASA Grants No.~20-LPS20-0011 and 21-ATP21-0010; by the John Templeton Foundation Grant 62840; and by the ITA-USA Science and Technology Cooperation program,
supported by the Ministry of Foreign Affairs of Italy (MAECI).
Richard Brito acknowledges financial support provided by FCT/Portugal under the Scientific Employment Stimulus -- Individual Call -- 2020.00470.CEECIND and under project No.~2022.01324.PTDC.
Marta Colleoni, Sascha Husa, Anna Heffernan and Pierre Mourier  are supported by the Spanish Agencia Estatal de Investigaci\`on grants PID2022-138626NB-I00, PID2019-106416GB-I00, IJC2019-041385, funded by MCIN/AEI/10.13039/501100011033; the MCIN with funding from the European Union NextGenerationEU/PRTR (PRTR-C17.I1); Comunitat Autonòma de les Illes Balears through the Direcci\'o General de Recerca, Innovaci\'o I Transformaci\'o Digital with funds from the Tourist Stay Tax Law (PDR2020/11 - ITS2017-006), the Conselleria d'Economia, Hisenda i Innovaci\'o grant numbers SINCO2022/18146 and SINCO2022/6719, co-financed by the European Union and FEDER Operational Program 2021-2027 of the Balearic Islands; the “ERDF A way of making Europe”;
Alvin J.~K.~Chua and Jonathan E.~Thompson acknowledge support from the NASA LISA Preparatory Science grant 20-LPS20-0005.
Katy Clough acknowledges funding from the UKRI Ernest Rutherford Fellowship (grant number ST/V003240/1).
David A.\ Nichols acknowledges support from the NSF Grants No.\ PHY-2011784 and No.\ PHY-2309021. Vasileios Paschalidis is supported by NSF Grant No.~PHY-2145421 and NASA Grant No.~80NSSC22K1605.
Stuart Shapiro is supported by NSF Grants No.~PHY-2006066 and PHY-2308242.
Scott Field acknowledges support from US National Science Foundation Grants Nos. PHY-2110496, DMS-2309609, and by UMass Dartmouth's Marine and Undersea Technology (MUST) Research Program funded by the Office of Naval Research (ONR) under Grant No.~N00014-23-1–2141.
Davide Gerosa is supported by ERC Starting Grant No.~945155--GWmining, Cariplo Foundation Grant No.~2021-0555, MUR PRIN Grant No.~2022-Z9X4XS, Leverhulme Trust Grant No.~RPG-2019-350, MSCA Fellowship No.~101064542--StochRewind, and the ICSC National Research Centre funded by NextGenerationEU.
Eliu Huerta acknowledges support from the U.S.\ Department
of Energy under Contract No.~DE-AC02-06CH11357, and from
the U.S. National Science Foundation (NSF) through award OAC-2209892.
Scott A. Hughes has been supported by NASA ATP Grant 80NSSC18K1091, NSF Grants PHY-1707549 and PHY-2110384, by MIT’s Margaret MacVicar Faculty Fellowship Program, and by the MIT Kavli Institute for Astrophysics and Space Research.
Chris Kavanagh acknowledges support from Science Foundation Ireland under Grant number 21/PATH-S/9610.
Gaurav Khanna acknowledges support from US National Science Foundation Grants No.~PHY-2307236 and DMS-2309609.
Larry Kidder acknowledges support from  NSF OAC-2209655, PHY-2308615 and Sherman Fairchild Foundation.
Pablo Laguna acknowledges support from US National Science Foundation Grants No.~PHY-2114582 and 2207780.
Georgios Lukes-Gerakopoulos has been supported by the fellowship Lumina Quaeruntur No.~LQ100032102 of the Czech Academy of Sciences.
Hyun Lim is supported by the LANL Laboratory Directed Research and Development program under project number 20220087DR. LANL is operated by Triad National Security, LLC, for the National Nuclear Security Administration of the U.S.DOE  (Contract No.~89233218CNA000001). This work is authorized for unlimited release under LA-UR-23-31548.
Tyson B. Littenberg is supported by the NASA LISA Study Office.
Carlos O.~Lousto gratefully acknowledge the National Science Foundation (NSF) for financial support from Grant No.~PHY- 1912632 and PHY-2207920.
Elisa Maggio acknowledges funding from the Deutsche Forschungsgemeinschaft (DFG) - project number: 386119226.
Richard O'Shaughnessy acknowledges support from NSF PHY-2012057, PHY-2309172, and AST-2206321.
Naritaka Oshita is supported by the Japan Society for the Promotion of Science (JSPS) KAKENHI Grant Number JP23K13111.
Rodrigo P.~Macedo and Maarten van de Meent acknowledge the financial support by the VILLUM Foundation (grant No.~VIL37766), the DNRF Chair program (grant No.~DNRF162) by the Danish National Research Foundation, and the European Union’s H2020 ERC Advanced Grant “Black holes: gravitational engines of discovery” grant agreement No.~Gravitas–101052587.
Adam Pound acknowledges the support of a Royal Society University Research Fellowship and a UKRI Frontier Research Grant under the Horizon Europe Guarantee scheme [grant number EP/Y008251/1].
Milton Ruiz acknowledges support from the Generalitat Valenciana Grant
CIDEGENT/2021/046 and the Spanish Agencia Estatal de Investigaci´on Grant PID2021-125485NB-C21.
Carlos F.~Sopuerta CFS  is supported by contracts PID2019-106515GB-I00 and PID2022-137674NB-I00 (MCIN/AEI/10.13039/501100011033) and partially supported by the program \textit{Unidad de Excelencia Mar\'{\i}a de Maeztu} CEX2020-001058-M (MCIN/AEI/10.13039/501100011033).
Stuart L.~Shapiro acknowledges support from NSF Grants No.~PHY-2006066
and No.~PHY-2308242.
Deirdre Shoemaker acknowledges support from NASA 22-LPS22-0023, 80NSSC21K0900 and NSF 2207780.
Antonios Tsokaros is supported by NSF Grants No.~PHY-2308242 and OAC-2310548.
Niels Warburton acknowledges support from a Royal Society~-~Science Foundation Ireland University Research Fellowship.
This publication has emanated from research conducted with the financial support of Science Foundation Ireland under Grant numbers 16/RS-URF/3428, 17/RS-URF-RG/3490 and 22/RS-URF-R/3825.
Helvi Witek acknowledges support provided by the National Science Foundation
under NSF Awards No.~OAC-2004879 and No.~PHY-2110416.
Huan Yang is supported by the Natural Sciences and Engineering Research Council of Canada and in part by Perimeter Institute for Theoretical Physics. 
Research at Perimeter Institute is supported in part by the Government of Canada through the Department of Innovation, Science and Economic Development Canada and by the Province of Ontario through the Ministry of Colleges and Universities.
Miguel Zilh\~ao acknowledges financial support by the Center for Research and Development in Mathematics and Applications (CIDMA) through the Portuguese Foundation for Science and Technology (FCT -- Funda\c{c}\~ao para a Ci\^encia e a Tecnologia) -- references UIDB/04106/2020 and UIDP/04106/2020 -- as well as FCT projects 2022.00721.CEECIND and 2022.04560.PTDC.

\section*{References}
\bibliography{WavWGWhitepaper,non-inspire-ads}

\end{document}